%% file: main.tex
\documentclass[10pt, twoside]{article}
\usepackage[paperwidth=170mm, paperheight=244mm]{geometry}
\usepackage{graphicx}
\usepackage{amsmath, amsfonts, amssymb, amsthm, bbm}
\usepackage{fancyhdr}
\usepackage{lettrine}
\usepackage[explicit]{titlesec}
\usepackage{watermark}
\usepackage{color}
\usepackage[final]{pdfpages}
\usepackage{setspace}
\usepackage{chngcntr}
\usepackage{enumitem}
\usepackage[colorlinks, citecolor=blue, linkcolor=black, urlcolor=blue]{hyperref}

\usepackage[T1]{fontenc}
\usepackage{lmodern}

\definecolor{grey}{rgb}{0.5,0.5,0.5}
\definecolor{l-grey}{rgb}{0.8,0.8,0.8}
\definecolor{white}{rgb}{1,1,1}
\definecolor{black}{rgb}{0,0,0}

\let\origdoublepage\cleardoublepage
\newcommand{\clearemptydoublepage}{%
  \clearpage
  {\pagestyle{empty}\origdoublepage}%
}

\let\cleardoublepage\clearemptydoublepage

\titleformat{\section}[display]{\vspace*{190pt} \bfseries\sffamily \Huge}
{\begin{picture}(0,0)\put(-60,-30){\textcolor{grey}{\thesection}}\end{picture}}
{0pt}
{#1}
[]
\titlespacing*{\section}{40pt}{10pt}{40pt}[40pt]

\titlespacing*{\subsection}{0pt}{30pt}{20pt}[0pt]
\titleformat{\subsection}[display]{\Large \sffamily}{}{0pt}{\thesubsection \ #1}[]

\begin{document}

\pagestyle{fancy}
\renewcommand{\headrulewidth}{0pt}
\fancyhead{}
\fancyfoot{}


\setcounter{page}{1}

\input{chap/title.tex}


This document is a PhD thesis developed during the period from January 2013 to September 2016 at QUTIS (Quantum Technologies for Information Science) group, led by Prof. Enrique Solano. This work was funded by the University of the Basque Country with a PhD fellowship. 

\vspace*{250pt}

\noindent \textcopyright 2016 by Julen Sim\'on Pedernales. All rights reserved. 

\bigskip

\noindent An electronic version of this thesis can be found at \href{http://www.qutisgroup.com}{www.qutisgroup.com}

\bigskip

\noindent Bilbao, October 2016

\vspace*{\fill}

{\setstretch{1.2}

\noindent This document was generated with the 2014  \LaTeX  \ distribution.

\noindent The \LaTeX \  template is adapted from a \href{https://gitlab.com/i-apellaniz/PhD_Thesis}{template} by Iagoba Apellaniz.

\noindent The bibliographic style was created by Sof\'ia Martinez Garaot.

}

\cleardoublepage


\vspace*{70pt}
\begin{flushright}
\emph{To my parents, \\ and to my siblings}
\end{flushright}

\cleardoublepage


\vspace*{350pt}
\hspace*{\fill}\begin{minipage}{\textwidth-90pt}
\emph{I'll make my report as if I told a story, for I was taught as a child on my homeworld that Truth is a matter of the imagination.}
\end{minipage}
\\
\vspace*{10pt}
\begin{flushright}
{\setstretch{2} -Ursula K. LeGuin, The Left Hand of Darkness}
\end{flushright}


\titleformat{\section}[display]
{\vspace*{200pt}
\bfseries\sffamily \Huge}
{\begin{picture}(0,0)\put(-64,-31){\textcolor{grey}{\thesection}}\end{picture}}
{0pt}
{\textcolor{black}{#1}}
[]
\titlespacing*{\section}{100pt}{10pt}{40pt}[40pt]

\textsf{\tableofcontents}


\section*{Abstract}
\pagenumbering{roman}
\fancyfoot[LE,RO]{\thepage}
\phantomsection
\addcontentsline{toc}{section}{Abstract}

It is believed that the ability to control quantum systems with high enough precision will entail a technological and scientific milestone for mankind, comparable to that of classical computation. Solving mathematical problems that are intractable with current technology, secure communication or unprecedented metrological precision are among the promises of quantum technologies. However, the control of quantum systems is still at its infancy, and the efforts to conquer this technological field are among the most exciting enterprises of the 21st century. Not only that, the endeavor of mastering quantum platforms is also a journey towards the understanding of the physics that governs them.

The quantum interaction between light and matter is at the core of almost every quantum platform, be it in systems where atoms and photons directly interact, like cavity QED, or in platforms where the same quantum optical models are used to describe analogous physics, like trapped ions or circuit QED. Light-matter interactions are of relevance in the initialization of the system, its evolution and the measurement process, which can be considered the three key steps of any quantum protocol. It is through these interactions that quantum systems get correlated, both in time and space, and in turn it is thanks to these correlations that quantum information processing, communication and sensing is possible. However, for almost all controllable quantum systems these interactions occur in physical regimes where the coupling strength is small as compared to the energies of the subsystems that interact. This restricts the number of dynamics that are accessible and as a consequence the correlations that can be generated. On the other hand, the extraction of these correlations from the system is generally nontrivial, in particular, time-correlation functions are known to be demanding to extract due to the invasiveness of the measurement process in quantum mechanics. Also, the correlation between different subsystems, the so called entanglement, is hard to quantify and to measure. 

We propose to use an ancillary system to efficiently extract time-correlation functions of arbitrary Hermitian observables in a given quantum system. Moreover, we show that these correlations can be used for the simulation of dissipative processes. We also report on an experimental demonstration of these ideas in an NMR platform in the laboratory of Professor Gui-Lu Long in Beijing. We continue to show that entanglement and its quantifiers can be efficiently extracted from a simulated dynamics, if we follow similar ancilla-based techniques. We describe a potential implementation of these ideas in trapped ions, which would be feasible with state-of-the-art technology. Indeed, two experiments have been realized with photons following our ideas, one in the lab of Professor Jian-Wei Pan in Hefei, and other in the lab of Professor Andrew White in Brisbane. In this thesis we will report on the latter, with a detailed description of the experiment. 

After exploring the quantum correlations, we show that the same models of light-matter interactions that were used to generate them can be simulated in a much broader regime of coupling strengths. The exploration of these models outside their native regimes is not only of fundamental interest, but will also allow us to generate nontrivial highly-correlated states, enhancing in this manner the complexity of the correlations accessible to quantum platforms. We will describe how these can be achieved in several quantum platforms with digital and analog simulation techniques, as well as a combination of them that we have dubbed digital-analog. It is noteworthy to mention that an experimental realization of these ideas was recently performed in the lab of Prof. Leonardo DiCarlo in TU Delft, where the simulation of models of light-matter interaction in the deep strong coupling regime was achieved, following digital-analog techniques. With this, a total of 4 experiments have been performed following ideas contained in this thesis.

Summarizing, we attack two main fronts that any quantum platform needs to master, namely the generation of correlations and the efficient measurement of them. In this sense, this thesis offers novel strategies for the extraction of the correlations present in controllable quantum systems, as well as for a full-fledged implementation of the models of light-matter interaction through which these correlations can be generated. We believe that this thesis will help reach a better understanding of light-matter interactions and their correlations in controllable quantum systems, in order to develop quantum technologies further, and to explore some of their applications.


\section*{Resumen}
\fancyfoot[LE,RO]{\thepage}
\phantomsection
\addcontentsline{toc}{section}{Resumen}

En la actualidad, el ser humano posee un dominio tecnológico suficiente para controlar de forma modesta fenómenos cuánticos. Estos fenómenos se dan en sistemas de laboratorio que pueden ser átomos, fotones individuales, o incluso sistemas mesoscópicos, que a temperaturas y presión adecuadas muestran coherencia cuántica. Por control entendemos que un conjunto de los parámetros que definen estos sistemas son manipulables, y que ello nos permite inducir en estos sistemas efectos cuánticos de interés, y no quedar restringidos a observar los efectos que se dan de forma natural. La teoría de la información cuántica predice que un control más profundo de estos sistemas supondrá la segunda revolución cuántica. La primera es aquella que ha permitido el nivel de desarrollo de los ordenadores tal y como los conocemos hoy. Las computadoras, desde los ordenadores de mesa hasta los teléfonos de bolsillo,  se basan en circuitos fabricados con transistores y materiales semiconductores que dependen de fenómenos cuánticos. Se podría decir que la tecnología de la información que domina el mundo hoy en día es un desprendimiento de los desarrollos teóricos que dieron lugar a la teoría de la mecánica cuántica. Sin embargo, en nuestros ordenadores la información es codificada en grados de libertad que se comportan de acuerdo a las leyes de la física clásica, es decir, que no muestran efectos cuánticos como la superposición o el carácter probabilístico de la medida. Por el contrario, la segunda revolución cuántica propone, no solo que la tecnología se valga de estos fenómenos cuánticos, sino que los grados de libertad en los cuales se codifica la información sean cuánticos también. Por ejemplo, que un bit pueda estar al mismo tiempo en el estado 1 y en el estado 0. Entre las promesas de la segunda revolución cuántica está el tener acceso a un poder computacional sin precedentes, el cual nos permitiría explorar los modelos matemáticos  que describen la naturaleza en regímenes y para un número de partículas fuera del alcance de las computadoras actuales. Esto podría permitir el desarrollo de nuevos materiales con propiedades muy diversas, como por ejemplo una alta eficiencia en la captación de luz en paneles solares, o el diseño de nuevos fármacos más eficaces en el tratamiento de enfermedades. Otra de las aplicaciones que se vislumbran, es la de la comunicación cuántica con métodos de seguridad infranqueables, o el desarrollo de sensores de alta precisión. Esta tesis se sitúa a la vanguardia de la tecnología cuántica actual para proponer escenarios en los cuales esta tecnología sería útil en el presente, así como para sugerir estrategias que empujen su frontera hacia delante, acercándola a las promesas de la información cuántica. 

Durante la segunda mitad del siglo XX, los campos de la óptica cuántica y de la óptica atómica han desarrollado experimentos cuánticos de forma controlada, y en consecuencia ha sido en sus laboratorios donde se han podido observar con mayor precisión los efectos predichos por las teorías cuánticas. En este sentido, una de las plataformas más relevantes es la de los iones atrapados, que consiste en átomos atrapados con potenciales eléctricos y que pueden ser manipulados con láseres. Por otro lado, también tenemos las cavidades ópticas o de microondas, que son capaces de atrapar fotones individuales entre dos espejos y hacerlos interaccionar con átomos  que vuelan a través de ellos. Por el control de estos sistemas los físicos David Wineland y Serge Haroche recibieron el premio Nobel de física en el año 2012. Estos sistemas que fueron motivados inicialmente por las teorías de la óptica cuántica y la óptica atómica, resultan ser sistemas cuánticos en los que se alcanza un grado de controlabilidad tal que sugieren que se podrían utilizar para implementar los modelos de computación de la teoría de la información cuántica. Por esto, en las últimas dos décadas la comunidad científica ha dedicado grandes esfuerzos a dominar estos sistemas con precisiones cada vez mayores. De ese esfuerzo han surgido nuevas plataformas como por ejemplo, circuitos superconductores aplicados a reproducir los modelos de electrodinámica cuántica, arreglos de fotónica lineal, espines nucleares controlados con técnicas de resonancia magnética nuclear, o defectos paramagnéticos en estructuras de diamante entre otros. Todas estas plataformas compiten por convertirse en la primera capaz de alcanzar algún resultado que vaya más allá de lo que las computadoras y los sensores actuales pueden ofrecer. 

La principal ventaja que ofrecen estos sistemas es el hecho de poder generar correlaciones cuánticas, es decir, correlaciones que sólo pueden ser descritas en el marco de la teoría cuántica. Estas correlaciones pueden ser temporales, o entre distintos subsistemas de la plataforma, lo que se conoce como  entrelazamiento cuántico.  Una vez generadas las correlaciones pueden ser explotadas para el procesamiento de información en modelos de computación cuántica, para comunicación en modelos de teleportación y para metrología de alta precisión. Las correlaciones que surgen en estos sistemas son consecuencia de las interacciones que pueden ser generadas entre distintas partes de los mismos, o de la propias dinámicas Hamiltonianas a las que pueden ser expuestos. En este sentido, los Hamiltonianos que rigen tanto las dinámicas como los tipos de interacción que se dan en estos sistemas, son Hamiltonianos derivados de la óptica cuántica que fueron desarrollados para describir cómo los átomos y la luz interaccionan. Típicamente, los átomos son reducidos a sistemas de dos niveles, dos de sus niveles electrónicos precisamente, los cuales tienen una diferencia energética similar a la de uno de los modos del campo electromagnético. Con todo, el sistema puede simplificarse a un modo electromagnético y un sistema de dos niveles. El Hamiltoniano que describe esta física es conocido como el Hamiltoniano cuántico de Rabi. Las interacciones que ocurren de forma natural siguen estos modelos en un régimen de acoplo muy concreto, que es aquel en el cual la fuerza de la interacción entre el átomo y la luz es mucho menor que la energía que tienen estos sistemas por separado.  En este régimen el Hamiltoniano puede simplificarse al conocido como Hamiltoniano de Jaynes-Cummings, que es analíticamente soluble. Históricamente, ha sido este último el que se ha estudiado tanto de forma teórica como en los laboratorios, ya que el modelo completo de Rabi carecía de realidad física. Sin embargo, con el desarrollo de las nuevas tecnologías cuánticas, ha sido posible inducir estos acoplos luz-materia con fuerzas mayores a las que se dan de forma natural, llegando al límite en el cual el simplificado modelo analítico de Jaynes-Cummings no describe correctamente las observaciones. Eso ha obligado a la comunidad científica a recuperar el modelo cuántico de Rabi en su totalidad. En 2011, el físico alemán Daniel Braak demostró que el modelo era integrable, algo que no se había conseguido demostrar desde que el modelo fuese propuesto por primera vez en los años~60. Con la capacidad de las nuevas plataformas para explorar el modelo cuántico de Rabi en regímenes de acoplo nunca antes observados se abre la puerta, no solo al análisis fundamental de las interacciones entre luz y materia, sino también a todo un mundo de correlaciones que pueden ser generadas en estas plataformas y después explotadas para provecho de la información cuántica. 

En esta tesis exploramos cómo el modelo cuántico de Rabi, y otros modelos derivados del mismo, pueden implementarse en plataformas cuánticas como los iones atrapados o los circuitos superconductores. Exploramos también cómo las correlaciones generadas en estos sistemas pueden ser extraídas y explotadas con fines de simulación cuántica. Introduciremos métodos novedosos para la simulación de estos modelos, combinando estrategias digitales y analógicas. En definitiva, esta tesis trata de explotar la puerta abierta por los nuevos regímenes de acoplo entre luz y materia que se dan, ya sea de forma directa o de forma simulada, en las modernas plataformas cuánticas. Y explora las posibles aplicaciones que surgen de las mismas. En una primera parte de esta tesis trataremos de forma intensiva la extracción así como el aprovechamiento de las correlaciones temporales y el entrelazamiento que se dan en las plataformas cuánticas. En una segunda parte estudiaremos los modelos de interacción luz-materia que dan lugar a estas correlaciones y discutiremos sobre cómo estas interacciones pueden ser generadas en regímenes nuevos y cómo puede hacerse que mantengan el ritmo de crecimiento de las plataformas.

Las correlaciones temporales han sido estudiadas comparativamente menos que el entrelazamiento. Esto tiene que ver con que la extracción o medida de correlaciones temporales está considerada en general complicada. Una de las principales razones es que extraer una correlación temporal necesita a priori la medida de un observable cuántico a dos tiempos distintos. Es bien sabido que en mecánica cuántica el proceso de medida hace colapsar el estado cuántico, que podría encontrarse en una superposición de estados, proyectándolo a uno y solo uno de ellos. Algunos protocolos han sido propuestos para resolver esta limitación, extrayendo las correlaciones de forma indirecta sin tener que medir el observable cuántico. Sin embargo, estos métodos requieren en general doblar el sistema, o en su defecto están restringidos a la medida de observables unitarios. Nosotros proponemos un método de medida, que utilizando un sistema auxiliar de dos niveles, permite la medida de correlaciones temporales de cualquier conjunto de observables. El único requisito es que nuestro sistema pueda seguir la evolución dictada por un Hamiltoniano que coincida con ese mismo observable que se quiere medir. Mostramos la eficiencia de nuestro método, el cual nunca requiere más de un sistema auxiliar y dos medidas. Además, demostramos que la medida de estas correlaciones es útil en la simulación de sistemas abiertos. Los sistemas abiertos siguen una dinámica que no es  unitaria. La simulación de estos sistemas tiene gran interés ya que cualquier sistema de estudio es en realidad un sistema abierto, un ejemplo claro sería el de una célula fotovoltaica expuesta a la radiación del sol. Comprender cómo estos sistemas funcionan desde un punto de vista cuántico, podría aumentar notablemente su eficiencia en la captación de luz. En esta tesis explicamos cómo la dinámica de un sistema abierto en la aproximación de Born-Markov puede ser codificada en las correlaciones temporales de un sistema cerrado que evoluciona de forma unitaria. Nuestros resultados son fácilmente extensibles a situaciones en las que la dinámica es no-Markoviana, algo que no es trivial para otros métodos de simulación. Nuestras ideas para la medida de correlaciones temporales, han sido demostradas experimentalmente con espines nucleares en una disolución de cloroformo en el laboratorio del Profesor Gui-Lu Long, en la ciudad de Beijing en China. En esta tesis damos una descripción detallada de este experimento, que ha logrado medir correlaciones temporales a dos tiempos para evoluciones bajo Hamiltonianos independientes del tiempo, y también para Hamiltonianos dependientes del tiempo, así como  correlaciones de ordenes superiores, incluyendo correlaciones entre 10 puntos temporales distintos. 

El entrelazamiento en un sistema cuántico es una correlación sin análogo clásico. Matemáticamente se define cómo un estado cuántico de dos o más sistemas que es  inseparable, es decir, que no puede escribirse como el producto de los estados cuánticos de cada uno de los sistemas. Esta negación de la separabilidad de un sistema es útil para identificar sistemas entrelazados. Sin embargo, determinar el nivel de entrelazamiento de un sistema es en general una tarea complicada, y uno de los retos actuales de la información cuántica, tanto a nivel teórico cómo experimental. Algunas propuestas, como las funciones monótonas de entrelazamiento, son capaces de cuantificar el entrelazamiento, pero no se conoce una forma eficiente de medirlas en un sistema cuántico. El procedimiento habitual es medir un conjunto completo de observables del sistema, de modo que su función de onda completa pueda ser reconstruida. Esta información después se utiliza para calcular el valor de las funciones monótonas de entrelazamiento. Sin embargo, este procedimiento se vuelve inviable cuando los sistemas empiezan a crecer, ya que el número de medidas necesarios para reconstruir la función de onda crece de forma exponencial con el tamaño del sistema. En esta tesis proponemos un método dentro del marco de la simulación cuántica, bajo el cual estas funciones podrían ser medidas de forma eficiente en un sistema simulado. El método consiste en incorporar un sistema auxiliar de dos niveles, e implementar la dinámica de una forma modificada que deje al descubierto estas funciones para ser extraídas eficientemente. Nuestro método aunque no mide el entrelazamiento real del sistema, ya que sólo lo hace para el del sistema simulado, podría ser útil en estudios fundamentales sobre el entrelazamiento, como por ejemplo conocer cuál es el comportamiento del entrelazamiento en sistemas de gran tamaño donde los ordenadores clásicos o los métodos analíticos no son útiles. A esta nueva generación de simuladores, diseñados para extraer de forma eficiente aspectos específicos de alto interés que quedan ocultos en las dinámicas naturales, los hemos llamado simuladores cuánticos embebidos. En esta tesis presentamos un ejemplo particular, pero el concepto es extensible a simuladores capaces de medir otro tipo magnitudes. Además del marco teórico, en esta tesis ofrecemos un protocolo concreto para su implementación en iones atrapados. Damos ejemplos de distintas dinámicas que generan entrelazamiento no-trivial y explicamos con detalle cómo podría ser extraído de nuestro simulador cuántico embebido. Además hacemos un análisis de las fuentes de error comunes en los sistemas de iones atrapados y cómo estos afectarían a nuestro simulador. Estas ideas se demostraron de forma experimental con fotones en arreglos de óptica lineal, en sendos experimento en el  laboratorio del Profesor Jian-Wei Pan, en la ciudad de Hefei en China,  y en el laboratorio del Profesor Andrew White, en la ciudad de Brisbane en Australia. En esta tesis damos una descripción detallada del experimento de Australia, donde se utilizaron tres fotones para simular el entrelazamiento de 2 qubits. En este experimento la función monótona de entrelazamiento para dos qubits, también llamada función de concurrencia, fue extraída con la medida de tan solo dos observables, frente a los 15 necesarios para reconstruir la función de onda completa.

En la segunda parte de la tesis nos centramos en las interacciones que pueden encontrarse en las plataformas cuánticas actuales, que son en definitiva el origen de las correlaciones. El modelo cuántico de Rabi describe la interacción más simple entre un átomo y un modo del campo electromagnético, cuando tanto el átomo como el modo son tratados de forma cuántica. Hoy en día, es posible atrapar iones en campos eléctricos y actuar sobre ellos con luz láser. El movimiento del ion en la trampa puede ser enfriado de forma que entre en el régimen cuántico, es decir, que su movimiento sea el de un oscilador armónico cuántico. Por otro lado, los niveles electrónicos del ion pueden reducirse a un sistema de dos niveles. El láser induce transiciones entre estos niveles electrónicos y dado que la longitud de onda del láser es comparable a la amplitud de las oscilaciones del ion, estas transiciones se vuelven dependientes de la posición del ion. De esta forma es posible inducir una interacción entre el movimiento del ion y sus grados de libertad internos. Dado que los grados de libertad del movimiento del ion son análogos a aquellos de un modo electromagnético, la interacción de los niveles internos del ion con su grado de libertad mecánico puede ser reinterpretada como una interacción del tipo luz-materia. En esta tesis explicamos cómo es posible hacer que esta interacción reproduzca el modelo cuántico de Rabi en todos sus regímenes. No solo eso, también mostramos la forma en la cual el régimen de la interacción puede ser modificado durante el propio proceso de interacción. Esto nos permite generar autoestados no-triviales en los regímenes de acoplo alto del modelo, modificando de forma adiabática el Hamiltoniano desde un régimen de interacción débil, donde los autoestados son conocidos, hasta un régimen de acoplo más intenso. Extendemos nuestros resultados a lo que se conoce como el modelo cuántico de Rabi de dos fotones, en el cual las interacciones entre el sistema de dos niveles y el modo electromagnético se dan a través del intercambio de dos excitaciones del campo electromagnético por cada una del átomo. Este modelo presenta varios aspectos exóticos desde un punto de vista matemático, como por ejemplo, el hecho de que su espectro discontinuo colapse a una banda continua para un valor especifico de la intensidad del acoplo. Al igual que para el modelo cuántico de Rabi, nuestro esquema presenta un alto grado de versatilidad en lo referente a los regímenes simulables, dando lugar a una herramienta de utilidad tanto para el estudio fundamental del modelo como para la generación de correlaciones en la plataforma. 

Por último, introducimos el concepto de simulación digital-analógica, el cual es una combinación de los métodos de simulación digital y analógicos. Los métodos de simulación digitales consisten en la descomposición de la dinámica en puertas lógicas que actúan sobre un registro de qubits, o sistemas cuánticos de dos niveles. Una simulación digital que ofrezca resultados no-triviales requeriría un número de qubits y una fidelidad de las operaciones que está lejos del alcance de ninguna plataforma actual. Las mejores simulaciones digitales hasta la fecha se reducen a una decena de qubits, y algunos  cientos de puertas lógicas sobre estos. Sin embargo, este método de simulación tiene la característica de ser universal, de modo que si alguna plataforma cuántica algún día llegara a tener el dominio necesario para implementar protocolos digitales suficientemente sofisticados, podría simular prácticamente cualquier modelo Hamiltoniano. Por otro lado, existen lo que se conoce como las simulaciones analógicas, las cuales no se restringen a un registro de qubits, ni a puertas lógicas, sino que explotan todos los grados de libertad que ofrece el sistema, como por ejemplo grados de libertad continuos. Las dinámicas no son necesariamente reducidas a una secuencia de puertas, sino que se utilizan dinámicas Hamiltonianas que son continuas en el tiempo.  Esto se consigue adaptando las dinámicas naturales de los sistemas a las dinámicas de interés. Sin embargo, esta adaptabilidad está obviamente limitada por las características propias del sistema. En consecuencia, el número de modelos simulables con técnicas analógicas es mucho más reducido que el número de modelos simulables con técnicas digitales. Sin embargo, estos modelos pueden ser simulados con las tecnologías actuales, ya que requieren un nivel de control muy inferior al que requieren los métodos digitales. En esta tesis proponemos combinar ambos, aplicando un número reducido de puertas lógicas de forma astuta sobre la evolución de un simulador analógico. Esto nos permite explotar el tamaño y la funcionalidad de los simuladores analógicos, y hacerlos más versátiles, de forma que puedan simular modelos mas allá de lo que es posible cuando se considera sólo su dinámica analógica. En esta tesis ejemplificamos este concepto con propuestas para la simulación de los modelos de Rabi y de Dicke en circuitos superconductores, y el modelo de espines de Heisenberg en iones atrapados. Nuestro enfoque de simulación garantiza que estos simuladores son escalables con la tecnología actual, sobrepasando así las barreras tecnológicas que tienen los modelos digitales, y las conceptuales que limitan a los métodos analógicos. 

En conclusión, esta tesis explora la generación, extracción y explotación de correlaciones cuánticas en las plataformas cuánticas actuales. Los modelos de interacción luz-materia responsables de generar estas interacciones son analizados desde un punto de vista fundamental, así como desde un punto de vista instrumental para la generación de correlaciones útiles en protocolos de computación. Nuestro análisis se ha mantenido siempre cercano a consideraciones experimentales realistas, que garantizan la viabilidad de los protocolos propuestos. Una buena muestra de ello es que esta tesis recoge dos experimentos, realizados en Beijing y en Brisbane, basados en las ideas aquí propuestas, y que otros dos experimentos basados en ideas aquí propuestas han sido realizados de forma paralela e independiente, en laboratorios de Hefei y Delft. Nuestras estrategias apuntan a garantizar la generación y extracción de las correlaciones cuando los sistemas crecen en tamaño, y son por lo tanto estrategias para la escalabilidad de las correlaciones. Estamos convencidos de que los resultados recogidos en esta tesis, no sólo aumentan las posibilidades de las tecnologías cuánticas actuales, sino que contribuirán al desarrollo de estas tecnologías en su intento de alcanzar las promesas de la información cuántica.


\section*{Acknowledgements}
\phantomsection
\addcontentsline{toc}{section}{Acknowledgements}

\emph{``The only people for me are the mad ones, the ones who are mad to live, mad to talk, mad to be saved, desirous of everything at the same time, the ones who never yawn or say a commonplace thing, but burn, burn, burn like fabulous yellow Roman candles exploding like spiders across the stars.''}\\
\begin{flushright} - Jack Kerouac, On The Road\end{flushright}
\vspace{20pt}

I am privileged. And it is my duty and my will to thank those who are responsible for that. There exists a risk (and it is the fear of many in my situation) of forgetting someone who deserves to be in this list. I apologize in advance, if this is the case, for it is surely not the fault of the forgotten ones but mine.

Little did I know when I started this journey with my great friends Mikel Palmero, with his good taste for controversy, and Aitor Aldama, who now pursues happiness at other latitudes, that I would meet so many amazing people in the way. I will start by mentioning the GNT group and its constituents that, everyday at lunchtime and occasionally in the bars, have helped me to unveil the contradictions of life. I want also to thank the C group and its constantly renovated member list of dreamers, for the good times, be it in the university, in the cinema or on the dance floor.

During these last years, I have visited several top-level research groups all around the world. Some of these visits have crystallized in research articles that are contained in this thesis, others have served to learn and inspire. I want to thank the QUBIT group at the Walther Meissner Institut, and specially his frontman Dr.~Frank~Deppe, the trapped ion group at ETH Zurich and his leader Prof.~JonathanHome, the people of IQOQI at Innsbruck, and specially Prof.~Rainer~ Blatt and Prof.~Gerhard Kirchmair for inviting me to their groups, Prof.~Ferdinand~Schmidt-Kaler for hosting me at Johannes Gutenberg Universität in Mainz, Prof.~Kihwan~Kim in Tsinghua University, Prof.~Adolfo~del~Campo at University of Massachusetts, and Prof.~Martin~Plenio in Ulm University. Upon returning from each of these trips, I was no longer the person I was when I left, which was indeed the reason for traveling.  

This thesis has been developed at the QUTIS group in the University of the Basque Country. Since I started here, I have witnessed the metamorphosis of the group, both in its constituents and in its spirit. I feel that parallel to my personal maturation process the group has also explored its possibilities and evolved accordingly. And I proudly think I have contributed to some aspects of that transformation. I want to thank present and past members of the QUTIS group for their company during our wandering. For reasons of economy of the language, I will only personally mention my closest collaborators. I want to thank my mentor during the first two years of my PhD, Dr. Jorge Casanova, for his immense creativity and combative soul. I am grateful to Dr. Roberto Di Candia, whom I consider to be a genius, for his mathematical lucidity and acid humor. I want also to mention Prof. Iñigo Egusquiza, who was the first person I met when I set foot in this university for the first time, more than 9 years ago, as an undergraduate student in the infamous ``curso cero''. For his infinite knowledge, he will always be the professor and I will always be the student, but incidentally I have also become a proud collaborator of him. I am grateful to him, and I have to admit that the shadow of his judgment has been present as I wrote this thesis, and if this document has any quality, it is also thanks to him. I want to mention Iñigo Arrazola, the youngest of my collaborators, with whom, since years, I maintain one, and only one, endless discussion. We invoke it every now and then, and it cuts through the fields of physics, music, cinema, literature, and in general any topic where we can sense and challenge our aesthetic tastes. In a world with an overdose of clones, Iñigo is by far one of the most genuine personalities I have ever met. From him I have learned, and with him I have laughed.

Of course, I want to thank my supervisors, firstly Dr. Lucas Lamata, the definition of an expert, knowledgeable and efficient, a balanced blend of rigorousness and creativity. He has taught me that victory is for those who resist, that success is the reward of the insistent, of the workers, of those who practice excellence every day and on every stage. Secondly, I am grateful to Prof. Enrique Solano, the creator of the QUTIS cosmogony, who has not only taught me about physics, but also how to communicate it, both in a written and in a spoken manner. He has taught me about scientific politics and economy as well. From him I have learned, that the virtues that took you from A to B will not take you from B to C, that betraying your past is not a symptom of weakness, but of progress, that coherence with yourself is only a pleasant delusion of certainty. Thanks to him I know that it is worth to fantasize, to slightly distort reality, for it is living according to our fantasies that we force the world to accommodate to them, and even if it is only slightly, we change it. 

And naturally, I want to thank my family and friends, everyone who confronts my opinions, and anyone who thinks different from me. Thanks to those that ever told me that I was wrong, I was able to evolve. It is them who purify me, who help to unmask my imposture.


\section*{List of Publications}
\phantomsection
\addcontentsline{toc}{section}{List of Publications}

This thesis is based on the following publications and preprints:
\\

{\bf Chapter 2: \nameref{sec2}}

\begin{enumerate}

\item {\color{magenta} J. S. Pedernales}, R. Di Candia, I. L. Egusquiza, J. Casanova, and E. Solano, \\
{\it Efficient Quantum Algorithm for Computing n-time Correlation Functions},\\
\href{http://dx.doi.org/10.1103/PhysRevLett.113.020505}{Physical Review Letters {\bf 113}, 020505 (2014).}

\item R. Di Candia, {\color{magenta} J. S. Pedernales}, A. del Campo, E. Solano, and J. Casanova,\\
{\it Quantum Simulation of Dissipative Processes without Reservoir Engineering},\\
\href{http://www.nature.com/articles/srep09981}{Scientific Reports {\bf 5}, 9981 (2015).}

\item T. Xin, {\color{magenta} J. S. Pedernales}, L. Lamata, Enrique Solano, and Gui-Lu Long,\\
{\it Measurement of Linear Response Functions in NMR},\\
\href{http://arxiv.org/abs/1606.00686}{arXiv preprint quant-ph/1606.00686 (2016).}

\end{enumerate}

{\bf Chapter 3: \nameref{sec3}}

\begin{enumerate}[resume]

\item {\color{magenta} J. S. Pedernales}, R. Di Candia, P. Schindler, T. Monz, M. Hennrich, J. Casanova,\\
and E. Solano,\\
{ \it Entanglement Measures in Ion-Trap Quantum Simulators without Full Tomography},\\
\href{http://journals.aps.org/pra/abstract/10.1103/PhysRevA.90.012327}{Physical Review A {\bf 90}, 012327 (2014).}

\item R. Di Candia, B. Mejia, H. Castillo, {\color{magenta} J. S. Pedernales}, J. Casanova, \\ and E. Solano,\\
{\it Embedding Quantum Simulators for Quantum Computation of Entanglement},\\
\href{http://journals.aps.org/prl/abstract/10.1103/PhysRevLett.111.240502}{Physical Review Letters {\bf 111}, 240502 (2013).}

\item J. C. Loredo, M. P. Almeida, R. Di Candia, {\color{magenta} J. S. Pedernales}, J. Casanova, \\ E. Solano, and A. G. White,\\
{\it Measuring Entanglement in a Photonic Embedding Quantum Simulator},\\
\href{http://journals.aps.org/prl/abstract/10.1103/PhysRevLett.116.070503}{Physical Review Letters {\bf 116}, 070503 (2016).}

\end{enumerate}

{\bf Chapter 4: \nameref{sec4}}

\begin{enumerate}[resume]

\item {\color{magenta} J. S. Pedernales}, I. Lizuain, S. Felicetti, G. Romero, L. Lamata, and E. Solano,\\
{\it Quantum Rabi Model with Trapped Ions},\\
\href{http://www.nature.com/articles/srep15472}{Scientific Reports {\bf 5}, 15472 (2015).}

\item S. Felicetti, {\color{magenta} J. S. Pedernales}, I. L. Egusquiza, G. Romero, L. Lamata, D. Braak, \\
and E. Solano,\\
{\it Spectral Collapse via Two-Phonon Interactions in Trapped Ions},\\
\href{http://journals.aps.org/pra/abstract/10.1103/PhysRevA.92.033817}{Physical Review A {\bf 92}, 033817 (2015).}

\end{enumerate}

{\bf Chapter 5: \nameref{sec5}}

\begin{enumerate}[resume]

\item A. Mezzacapo, U. Las Heras, {\color{magenta} J. S. Pedernales}, L. DiCarlo, E. Solano,\\ and L. Lamata,\\
{\it Digital Quantum Rabi and Dicke Models in Superconducting Circuits},\\
\href{http://www.nature.com/articles/srep07482}{Scientific Reports {\bf 4}, 7482 (2014).}

\item I. Arrazola, {\color{magenta} J. S. Pedernales}, L. Lamata, and E. Solano\\
{\it Digital-Analog Quantum Simulation of Spin Models in Trapped Ions},\\
\href{http://www.nature.com/articles/srep30534}{Scientific Reports {\bf 6}, 30534 (2016).}

\end{enumerate}
\vspace{\fill}

Other publications and preprints not included in this thesis:

\begin{enumerate}[resume]

\item {\color{magenta} J. S. Pedernales}, R. Di Candia, D. Ballester, and E. Solano,\\
{\it Quantum Simulations of Relativistic Quantum Physics in Circuit QED},\\
\href{http://iopscience.iop.org/article/10.1088/1367-2630/15/5/055008/meta}{New Journal Physics {\bf 15}, 055008 (2013).}

\item X.-H. Cheng, I. Arrazola, {\color{magenta} J. S. Pedernales}, L. Lamata, X. Chen, and E. Solano,\\
{\it Switchable Particle Statistics with an Embedding Quantum Simulator},\\
\href{http://arxiv.org/abs/1606.04339}{arXiv preprint quant-ph/1606.04339 (2016).}

\item R. L. Taylor, C. D. B. Bentley, {\color{magenta} J. S. Pedernales}, L. Lamata, E. Solano, \\ A. R. R. Carvalho, and J. J. Hope,\\
{\it Fast Gates Allow Large-Scale Quantum Simulation with Trapped Ions},\\
\href{http://arxiv.org/abs/1601.00359}{arXiv preprint quant-ph/1601.00359 (2016).}

\end{enumerate}

\cleardoublepage


\phantomsection
\listoffigures
\addcontentsline{toc}{section}{List of Figures}
\counterwithin{figure}{section}

\cleardoublepage


\section*{List of Abbreviations}
\phantomsection
\addcontentsline{toc}{section}{List of Abbreviations}

\begin{enumerate}[leftmargin=5cm]
\item [\bf AJC]{Anti-Jaynes-Cummings}
\item [\bf APD]{Avalanche Photodiode}
\item [\bf BBO]{Beta-Barium Borate}
\item [\bf COM]{Center of Mass}
\item [\bf cQED]{circuit Quantum ElectroDynamics}
\item [\bf CQED]{Cavity Quantum ElectroDynamics}
\item [\bf DAQS]{Digital-Analog Quantum Simulation}
\item [\bf DSC]{Deep Strong Coupling}
\item [\bf EQS]{Embedding Quantum Simulator}
\item [\bf GRAPE]{GRadient Ascending Pulse Engineering}
\item [\bf GT]{Glan Taylor}
\item [\bf HWP]{Half-Wave Plate}
\item [\bf JC]{Jaynes-Cummings}
\item [\bf LOCC]{Local Operations and Classical Communication}
\item [\bf MS]{M\o lmer-S\o rensen}
\item [\bf NMR]{Nuclear Magnetic Resonance}
\item [\bf PPBS]{Partially Polarizing Beam Splitter}
\item [\bf PPS]{Pseudo-Pure State}
\item [\bf QRM]{Quantum Rabi Model}
\item [\bf QST]{Quantum State Tomography}
\item [\bf QWP]{Quarter-Wave Plate}
\item [\bf RWA]{Rotating Wave Approximation}
\item [\bf SC]{Strong Coupling}
\item [\bf USC]{UltraStrong Coupling}
\end{enumerate}

\cleardoublepage


\renewcommand{\headrulewidth}{0.5pt}
\fancyfoot[LE,RO]{\thepage}
\fancyhead[LE]{\rightmark}
\fancyhead[RO]{\leftmark}


\titleformat{\section}[display]
{\vspace*{190pt}
\bfseries\sffamily \huge}
{\begin{picture}(0,0)\put(-50,-25){\textcolor{grey}{\thesection}}\end{picture}}
{0pt}
{\textcolor{white}{#1}}
[]
\titlespacing*{\section}{80pt}{10pt}{50pt}[20pt]


\pagenumbering{arabic}
\section{Introduction}
\thiswatermark{\put(1,-280){\color{l-grey}\rule{70pt}{42pt}}
\put(70,-280){\color{grey}\rule{297pt}{42pt}}}

\input{chap/introduction.tex}


\section{Quantum Correlations in Time} \label{sec2}
\thiswatermark{\put(1,-280){\color{l-grey}\rule{70pt}{42pt}}
\put(70,-280){\color{grey}\rule{297pt}{42pt}}}

\input{chap/chapter2.tex}


\titleformat{\section}[display]
{\vspace*{190pt}
\bfseries\sffamily \LARGE}
{\begin{picture}(0,0)\put(-50,-35){\textcolor{grey}{\thesection}}\end{picture}}
{0pt}
{\textcolor{white}{#1}}
[]
\titlespacing*{\section}{80pt}{10pt}{40pt}[20pt]


\section[Quantum Correlations in Embedding Quantum Simulators]{Quantum Correlations in \\ Embedding Quantum Simulators} \label{sec3}
\thiswatermark{\put(1,-302){\color{l-grey}\rule{70pt}{60pt}}
\put(70,-302){\color{grey}\rule{297pt}{60pt}}}

\input{chap/chapter3.tex}

\section[Analog Quantum Simulations of Light-Matter Interactions]{Analog Quantum Simulations \\ of Light-Matter Interactions} \label{sec4}
\thiswatermark{\put(1,-302){\color{l-grey}\rule{70pt}{60pt}}
\put(70,-302){\color{grey}\rule{297pt}{60pt}}}
\fancyhead[RO]{\small \leftmark}

\input{chap/chapter4.tex}


\section[Digital-Analog Generation of Quantum Correlations]{Digital-Analog Generation of \\ Quantum Correlations} \label{sec5}
\thiswatermark{\put(1,-302){\color{l-grey}\rule{70pt}{60pt}}
\put(70,-302){\color{grey}\rule{297pt}{60pt}}}
\fancyhead[RO]{\leftmark}

\input{chap/chapter5.tex}


\titleformat{\section}[display]
{\vspace*{190pt}
\bfseries\sffamily \huge}
{\begin{picture}(0,0)\put(-50,-25){\textcolor{grey}{\thesection}}\end{picture}}
{0pt}
{\textcolor{white}{#1}}
[]
\titlespacing*{\section}{80pt}{10pt}{50pt}[20pt]

\section{Conclusions}
\fancyhead[LE]{\rightmark}

\thiswatermark{\put(1,-280){\color{l-grey}\rule{70pt}{42pt}}
\put(70,-280){\color{grey}\rule{297pt}{42pt}}}

\input{chap/conclusions.tex}


\titleformat{\section}[display]
{\vspace*{150pt}
\bfseries\sffamily \Huge}
{\begin{picture}(0,0)\put(-64,-31){\textcolor{grey}{\thesection}}\end{picture}}
{0pt}
{\textcolor{black}{#1}}
[]
\titlespacing*{\section}{80pt}{10pt}{50pt}[0pt]

\section*{Appendices}
\phantomsection
\addcontentsline{toc}{section}{Appendices}

\fancyhead[RO]{APPENDICES}

\cleardoublepage

\appendix

\fancyhead[RO]{\leftmark}

\input{chap/appendix.tex}


\gdef\thesubsection{}
  
\section*{Bibliography}
\fancyhead[RO]{BIBLIOGRAPHY}
\fancyhead[LE]{}
\phantomsection
\addcontentsline{toc}{section}{Bibliography}

\begin{flushright}
{\it Nothing comes from nothing.}

-Parmenides
\end{flushright}

\renewcommand{\refname}{}

\bibliographystyle{bib/sofia}

\let\oldbibliography\bibliography

\renewcommand{\bibliography}[1]{{%
\let\section\subsection
\oldbibliography{#1}}}
  
\bibliography{bib/library}

\end{document}

%% file: chap/title.tex
\begin{center}

\hrule 

\vspace{16pt}
{\huge Quantum Correlations\\ of Light-Matter Interactions}
\vspace{16pt}

\hrule

\vspace{40pt}

{\Large {\bf Julen S. Pedernales } }

\vspace{40pt}

\emph{Supervised by} \\

\vspace{20pt}

{\large 

Prof. Enrique Solano\\ \vspace{8pt} and\\ \vspace{8pt} Dr. Lucas Lamata

}

\vspace{50pt}

\includegraphics[height=2.5cm]{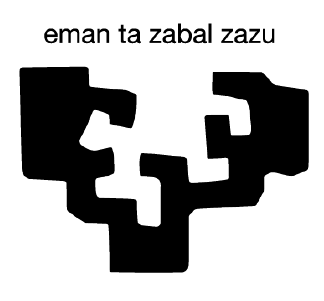}

\vspace{40pt}

Departamento de Química Física\\
y\\
\vspace{5pt}
Departamento de Física Teórica e Historia de la Ciencia \\
\vspace{5pt}
Facultad de Ciencia y Tecnología\\
Universidad del País Vasco

\vspace{20pt}

{\large October 2016}

\end{center}

%% file: chap/introduction.tex
\lettrine[lines=2, findent=3pt,nindent=0pt]{I}{n} Plato's myth of the cave~\cite{Plato}, a group of people lives in captivity since their childhood inside of a cave. They have their heads and legs chained so that they are forced to face a blank wall in front of them. At their backs, a fire projects shadows of anything passing between it and the prisoners onto the wall. When one of the captives manages to get liberated, he learns that what he considered to be reality were just shadows of objects that before were inaccessible to him. One could argue that similarly modern scientists have access to certain aspects of reality from which they try to infer a mathematical model. This model, being an idealization of reality, can be considered to belong to the platonic world of ideas. In this sense, science is a journey of abstraction, from the particular to the general, from the shadows to the objects. However, not all models are distilled from nature, the scientist, as a  creative being, can fabricate its own models and ask for the possibility of their implementation in nature. In this sense, the path towards the modeling of nature shall be walked in both directions: from reality to the model, and from the model to reality. The first starts from experimentation and observation and ends up in a mathematical model, the second begins with a model and ends up in a particular implementation of it. When this implementation occurs in a system different than that from which the model was originated we refer to it as a simulation.

A simulation is a specific experimental realization of a model, which is assumed to preserve some of the generalities ascribed to the model. Let me illustrate this with a somewhat speculative example: early humans would have noticed that when you gather a set of three stones with a set of two you end up with five stones, and that this was true as well for sticks, bones or apples, but not for water or fire, where the unit was not defined. From this observation, they would have developed a simple arithmetic model for addition, they would have performed the journey from the particular of the bones and the stones to the abstract of the arithmetics of natural numbers. Later, when they gained control over their environment, when they developed a minimal technology, they would have managed to reverse the journey and fabricate a physical counterpart of their model: the abacus. The abacus preserves some of the abstractions of the model in that it serves to describe the addition of a plethora of objects in nature. Moreover, it becomes a tool to explore the model itself, eventually upgrading it to include subtraction, or multiplication, which is already a product of the model with no counterpart in nature. A plethora of examples of simulation exist in the history of mankind: orreries are mechanical simulations of the solar system, which have served to predict the positions of the planets and the moons, eclipses, the seasons, etc. Tide predictors built of pulleys and wires were capable of predicting tide levels and were very useful for navigation. Gun directors were commonly used in 20th century's warships to quickly calculate the best firing parameters. These devices could simulate projectile shootings for a number of time varying conditions, like the target position and the speed of the wind. More recently, wind tunnels are used to simulate aerodynamic phenomena and serve to study a flying object with it being still. It is clear, that as models get more sophisticated, a more developed technology is needed for their simulation. In this sense, technology can be considered a tool for the physical fabrication of ideas, be it mathematical models, engineering solutions, or arts. Indeed, one could argue that technology is an extension of the mind, and therefore, that its evolution is as well the evolution of the human intellect. 

From this point of view, a simulation is a map from the simulated model, the idea, to the simulating system, the technological platform. Therefore, it is not possible to simulate something if the simulator itself is not well characterized, that is to say, if we lack a faithful model of the simulator system. In this sense, our simulator, which we can consider to be the exponent of our technology, needs to be deeply understood and mastered. Then, the simulation is nothing but the link from each element of the model that one wants to simulate to an element of the model of the simulator. For instance, in the case of the abacus, natural numbers are classified as ones, tens, hundreds and so on, and linked to different elements of the abacus, a mechanical procedure then implements the logic operation of summation. The arithmetic model is mapped onto the mechanical one. So far, all the examples of simulators we have given seem to be individually designed to accommodate a particular model. It was not until the first half of the 20th century that the ideas of simulation, and computation in general, started to be formalized. Alan Turing proposed a model for computation to which any problem, as long as it could be written as an algorithm, could be mapped~\cite{Turing1937}. The original work of Turing inspired other computational models~\cite{VonNeumann1993} that later have resulted in computers as we know them today. Modern digital computers are technological devices capable of implementing universal models of computation, and therefore capable of simulating almost every physical model. In this sense, computers are universal simulators. They are typically fabricated from transistors and semiconductors, which heavily rely on quantum effects. A deep technological revolution that stemmed from the quantum theory was crucial for the development of computers as we know them today. Indeed this technological milestone is typically referred to as the first quantum revolution. 

A given model might be computable, in the sense that it can be mapped onto an algorithm that is then fed to a computer. However, it might not be efficient, in the sense that the computer would take too much time to yield a solution, or that it would require an unreasonable number of constituents to be implemented. Therefore, the matter of efficiency is a major concern when making a computer simulation. To date, there exist a number of interesting problems for which efficient algorithms are not known. For example, an efficient algorithm for factorizing large numbers into a product of prime numbers is not known. More interesting to this thesis, efficient algorithms for the simulation of general quantum mechanical systems are not known. This inefficiency is believed to have its origin in the size of the mathematical machinery used to describe quantum systems. The exponential growth of the dimensionality of the Hilbert space with the number of constituents of a quantum mechanical system is a major obstacle for its simulation with classical devices. It was suggested by Richard Feynman in the early 80's that this difficulty could be understood in terms of a conflict between the quantum character of the simulated model, and the classical character of the one describing the simulator. Feynman suggested that it should be possible to aid this discrepancy if instead the simulated quantum model was mapped onto another quantum model, that is to say, that the simulator system itself behaved under the rules of quantum mechanics~\cite{Feynman1982}. These ideas initiated the quantum theory of information, which  analogously to the classical theory of information investigates and formalizes the processing of information under quantum mechanical models~\cite{NielsenChuang}. How can the models of interest be mapped onto a universal language based on quantum mechanics? Which are the necessary resources to implement such a model of computation? How can these resources be quantified?, etc. However, just like classical computers have needed the control of circuits and transistors, a comparable control of quantum mechanical systems will be required to successfully implement a quantum mechanical model of computation. This technological paradigm is typically referred to as the second quantum revolution. 

In 2012, the Nobel prize in physics was awarded to Serge Haroche and David Wineland ``for ground-breaking experimental methods that enable measuring and manipulation of individual quantum systems''~\cite{Haroche2013, Wineland2013}. These individual quantum systems were single atoms trapped in time-varying electric potentials in the case of David Wineland, and single photons trapped in cavities that were made to interact with single atoms, in the case of Serge Haroche. The field of quantum optics has achieved a great control over individual quantum systems, with many different purposes, among them the study of light matter interactions, the generation of quantum states of light or mass spectrometry of atoms. As an unexpected consequence of this effort, the physical control over individual quantum systems has opened the door to physical the implementation of quantum computational models~\cite{Lloyd1993, Barenco1995, Sleator1995, Cirac1995}. A plethora of quantum platforms have proliferated in the last two decades, creating a playground where the theory of quantum information finds a natural scenario for the implementation of models of quantum information processing. One of the central features of the quantum mechanical description of nature is that it allows for a superposition of states. That is, systems can be in several of the states accessible to them at the same time, unlike classical systems that need to be in one and only one of them. A direct consequence of the superposition principle is the emergence of entanglement. Entanglement is a non-classical correlation among quantum systems, which can be understood as a state of a composite system that cannot be described by the states of each subsystem independent of the others, even when these subsystems are space-like separated. Entanglement has been identified as a central resource for the theory of quantum information and communication. And it is therefore of great interest to develop platforms where the different elements can get arbitrarily entangled, and where these correlations are accessible. Given that two initially uncorrelated systems need to interact in order to get entangled, one could shift the focus and say that quantum interactions are the central resource for quantum computation and simulation. In this sense, the main models characterizing the interactions in almost every quantum platform are models of quantum and atom optics, more specifically  models of light-matter interaction.  Light typically understood as an electromagnetic field and matter as atoms, which are typically reduced to two level systems. Even for platforms where the degrees of freedom do not correspond to light or atoms, these models are used to describe the physics of the platforms. Therefore, the study of the interactions of light and matter, and the correlations that can be created, and how these can be detected and exploited seems a key area of study for a full characterization and understanding of the capabilities of these quantum platforms. Entanglement is challenging to quantify even from a theoretical point of view. Its experimental quantification seems also a very inefficient task. Other quantum correlations, like time correlations, are also demanding to measure from a quantum mechanical system and it is not clear what their measurement could be useful for. On the other hand, the interactions of light and matter that generate these correlations in quantum platforms are restricted to very specific coupling regimes, which in turn obstruct the complexity of correlations that can be generated. These are some of the challenges of this field, and we will try to attack them in this thesis.

Hopefully, by now, the reader has situated the relevance of simulation and computation in the scientific endeavor. The reader has probably also understood that the development of technology cannot be detached from this enterprise. Present computers and simulators cannot follow the demands of modern science, that is to say, they cannot simulate the quantum mechanical models that describe nature. In order to breach this barrier, the control of quantum mechanical systems seems unavoidable. This is a cutting-edge technological problem that lies at the frontier of the human understanding of the universe.  

\subsection{What you will find in this thesis}

In this thesis, we explore how correlations in quantum platforms can be generated, extracted and used for quantum simulation and quantum computing purposes. In the cases where efficient methods to extract these correlations are not available, we show that they can be simulated. We also explore how quantum platforms, like trapped ions and circuit QED, can simulate the models of light-matter interaction that are behind the generation of these correlations. We specially explore the ability of these platforms to simulate quantum optical models outside the physical regimes that they can naturally reach, opening the door to the generation of more complex correlations, as well as to the fundamental study of these models. This thesis is divided in 5 chapters, this introduction being the first one.

In chapter \ref{sec2}, we provide an efficient method for the extraction of time-correlation functions from a controllable quantum system evolving under an arbitrary evolution. We show that unlike previous methods, time correlations of generic Hermitian operators can be measured. Moreover, we will show that these time-correlation functions are useful in the simulation of open system quantum dynamics. In this chapter, we will also report on the experimental implementation of a proof-of-principle demonstration of such a protocol in NMR technologies, which was carried out in Beijing in the lab of Professor Gui-Lu Long.

In chapter \ref{sec3}, we explain how the entanglement generated during a given Hamiltonian evolution can be efficiently quantified in a quantum simulation. The quantification of entanglement is a hard task in general, and its extraction from a quantum system is inefficient. In this chapter, we show that in the framework of a quantum simulator, however, it is posible to quantify the entanglement of the simulated system efficiently. We show that the interactions available in trapped-ion setups suit well for this kind of simulations and we propose an experimental implementation. Not only that, we describe a proof-of-principle experiment of these ideas with photonic systems, that was performed in the laboratory of Professor Andrew White in  Brisbane. 

In chapter \ref{sec4}, we will focus on the simulation of Hamiltonians of the Rabi class in trapped ions. These Hamiltonians, namely the quantum Rabi Hamiltonian, the two-photon quantum Rabi Hamiltonian and the Dicke Hamiltonian, are ubiquitous in quantum platforms, and their understanding is of fundamental interest, as well as important for the generation of nontrivially correlated states in these platforms. We will show how an ion trapped in an electric potential can simulate these models beyond the parameter regimes provided by nature.

In chapter \ref{sec5}, we introduce the concept of digital-analog quantum simulation, which is of relevance for the simulation of Rabi kind Hamiltonians, and for quantum simulations in general. We describe two experimental proposals based on these techniques, one for the simulation of the Rabi and the Dicke models in superconducting circuits, and the other one for the simulation of spin Hamiltonians in trapped ion setups.

Furthermore, we dedicate a chapter to discuss the overall conclusions of this thesis work, and we provide an appendix section with additional material to complement the discussions held during the text. Finally, a complete bibliography can be found at the end of this document.

%% file: chap/chapter2.tex
\lettrine[lines=2, findent=3pt,nindent=0pt]{I}{f} the evaluation of a quantity that randomly fluctuates in time serves, with certain probability, as a predictor of the outcome of another random quantity measured at a different time, these two quantities are said to be correlated in time. Equivalently,  two quantities that have no potential of predicting each other are said to be uncorrelated. When the correlation corresponds to the same dynamical variable evaluated at different moments, we talk about an auto-correlation. In the theory of quantum mechanics a two-time quantum correlation function is defined as
\begin{equation}
C_{A, B} (t)=\langle \Psi | A (0) B(t) | \Psi \rangle
\end{equation}
and gives the value of the time correlation as a function of the distance in time between the evaluation of observables $A$ and $B$, for a system in the initial state $|\Psi \rangle$. Here, we have adopted the Heisenberg picture, so operators depend on time, while states do not. For a closed quantum system following the evolution dictated by a Hamiltonian $H$, the time dependence of observable $B$ can be explicitly given as $B(t)=e^{i H t} B e^{-i H t}$. The concept can be naturally extended to correlations of arbitrary order $n$ of the form
\begin{equation}
C_{A,B,C..., \alpha} (t_1, t_2, ..., t_n) =\langle \Psi | A (0) B (t_1) C(t_2) ... \alpha (t_n)) | \Psi \rangle.
\end{equation}

From a physical point of view, time-correlation functions have a plethora of applications. In the theory of statistical mechanics, time-correlation functions become a tool for the analysis of dynamical processes that could be compared to the value of the partition function for a system in equilibrium~\cite{Zwanzig1965}, in the sense that once one of these is known, all the relevant quantities of the system are accessible. In the linear response theory introduced by Kubo~\cite{Kubo1957}, it is shown that the linear response of a system to a perturbation can be computed in terms of time-correlation functions of microscopic degrees of freedom of the unperturbed system. Consider that a system in a thermal state with respect to a reference Hamiltonian $H_0$ is perturbed with a time dependent force $F(t)$, represented by the Hamiltonian term $H'(t)= A F(t)$. Here $A$ describes the form of the perturbation, such that the time-dependent Hamiltonian is now $H(t)= H_0 + A F(t)$. In such a scenario, the expectation value of a given observable $B$ is
\begin{equation}
\label{Kubo formula}
\langle B(t) \rangle = \langle B_0 \rangle + \int_{-\infty}^t \phi_{ab}(t-t') F(t') dt',
\end{equation}
where $\phi(t-t')$ is the so called {\it response function}, and $\langle B_0 \rangle$ is the expectation value of $B$ in the absence of the perturbation. The response function reads
\begin{equation}
\label{Response function}
\phi_{ab}(t)= \langle [A, B(t)] \rangle / (i \hbar),
\end{equation}
where the averaging is done over a thermal equilibrium ensemble of the $H_0$ Hamiltonian, and it is therefore completely defined by time-correlation functions of the unperturbed system. A text book example of the application of this theory is that of the magnetic susceptibility, which gives a measure of the degree of magnetization of a material in response to an applied magnetic field, which can be originated for example from an electromagnetic wave impacting on the material. If we consider the case of a periodic force $F(t)=F_0 e^{i \omega t}$, one can reorganize Equation~(\ref{Kubo formula}) to explicitly show the frequency dependent admittance $\chi^\omega_{ab}$,
\begin{equation}
\langle B(t) \rangle = \langle B_0 \rangle + \chi_{ab}^\omega F_0 e^{i \omega t },
\end{equation}
where 
\begin{equation}
\label{Susceptibility}
\chi_{ab}^\omega= \int_{-\infty}^t \phi_{ab}(t-s)e^{i w(t-s)} ds.
\end{equation} 
For the particular case where B is the magnetization and A a magnetic field, the admittance $\chi_{ab}^\omega$ corresponds to the frequency dependent susceptibility. 

As we can see, time-correlation functions play a central role in the theory of both classical and quantum statistical mechanics. However, their physical relevance is not restricted to that, time-correlation functions have a similar importance in quantum optics, more precisely they are at the core of the theory of quantum optical coherence developed by Glauber~\cite{Glauber1963}. In his seminal work, Glauber extended the classical theory of optical coherence to the concept of an arbitrary $n$-th order coherence, where $n$ is the order of the quadrature. The very well known coherence functions $g^{(1)}(\tau)$ and $g^{(2)}(\tau)$ are time-correlation functions of the electric-field amplitude and electric-field intensity, respectively. They offer a classification of the light depending on its degree of coherence, and serve as identifiers of quantum states of light. In spectroscopy, Fourier transforms of time correlation functions yield the energy spectrum of a system. In quantum field theories, Feynman propagators are generally defined as time-correlation functions. Despite the ubiquity of time-correlation functions across the different theories in physics, it turns out that the measurement of time-correlation functions in a quantum system is challenging.

Let us consider a two-time correlation function $\langle A(t) B(0) \rangle$ where we define $A(t)= U^{\dagger}(t) A(0) U(t)$, $U(t)$ being a given unitary operator, while $A(0)$ and $B(0)$ are both Hermitian. Remark that, generically,  $A(t)B(0)$ will not be Hermitian. However, one can always construct two self-adjoint operators $C(t) = \frac{1}{2} \{ A(t), B(0) \}$ and $D(t) = \frac{1}{2i} [ A(t) , B(0) ]$ such that ${\langle A(t) B(0) \rangle=\langle C(t) \rangle +  i \langle D(t) \rangle}$. According to the quantum mechanical postulates, there exist two measurement apparatus associated with observables $C(t)$ and $D(t)$.  In this way, we may formally compute $\langle A(t) B(0) \rangle$ from the measured $\langle C(t) \rangle$ and $\langle D(t) \rangle$. However, the determination of $\langle C(t) \rangle$ and $\langle D(t) \rangle$ depends non trivially on the correlation times and on the complexity of the specific time evolution operator $U(t)$. Furthermore, we point out that the computation of $n$-time correlations, as $\langle \Psi | \Psi' \rangle = \langle \Psi | U^{\dagger}(t) A U(t) B | \Psi \rangle$, is not a trivial task even if one has access to full state tomography, due to the ambiguity of the global phase of state $| \Psi' \rangle = U^{\dagger}(t) A U(t) B | \Psi \rangle$. Therefore, we are confronted with a cumbersome problem: the design of measurement apparatus depending on the system evolution for determinating $n$-time correlations of generic Hermitian operators of a system whose evolution may not be accessible.

This chapter is divided in three sections. In section~\ref{sec:TimeCor}, we will first introduce an algorithm that will be useful for the extraction of arbitrary order time-correlation functions of generic Hermitian operators. In section~\ref{sec:SimDiss}, we will introduce a simulation technique that profiting from the algorithm introduced in the previous section will be able to simulate open quantum dynamics in a controllable quantum platform. And finally, in section~\ref{sec:NMRExp} we will report on an experimental realization of the algorithm with NMR technologies in the laboratory of Prof. Gui-Lu Long in Beijing.

\subsection{An algorithm for the measurement of time-correlation functions} \label{sec:TimeCor}

The computation of time correlation functions for propagating signals is at the heart of quantum optical methods~\cite{MandelWolf}, including the case of propagating quantum microwaves~\cite{Bozyigit2011,Menzel2012,DiCandia2014}. However, these methods are not necessarily easy to export to the case of spinorial, fermionic and bosonic degrees of freedom of massive particles. In this sense, recent methods have been proposed for the case of two-time correlation functions associated to specific dynamics in optical lattices~\cite{Knap2013}, as well as in setups where post-selection and cloning methods are available~\cite{Buscemi2013}.  On the other hand, in quantum computer science the SWAP test~\cite{Buhrman2001} represents a standard way to access $n$-time correlation functions if a quantum register is available that is, at least,  able to store two copies of a  state,  and to perform a generalized-controlled swap gate~\cite{Wilmott2012}. However, this could be demanding if the system of interest is large, for example, for an $N$-qubit system the SWAP test requires the quantum control of a system of more than $2N$ qubits. Another possibility corresponds to the Hadamard test~\cite{Somma2002}, which exploits an ancillary qubit and controlled operations to extract time-correlation functions of unitary operators. Following similar routs, in this section, we propose a method for computing $n$-time correlation functions of arbitrary spinorial, fermionic, and bosonic operators, consisting of an efficient quantum algorithm that encodes these correlations in an initially added ancillary qubit for probe and control tasks. For spinorial and fermionic systems, the reconstruction of arbitrary $n$-time correlation functions requires the measurement of two ancilla observables, while for bosonic variables time derivatives of the same observables are needed. Finally, we provide examples applicable to different quantum platforms in the frame of the linear response theory.

\begin{figure}[h]
\begin{center}
\includegraphics [width= 0.9\columnwidth]{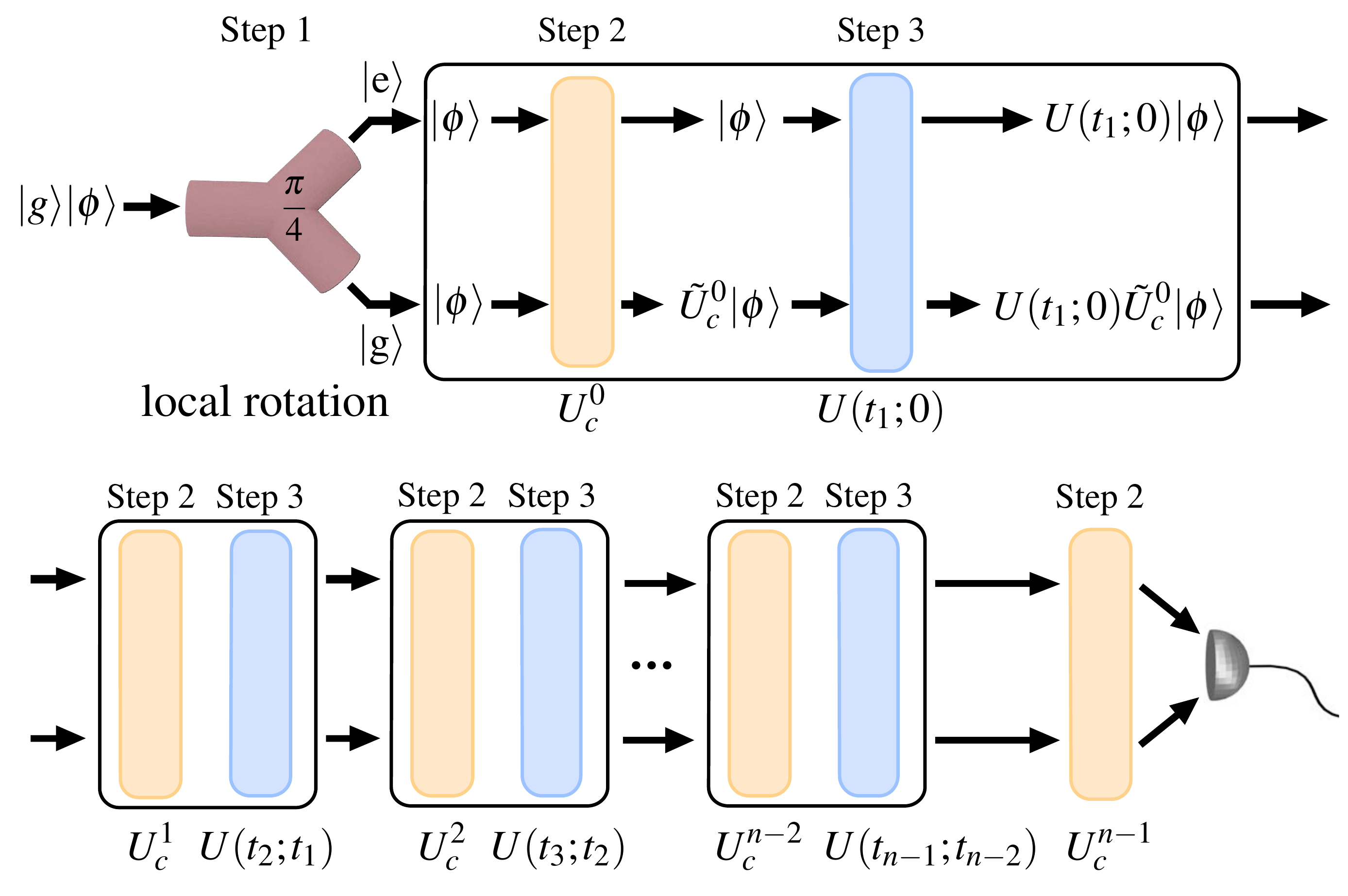}
\end{center}
\caption[Quantum algorithm for computing $n$-time correlation functions]{\footnotesize{{\bf Quantum algorithm for computing $n$-time correlation functions.} The ancilla state $\frac{1}{\sqrt 2}(|{\rm e}\rangle + |{ \rm g }\rangle)$ generates the $|{\rm e}\rangle$ and $|{ \rm g }\rangle$ paths, step $1$, for the ancilla-system coupling. After that, controlled gates $U_c^m$ and unitary evolutions $U(t_m; t_{m-1})$ applied to our system, steps $2$ and $3$,  produce the final state $\Phi$. Finally, the measurement of the ancillary spin operators $\sigma_x$ and $ \sigma_y$ leads us to $n$-time correlation functions.}}\label{figscheme}
\end{figure}

The protocol works under the following two assumptions.  First, we are provided with a controllable quantum system undergoing a given quantum evolution described by the Schr\"odinger equation
\begin{equation}\label{scho}
i\hbar\partial_t |\phi\rangle = H |\phi\rangle.
\end{equation}
And second, we require the ability to perform entangling operations, for example M\o lmer-S\o rensen~\cite{Sorensen1999}  or equivalent controlled gates~\cite{NielsenChuang},  between  some part of the system and the ancillary qubit. More specifically, and as it is discussed in appendix~\ref{app:efficiency}, we require a number of entangling gates that grows linearly  with the order $n$ of the $n$-time correlation function and that remains fixed with increasing system-size. This protocol  will provide us with  the efficient measurement of  generalized $n$-time correlation functions  of the form ${\langle\phi| O_{n-1}(t_{n-1}) O_{n-2}(t_{n-2}) ... O_1(t_1) O_0(t_0)|\phi\rangle}$, where $O_{n-1}(t_{n-1}) ... O_0(t_0)$ are certain operators evaluated at different times, e.g. $O_k(t_k) = U^{\dag}(t_k; t_0) O_k \ U(t_k; t_0)$, $U(t_k; t_0)$ being the unitary operator evolving the system from $t_0$ to $t_k$. For the case of dynamics governed by time-independent Hamiltonians, $U(t_k; t_0)=U(t_k - t_0) = e^{-\frac{i}{\hbar}H(t_k - t_0)}$. However, our method applies also to the case where $H=H(t)$, and can be sketched as follows. First, the ancillary qubit  is prepared in state ${\frac{1}{\sqrt 2}(|{\rm e} \rangle + |{ \rm g } \rangle)}$ with ${|{ \rm g }\rangle}$ its ground state, as in step $1$ of Fig.~\ref{figscheme}, so that the whole ancilla-system quantum state is ${\frac{1}{\sqrt 2}(|{\rm e} \rangle + |{ \rm g }\rangle) \otimes |\phi\rangle}$, where $|\phi\rangle$ is the  state of the system. Second, we apply  the controlled quantum gate $U^0_c = \exp{( - \frac{i}{\hbar}|{ \rm g }\rangle\langle { \rm g }| \otimes  H_0 \tau_0 ) }$, where, as we will see below, $H_0$ is a Hamiltonian related to the operator $O_0$, and $\tau_0$ is the gate time. As we point out in the appendix~\ref{app:MolmerSorensen}, this entangling gate can be implemented efficiently with M\o lmer-S\o rensen gates for operators $O_0$ that consist in a tensor product of Pauli matrices~\cite{Sorensen1999}. This operation entangles the ancilla with the system generating the state $ \frac{1}{\sqrt 2} (|{\rm e}\rangle \otimes |\phi\rangle + |{ \rm g }\rangle \otimes \tilde{U}_c^0|\phi\rangle)$, with $\tilde{U}_c^0= e^{- \frac{i}{\hbar}H_0 \tau_0}$,  step $2$ in Fig.~\ref{figscheme}. Next, we switch on the dynamics of the system governed by  Eq.~(\ref{scho}). For the sake of simplicity let us assume $t_0=0$. The effect on the ancilla-system wavefunction is to produce the state ${\frac{1}{\sqrt{2}}\big(|{\rm e} \rangle \otimes U(t_1; 0) |\phi \rangle + |{ \rm g }\rangle \otimes U(t_1; 0) \tilde{U}_c^0 |\phi\rangle \big)}$, step $3$ in Fig.~\ref{figscheme}. Note that, remarkably, this last step does not require an interaction between the system and the ancillary-qubit degrees of freedom nor any knowledge of the Hamiltonian $H$. These techniques, as will be evident below, will find a natural playground in the context of quantum simulations, preserving its analogue or digital character. If we iterate $n$ times step~$2$ and step~$3$ with a suitable choice of  gates and evolution times, we obtain the state $\Phi = \frac{1}{\sqrt{2}}(|{\rm e}\rangle \otimes U(t_{n-1}; 0 )  |\phi\rangle + |{ \rm g }\rangle   \otimes \tilde{U}_c^{n-1}   U(t_{n-1}; t_{n-2})   ... \ U(t_{2}; t_{1})  \tilde{U}_c^1 U(t_1; 0)   \tilde{U}_c^0 |\phi\rangle )$. Now, we target the quantity  ${\rm Tr}( |{\rm e}\rangle\langle { \rm g } |  | \Phi\rangle\langle\Phi|)$ by measuring the $\langle \sigma_x \rangle$ and $\langle \sigma_y \rangle$ corresponding to the ancillary degrees of freedom. Simple algebra leads us to 
\begin{eqnarray}
\label{near}
&&{\rm Tr}( |{\rm e} \rangle\langle { \rm g } |  | \Phi\rangle\langle\Phi|) = \frac{1}{2}\left(\langle\Phi|\sigma_x| \Phi\rangle + i \langle\Phi|\sigma_y| \Phi\rangle\right)  =\nonumber \\ 
&&\frac{1}{2}\langle\phi| U^{\dag}(t_{n-1}; 0)  \tilde{U}_c^{n-1}   U(t_{n-1}; t_{n-2})     ... \ U(t_2; t_1)  \tilde{U}_c^1 U(t_1; 0)   \tilde{U}_c^0 |\phi\rangle. \nonumber \\
\end{eqnarray}
It is not difficult to see that,  by using the composition property $U(t_k; t_{k-1}) = U(t_k; 0) U^{\dag}(t_{k-1}; 0)$, Eq.~(\ref{near}) corresponds to a general construction relating $n$-time correlations of system operators $\tilde{U}_c^{k}$ with two one-time ancilla measurements. In order to explore its depth, we shall examine several classes of systems and suggest concrete realizations of the proposed algorithm. The crucial point is  establishing a connection that associates the $\tilde{U}_c^k$ unitaries with $O_k$ operators. 

Starting with  the discrete variable case, e.g. spin systems, and profiting from the fact that Pauli matrices are both Hermitian and unitary, it follows that
\begin{equation}
\label{linear}
\tilde{U}_c^m\big|_{\Omega \tau_m =\pi/2 }= \exp{( - \frac{i}{\hbar} H_m \tau_m)}\big|_{\Omega\tau_m =\pi/2 } = -i O_m,
\end{equation}
where $H_m = \hbar \Omega O_m$, $\Omega$ is a coupling constant, and  $O_m$ is a tensor product of Pauli matrices of the form ${O_m = \sigma_{i_m} \otimes \sigma_{j_m} ... \sigma_{k_m}}$ with $i_m, j_m, ..., k_m \in 0, x, y, z$, and $\sigma_0=\mathbb{I}$. In consequence, the controlled quantum gates in step 2 correspond to $U_c^m\big|_{\Omega \tau_m =\pi/2 }= \exp{(-i |{ \rm g }\rangle\langle { \rm g } | \otimes \Omega O_m \tau_m)}$, which can be implemented efficiently, up to local rotations, with four M\o lmer-S\o rensen gates~\cite{Sorensen1999, Muller2011, Lanyon2011, Casanova2012}. In this way, we can write the second line of Eq.~(\ref{near}) as 
\begin{eqnarray}\label{ntimespin}
(-i)^{n}\langle\phi| O_{n-1}(t_{n-1}) O_{n-2}(t_{n-2}) ... O_0(0)|\phi\rangle,
\end{eqnarray}
which amounts to the measured \( n \)-time correlation function of Hermitian and unitary operators $O_k$. We can also apply these ideas to the case of non-Hermitian operators, independent of their unitary character, by considering  linear superpositions of the Hermitian objects appearing in Eq.~(\ref{ntimespin}).

We show now how to apply this result to the case of fermionic systems. In principle, the previous proposed steps would apply straightforwardly if we had access to the corresponding fermionic operations. In the case of quantum simulations, a similar result is obtained via the Jordan-Wigner mapping of fermionic operators to tensorial products of Pauli matrices, $b^\dag_p\rightarrow\Pi_{r=1}^{p-1}\sigma_{+}^p\sigma_z^r$~\cite{Jordan1928}. Here, $b^{\dag}_p$ and $b_q$ are creation and annihilation fermionic operators obeying anticommutation relations, $\{b^{\dag}_p, b_q \} = \delta_{p, q}$. For trapped ions, a quantum algorithm for the efficient implementation of fermionic models has been recently proposed~\cite{Casanova2012, Lamata2014, Yung2014}. Then, we code $\langle b_p^{\dag}(t) b_q(0)\rangle = \langle\Phi|(\sigma^p_+\otimes\sigma^{p-1}_z ... \sigma_z^1)_{t} \  \sigma^q_-\otimes\sigma^{q-1}_z ... \sigma_z^1|\Phi\rangle$, where $(\sigma^p_+\otimes\sigma^{p-1}_z ... \sigma_z^1)_{t} = e^{\frac{i}{\hbar}H t} \sigma^p_+\otimes\sigma^{p-1}_z ... \sigma_z^1 e^{- \frac{i}{\hbar}H t}$. Now, taking into account that $\sigma_{\pm} = \frac{1}{2}(\sigma_x \pm i \sigma_y)$, the fermionic correlator $\langle b_p^{\dag}(t) b_q(0)\rangle$ can be written as the sum of four terms of the kind appearing in Eq.~(\ref{ntimespin}). This result extends naturally to multitime correlations of fermionic systems.

The case of bosonic $n$-time correlators requires a variant in the proposed method, due to the nonunitary character of the associated bosonic operators. In this sense, to reproduce a linearization similar to that of Eq.~(\ref{linear}), we can write
\begin{equation}
\partial_{\Omega \tau_m}\tilde{U}_c^m\big|_{\Omega \tau_m =0} =\partial_{\Omega \tau_m}\exp{(- \frac{i}{\hbar} H_m \tau_m)}\big|_{\Omega\tau_m =0 } =-i O_m\,,
\end{equation}
with $H_m = \hbar \Omega O_m$. Consequently, it follows that
\begin{eqnarray}
\label{nolabel}
&&\partial_{\Omega \tau_j} ... \partial_{\Omega\tau_k} {\rm Tr}( |{\rm e}\rangle\langle { \rm g } |  | \Phi\rangle\langle\Phi|)\big|_{\Omega(\tau_\alpha ... \tau_\beta) = \frac{\pi}{2}, \ \Omega(\tau_j ... \tau_k) = 0} =\nonumber\\
&&\ \ \ \ \ (-i)^{n}\langle\phi| O_{n-1}(t_{n-1}) O_{n-2}(t_{n-2}) ... O_0(0)|\phi\rangle \, ,
\end{eqnarray}
where the label $(\alpha, ..., \beta)$ corresponds to spin operators and $(j,...,k)$ to spin-boson operators. The right-hand side is a correlation of Hermitian operators, thus substantially extending our previous results. For example, $O_m$ would include spin-boson couplings as $O_m = \sigma_{i_m} \otimes \sigma_{j_m} ... \sigma_{k_m} (a + a^{\dag})$. The way of generating the associated evolution operator ${\tilde{U}_c^m = \exp(-i \Omega O_m \tau_m)}$ has been shown in~\cite{Casanova2012, Lamata2014, Mezzacapo2012}, see also appendix~\ref{app:MolmerSorensen}. Note that, in general, dealing with discrete derivatives of experimental data is an involved task~\cite{Lougovski2006, Casanova2012a}. However, recent experiments in trapped ions~\cite{Gerritsma2010, Gerritsma2011, Zahringer2010} have already succeeded in the extraction of precise information from data associated to first and second-order derivatives. 

The method presented here works as well when the system is prepared in a mixed-state $\rho_0$, e.g. a state in thermal equilibrium~\cite{Kubo1957, Zwanzig1965}.  Accordingly, for the case of spin correlations, we have 
\begin{eqnarray}
\nonumber
{\rm Tr}(|{\rm e}\rangle\langle { \rm g } | \tilde{\rho}) =(-i)^{n}{\rm Tr}( O_{n-1}(t_{n-1})O_{n-2}(t_{n-2}) ... O_0(0) \rho_0) , \\
\end{eqnarray}
with
\begin{eqnarray}
\tilde{\rho} =  \big(... U(t_2; t_1)U_c^1U(t_1; 0)U_c^0\big)  \tilde{\rho}_0  \big(U_c^{0 \dag} U(t_1; 0)^{\dag} U_c^{1\dag}U(t_2; t_1)^{\dag} ... \big) \nonumber \\
\end{eqnarray}
and $\tilde{\rho}_0 = \frac{1}{2}(|{\rm e}\rangle+|{ \rm g } \rangle)(\langle {\rm e}| + \langle { \rm g } |)\otimes \rho_0$. If bosonic variables are involved, the analogue to Eq.~({\ref{nolabel}) reads
\begin{eqnarray}
&&\partial_{\Omega \tau_j} ... \partial_{\Omega\tau_k}{\rm Tr}(|{\rm e}\rangle\langle { \rm g } | \tilde{\rho})\big|_{\Omega(\tau_\alpha ... \tau_\beta) = \frac{\pi}{2}, \ \Omega(\tau_j ... \tau_k) = 0} =\nonumber\\&&\ \ \ \ \ \ \ (-i)^n{\rm Tr}( O_{n-1}(t_{n-1})O_{n-2}(t_{n-2}) ... O_0(0) \rho_0) .
\end{eqnarray}

We will exemplify the introduced formalism with the case of quantum computing of spin-spin correlations of the form 
\begin{equation}\label{ss}
\langle\sigma_i^k(t) \sigma_{j}^{l}(0)\rangle,
\end{equation}
where $k, l = x, y, z$, and $i,j = 1,... ,N$, $N$ being the number of spin-particles involved. In the context of spin lattices, where several quantum models can be simulated in different quantum platforms as trapped ions~\cite{Britton2012, Porras2004, Friedenauer2008, Kim2009, Kim2010, Richerme2013}, optical lattices~\cite{GarciaRipoll2004, Simon2011, Greiner2002}, and circuit QED~\cite{GarciaRipoll2008, Tian2010, Zhang2014, Viehmann2013},  correlations like~(\ref{ss}) are a crucial element in the computation of, for example, the magnetic susceptibility~\cite{Kubo1957, Zwanzig1965, Forster}. In particular, with our protocol, we have access to  the frequency-dependent susceptibility  $\chi_{\sigma, \sigma}^\omega$ that quantifies the linear response of a spin-system when it is driven by a monochromatic field. This situation is described by  the Schr\"odinger equation  $i\hbar \partial_t | \psi \rangle = (H + f_{\omega} \sigma_j^l e^{i\omega t}) | \psi \rangle$, where, for simplicity, we assume $H\neq H(t)$. With a perturbative approach, and following the Kubo relations~\cite{Kubo1957, Zwanzig1965}, one can calculate the first-order effect of a magnetic  perturbation acting on the  \( j \)-th spin in the polarization of the \( i \)-th spin as
\begin{equation}\label{lr}
\langle\sigma_i^{k} (t) \rangle = \langle\sigma_i^{k} (t) \rangle_0+\chi_{\sigma, \sigma}^\omega  f_\omega e^{i\omega t}.
\end{equation}
Here, $\langle\sigma_i^{k} (t) \rangle_0$ corresponds to the value of  the observable $\sigma_i^k$ in the absence of perturbation, and the frequency-dependent susceptibility $\chi_{\sigma, \sigma}^\omega$ is
\begin{equation}
\chi_{\sigma, \sigma}^\omega=\int_0^t ds \ \phi_{\sigma, \sigma}(t-s) e^{i\omega (s-t)}
\end{equation}
where $ \phi_{\sigma, \sigma}(t-s)$ is called the response function, which can be written in terms of two-time correlation functions,
\begin{equation}\label{aef}
\phi_{\sigma, \sigma}(t-s) = \frac{i}{\hbar}\langle [\sigma_i^k(t-s),\sigma_j^l(0)]\rangle =  \frac{i}{\hbar}{\rm Tr}\big( [\sigma_i^k(t-s),\sigma_j^l(0)] \rho\big), 
\end{equation}
with $\rho = U(t)\rho_0 U^{\dag}(t)$, $\rho_0$ being the initial state of the system and $U(t)$ the perturbation-free time-evolution operator ~\cite{Kubo1957}. Note that for thermal states or energy eigenstates, we have  $\rho = \rho_0$. According to our proposed method, and assuming for the sake of simplicity $\rho=|\Phi\rangle\langle\Phi|$, the measurement of the commutator in Eq.~(\ref{aef}), corresponding to the imaginary part of ${\langle \sigma_i^k(t-s) \sigma_j^l(0) \rangle }$,  would require the following  sequence of interactions: ${| \Phi \rangle \rightarrow U_c^1 U(t-s)U_c^0 | \Phi \rangle}$, where ${U_c^0=e^{-i | { \rm g } \rangle \langle { \rm g } | \otimes \sigma_j^l \Omega \tau}}$, ${U(t-s)=e^{ -\frac{i}{\hbar} H (t-s)}}$, and ${U_c^1=e^{-i | { \rm g } \rangle \langle { \rm g } | \otimes \sigma_i^k \Omega \tau}}$, for $\Omega \tau = \pi /2$. After such a gate sequence, the expected value in Eq. (\ref{aef}) corresponds to $-1/2  \langle \Phi | \sigma_y | \Phi \rangle $. In the same way, Kubo relations allow the computation of higher-order corrections of the perturbed dynamics in terms of higher-order  time-correlation functions. In particular, second-order corrections to the linear response of Eq.~(\ref{lr}) can be calculated through the computation of three-time correlation functions of the form $\langle \sigma_i^k(t_2) \sigma_j^l(t_1) \sigma_j^l(0)\rangle$. Using the method introduced in this section, to measure such a three-time correlation function one should perform the evolution $| \Phi \rangle \rightarrow U_c^1U(t_2 - t_1)U_c^0 U(t_1)U_c^0 | \Phi \rangle$, where $U_c^0=e^{-i | { \rm g } \rangle \langle { \rm g } | \otimes \sigma_j^l \Omega \tau}$, $U(t)=e^{-\frac{i}{\hbar} H t}$ and $U_c^1=e^{-i | { \rm g } \rangle \langle { \rm g } | \otimes \sigma_i^k \Omega \tau}$ for $\Omega \tau = \pi /2$. The searched time correlation then corresponds to the quantity $1/2(i \langle \Phi | \sigma_x | \Phi \rangle -  \langle \Phi | \sigma_y | \Phi \rangle  )$.

Our method is not restricted to corrections of observables that involve the spinorial degree of freedom. Indeed, we can show how the method applies when one is interested in the effect of the perturbation onto the motional degrees of freedom of the involved particles. According to the linear response theory, corrections to observables involving the motional degree of freedom enter in the response function, $\phi_{a+a^{\dag}, \sigma}(t-s)$, as time correlations of the type $\langle  (a_i + a^{\dag}_i)_{(t-s)}  \sigma_j^l\rangle$, where ${(a_i + a^{\dag}_i)_{(t-s)}=e^{\frac{i}{\hbar}H(t-s)}(a_i + a^{\dag}_i) e^{-\frac{i}{\hbar}H(t-s)}}$. The response function can be written as in Eq.~(\ref{aef}) but replacing the operator $\sigma_i^k(t-s)$ by $(a_i + a^{\dag}_i)_{(t-s)}$. The corrected expectation value is now
\begin{equation}\label{lrb}
\langle(a_i + a_i^{\dag})_{t} \rangle = \langle (a_i + a_i^{\dag})_{t} \rangle_0+\chi_{a+a^{\dag}, \sigma}^\omega  f_\omega e^{i\omega t}.
\end{equation}
In this case, the gate sequence for the measurement of the associated correlation function $\langle  (a_i + a^{\dag}_i)_{(t-s)}  \sigma_j^l\rangle$ reads $| \Phi \rangle \rightarrow U_c^1 U(t-s)U_c^0 | \Phi \rangle$, where ${U_c^0=e^{-i | { \rm g } \rangle \langle { \rm g } | \otimes \sigma_j^l \Omega_0 \tau_0}}$, ${U(t-s)=e^{-\frac{i}{\hbar} H (t-s)}}$, and ${U_c^1=e^{-i | { \rm g } \rangle \langle { \rm g } | \otimes (a_i + a^{\dag}_i) \Omega_1 \tau_1}}$, for ${\Omega_0 \tau_0 = \pi /2}$. The time correlation is now obtained through the first derivative of the expectation values of Pauli operators as $ -1/2 \partial_{\Omega_1 \tau_1} (\langle \Phi | \sigma_x | \Phi \rangle + i  \langle \Phi | \sigma_y | \Phi \rangle) |_{\Omega_1 \tau_1 = 0}$.

Equations~(\ref{lr}) and (\ref{lrb}) can be extended to describe the effect on the system of light pulses containing frequencies in a certain interval $(\omega_0, \omega_0+\delta)$. In this case, Eqs.~(\ref{lr}) and (\ref{lrb}) read}
\begin{equation}\label{lrmf}
\langle\sigma_i^{k} (t) \rangle = \langle\sigma_i^{k} (t) \rangle_0+\int_{\omega_0}^{\omega_0 + \delta}\chi_{\sigma, \sigma}^\omega  f_\omega e^{i\omega t} d\omega,
\end{equation}
and  
\begin{equation}\label{lrbmf}
\langle(a_i + a_i^{\dag})_{t} \rangle = \langle (a_i + a_i^{\dag})_{t} \rangle_0 + \int_{\omega_0}^{\omega_0+\delta}\chi_{a+a^{\dag}, \sigma}^\omega  f_\omega e^{i\omega t} d\omega.
\end{equation}
Note that despite the presence of many frequency components of the light field in the integrals of Eqs.~(\ref{lrmf}, \ref{lrbmf}), the computation of the susceptibilities, $\chi_{\sigma, \sigma}^\omega $ and $\chi_{a+a^{\dag}, \sigma}^\omega$, just requires the knowledge of the time correlation functions $\langle [\sigma_i^k(t-s),\sigma_j^l(0)]\rangle$ and $\langle [(a+a^{\dag})_{(t-s)},\sigma_j^l(0)]\rangle$, which can be efficiently calculated with the protocol described in  Fig~\ref{figscheme}. In this manner, we provide an efficient quantum algorithm to characterize the response of different quantum systems to external perturbations. Our method may be related to the quantum computation of transition probabilities ${|\alpha_{{\rm f},{\rm i}}(t)|^2=|\langle { \rm f }| U(t) | {\rm i } \rangle|^2=\langle {\rm i} | P_{\rm f} (t) | {\rm i} \rangle }$, between initial and final states,  $ | {\rm i} \rangle$ and $| {\rm f} \rangle$, with $P_{\rm f} (t)=U(t)^{\dag}| {\rm f}\rangle\langle { \rm f} | U(t)$, and to transition or decay rates $\partial_t |\alpha_{{\rm f},{\rm i }}(t)|^2$ in atomic ensembles. These questions are of general interest for evolutions perturbed by external driving fields or by interactions with other quantum particles. 

Summarizing, in this section we have presented a quantum algorithm to efficiently compute  arbitrary $n$-time correlation functions. The protocol requires the initial addition of a single probe and control qubit and is valid for arbitrary unitary evolutions. Furthermore, we have applied this method to interacting fermionic, spinorial, and bosonic systems,  showing how to compute second-order effects beyond the linear response theory. Moreover, if used in a quantum simulation, the  protocol preserves the analogue or digital character of the associated dynamics. We believe that the proposed concepts pave the way for making accessible a wide class of $n$-time correlators in a wide variety of physical systems. 


\subsection{Simulating open quantum dynamics with time-correlation functions}\label{sec:SimDiss}

While every physical system is indeed coupled to an environment~\cite{BreuerPetruccione, Rivas2012}, modern quantum technologies have succeeded in isolating systems to an exquisite degree in a variety of platforms~\cite{Leibfried2003, Devoret2013a, Bloch2005, OBrien2009}. In this sense, the last decade has witnessed great advances in testing and  controlling the quantum features of these  systems, spurring the quest for the development of quantum simulators \cite{Feynman1982, Lloyd1996a, Cirac2012,Georgescu2014}. These efforts are guided by the early proposal of using a highly tunable quantum device to mimic  the behavior of another quantum system of interest,  being  the latter complex enough to render its description by classical means intractable. By now, a series of proof-of-principle experiments have  successfully demonstrated the basic tenets of quantum simulations revealing quantum technologies as  trapped ions \cite{Schneider2012}, ultracold quantum gases \cite{Bloch2012}, and superconducting circuits \cite{Houck2012}    as  promising candidates to harbor  quantum simulations beyond the computational capabilities of classical devices.

It was soon recognised that this endeavour should not  be limited to simulating the dynamics of isolated complex  quantum systems, but should more generally aim at the emulation of  arbitrary physical processes, including the open quantum dynamics of  a system coupled to an environment. Tailoring the complex nonequilibrium dynamics of an open system has the potential to uncover a plethora of technological and scientific applications. A remarkable instance results from the understanding of the role played by quantum effects in the open dynamics of photosynthetic processes in biological systems \cite{Huelga2013,Mostame2012}, recently used in the design of artificial light-harvesting nanodevices \cite{Scully2011,Dorfman2013,Creatore2013}. At a more fundamental level, an open-dynamics quantum simulator would be invaluable to shed new light on core issues of foundations of physics, ranging from the quantum-to-classical transition  and quantum measurement theory \cite{Zurek2003} to the characterization of Markovian and non-Markovian systems \cite{Breuer2009,Rivas2010,Liu2011}. Further motivation arises at the forefront of quantum technologies.  As  the available resources increase,  the verification with classical computers of quantum annealing devices \cite{Boixo2013,Boixo2014}, possibly operating with a  hybrid quantum-classical performance, becomes a daunting task. The comparison between different experimental implementations of quantum simulators is required to establish a confidence level, as customary with other quantum technologies, e.g., in the use of atomic clocks for time-frequency standards. In addition, the knowledge and control of dissipative processes can be used as well as a resource for quantum state engineering~\cite{Verstraete2009}. 

Facing  the high dimensionality of the Hilbert space of the composite system made of a quantum device embedded in an environment, recent developments have been focused on the reduced  dynamics of the system that emerges after  tracing out the environmental degrees of freedom. The resulting nonunitary dynamics is governed by a dynamical map,  or equivalently,  by a master equation~\cite{BreuerPetruccione,Rivas2012}. In this respect, theoretical~\cite{Lloyd2001, Kliesch2011, Wang2011} and experimental~\cite{Barreiro2011} efforts in the simulation of open quantum systems have exploited the combination of coherent quantum operations with controlled dissipation. Notwithstanding, the experimental complexity required to simulate an arbitrary open quantum dynamics is recognised to substantially surpass that needed in the case of  closed systems, where a smaller number of generators suffices to design a general time-evolution. Thus, the quantum simulation of open systems remains a challenging task.

In this section, we propose a quantum algorithm to simulate finite dimensional Lindblad master equations, corresponding to Markovian or non-Markovian processes. Our protocol shows how to reconstruct, up to an arbitrary finite error, physical observables that evolve according to a dissipative dynamics, by evaluating multi-time correlation functions of its Lindblad operators. We show that the latter requires the implementation of the unitary part of the dynamics in a quantum simulator, without the necessity of physically engineering the system-environment interactions.  Moreover, we demonstrate  how these multi-time correlation functions can be computed  with a reduced number of measurements. We further show that our method can be applied as well to the simulation of  processes associated with non-Hermitian Hamiltonians. Finally, we provide specific error bounds to estimate the accuracy of our approach. 

Consider a quantum system coupled to an environment whose dynamics is described by  the von Neumann equation  $ i\frac{d{\bar{\rho}}}{dt} = [\bar H, \bar{\rho}]$. Here, $\bar{\rho}$ is the system-environment density matrix, $\bar{H}= H_s + H_e + H _I$,   where $H_s$ and $H_e$  are the  system and environment Hamiltonians, while  $H_I$ corresponds to  their interaction. Assuming weak coupling and short time-correlations between the system and the environment, after tracing out the environmental degrees of freedom we obtain the Markovian master equation
\begin{equation}\label{master}
\frac{d\rho}{dt} =\mathcal{L}^t_{}\rho,
\end{equation}
being $\rho={\rm Tr}_e(\bar{\rho})$ and $\mathcal{L}^t$  the time-dependent superoperator governing the dissipative dynamics~\cite{BreuerPetruccione, Rivas2012}.  Notice that there are different ways to recover Eq.~\eqref{master} \cite{Alicki2006}. Nevertheless, Eq.~\eqref{master} is our starting point, and in the following we show how to simulate this equation regardless of its derivation. Indeed, our algorithm does not need control any of the approximations done to achieve this equation. We can decompose  $\mathcal{L}^t$ into $\mathcal{L}^t=\mathcal{L}_{H}^t+\mathcal{L}_D^t$. Here, $\mathcal{L}_H^t$ corresponds to a unitary part, i.e. $\mathcal{L}_{H}^t\rho\equiv-i[H(t),\rho]$, where $H(t)$ is defined by $H_s$ plus a term due to the lamb-shift effect and it may depend on time. Instead, $\mathcal{L}_D^t$ is the dissipative contribution and it follows the Lindblad form~\cite{Lindblad1976} $\mathcal{L}_D^t\rho\equiv\sum_{i=1}^N\gamma_i(t)\left(L_i\rho L_i^\dag-\frac{1}{2}\{L_i^{\dag}L_i, \rho\}\right)$, where $L_i$ are the Lindblad operators modelling the effective interaction of the system with the bath that may depend on time, while $\gamma_i(t)$ are nonnegative parameters.  Notice that, although the standard derivation of Eq.~\eqref{master} requires the Markov approximation, a non-Markovian equation can have the same form. Indeed, it is known that if $\gamma_i(t)<0$ for some $t$ and $\int_0^tdt'\;\gamma_i(t')>0$ for all $t$, then Eq.~\eqref{master} corresponds to a completely positive non-Markovian channel~\cite{Rivas2014}. Our approach can deal also with non-Markovian processes of this kind, keeping the same efficiency as the Markovian case. While we will consider the general case $\gamma_i=\gamma_i(t)$, whose sign distinguishes the Markovian processes by the non-Markovian ones, for the sake of simplicity we will consider the case $H\neq H(t)$ and $L_i\neq L_i(t)$ (in the following, we will denote $\mathcal{L}_H^t$ simply as $\mathcal{L}_H$). However, the inclusion in our formalism of  time-dependent Hamiltonians and Lindblad operators is straightforward.

One can integrate Eq.~\eqref{master} obtaining a Volterra equation~\cite{Bellman}  
\begin{align}\label{Volterra}
\rho(t)=e^{t\mathcal{L}_{H}}\rho(0)+\int_0^tds\;e^{(t-s)\mathcal{L}_{H}}\mathcal{L}_D^s\ \rho(s),
\end{align}
where $e^{t\mathcal{L}_{H}}\equiv\sum_{k=0}^\infty t^k\mathcal{L}_{H}^k/k!$.  The first term at the right-hand-side of Eq.~(\ref{Volterra}) corresponds to the  unitary evolution of $\rho(0)$ while the second term gives rise to the dissipative correction. Our goal is to find a perturbative expansion of Eq.~\eqref{Volterra} in the $\mathcal{L}_D^t$ term, and to provide with a  protocol to measure the resulting expression in a unitary way. In order to do so, we consider the iterated solution of Eq.~(\ref{Volterra}) obtaining 
\begin{align}
\rho(t)\equiv  \sum_{i=0}^\infty \rho_i(t).\label{series}
\end{align}
\sloppy Here, $\rho_0(t)=e^{t\mathcal{L}_{H}}\rho(0)$, while, for $i\geq1$, $\rho_i(t)$ has the following general structure: ${\rho_i(t)= \Pi_{j=1}^i  \Phi_j  \ e^{s_i \mathcal{L}_{H}} \rho(0)}$, $\Phi_j$ being a superoperator acting on an arbitrary matrix $\xi$ as  ${\Phi_j \xi = \int_0^{s_{j-1}} ds_{j} \ e^{(s_{j-1} - s_{j}) \mathcal{L}_{H}}\mathcal{L}_D^{s_{j}}\, \xi}$, where $s_0\equiv t$. For instance,  $\rho_2(t)$ can be written as 
\begin{eqnarray}
\nonumber \rho_2(t) &=& \Pi_{j=1}^2 \Phi_j   e^{s_2 \mathcal{L}_{H}} \rho(0) = \Phi_1\Phi_2\ e^{s_2 \mathcal{L}_{H}} \rho(0) \\ &=& \int_0^{t}ds_1e^{(t-s_1)\mathcal{L}_{H}}\mathcal{L}_D^{s_1}\int_{0}^{s_1}ds_2 e^{(s_1-s_2)\mathcal{L}_{H}}\mathcal{L}_D^{s_2}  e^{s_2\mathcal{L}_{H}}\rho(0).\nonumber
\end{eqnarray}
In this way, Eq.~(\ref{series}) provides us with a general and useful expression of the solution of Eq.~\eqref{master}. Let us consider the truncated series in Eq.~(\ref{series}), that is $\tilde\rho_{n}(t) = e^{t\mathcal{L}_{H}}\rho(0) +  \sum_{i=1}^{n} \rho_i(t)$, where $n$  corresponds to the order of the approximation. We will prove that an expectation value $\langle O \rangle_{\rho(t)}\equiv\text{Tr}\,[O\rho(t)]$ corresponding to a dissipative dynamics can be well approximated as
\begin{equation}\label{apprO}
\langle O\rangle_{\rho(t)}\approx {\rm Tr}[O e^{t\mathcal{L}_{H} } \rho(0)]  + \sum_{i=1}^{n} {\rm Tr}[O \rho_i(t)].
\end{equation}

In the following, we will supply with a quantum algorithm based on single-shot random measurements to compute each of the terms appearing in Eq.~(\ref{apprO}), and we will derive specific upper-bounds quantifying the accuracy of  our method. Notice that the first term at the right-hand-side of  Eq.~(\ref{apprO}), i.e. ${\rm Tr}[O e^{t\mathcal{L}_{H} } \rho(0)]$,  corresponds to the expectation value of the operator $O$ evolving under a unitary dynamics, thus it can be measured directly in a unitary quantum simulator where the dynamic associated with the Hamiltonian $H$ is implementable. However, the successive terms of the considered series, i.e. ${\rm Tr}[O \rho_i(t)]$ with $i\geq1$, require a specific development because they involve  multi-time correlation functions of the Lindblad operators and the operator $O$. 

Let us consider the first order term of the series in Eq.~\eqref{apprO}
\begin{eqnarray}\label{first}
\langle O\rangle_{\rho_1(t)}&=&\int_0^tds_1\; \text{Tr}\,[Oe^{(t-s_1)\mathcal{L}_{H}}\mathcal{L}_D^{s_1} \rho_0(s_1)]  \\
&=&\sum_{i=1}^N\int_0^tds_1\;\gamma_i(s_1)\bigg[\langle L_i^\dag (s_1)O(t)L_i (s_1)\rangle-\frac{1}{2}\langle \left\{O(t), L_i^\dag L_i(s_1)\right\}\rangle\bigg], \nonumber
\end{eqnarray}
where $\xi(s)\equiv e^{iH s}\xi e^{-iH s}$ for a general operator $\xi$ and time $s$, and all the expectation values are computed in the state $\rho(0)$. Note that the average values  appearing in the second and third lines of  Eq.~\eqref{first} correspond to time correlation functions of the operators $O$, $L_i^{}$, $L^{\dag}_i$, and $L_i^\dag L_i$. In the following, we consider a basis $\{Q_{j}\}_{j=1}^{d^2}$, where $d$ is the system dimension and $Q_{j}$  are Pauli-kind operators, i.e. both unitary and Hermitian \footnote{For a discussion on the decomposition of operators in a unitary basis see appendix~\ref{app:decompPauli}.}. The operators $L_i$ and $O$ can be decomposed as $L_i=\sum_{k=1}^{M_i} q^i_{k}Q^i_{k}$ and $O=\sum_{k=1}^{M_O} q^O_{k}Q^O_{k}$, with $q^{i,O}_{k}\in\mathbb{C}$, $Q^{i,O}_{k}\in \{ Q_j\}_{j=1}^{d^2}$, and $M_i,M_O\leq d^2$. We obtain then
\begin{equation}\label{decomp}
\langle L_i^\dag(s_1) O(t)L_i(s_1) \rangle=\sum_{l=1}^{M_{O}}\sum_{k,k'=1}^{M_i}q_l^{O}q_{k}^{i\,*}q_{{k'}}^i\langle Q^i_{k}(s_1) Q_l^O(t)Q_{{k'}}^i(s_1)\rangle,
\end{equation}
that is a sum of correlations of unitary operators. The same argument applies to the terms including $L_i^\dag L_i$ in  Eq.~\eqref{first}. Accordingly, we have seen that the problem of estimating the first-order correction is moved to the  measurement of some specific multi-time correlation functions involving the $Q_k^{i,O}$ operators. The argument can be easily extended to higher-order corrections. Indeed, for the $n$-th order, we have to evaluate the quantity
\begin{eqnarray}\label{nquant}
\langle O\rangle_{\rho_n(t)}&=&\int dV_n\;\text{Tr}[Oe^{(t-s_1)\mathcal{L}_{H}}\mathcal{L}_D^{s_1} \dots  \mathcal{L}_D^{s_n}e^{s_n\mathcal{L}_{H}}\rho(0)] \nonumber \\ &\equiv& \sum_{i_1,\dots,i_n=1}^N\int dV_n\;\langle A_{[i_1,\cdots,i_n]}(\vec{s})\rangle.
\end{eqnarray}
Here,
\begin{align}
A_{[i_1,\dots,i_n]}(\vec{s})&\equiv  e^{s_n\mathcal{L}_{H}^\dag}\mathcal{L}_D^{s_n,i_n\dag}\dots \mathcal{L}_D^{s_2,i_2\dag}e^{(s_1-s_2)\mathcal{L}^\dag_{H}} \mathcal{L}_D^{s_1,i_1\dag}e^{(t-s_1)\mathcal{L}^\dag_{H}}O,\nonumber
\end{align}
where $\mathcal{L}_{D}^{s,i} \xi\equiv \gamma_{i}^{}(s)\left(L_{i}\xi L_{i}^{\dag}-\frac{1}{2}\{ L^{\dag}_{i} L_{i}, \xi\}\right)$,  $\vec{s} = (s_1, \dots, s_n)$, $\int dV_n = \int_0^t \dots \int_0^{s_{n-1}} ds_1\dots ds_{n}$, and $\mathcal{L}^\dag\xi\equiv(\mathcal{L}\xi)^\dag$ for a general superoperator $\mathcal{L}$. As  in Eq.~(\ref{first}), the above expression contains multi-time correlation functions of the Lindblad operators $L_{i_1},\dots, L_{i_n}$ and the observable $O$, that have to be evaluated in order to compute  each  contribution  in Eq.~(\ref{apprO}). 

Our next step is to provide a method to evaluate general terms as the one appearing in Eq.~\eqref{nquant}. The standard approach to estimate this kind of quantities corresponds to measuring the expected value $\langle A_{[i_1,\cdots,i_n]}(\vec{s})\rangle$ at different random times $\vec{s}$ in the integration domain, and then calculating the average. Nevertheless, this strategy involves a huge number of measurements, as we need to estimate an expectation value at each chosen time. Our technique, instead, is based on single-shot random measurements and, as we will see below, it leads to an accurate estimate of  Eq.~(\ref{nquant}). More specifically, we will prove that 
\begin{equation}\label{casilla}
\sum_{i_1,\dots,i_n=1}^N\int dV_n\;\langle A_{[i_1,\cdots,i_n]}(\vec{s})\rangle\approx \frac{N^n|V_n|}{|\Omega_n|} \sum_{\Omega_n}\tilde A_{\vec{\omega}}(\vec{t}),
\end{equation}
where $\tilde A_{\vec{\omega}}(\vec{t})$ corresponds to a single-shot measurement of $A_{\vec{\omega}}(\vec{t})$, being  $[\vec{\omega},\vec{t}]\in\Omega_n\subset\{[\vec{\omega},\vec{t}]\;|\; \vec{\omega}=[i_1,\dots, i_n], i_k\in[1,N], \vec{t}\in V_n \}$, $|\Omega_n|$ is the size of $\Omega_n$, and $[\vec{\omega},\vec{t}]$ are sampled uniformly and independently. As already pointed out, the integrand in Eq.~\eqref{nquant} involves multi-time correlation functions. Indeed, in section~\ref{sec:TimeCor} we have shown how, by adding only one ancillary qubit to the simulated system,  general time-correlation functions are accessible by implementing only unitary evolutions of the kind $e^{t\mathcal{L}_{H}}$, together with entangling operations between the ancillary qubit and the system. It is noteworthy to mention that these operations have already experimentally demonstrated in quantum systems as trapped ions~\cite{Lanyon2011} or quantum optics~\cite{OBrien2009}, and have been recently proposed for cQED architectures~\cite{Mezzacapo2014b}. Moreover, the same quantum algorithm allows us to measure single-shots of the real and imaginary part of these quantities providing, therefore, a way to compute the term at the right-hand-side of Eq.~(\ref{casilla}). Notice that the evaluation of each term $\langle A_{[i_1,\cdots,i_n]}(\vec{s})\rangle$ in Eq.~\eqref{nquant}, requires a number of measurements that depends on the observable decomposition, see Eq.~\eqref{decomp}. After specifying it, we measure the real and the imaginary part of the corresponding correlation function. Finally, in appendix~\ref{app:proofEqnumer1} we prove that
\begin{align}\label{numer1}
\left|\sum_{i_1,\dots,i_n=1}^N\int dV_n\;\langle A_{[i_1,\cdots,i_n]}(\vec{s})\rangle-\frac{(Nt)^n}{n!|\Omega_n|}\sum_{\Omega_n}\tilde A_{\vec{\omega}}(\vec{t})\right|\leq\delta_n,
\end{align}
with probability higher than $1-e^{-\beta}$, provided that $|\Omega_n|>\frac{36M_O^2(2+\beta)}{\delta_n^2}\frac{(2\bar{\gamma} MN t)^{2n}}{n!^2}$, where $\bar{\gamma}\equiv\max_{i,s\in[0,t]}|\gamma_i(s)|$ and $M\equiv\max_i M_i$. Equation~\eqref{numer1} means that  the quantity in Eq.~\eqref{nquant} can be estimated with arbitrary precision by random single-shot measurements of $A_{[i_1,\cdots,i_n]}(\vec{s})$, allowing, hence, to dramatically reduce  the resources required by our quantum simulation algorithm. Notice that the required number of measurements to evaluate the order $n$ is bounded by $3^n|\Omega_n|$, and the total number of measurements needed to compute the correction to the expected value of an observable up the order $K$ is bounded by $\sum_{n=0}^K3^n|\Omega_n|$. In the following, we discuss at which order we need to truncate in order to have a certain error in the final result.
 
So far, we have proved that we can compute, up to an arbitrary order in $\mathcal{L}_D^t$, expectation values corresponding to dissipative dynamics with a unitary quantum simulation. It is noteworthy  that our method does not require to physically engineer  the system-environment interaction. Instead, one only needs to implement the system Hamiltonian $H$. In this way we are opening a new avenue for the quantum simulation of open quantum dynamics in situations where the complexity on the design of the dissipative terms excedes the capabilities of quantum platforms. This covers a wide range of physically relevant situations. One example corresponds to the case of fermionic theories where the encoding of the fermionic behavior in the  degrees of freedom of the quantum simulator gives rise to highly delocalized operators~\cite{Jordan1928, Casanova2012}. In this case a reliable dissipative term should act on these non-local operators instead of on the individual qubits of the system. Our protocol solves this problem because it avoids the necessity of implementing the Lindblad superoperator. Moreover, the scheme allows one to simulate at one time a class of master equations corresponding to the same Lindblad operators, but with different choices of $\gamma_i$, including the relevant case when only a part of the system is subjected to dissipation, i.e. $\gamma_i=0$ for some values of $i$.

We shall next quantify the quality of our method. In order to do so, we will find an error bound certifying how the truncated series in Eq.~\eqref{series} is close to the solution of Eq.~\eqref{master}. This error bound will depend on the system parameters, i.e. the time $t$ and the dissipative parameters $\gamma_i$. As figure of merit we choose the trace distance, defined by
\begin{equation}
D_1(\rho_1,\rho_2)\equiv \frac{\|\rho_1-\rho_2\|_1}{2},
\end{equation}
where $\|A\|_1\equiv\sum_i\sigma_i(A)$, being $\sigma_i(A)$ the  singular values of $A$~\cite{Watrous2004}. Our goal is to find a bound for $D_1(\rho(t),\tilde {\rho}_n(t))$, where $\tilde \rho_n(t)\equiv \sum_{i=0}^n\rho_i(t)$ is the series of Eq. \eqref{series} truncated at the $n$-th order. We note that the the following recursive relation holds 
\begin{equation}\label{generaldif}
{\tilde {\rho}}_n(t)=e^{t\mathcal{L}_H}\rho(0)+\int_0^tds\;e^{(t-s)\mathcal{L}_{H}}\mathcal{L}_D^s\tilde{\rho}_{n-1}(s).
\end{equation}
From Eq. \eqref{generaldif}, it follows that
\begin{eqnarray}\label{induce}
\nonumber D_1(\rho(t),\tilde{\rho}_n(t))&=&\frac{1}{2}\left\|\int_0^tds\;e^{(t-s)\mathcal{L}_{H}} \mathcal{L}_D^s(\rho(s)-\tilde{\rho}_{n-1}(s))\right\|_1\\ &\leq& \int_0^tds\;\|\mathcal{L}_D^s\|_{1\rightarrow1}D_1(\rho(s),\tilde{\rho}_{n-1}(s)),
\end{eqnarray}
where we have introduced the induced superoperator norm $\| \mathcal{A}\|_{1\rightarrow 1}\equiv\sup_\sigma\frac{\|\mathcal{A}\sigma\|_1}{\|\sigma\|_1}$~\cite{Watrous2004}. For $n=0$, i.e. for $\tilde \rho_n(t)\equiv \tilde{\rho}_0(t) = e^{t\mathcal{L}_{H}}\rho(0)$, we obtain the following bound\footnote{For a complete derivations of this bound see appendix~\ref{app:proofbounds}.}
\begin{align}\label{D1zero}
D_1(\rho(t),\tilde{\rho}_0(t))&\leq \frac{1}{2}\int_0^tds\; \|\mathcal{L}_D^s\|_{1\rightarrow1}\|\rho(s)\|_1\leq \sum_{i=1}^N |\gamma_i(\epsilon_i)|\|L_i\|_\infty^2t,
\end{align}
where $0\leq \epsilon_{i}\leq t$, and $\|A\|_\infty \equiv\sup_i\sigma_i(A)$. Notice that, in finite dimension, one can always renormalize $\gamma_i$ in order to have $\|L_i\|_{\infty}=1$,  i.e. if we transform $L_i\rightarrow L_i / \|L_i\|_\infty$, $\gamma_i\rightarrow \|L_i\|_\infty \gamma_i$, the master equation remains invariant. Using Eq.~\eqref{induce}-\eqref{D1zero}, one can shown by induction that for the general $n$-th order the following bound holds 
\begin{equation}\label{bounds}
D_1(\rho(t),\tilde{\rho}_n(t))\leq\prod_{k=0}^n\bigg[2\sum_{i_k=1}^N |\gamma_{i_k}(\epsilon_{i_k})|\bigg]\frac{t^{n+1}}{2(n+1)!}\leq\frac{(2\bar \gamma N t)^{n+1}}{2(n+1)!},
\end{equation}
where $0\leq \epsilon_{i_k}\leq t$ and we have set $\| L_i\|_\infty=1$. From Eq.~\eqref{bounds}, it is clear that the series converges uniformly to the solution of Eq.~\eqref{master} for every finite value of $t$ and choices of $\gamma_i$. As a result, the number of measurements needed to simulate a certain dynamics at time $t$ up to an error $\varepsilon<1$ is\footnote{For a discussion on the required number of measurements see appendix~\ref{app:numofmeas}.} $O\left(\left(\bar{t}+\log\frac{1}{\varepsilon}\right)^2\frac{e^{12M\bar{t}}}{\varepsilon^2}\right)$, where $\bar{t}=\bar \gamma N t$. Here, a discussion on the efficiency of the method is needed. From the previous formula, we can say that our method performs well when $M$ is low, i.e. in that case where each Lindblad operators can be decomposed in few Pauli-kind operators. Moreover, as our approach is perturbative in the dissipative parameters $\gamma_i$, it is reasonable that the method is more efficient when $|\gamma_i|$ are small. Notice that analytical perturbative techniques are not available in this case, because the solution of the unperturbed part is assumed to be not known. Lastly, it is evident that the algorithm is efficient for a certain choices of time, and the relevance of the simulation depends on the particular cases. For instance, a typical interesting situation is a strongly coupled Markovian system. Let us assume  with site-independent couple parameter $g$ and dissipative parameter $\gamma$. We have that $e^{12M\bar t}\leq 1+12eM\bar t$ if $t\leq \frac{1}{12M\gamma N}\equiv t_c$. In this period, the system oscillates typically $C\equiv gt_c=\frac{g/\gamma}{12MN}$ times, so the simulation can be considered efficient for $N\sim g/\gamma C$, which, in the strong coupling regime, can be of the order of $10^3/C$. Notice that, in most relevant physical cases, the number of Lindblad operators $N$ is of the order of the number of system parties~\cite{Kliesch2011}.

All in all, our method is aimed to simulate a different class of master equations with respect the previous approaches, including non-Markovian quantum dynamics, and it is efficient in the range of times where the exponential $e^{M\bar t}$ may be truncated at some low order. A similar result is achieved by the authors of Ref.~\cite{Kliesch2011}, where they simulate a Lindblad equation via Trotter decomposition. They show that the Trotter error is exponentially large in time, but this exponential can be truncated at some low order by choosing the Trotter time step $\Delta t$ sufficiently small. Our method is qualitatively different, and it can be applied also to analogue quantum simulators where suitable entangled gates are available.

Lastly, we note that this method is also applicable to simulate dynamics under a non-Hermitian Hamiltonian $J=H-i\Gamma$, with $H=H^\dag$, $\Gamma=\Gamma^\dag$. This type of generator  emerges as an effective Hamiltonian in the Feshbach partitioning formalism~\cite{Muga2004}, when one looks for the evolution of the density matrix projected onto a subspace. The new Schr\"odinger equation reads
\begin{equation}
\frac{d\rho}{dt}=-i[H,\rho]+\{\Gamma,\rho\},
\end{equation}
This kind of equation is useful in understanding several phenomena, e.g. scattering processes~\cite{Moiseyev1998} and dissipative dynamics~\cite{Plenio1998}, or in the study of $PT$-symmetric Hamiltonian~\cite{Bender2007}.  Our method consists in considering the non-Hermitian part as a perturbative term. As in the case previously discussed, similar bounds can be easily found\footnote{See appendix~\ref{app:nonHermitian} for a discussion on the bounds of the {non-Hermitian} case.}, and this proves that the method is reliable also in this situation.

In conclusion, we have proposed a method to compute expectation values of observables that evolve according to a generalized Lindblad master equation, requiring only the implementation of its unitary part.  Through the quantum computation of $n$-time correlation functions of the Lindblad operators, we are able to reconstruct the corrections of the dissipative terms to the  unitary quantum evolution without reservoir engineering techniques. We have provided a complete recipe that combines quantum resources and specific theoretical developments   to compute these corrections, and error-bounds quantifying the accuracy of the proposal and defining the cases when the proposed method is efficient. Our technique can be also applied, with small changes, to the quantum simulation of  non-Hermitian Hamiltonians. The presented method provides a general strategy to perform quantum simulations of open systems, Markovian or not,  in a variety of quantum platforms.


\subsection{An experimental demonstration of the algorithm in NMR} \label{sec:NMRExp}

In this section we report on the  measurement of multi-time correlation functions of a set of Pauli operators on a two-level system, which can be used to retrieve its associated linear response functions. The two-level system is an effective spin constructed from the nuclear spins of $^{1}$H atoms in a solution of $^{13}$C-labeled chloroform. Response functions characterize the linear response of the system to a family of perturbations, allowing us to compute physical quantities such as the magnetic susceptibility of the effective spin. We use techniques introduced in section~\ref{sec:TimeCor} to measure time correlations on the two-level system. This approach requires the use of an ancillary qubit encoded in the nuclear spins of the $^{13}$C atoms and a sequence of controlled operations. Moreover, we demonstrate the ability of such a quantum platform to compute time-correlation functions of arbitrary order, which relate to higher-order corrections of perturbative methods. Particularly, we show three-time correlation functions for arbitrary times, and we also measure time correlation functions at fixed times up to tenth order.

\begin{figure}[htb!]
\begin{center}
\includegraphics[width= 0.8\columnwidth]{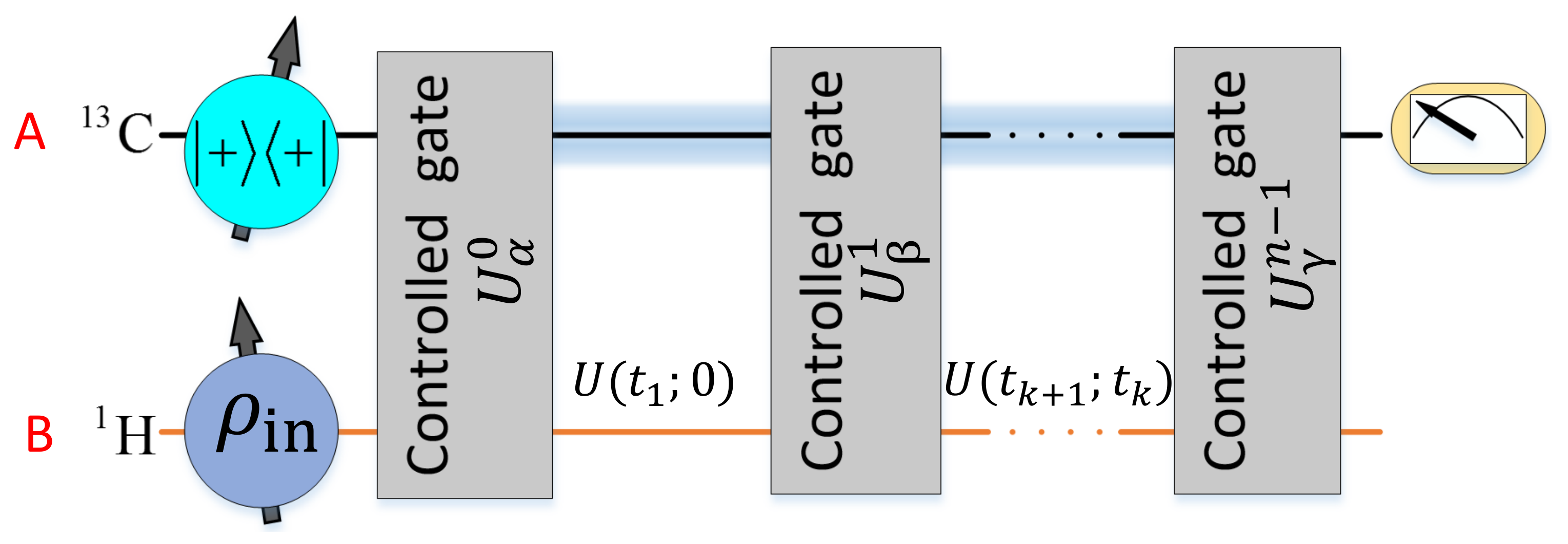}
\end{center}
\setlength{\abovecaptionskip}{-0.00cm}
\caption[Two-qubit quantum circuit for measuring general $n$-time correlation functions in a solution of $^{13}$C-labeled chloroform]{\footnotesize{ {\bf Two-qubit quantum circuit for measuring general $n$-time correlation functions.} Qubit A is the ancilla (held by the nuclear spin of $^{13}$C), and qubit B is the system qubit (held by the nuclear spin of $^{1}$H). The blue zone between the different controlled gates $U^k_\alpha$ on the line of qubit A represents the decoupling of the $^{13}$C nucleus from the nuclear spin of $^{1}$H, while the latter evolves according to $U(t_{k+1};t_{k})$. The measurement of the quantities  $\langle \sigma_x \rangle$ and $\langle \sigma_y\rangle$ of the ancillary qubit at the end of the circuit will directly provide the real and imaginary values of the $n$-time correlation function for the initial state $\rho_{\rm in}= | \phi \rangle \langle \phi | $.} } \label{circuit_total}
\end{figure}

We will follow the algorithm introduced in section~\ref{sec:TimeCor} to extract  $n$-time correlation functions of the form $f(t_1, ... , t_{n-1})=\langle \phi | \sigma_\gamma (t_{n-1})...\sigma_\beta(t_1)\sigma_\alpha (0) | \phi \rangle$ from a two-level quantum system, with the  assistance of one ancillary qubit. Here, $ | \phi \rangle$ is the quantum state of the system and $\sigma_\alpha(t)$ is a time-dependent Pauli operator in the Heisenberg picture, defined as ${\sigma_\alpha(t)=U^{\dagger}(t;0)\sigma_\alpha U(t;0)}$, where ${\alpha=x,y,z}$, and $U(t_j; t_i)$ is the evolution operator from time $t_i$ to $t_j$.  The considered algorithm is depicted in Fig.~(\ref{circuit_total}), for the case where qubit A and B respectively encode the ancillary qubit and the two-level quantum system, and consists of the following steps: \\
$(i)$ The input state of the probe-system qubits is prepared in $\rho^{\rm AB}_{\rm in}= | + \rangle \langle + | \otimes\rho_{\rm in}$, with $ | + \rangle = ( | 0 \rangle + | 1 \rangle )/\sqrt{2}$ and $\rho_{\rm in}=| \phi \rangle \langle \phi |$. \\
$(ii)$ The controlled quantum gate $U_\alpha^k=|1 \rangle \langle 1 | \otimes S_\alpha + | 0 \rangle \langle 0 | \otimes \mathbb{I}_2$ is firstly applied on the two qubits,  with $S_x=\sigma_x$, $S_y=-i\sigma_y$ and $S_z=i\sigma_z$. $\mathbb{I}_2$ is a $2 \times 2$ identity matrix. \\
$(iii)$ It follows a unitary evolution of the system qubit from $t_k$ to time $t_{k+1}$, $U(t_{k+1};t_k)$, which needs not be known to the experimenter. In our setup, we engineer this dynamics by decoupling qubit A and B, such that only the system qubit evolves under its free-energy Hamiltonian. An additional evolution can also be imposed on the system qubit according to the considered problem. Steps $(ii)$ and $(iii)$ will be iterated $n$ times, taking $k$ from $0$ to $n-1$ and avoiding step $(iii)$ in the last iteration. With this, all $n$ Pauli operators will be interspersed between evolution operators with the time intervals of interest $\{t_{k}, t_{k+1} \}$. \\
$(iv)$ Finally, the time correlation function is extracted as a non-diagonal operator of the ancilla, Tr($ | 0 \rangle \langle 1 | \varphi_{\rm out} \rangle \langle \varphi_{\rm out} | $), where $ | \varphi_{\rm out} \rangle = ( | 0 \langle \otimes U(t_{n-1};0) | \phi \rangle + | 1 \rangle \otimes S_\gamma U(t_{n-1};t_{n-2})...U(t_2;t_1)S_\beta U(t_1;0) S_\alpha | \phi \rangle)/\sqrt{2}$. We recall here that $| 0 \rangle \langle 1 | =( \sigma_x + i \sigma_y)/2$, such that ${f(t_1, ..., t_{n-1})=  c_y c_z (\langle \sigma_x \rangle + i \langle \sigma_y \rangle})$, which is in general a complex magnitude.  The additional factors $c_y=i^r$ and $c_z=(-i)^l$, where integers $r$ and $l$ are the occurrence numbers of Pauli operators $\sigma_y$ and $\sigma_z$ in $f(t_1, ..., t_{n-1})$. 

As already explained in the introduction to this chapter, the measurement of $n$-time correlation functions plays a significant role in the linear response theory. For instance, we can microscopically derive useful quantities such as the conductivity and the susceptibility of a system, with the knowledge of 2-time correlation functions. As an illustrative example, we study the case of a spin-$1/2$ particle in a uniform magnetic field of strength $B$ along the z-axis, which has a natural Hamiltonian $\mathcal{H}_{\rm 0}= -\gamma B \sigma_z$, where $\gamma$ is the gyromagnetic ratio of the particle. We assume now that a magnetic field with a sinusoidal time dependence $B'_0e^{-i\omega t}$ and arbitrary direction $\alpha$ perturbs the system. The Hamiltonian representation of such a situation is given by $\mathcal{H}=\mathcal{H}_{\rm 0}- \gamma B'_0 \sigma_\alpha e^{-i \omega t}$, with ${\alpha= x, y, z}$. The magnetic susceptibility of the system is the deviation of the magnetic moment from its thermal expectation value as a consequence of such a perturbation. For instance, the corrected expression for the magnetic moment in direction $\beta\  ( \mu_\beta= \gamma \sigma_\beta)$ is given by $\mu_\beta(t)= \mu_\beta(0) + \chi_{\alpha, \beta}^\omega e^{-i \omega t}$, where $\chi_{\alpha, \beta}^\omega$ is the frequency-dependent susceptibility. From linear response theory, we learn that the susceptibility can be retrieved integrating the linear response function as ${\chi_{\alpha, \beta}^\omega= \int_{-\infty}^{t} \phi_{\alpha, \beta}(t-s) e^{-i \omega (t-s)} ds}$. Moreover, the latter can be given in terms of time-correlation functions of the measured and perturbed observables, ${\phi_{\alpha, \beta}(t)=\langle [\gamma B \sigma_\alpha,  \gamma \sigma_\beta (t)] \rangle/(i \hbar)}$, where ${\sigma_\beta (t)= e^{i/\hbar H_0 t} \sigma_\beta e^{-i/\hbar H_0 t}}$, and the averaging is made over a thermal equilibrium ensemble. Notice that for a two level system, the thermal average can easily be reconstructed from the expectation values of the ground and excited states. So far, the response function can be retrieved by measuring the 2-time correlation functions of the unperturbed system $\langle \sigma_\alpha \sigma_\beta (t) \rangle $ and $\langle  \sigma_\beta (t) \sigma_\alpha\rangle$. Notice that when $\alpha=\beta$, ${\langle \sigma_\alpha (t) \sigma_\alpha \rangle^*=\langle \sigma_\alpha \sigma_\alpha(t) \rangle}$, and it is enough to measure one of them. All in all, measuring two-time correlation functions from an ensemble of two level systems is not merely a computational result, but an actual measurement of the susceptibility of the system to arbitrary perturbations. Therefore, it gives us insights about the behavior of the system, and helps us to characterize it. In a similar fashion, further corrections to the expectation values of the observables of the system will be given in terms of higher-order correlation functions. In this experiment, we will not only measure two-time correlation functions that will allow us to extract the susceptibility of the system, but we will also show that higher-order correlation functions can be obtained.

We will measure $n$-time correlation functions of a two-level quantum system with the  assistance of one ancillary qubit by implementing the quantum circuit shown in Fig.~(\ref{circuit_total}). Experiments are carried out using NMR \cite{Cory2000,Havel2002,Suter2008}, where the sample used is $^{13}$C-labeled chloroform. Nuclear spins of $^{13}$C and $^{1}$H encode the ancillary qubit and the two-level quantum system, respectively.  With the weak coupling approximation, the internal Hamiltonian of $^{13}$C-labeled chloroform is
\begin{equation}
\mathcal{H}_{\rm int}=-\pi (\nu _1-\omega_1)\sigma_z^1-\pi (\nu _2-\omega_2)\sigma_z^2+0.5 \pi J_{12} \sigma_z^1 \sigma_z^2,
\label{hamiltonian}
\end{equation}
where $\nu_j$ ($j=1,2$) is the chemical shift, $\emph{J}_{12}$ is the $J$-coupling strength as illustrated in appendix~\ref{app:NMRplatform}, while $\omega_1$ and $\omega_2$ are reference frequencies of $^{13}$C and $^{1}$H, respectively.  We set $\nu _1=\omega_1$ and $\nu _2-\omega_2=\bigtriangleup\omega$ such that the natural Hamiltonian of the system qubit is $\mathcal{H}_{0}=-\pi \bigtriangleup \omega \sigma_z$.  The detuning frequency $\bigtriangleup\omega$ is chosen as hundreds of Hz to assure the selective excitation of different nuclei via hard pulses.  All experiments are carried out on a Bruker AVANCE 400MHz spectrometer at room temperature. 

It is widely known that the thermal equilibrium state of an NMR ensemble is a highly-mixed state with the following structure~\cite{Cory2000}
\begin{equation}
\mathcal{\rho}_{eq}\approx \frac{1-\epsilon}{4}\mathbb{I}_4+\epsilon(\frac{1}{4}\mathbb{I}_4+\sigma^1_z+4\sigma^2_z ).
\end{equation}
Here, $\mathbb{I}_4$ is a $4 \times 4$ identity matrix and $\epsilon\approx10^{-5}$ is the polarization at room temperature. Given that $\mathbb{I}_4$ remains unchanged and that it does not contribute to the NMR spectra, we consider the deviation density matrix $\bigtriangleup \rho=0.25\mathbb{I}_4+\sigma^1_z+4\sigma^2_z $ as the effective density matrix describing the system. The deviation density matrix can be initialized in the pure state $ | 00 \rangle \langle 00 |$ by the spatial averaging technique \cite{Cory1997,Cory1998,Knill1998}, transforming the system into a so-called pseudo-pure state (PPS). Input state $\rho^{\rm CH}_{\rm in}$ can be easily created by applying local single-qubit rotation pulses after the preparation of the PPS.

In Step $(ii)$, all controlled quantum gates $U^k_\alpha$ can be easily realized by using single-qubit rotation pulses and a $J$-coupling evolution \cite{Xin2015,Vandersypen2004}. All controlled operations are chosen from the set of gates $\{C-R^2_z(-\pi), C-iR^2_x(\pi), C-R^2_y(\pi) \}$. The notation $C-U$ means operator $U$ will be applied on the system qubit only if the ancilla qubit is in state $| 1 \rangle \langle 1 |$, while $R^j_{\hat{n}}(\theta)$ represents a single-qubit rotation on qubit $j$ along the $\hat{n}$-axis with the rotation angle $\theta$. In an NMR platform, we decompose gates $U^k_\alpha$ in the following way,
\begin{eqnarray}
&& C-R^2_z(-\pi) = U(\frac{1}{2J})R^2_z(-\frac{\pi}{2}) , \nonumber \\
&& C-iR^2_x(\pi) = \sqrt{i}R^1_z(\frac{\pi}{2})R^2_z(-\frac{\pi}{2})R^2_x(\frac{\pi}{2})U(\frac{1}{2J})R^2_y(\frac{\pi}{2}) , \nonumber \\
&& C-R^2_y(\pi) = R^2_x(\frac{\pi}{2})U(\frac{1}{2J})R^2_x(-\frac{\pi}{2})R^2_y(\frac{\pi}{2}).
\label{decompose}
\end{eqnarray}
Here, $U(\frac{1}{2J})$ is the $J$-coupling evolution $e^{-i\pi\sigma^1_z\sigma^2_z/4}$. Moreover, any $z$-rotation $R_z(\theta)$ can be decomposed in terms of rotations around the $x$ and $y$ axes, $R_z(\theta)=R_y(\pi/2)R_x(-\theta)R_y(-\pi/2)$.  On the other hand, the decoupling of the interaction between $^{13}$C and $^{1}$H nuclei can be realized by using refocusing pulses \cite{RobinBendall1983} or the Waltz-4 sequence \cite{Widmaier1998,Shaka1983,Shaka1983a}. 

We will now apply the described algorithm to a collection of situations of physical interest. The detailed NMR sequences for all considered experiments can be found in appendix~\ref{app:detailsequence}. In Fig.~(\ref{classA}), we extract the time-correlation functions of a two-level system evolving under Hamiltonian $\mathcal{H}_{0}=-100\pi\sigma_z$ for a collection of $\alpha$ and $\beta$, and different initial states. These correlation functions are enough to retrieve the response function for a number of physical situations corresponding to different magnetic moments and applied fields. In this experiment, $\mathcal{H}_{0}$ is realized by setting $\nu _1=\omega_1$ and $\nu _2-\omega_2=100$ Hz in Eq.~(\ref{hamiltonian}). A rotation pulse $R^1_y(\pi/2)$ is applied on the first qubit after the PPS preparation to create $\rho^{\rm CH}_{\rm in}= | + \rangle \langle + | \otimes | 0 \rangle \langle 0 |$. Similarly, a $\pi$ rotation on the second qubit is additionally needed to prepare $\rho^{\rm CH}_{\rm in}= | + \rangle \langle + | \otimes | 1 \rangle \langle 1 |$ as the input state of the ancilla-system compound, or alternatively a $R_y^2(\pi/2)$ rotation to generate the initial state $\rho_{\rm in}^{\rm CH}= | + \rangle \langle + | \otimes | + \rangle \langle + |$. Extracting correlation functions for initial states $| 0 \rangle$ and $| 1 \rangle $ will allow us to reconstruct such correlation functions for a thermal state of arbitrary temperature.

\begin{figure}[htb]
\begin{center}
\includegraphics[width= 0.8\columnwidth]{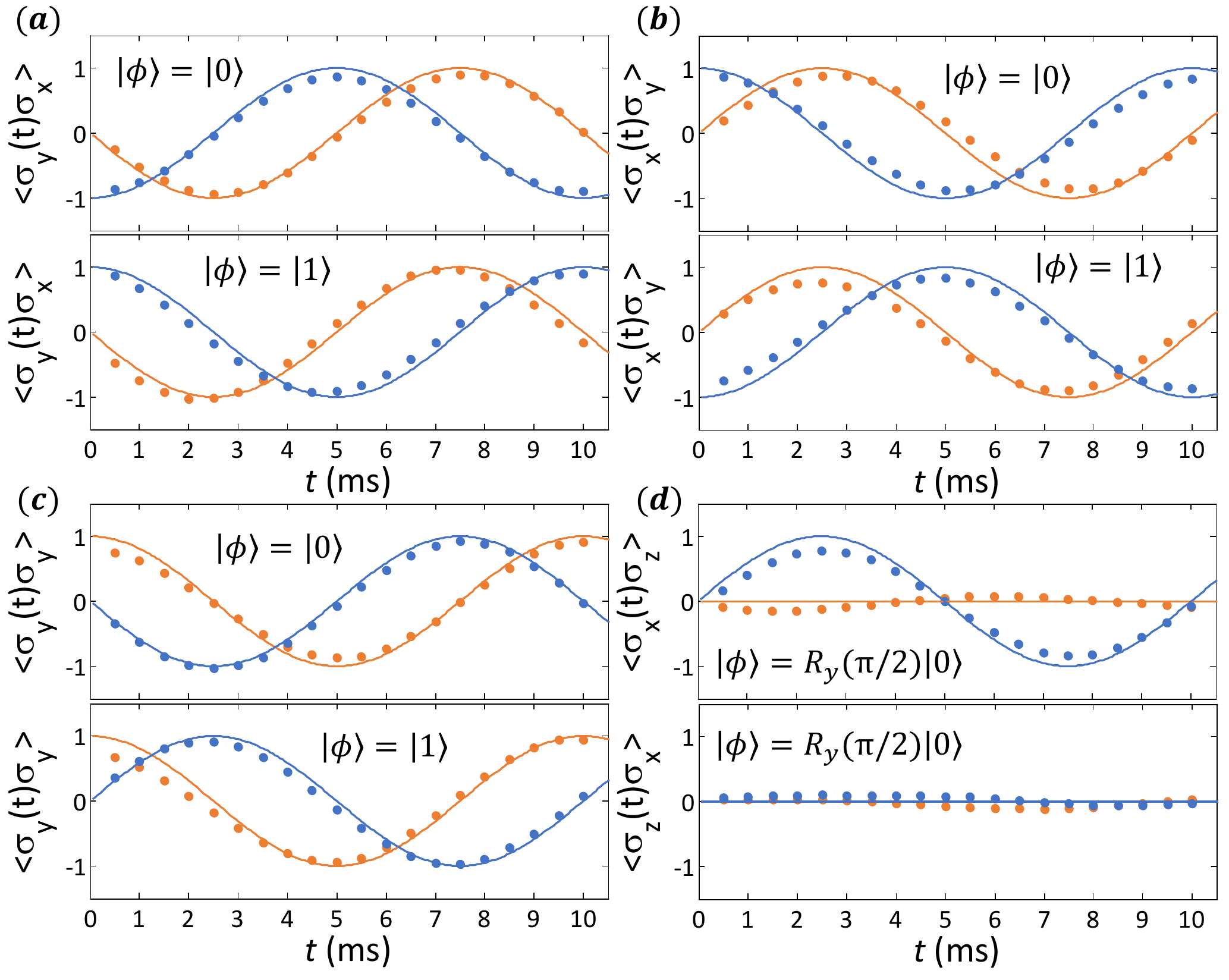}
\end{center}
\setlength{\abovecaptionskip}{-0.00cm}
\caption[Experimental measurement of a 2-time correlation function of $^1$H nuclei evolving under a time-independent Hamiltonian]{ \footnotesize{{\bf Experimental results (dots) for 2-time correlation functions.} In this case, only two controlled quantum gates $U^0_\alpha$ and $U^1_\beta$ are applied  with an interval of $t$. For example, $U^0_\alpha$ and $U^1_\beta$ should be chosen as $C-iR^2_x(\pi)$ and $C-R^2_y(\pi)$, respectively, to measure the 2-time correlation function $\langle\sigma_y(t)\sigma_x\rangle$. $t$ is swept from $0.5$ms to $10$ms with a $0.5$ms increment. The input state of $^{1}$H nuclei $\rho_{\rm in}=| \phi \rangle \langle\phi |$ is shown on each diagram. All experimental results are directly obtained from measurements of the expectation values of  $\langle\sigma_x\rangle$ and $\langle\sigma_y\rangle$ of the ancillary qubit. }} \label{classA}
\end{figure}

\begin{figure}[htb]
\begin{center}
\includegraphics[width= 0.8\columnwidth]{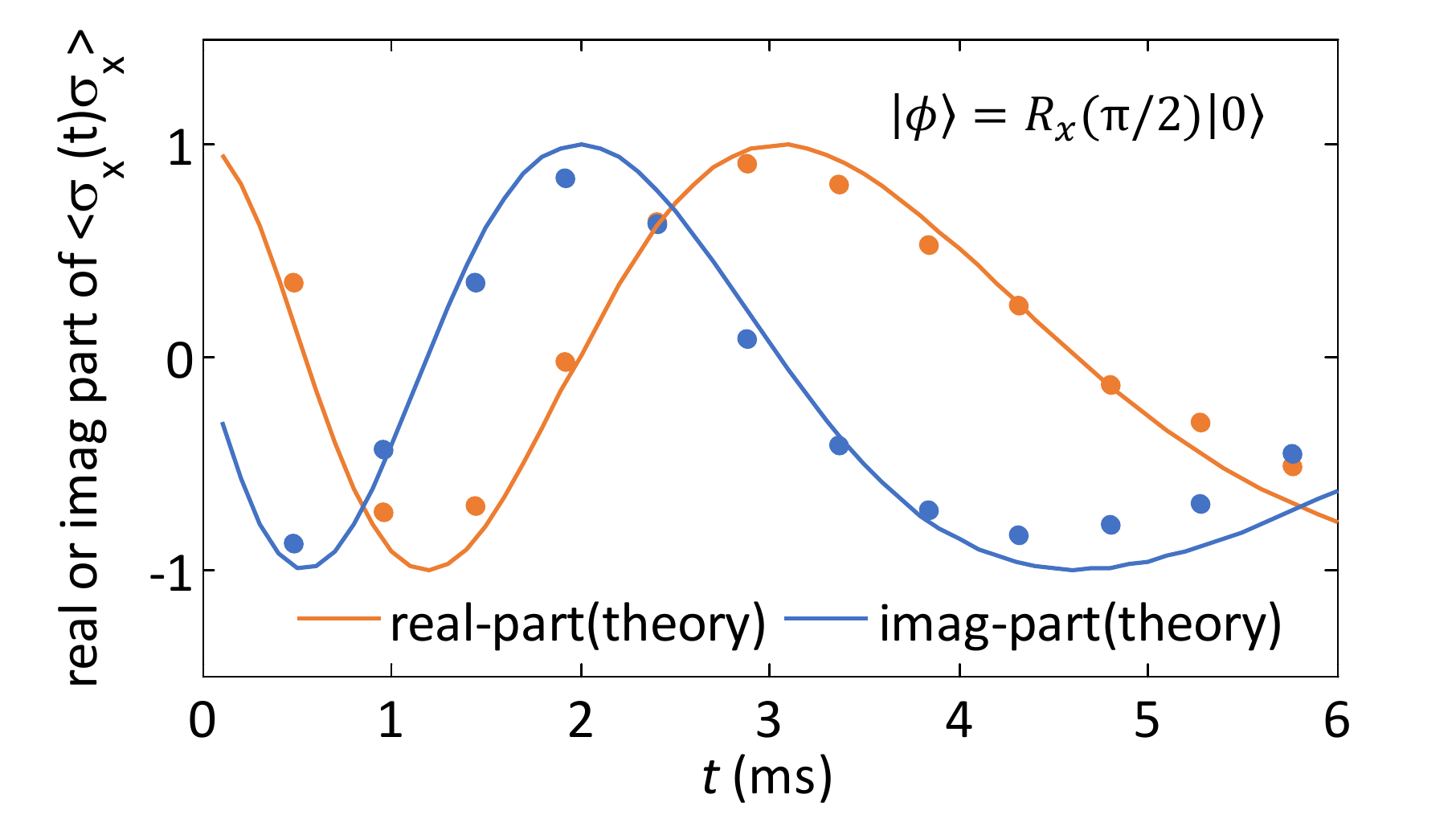}
\end{center}
\setlength{\abovecaptionskip}{-0.00cm}
\caption[Experimental measurement of a 2-time correlation function of $^1$H nuclei evolving under a time-dependent Hamiltonian]{\footnotesize{ {\bf Experimental results (dots) for a 2-time correlation function of the $^1$H nuclei evolving under a time-dependent Hamiltonian.} For this experiment, the $^1$H nuclei have a natural Hamiltonian $\mathcal{H}_{0}=0$ and an initial state $| \phi \rangle=R_x(\pi/2) | 0 \rangle$. An evolution $U(t;0)$ between $U^0_x$ and $U^1_x$ is applied on the system, which is described by the evolution operator $e^{-i\int_{0}^t \mathcal{H'}(s) ds}$ with $\mathcal{H'}(s)=500e^{-300s}\pi\sigma_y$. $t$ is changed from $0.48$ms to $5.76$ms with a $0.48$ms increment per step.}} \label{classD}
\end{figure}

We now consider a more involved situation where the system is in a magnetic field with an intensity that is decaying exponentially in time, that is, the unperturbed system Hamiltonian turns now into a time-dependent $\mathcal{H'}= \gamma B_0 e^{-a t} \sigma_y$. In this case, the response function needs to be computed in terms of time-correlation functions of the system observables evolving under this new Hamiltonian. In Fig.~(\ref{classD}), we show such a case, for the time-correlation function $\langle\sigma_x (t) \sigma_x\rangle$ and the initial state $(| 0 \rangle-i| 1 \rangle)/\sqrt{2}$. For this, we set $\nu _1=\omega_1$ and $\nu _2=\omega_2$ in Eq.~(\ref{hamiltonian}), making the system free Hamiltonian $\mathcal{H}_{0}=0$. The initial state $ | \phi \rangle=R_x(\pi/2) | 0 \rangle$ can be prepared by using a rotation pulse $R_x(\pi/2)$ on the initial PPS. Two controlled quantum gates $U^0_x=C-iR^2_x(\pi)$ and $U^1_x=C-iR^2_x(\pi)$ are applied  with a time interval $t$. A decoupling sequence Waltz-4 \cite{Widmaier1998} is used to cancel the interaction between the $^{13}$C and $^{1}$H nuclei during the evolution between the controlled operations. During the decoupling period, a time-dependent radio-frequency pulse is applied on the resonance of the system qubit $^1$H to create the Hamiltonian $\mathcal{H'}(t)=500e^{-300t}\pi\sigma_y$.

\begin{figure}[htb]
\begin{center}
\includegraphics[width= 0.8 \columnwidth]{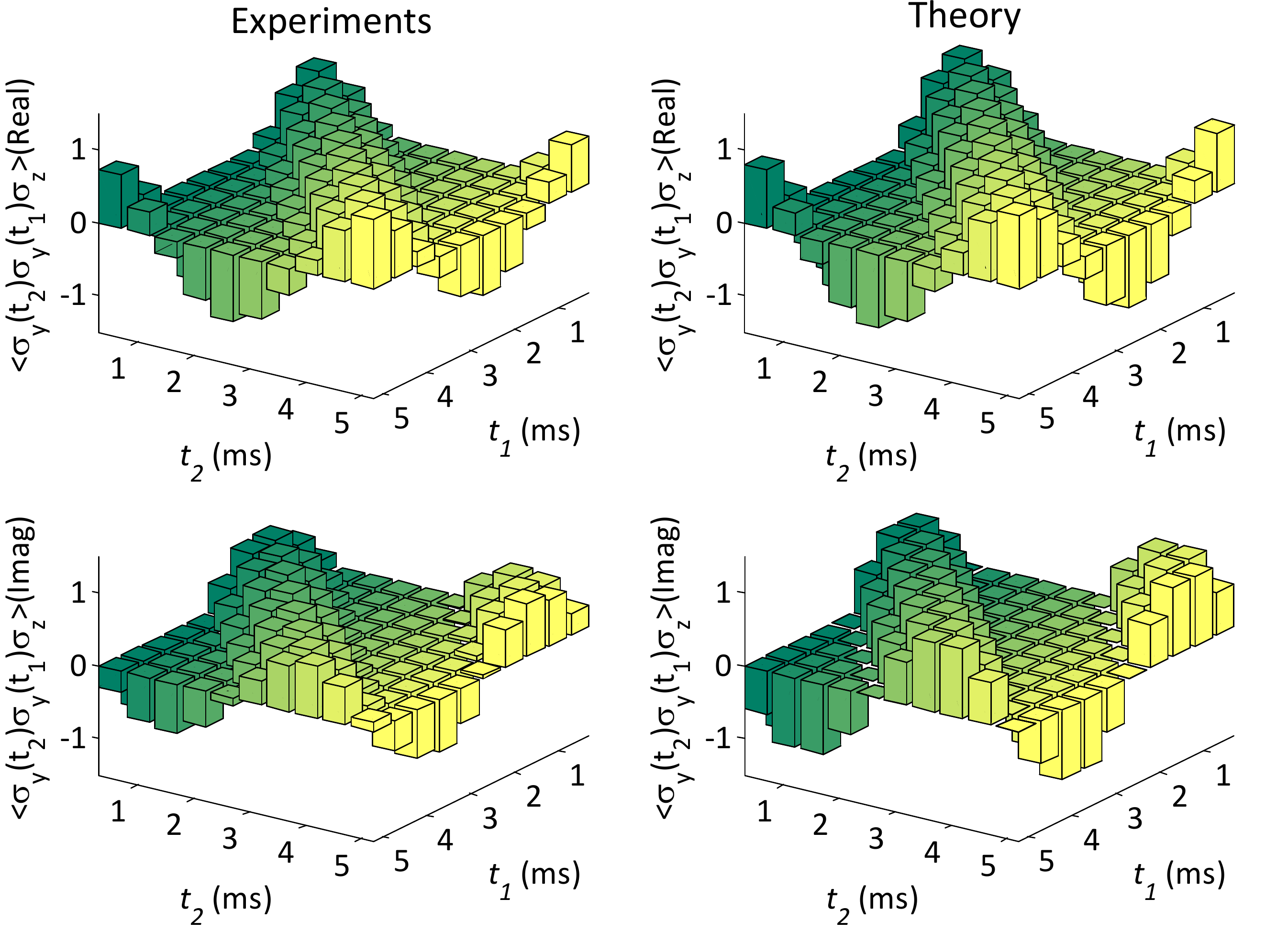}
\end{center}
\setlength{\abovecaptionskip}{-0.00cm}
\caption[Experimental measurement of $3$-time correlation functions] { \footnotesize{{\bf Experimental results for the $3$-time correlation functions.} We plot $\mathcal{M}^3_{zyy} = \langle\sigma_y(t_2)\sigma_y(t_1)\sigma_z\rangle$ for $t_1$ and $t_2$ going from 0.5ms to 5ms with 0.5ms time step, showing the agreement of experimental results with theoretical predictions. The quantum circuit for measuring $\mathcal{M}^3_{zyy}$ includes three controlled quantum gates $U^0_z$,  $U^1_y$ and $U^2_y$, which are decomposed according to Eq.~(\ref{decompose}) and experimentally implemented by hard pulses. }} \label{classB}
\end{figure}

When the perturbation is not weak enough, for instance when the radiation field applied to a material is of high intensity, the response of the system might not be linear. In such situations,  higher-order response functions, which depend in higher-order time-correlation functions, will be needed to account for the non-linear corrections~\cite{Kubo1957, Peterson1967}. For example, the second order correction to an observable $B$ when the system suffers a perturbation of the type $H(t)= H_0 + A F(t)$ would be given by ${\Delta B^{(2)}= \int_{-\infty}^t \int_{-\infty}^{t_1} \langle [B(t), [A(t_1), A(t_2)] \rangle F(t_1) F(t_2) dt_1 dt_2 }$.  In Fig.~(\ref{classB}), we show real and imaginary parts of 3-time correlation functions as compared to their theoretically expected values. We measure the 3-time correlation function $\langle\sigma_y(t_2)\sigma_y(t_1)\sigma_z\rangle$ versus $t_1$ and $t_2$. In this case, we simulate the system-qubit free Hamiltonian $\mathcal{H}_{0}=-200\pi\sigma_z$ and an input state $\rho_{\rm in}= | 0 \rangle \langle 0 |$. For this, we set $\nu _1=\omega_1$ and $\nu _2-\omega_2=200$ Hz in Eq.~(\ref{hamiltonian}). The $J$-coupling term of Eq.~(\ref{hamiltonian}) will be canceled by using a refocusing pulse in the circuit.  Three controlled quantum gates $U^0_\alpha$,  $U^1_\beta$ and $U^2_\gamma$ should be chosen as $C-R^2_z(-\pi)$, $C-R^2_y(\pi)$ and $C-R^2_y(\pi)$. The free evolution of the $^{1}$H nuclei between $U^0_\alpha$ and $U^1_\beta$ is given by the evolution operator $e^{-i\mathcal{H}_{0}t_1}$. Accordingly,  the free evolution of the $^{1}$H nuclei between $U^1_\beta$ and $U^2_\gamma$ is given by $e^{-i\mathcal{H}_{0}(t_2-t_1)}$. However, when $t_2$<$t_1$, we perform the evolution $e^{-i(-\mathcal{H}_{0})(t_1-t_2)}$ by inverting the phase of the Hamiltonian $\mathcal{H}_{0}$, which is realized by using double $\pi$ pulses at the beginning and at the end of the evolution~\cite{Vandersypen2004}. The data shown in Fig.~(\ref{classB}) demonstrates the power of this algorithm, with full range 3-time correlation functions, for all combinations of $t_1$ and $t_2$.\\

\begin{figure}[htb]
\begin{center}
\includegraphics[width= 0.8\columnwidth]{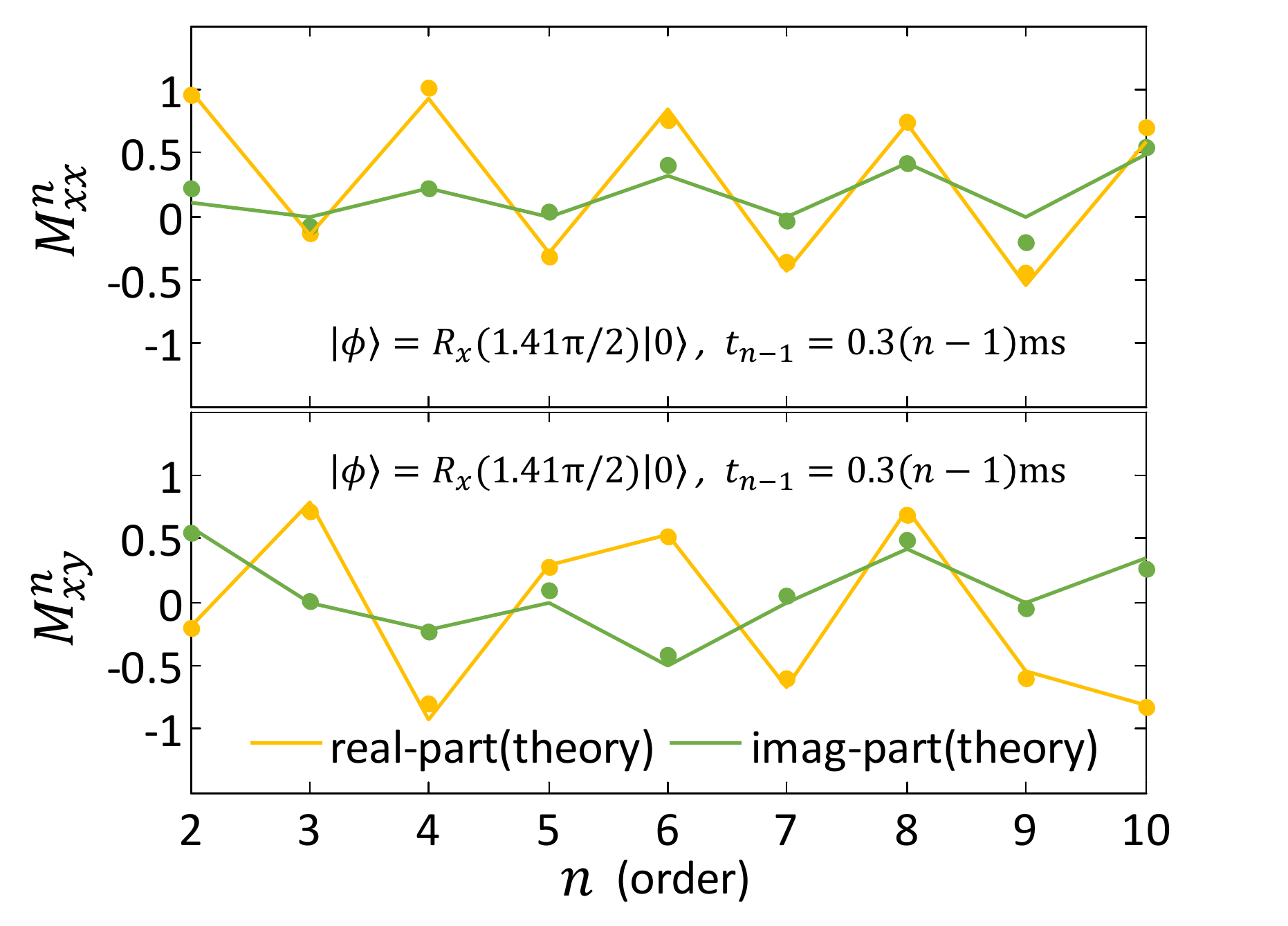}
\end{center}
\setlength{\abovecaptionskip}{-0.00cm}
\caption[Experimental measurement of high-order time-correlation functions]{ \footnotesize{{ \bf Experimental results (dots) for high-order time-correlation functions $\mathcal{M}^n_{xx}$ and $\mathcal{M}^n_{xy}$ ($n=2,3,...,10$).} $n$ controlled quantum gates $U^0_\alpha$,$U^1_\beta$,...,$U^{n-1}_\gamma$ are sequentially applied  with a time interval $\bigtriangleup t=0.3$ms. Refocusing pulses are used to decouple the interaction between the nuclei of $^{13}$C and $^{1}$H, during the time intervals $\bigtriangleup t$ between gates. For this experiment, we use the GRAPE pulsed technique to implement the quantum circuit. Using hard pulses like in the previous experiments would result in poor quality of the measured data due to the high number of pulses required and the cumulative effect of their imperfections.}} \label{classC}
\end{figure}

For testing scalability, we also measure higher-order time-correlation functions, up to $n=10$. In this case, we consider the free Hamiltonian ${\mathcal{H}_{0}=-100\pi\sigma_z}$ and the input state $| \phi \rangle=R_x(1.41\pi/2) | 0 \rangle$, and we measure the following high-order correlation functions as a function of the correlation order $n$, with time intervals $t_{n-1}=0.3(n-1)$ ms,
\begin{equation}
\begin{array}{l}
\mathcal{M}^n_{xx}=\langle\sigma_x(t_{n-1})\sigma_x(t_{n-2})...\sigma_x(t_{1})\sigma_x\rangle,\\
\mathcal{M}^n_{xy}=\langle ... \sigma_x(t_{2m})\sigma_y(t_{2m-1})...\sigma_y(t_{1})\sigma_x\rangle.
\end{array}
\end{equation}

Here, the superscripts and subscripts of $\mathcal{M}$ are the order of the correlation and the involved Pauli operators, respectively. Index $m$ runs from $1$ to $(n-1)/2$ for odd $n$, and to $n/2$ for even $n$. The quantum circuit used to measure $\mathcal{M}^n_{xx}$ and $\mathcal{M}^n_{xy}$ is based in the gradient ascent pulse engineering (GRAPE) technique \cite{Khaneja2005,Ryan2008}, which is designed to be robust to the static field distributions ($T_2^{*}$ process) and RF inhomogeneities. High-order correlation functions are related to the system susceptibilities beyond the linear response. In Fig. \ref{classC}, we show the measured results for this case, demonstrating the scalability of the technique and the high accuracy even for a $10$-time correlation function.
 
For all cases here discussed, experimental data shows a high degree of agreement with the theoretical predictions. Error bars are not shown, as they are always smaller than the used dots themselves. In our setup, the sources of errors are related to the initialization of the PPS, and imperfections in the width of the employed hard pulses. Moreover, the latter effect is cumulative and can result in a snowball effect. Additionally, factors such as $T_2^*$ processes, RF inhomogeneities, and measurement errors, bring in a signal loss.

Summarizing, we have shown that the measurement of time-correlation functions of arbitrary order in NMR is an efficient task, and can be used to obtain the linear response function of the system. We have demonstrated that such magnitudes can be experimentally retrieved with high accuracy, opening the door to the quantum simulation of physical models where time correlations play a central role. These experimental techniques are platform independent and may be extended to systems of arbitrary size, where a single ancillary qubit will always suffice.

%% file: chap/chapter3.tex
\lettrine[lines=2, findent=3pt,nindent=0pt]{E}{ntanglement} is considered  one of the most remarkable features of quantum mechanics~\cite{NielsenChuang, Horodecki2009a}, stemming from bipartite or multipartite correlations without classical counterpart. Firstly revealed by Einstein, Podolsky, and Rosen as a possible drawback of quantum theory~\cite{Einstein1935}, entanglement was subsequently identified as a fundamental resource for quantum communication~\cite{Ekert1991, Bennett1993} and quantum computing purposes~\cite{Shor1995, Chuang1999}. Beyond considering entanglement  as a purely theoretical feature, the development of quantum technologies has allowed us to create, manipulate, and detect entangled states in different quantum platforms. Among them, we can mention trapped ions, where eight-qubit W and fourteen-qubit GHZ states have been created~\cite{Haffner2005, Monz2011}, circuit QED (cQED) where  seven superconducting elements have been entangled~\cite{Mariantoni2011}, superconducting circuits where continuous-variable entanglement has been realized in propagating quantum microwaves~\cite{Menzel2012}, and bulk-optic based setups  where entanglement between eight photons has been achieved~\cite{Gao2010}.

Quantifying entanglement is considered a particularly difficult task, both from theoretical and experimental viewpoints. In fact, it is challenging to define entanglement measures for an arbitrary number of parties~\cite{Wong2001, Barnum2004}. Moreover, the existing entanglement monotones~\cite{Vidal2000} do not correspond directly to the expectation value of a Hermitian operator~\cite{Wootters1998}. Accordingly,  the computation of many entanglement measures, see Ref.~\cite{Guhne2007} for lower bound estimations, requires  previously the reconstruction of the full quantum state, which could be a cumbersome problem if the size of the associated Hilbert space is large. If we consider, for instance, an $N$-qubit system, quantum tomography techniques become already experimentally unfeasible for $N\sim10$ qubits. This is because the dimension of the Hilbert space grows exponentially with $N$, and the number of observables needed to reconstruct the quantum state scales~as~$2^{2N}-1$.

In this chapter we will introduce the concept of embedding quantum simulators (EQS) for the efficient measurement of entanglement monotones in simulation schemes. In section~\ref{sec:EQS}, we will establish the theoretical framework of EQS, for later in section~\ref{sec:EQSIons}, give a detailed recipe for their implementation in trapped ions. Finally, in section~\ref{sec:EQSPhotons}, we report on an actual experimental realization of the concept with photons in the laboratory of Prof. Andrew White.


\subsection{Embedding quantum simulators} \label{sec:EQS}

\begin{figure}[h]
\begin{center}
\vspace{0.5cm}
\hspace{-0.3cm}
\includegraphics[width= 0.97 \columnwidth]{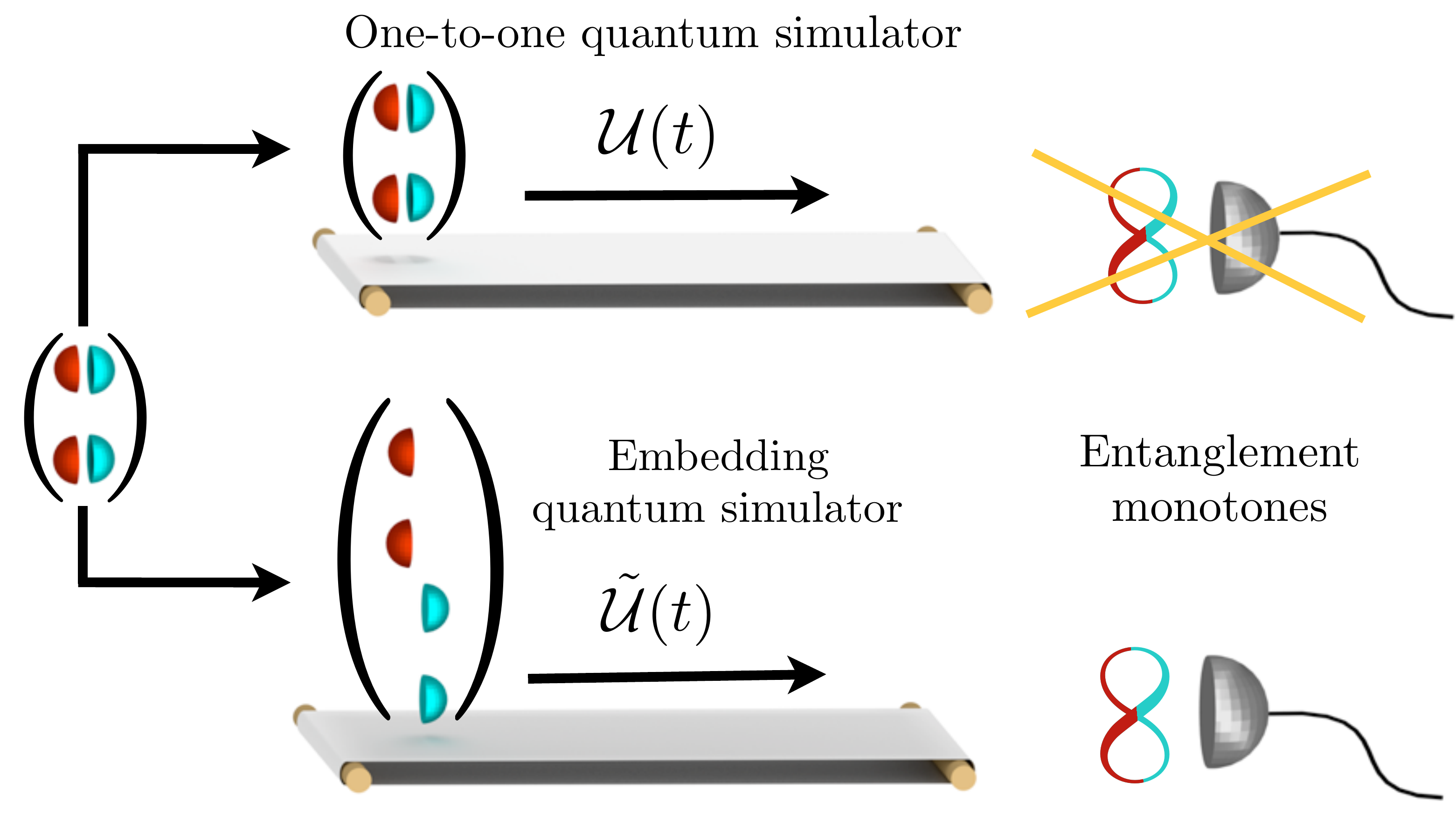}
\end{center}
\caption[Pictorical representation of a one-to-one quantum simulator versus an EQS.]{\footnotesize{{\bf Pictorical representation of a one-to-one quantum simulator versus an EQS.} The conveyor belts represent the dynamical evolution of the quantum simulators. The real (red) and imaginary (blue) parts of the simulated wave vector components are split in the embedding quantum simulator, allowing the efficient computation of entanglement monotones.}}\label{figembedding}
\end{figure}

From a general point of view, a {\it standard quantum simulation} is meant to be implemented in a {\it one-to-one quantum simulator} where, for example, a two-level system in the simulated dynamics is directly represented by another two-level system in the simulator. In this section, we introduce the concept of {\it embedding quantum simulators}, allowing the efficient computation of a wide class of entanglement monotones~\cite{Vidal2000}. This method can be applied at any time of the evolution of a simulated bipartite or multipartite system, with the prior knowledge of the Hamiltonian $H$ and the corresponding initial state $|\psi_0\rangle$. The efficiency of the protocol lies in the fact that, unlike standard quantum simulations, the evolution of  the state $|\psi_0\rangle$ is embedded in an {\it enlarged Hilbert space} dynamics (see Fig.~\ref{figembedding}). Note that enlarged-space structures have been previously considered for different purposes in Refs.~\cite{Rudolph2002, Fernandez2003, McKague2009}. In our case, antilinear operators associated with a certain class of entanglement monotones can be efficiently encoded into physical observables, overcoming the necessity of full state reconstruction. The simulating quantum dynamics, which embeds the desired quantum simulation, may be implemented in different quantum technologies with analog and digital simulation methods.

\subsubsection{Complex conjugation and entanglement monotones}
An entanglement monotone is a function of the quantum state, which is zero for all separable states and does not increase on average under local quantum operations and classical communication ~\cite{Vidal2000}. There are several functions satisfying these basic properties, as concurrence~\cite{Wootters1998} or three-tangle~\cite{Dur2000}, extracting information about a specific feature of entanglement. For pure states, an entanglement monotone $E(|\psi\rangle)$ can be defined univocally, while the standard approach for mixed states requires the computation of the convex roof 
\begin{equation}\label{roof}
E(\rho)=\min_{\{p_i,|\psi_i\rangle\}}\;\sum_{i}p_iE(|\psi_i\rangle).
\end{equation}
Here, $\rho=\sum_{i}p_i|\psi_i\rangle\langle\psi_i|$ is the density matrix describing the system, and the minimum in Eq.~(\ref{roof}) is taken over all possible pure-state decompositions~\cite{Horodecki2009a}. 

A systematic procedure to define entanglement monotones for pure states involves  the complex-conjugation operator $K$~\cite{Osterloh2005, Uhlmann2000}. For instance, the concurrence for  two-qubit pure states~\cite{Wootters1998} can be written as
\begin{equation}
\label{Concurrence}
C(|\psi\rangle)\equiv|\langle\psi|\sigma_y\otimes\sigma_y K|\psi\rangle | .
\end{equation}
Note that $\sigma_y\otimes\sigma_yK$,  where $K | \psi \rangle \equiv | \psi^* \rangle$, is an antilinear operator that cannot be associated with a physical observable. In general, we can construct entanglement monotones for  $N$-qubit systems  combining three operational building blocks: $K$,  $\sigma_y$, and $g^{\mu\nu}\sigma_\mu\sigma_\nu$, with $g^{\mu\nu}=\text{diag}\{-1,1,0,1\}$, $\sigma_0=\mathbb{I}_2$, $\sigma_1=\sigma_x$, $\sigma_2=\sigma_y$, $\sigma_3=\sigma_z$,   where we assume the repeated index summation convention~\cite{Osterloh2005}. For a two-qubit system, $N=2$, we can define $|\langle\psi|\sigma_y\otimes\sigma_yK|\psi\rangle|$ and  $|g^{\mu\nu}g^{\lambda\tau}\langle\psi|\sigma_\mu\otimes\sigma_\lambda K|\psi\rangle\langle\psi|\sigma_\nu\otimes\sigma_\tau K|\psi\rangle|$  as entanglement monotones. The first expression corresponds to the concurrence and the second one is a second-order monotone defined in Ref.~\cite{Osterloh2005}. For $N=3$ we have $|g^{\mu\nu}\langle\psi|\sigma_\mu\otimes\sigma_y\otimes\sigma_yK|\psi\rangle\langle\psi|\sigma_\nu\otimes\sigma_y\otimes\sigma_yK|\psi\rangle|$, corresponding to the $3$-tangle~\cite{Dur2000}, and so on. 

To evaluate the above class of entanglement monotones in a one-to-one quantum simulator, we would need to perform full tomography on the system. This is because  terms like $\langle\psi|OK|\psi\rangle\equiv \langle\psi|O|\psi^*\rangle$, with $O$ Hermitian, do not correspond to  the expectation value of a physical observable, and they have to be computed classically once each complex component of $|\psi\rangle$ is known. We will explain now how to compute efficiently quantities as $\langle\psi| O K|\psi\rangle$ in our proposed embedding quantum simulator, via the measurement of a reduced number of observables.

Consider a pure quantum state $|\psi\rangle$ of an $N$-qubit system $\in \mathbb{C}_{2^N}$, whose evolution is governed  by the Hamiltonian $H$ via the Schr\"odinger equation ($\hbar=1$)
\begin{equation}\label{Schro}
(i  \partial_{t} -H )|\psi(t)\rangle=0.
\end{equation}
The quantum dynamics associated with the Hamiltonian $H$ can be implemented in a one-to-one quantum simulator~\cite{Feynman1982, Lloyd1996a} or, alternatively, it can be encoded in an embedding quantum simulator, where $K$  may become a physical quantum operation~\cite{Casanova2011}. The latter can be achieved according to the following rules.

{\it Embedding quantum simulator.---}  We define a mapping $\mathcal{M}:\mathbb{C}_{2^N}\rightarrow\mathbb{R}_{2^{N+1}}$ in the following way:
\begin{equation}\label{map}
|\psi\rangle=\left( \begin{array}{c} \psi_{\rm{re}}^1+i\psi_{\rm{im}}^1\\ \psi_{\rm{re}}^2+i\psi_{\rm{im}}^2\\ \psi_{\rm{re}}^3+i\psi_{\rm{im}}^3\\  \vdots\end{array} \right) {\huge\xrightarrow {\mathcal{M}}} \,\, |\tilde \psi\rangle=\left( \begin{array}{c} \psi_{\rm{re}}^1\\ \psi_{\rm{re}}^2\\ \psi_{\rm{re}}^3\\  \vdots\\ \psi_{\rm{im}}^1 \\  \psi_{\rm{im}}^2 \\  \psi_{\rm{im}}^3 \\  \vdots\end{array} \right).
\end{equation}
Hereafter, we will call $\mathbb{C}_{2^N}$ the {\it simulated space}  and $\mathbb{R}_{2^{N+1}}$ the {\it simulating space} or the {\it enlarged space}. We note that the resulting vector  $|\tilde{\psi}\rangle$ has only real components (see refs.~\cite{Rudolph2002, Fernandez2003, McKague2009} for other developments involving real Hilbert spaces), and that the reverse mapping is  $|\psi\rangle=M|\tilde\psi\rangle$, with $M=\left(1\;,\; i\right)\otimes\mathbb{I}_{2^N}$. It is noteworthy to mention that, for an unknown initial state, the mapping $\mathcal M$ is not physically implementable. However, according to Eq.~(\ref{map}), the knowledge of the initial state in the simulated space determines completely the possibility of initializing the state in the enlarged space. Furthermore, it can be easily checked that the inverse mapping $M$ can always be completed to form a unitary operation.

Now, we can write
\begin{equation}\label{relation}
K|\psi\rangle \equiv|\psi^*\rangle=M|\tilde\psi^*\rangle= M(\sigma_z\otimes\mathbb{I}_{2^N})|\tilde\psi\rangle\equiv M\tilde K|\tilde\psi\rangle ,
\end{equation}
which, despite its simple aspect, has important consequences. Basically, Eq.~(\ref{relation})  tells us that while $|\psi\rangle$ and $|\psi^*\rangle$ are connected by the unphysical operation $K$ in the simulated space, the relation between their images in the enlarged space, $|\tilde{\psi}\rangle$ and $|\tilde{\psi}^*\rangle$, is a physical quantum gate $\tilde{K}\equiv(\sigma_z\otimes\mathbb{I}_{2^N})$. In this way, we obtain that
\begin{equation}\label{obs}
\langle\psi|OK|\psi\rangle=\langle\tilde\psi|M^\dag O M (\sigma_z\otimes\mathbb{I}_{2^N})|\tilde \psi\rangle,
\end{equation}
where we can prove that
\begin{equation}
M^\dag  O M (\sigma_z\otimes\mathbb{I}_{2^N}) = (\sigma_z-i\sigma_x)\otimes O.
\end{equation}
Note that $M^\dag  O M (\sigma_z\otimes\mathbb{I}_{2^N})$ is a linear combination of Hermitian operators $\sigma_z\otimes O$ and $\sigma_x\otimes O$. Hence, its expectation  value can be efficiently computed via the measurement  of these two observables in the enlarged space.

\begin{figure}[h]
\begin{center}
\vspace{0.5cm}
\hspace{-0.3cm}
\includegraphics [width= 0.8 \columnwidth]{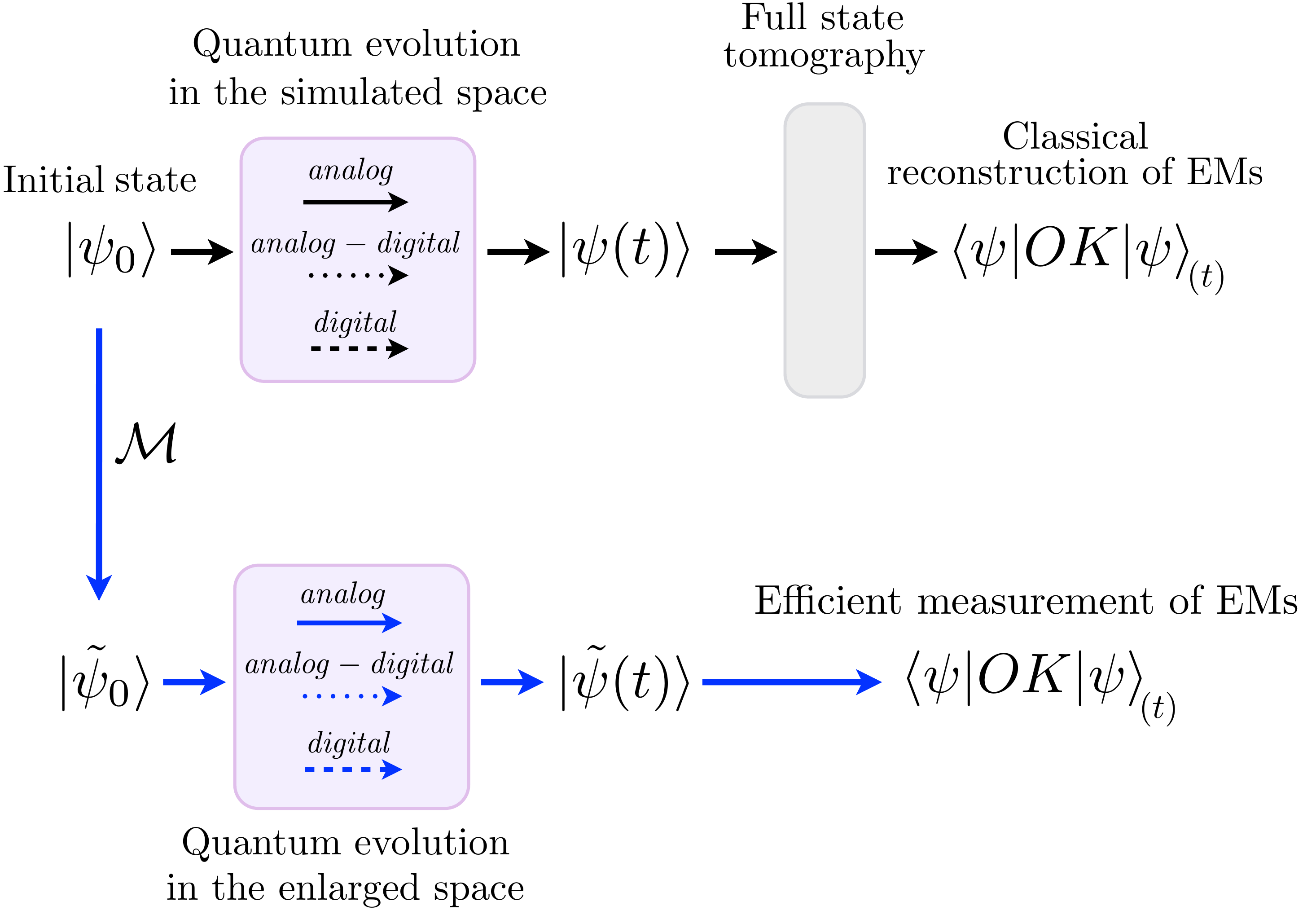}
\end{center}
\caption[Protocols for computing entanglement monotones.]{\footnotesize{{\bf Protocols for computing entanglement monotones.} In the lower part the enlarged space formalism  (blue arrows) is depicted, while in the upper we schematize the usual protocol  (black arrows). For any initial state $|\psi_0\rangle$,  we can construct throught the mapping $\mathcal{M}$ its image $|\tilde{\psi}_0\rangle$ in the enlarged space. The evolution will be implemented using analog or digital techniques giving rise to the state $|\tilde{\psi}(t)\rangle$. The subsequent  measure of a reduced number of observables  will provide us with the EMs.}}\label{chap3figscheme}
\end{figure}

So far, we have found a mapping for quantum states and expectation values between the simulated space and the simulating space. If we also want to consider an associated quantum dynamics, we would need to map the Schr\"odinger equation~(\ref{Schro}) onto another one in the enlarged space. In this sense, we look for a wave equation
\begin{equation}\label{enlarSchro}
(i\partial_t-\tilde H)|\tilde\psi(t)\rangle=0 ,
\end{equation}
whose solution  respects $|\psi(t)\rangle=M|\tilde\psi(t)\rangle$ and  $|\psi^*(t)\rangle=M\tilde{K}|\tilde\psi(t)\rangle$, thereby assuring that  the complex conjugate operation can be applied at any time $t$ with the same single qubit gate. If we define in the enlarged space a (Hermitian) Hamiltonian $\tilde H$ satisfying $M\tilde H=HM$, while applying $M$ to both sides of Eq.~\eqref{enlarSchro}, we arrive to equation $(i\partial_t-H)M|\tilde\psi(t)\rangle=0$. It follows that if $|\tilde \psi (t)\rangle$ is solution of Eq.~\eqref{enlarSchro} with initial condition $|\tilde\psi_0\rangle$, then $M|\tilde\psi(t)\rangle$ is solution of the original Schr\"odinger equation~\eqref{Schro} with initial condition $M|\tilde{\psi}_0\rangle$. Thus, if $|\psi_0\rangle=M|\tilde\psi_0\rangle$, then $|\psi(t)\rangle=M|\tilde\psi(t)\rangle$, as required. The Hamiltonian $\tilde H$ satisfying $HM=M\tilde H$ reads 
\begin{equation}\label{H}
\tilde H=\left(
\begin{array}{cc}
 i B & i A \\
 -i A & i B \\
\end{array}
\right)\equiv \big[i \mathbb{I}_2\otimes B-\sigma _y\otimes A\big],
\end{equation} 
where $H=A+iB$, with $A=A^\dag$ and $B=-B^\dag$  real matrices,  corresponds to the original Hamiltonian in the simulated space.  We note that $\tilde H$ is a Hermitian  imaginary  matrix,  e.g.   $H=\sigma_x\otimes \sigma_y+\sigma_x\otimes\sigma_z $ is mapped into $\tilde H=\mathbb{I}_2\otimes\sigma_x\otimes\sigma_y-\sigma_y\otimes\sigma_x\otimes \sigma_z$ which is Hermitian and imaginary. In this sense, $|\tilde\psi_0\rangle$ with real entries implies the same character for $|\tilde\psi(t)\rangle$, given that the Schr\"odinger equation is a first order differential equation with real coefficients. In this way, the complex-conjugate operator in the enlarged space $\tilde K=\sigma_z\otimes\mathbb{I}_{2^N}$ is the same at any time $t$.

On one hand, the implementation of the dynamics of Eq.~(\ref{enlarSchro}) in a quantum simulator will turn the computation of entanglement monotones into an efficient process, see Fig.~\ref{chap3figscheme}. On the other hand, the evolution associated to Hamiltonian $\tilde{H}$ can be implemented efficiently in different quantum simulator platforms, as is the case of trapped ions or superconducting circuits~\cite{Lanyon2011,Devoret2013a}. We want to point out that, in the most general case, the dynamics of a simulated system involving n-body interactions will require an embedding quantum simulator with (n+1)-body couplings. This represents, however, a small overhead of experimental resources. It is noteworthy to mention that the implementation of many-body spin interactions have already been realized experimentally in digital quantum simulators in trapped ions~\cite{Lanyon2011}. Concluding, quantum simulations in the enlarged space require the quantum control of {\it only one additional qubit}.

\subsubsection{Efficient computation of entanglement monotones} A general entanglement monotone constructed with $K$, $\sigma_y$, and $g^{\mu\nu}\sigma_\mu\sigma_\nu$, contains at most $3^k$ terms of the form $\langle\psi| O K|\psi\rangle$, $k$ being the number of times that $g^{\mu\nu}\sigma_\mu\sigma_\nu$ appears. Thus, to evaluate the most general set of entanglement monotones, we  need to measure $2\cdot 3^k$ observables, in contrast with the $2^{2N}-1$ required for full tomography.

We present now examples showing how our protocol minimizes the required experimental resources. 

{\it i)} {\it The concurrence.---} This two-qubit entanglement monotone defined in Eq.~(\ref{Concurrence})  is built using  $\sigma_y$ and $K$, and it can be evaluated with the measurement of  $2$ observables  in the enlarged space, instead of the $15$ required for  full tomography. Suppose we know $|\psi_0\rangle$ and want to compute $C(|\psi(t)\rangle)$, where $|\psi(t)\rangle \equiv e^{-iHt}|\psi_0\rangle$. We first initialize the quantum simulator with the state $|\tilde\psi_0\rangle$ using the mapping of Eq.~(\ref{map}). Second, this state evolves according to Eq.~(\ref{enlarSchro}) for a time $t$. Finally, following Eq.~\eqref{obs} with $O=\sigma_y\otimes\sigma_y$, we compute the quantity 
\begin{equation}\label{defen}
\langle\tilde \psi(t)|\sigma_z\otimes\sigma_y\otimes\sigma_y-i\sigma_x\otimes\sigma_y\otimes\sigma_y|\tilde\psi(t)\rangle ,
\end{equation}
by measuring the  observables $\sigma_z\otimes\sigma_y\otimes\sigma_y$ and $\sigma_x\otimes\sigma_y\otimes\sigma_y$ in the enlarged space.

{\it ii) The $3$-tangle.---} The $3$-tangle~\cite{Dur2000} is a $3$-qubit entanglement monotone defined as $\tau_3(|\psi\rangle)=|g^{\mu\nu}\langle\psi|\sigma_\mu\otimes\sigma_y\otimes\sigma_yK|\psi\rangle\langle\psi|\sigma_\nu\otimes\sigma_y\otimes\sigma_yK|\psi\rangle|$. It is built using  $g^{\mu\nu}\sigma_\mu\sigma_\nu$ and $K$, so the computation of $\tau_3$ in the enlarged space requires $6$ measurements  instead of the $63$ needed for full-tomography. The evaluation of $\tau_3(|\psi(t)\rangle)$ can be achieved  following  the same steps explained in the previous example, but now computing the quantity
\begin{eqnarray}
\big|&&-\langle\tilde\psi(t)|\sigma_z\otimes\mathbb{I}_2\otimes\sigma_y\otimes\sigma_y-i\sigma_x\otimes\mathbb{I}_2\otimes\sigma_y\otimes\sigma_y|\tilde\psi(t)\rangle^2  \nonumber \\ 
&&+\langle\tilde\psi(t)|\sigma_z\otimes\sigma_x\otimes\sigma_y\otimes\sigma_y-i\sigma_x\otimes\sigma_x\otimes\sigma_y\otimes\sigma_y|\tilde\psi(t)\rangle^2  \nonumber \\
&&+\langle\tilde\psi(t)|\sigma_z\otimes\sigma_z\otimes\sigma_y\otimes\sigma_y-i\sigma_x\otimes\sigma_z\otimes\sigma_y\otimes\sigma_y|\tilde\psi(t)\rangle^2 \ \ \big| , \nonumber \\
\end{eqnarray}
with  the corresponding measurement of  observables in the enlarged space.

{\it iii) N-qubit monotones.---}
In this case, the simplest entanglement monotone is $|\langle\psi|\sigma_y^{\otimes N}K|\psi\rangle|$ if $N$ is even (expression that is identically zero if $N$ is odd), and $|g^{\mu\nu}\langle\psi|\sigma_\mu\otimes\sigma_y^{\otimes N-1}K|\psi\rangle\langle\psi|\sigma_\nu\otimes\sigma_y^{\otimes N-1}K|\psi\rangle|$ if $N$ is odd. The first entanglement monotone needs $2$ measurements, while the second one needs $6$. This minimal requirements have to be compared with the $2^{2N}-1$ observables  required for full quantum tomography.
	
{\it iv) The mixed-state case.---}
Once we have defined  $E(|\psi\rangle)$ for the pure state case, we can extend our method to the mixed state case via the convex roof construction, see Eq. (\ref{roof}). Such a definition  is needed  because the possible  pure state decompositions of $\rho$ are infinite, and each of them brings a different value of $\sum_{i}p_iE(|\psi_i\rangle)$.  By considering its minimal value, as in Eq.~(\ref{roof}), we eliminate this ambiguity preserving the properties that define an entanglement monotone. To decide when $E(\rho)$ is zero  is called {\it separability problem}, and it is proven to be NP-hard for states close enough  to the border between the sets of entangled and separable states~\cite{Gurvits2003, Gharibian2010}. However, there exist useful classical algorithms\footnote{We stress that this is not equivalent to solving the separability problem, as the classical algorithm will always give a strict upper bound of $E(\rho)$.} able to find an estimation of $E(\rho)$ up to a finite error~\cite{Rothlisberger2009, Rothlisberger2012}.

Our approach for mixed states involves a hybrid quantum-classical algorithm, working well in cases in which $\rho$ is approximately a low-rank state. We restrict our study to the case of unitary evolutions acting on mixed-states, given that the inclusion of dissipative processes would require an independent development. Let us consider a state with rank $r$ and assume that the pure state decomposition solving Eq. \eqref{roof} has $c$ additional terms. That is, $k=r+c$, with $k$ being the number of terms in the optimal decomposition, while $c$ is assumed to be low. An algorithm that solves Eq.~\eqref{roof}, see for example~\cite{Rothlisberger2009, Rothlisberger2012}, evaluates at each step the quantity $\sum_{i=1}^kp_iE(|\psi_i\rangle)$ and, depending on the result, it changes $\{p_i,|\psi_i\rangle\}$ in order to find the minimum. Our method consists in inserting an embedded quantum simulation protocol in the evaluation of each $E(|\psi_i\rangle)$, which can be done with few measurements in the enlarged space. We gain in efficiency with respect to full tomography if  $k\cdot l\cdot m<2^{2N}-1$, where $l$ is the number of iterations of the algorithm and $m$ is the number of measurements to evaluate the specific entanglement monotone. We note that $m$ is a constant that can be low, depending on the choice of $E$, and, if $\rho$ is low rank, $k$ is a low constant too. With this approach, the performance of the computation of entanglement monotones, $E ( \rho)$, can be cast in two parts: while the quantum computation of $\sum_{i=1}^kp_iE(|\psi_i\rangle)$ can be efficiently implemented, the subsequent minimization remains a difficult~task.

In this section we have presented a paradigm for the efficient computation of a class of entanglement monotones requiring minimal experimental added resources. The proposed framework consists in the adequate embedding of  a quantum dynamics in the degrees of freedom of an enlarged-space quantum simulator.  In this manner, we have proposed novel concepts merging the fundamentals of quantum computation with those of quantum simulation. We believe that this novel embedding framework for quantum simulators will enhance the capabilities of one-to-one quantum simulations.


\subsection{How to implement an EQS with trapped ions} \label{sec:EQSIons}

In this section, we provide an experimental quantum simulation recipe to efficiently compute entanglement monotones involving antilinear operations, developing the EQS concepts for an ion-trap based quantum computer. The associated quantum algorithm is composed of two steps. First, we embed the $N$-qubit quantum dynamics of interest into a larger Hilbert space involving only one additional ion qubit and stroboscopic techniques. Second, we extract the corresponding entanglement monotones with a protocol requiring only the measurement of the additional single qubit. Finally, we show how to correct experimental imperfections induced by our quantum algorithm.

Trapped-ion systems are among the most promising technologies for quantum computation and quantum simulation protocols~\cite{Haffner2008}. In such systems,  fidelities of state preparation, two-qubit gate generation, and qubit detection, exceed values of $99\%$~\cite{Schindler2013}. With current technology, more than $140$ quantum gates including many body interactions  have been performed~\cite{Lanyon2011}. In this sense, the technology of trapped ions becomes a promising quantum platform to host the  embedded quantum algorithm described in section~\ref{sec:EQS}. In the following analysis, we will rely only on a set of operations involving local rotations and global entangling M{\o}lmer-S{\o}rensen (MS) gates~\cite{Molmer1999, Schindler2013}. In this sense, our method is not only applicable to trapped-ion systems. In general, it can be used in any platform where MS gates, or other long-range entangling interactions, as well as local rotations and qubit decoupling are available. Among such systems, we can mention cQED~\cite{Devoret2013a} where an implementation of MS gates has been recently proposed~\cite{Mezzacapo2014b}, or quantum photonics where MS interactions are available after a decomposition in controlled NOT gates~\cite{OBrien2009}.

The embedded dynamics of an interacting-qubit system is governed by the Schr\"odinger equation $i\hbar\partial_t \Psi = \tilde{H} \Psi$, where the Hamiltonian $\tilde{H}$ is $\tilde{H}=\sum_j \tilde{H}_j$ and each $\tilde{H}_j$ operator corresponds to a tensorial product of Pauli matrices. In this way, an embedded $N$-qubit dynamics can be implemented in two steps. First, we decompose the evolution operator using standard Trotter techniques~\cite{Lloyd1996a,Trotter1959},
\begin{equation}
\label{Trotterexpansion}
U_t = e^{-\frac{i}{\hbar}\sum_{j}\tilde{H}_jt}\approx \left(\Pi_je^{-i\tilde{H}_j t/n}\right)^n,
\end{equation}
where $n$ is the number of Trotter steps. Second, each exponential
$e^{-\frac{i}{\hbar}\tilde{H}_j t/n}$ can be implemented with a sequence of two M\o
lmer-S\o rensen gates~\cite{Molmer1999}  and a
single qubit rotation between them~\cite{Muller2011, Casanova2012}.  These
three quantum gates generate the evolution operator\footnote{For a detailed description see appendix~\ref{app:MolmerSorensen}.}
\begin{eqnarray}\label{exp}
e^{{i \varphi \sigma^z_1 \otimes \sigma^x_2 \otimes \sigma^x_3 \ldots \otimes \sigma^x_N}},
\end{eqnarray}
where $\varphi=gt$, $g$ being the coupling constant of the single qubit rotation~\cite{Muller2011}. In Eq.~(\ref{exp}), subsequent local rotations will produce any combination of Pauli matrices.
\begin{figure}[h]
\begin{center}
\vspace{0.5cm}
\hspace{-0.3cm}
\includegraphics[width= 0.8 \columnwidth]{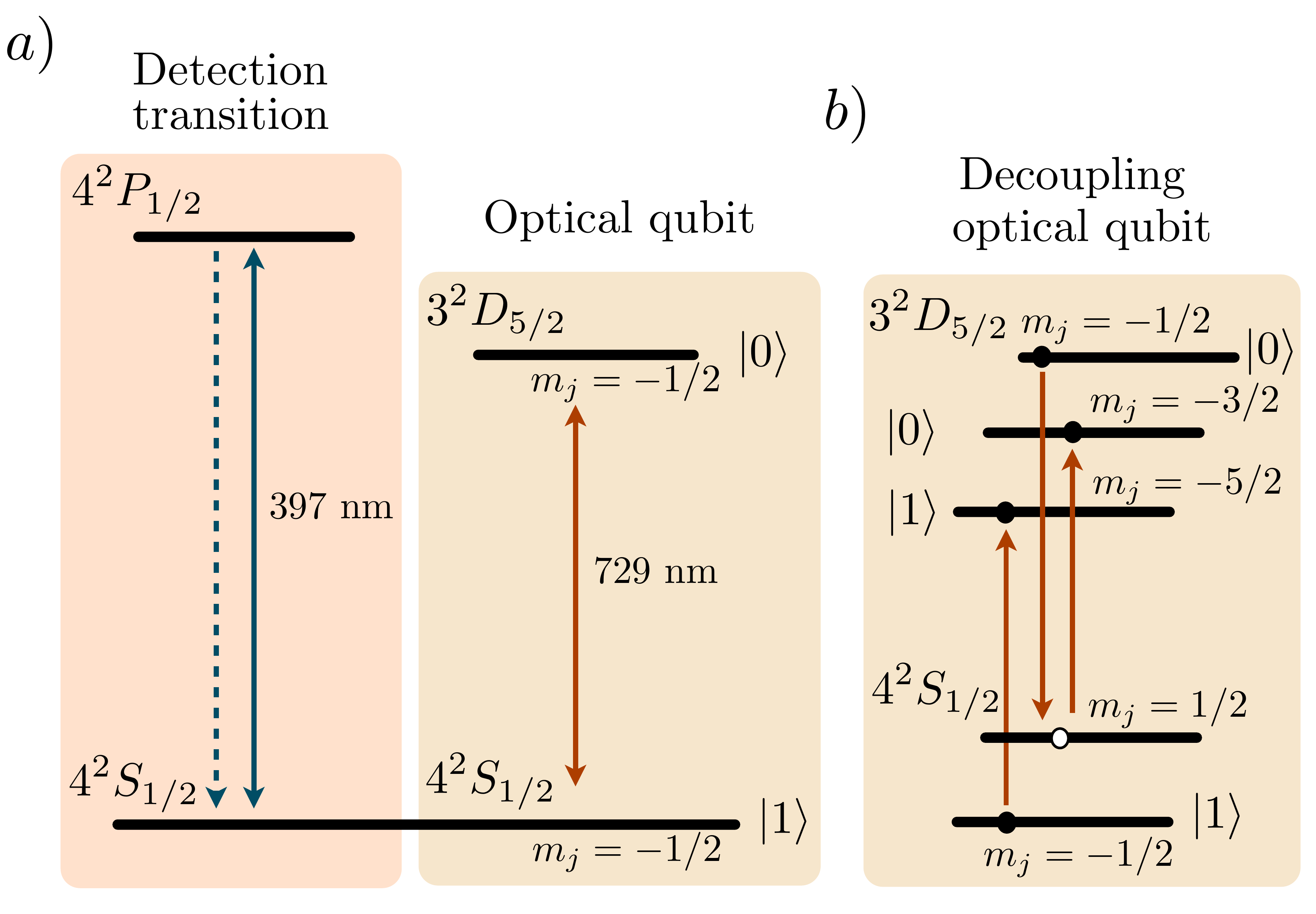}
\end{center}
\caption[Level scheme of $^{40}$Ca$^+$ ions.]{\footnotesize{{\bf Level scheme of $^{40}$Ca$^+$ ions.} a) The standard optical qubit is encoded in the $m_j=-1/2$ substates of the   $3D_{5/2}$ and $4S_{1/2}$ states. The measurement is performed via fluorescence detection exciting the  $4^2 S_{1/2} \leftrightarrow 4^{2} P_{1/2}$ transition.  b) The qubit can be  spectroscopically decoupled from the evolution by shelving the information in the
 $m_j=-3/2,-5/2$ substates of the $3D_{5/2}$ state.}} \label{figures levelscheme}
\end{figure}
As it is the case of quantum models involving Pauli operators, there exist different representations of the same dynamics. For example, the physically equivalent Ising Hamiltonians, $H_1=\omega_1 \sigma^x_1  + \omega_2  \sigma^x_2 + g \sigma^y_1 \otimes \sigma^y_2$ and $H_2 = \omega_1 \sigma^y_1 + \omega_2  \sigma^y_2 + g \sigma^x_1 \otimes \sigma^x_2$, are mapped onto the enlarged space as $\tilde{H}_1=-\omega_1 \sigma^y_0 \otimes \sigma^x_1 - \omega_2 \sigma^y_0 \otimes \sigma^x_2 - g \sigma^y_0 \otimes \sigma^y_1 \otimes \sigma^y_2$ and  $\tilde{H}_2= \omega_1  \sigma^y_1 + \omega_2 \sigma^y_2 -  g \sigma^y_0 \otimes \sigma^x_1 \otimes \sigma^x_2$. In principle, both Hamiltonians $\tilde{H}_1$ and $\tilde{H}_2$ can be implemented in trapped ions. However, while $\tilde{H}_1$ requires two- and three-body interactions, $\tilde{H}_2$ is implementable with a collective rotation applied to the ions $1$ and $2$ for the implementation of the free-energy terms, and MS gates for the interaction term. In this sense, $\tilde H_2$ requires less experimental resources for the implementation of the EQS dynamics. Therefore, a suitable choice of the system representation can considerably enhance  the performance of the simulator.

\subsubsection{Measurement protocol}  We want to measure correlations of the form 
\begin{eqnarray}\label{anti}
\langle\psi|\Theta|\psi^*\rangle &=& \langle\Psi|M^{\dag} \Theta M (\sigma^z \otimes \mathbb{I}_{2^N})|\Psi\rangle\nonumber\\
 &=& \langle\Psi|(\sigma^z-i\sigma^x) \otimes \Theta |\Psi\rangle,
\end{eqnarray}
with $\Theta$ a linear combination of tensorial products of Pauli matrices and identity operators. This information can be encoded in the expectation value $\langle \sigma_a^\alpha \rangle$ of one of the ions in the chain after performing two evolutions of the form of Eq.~(\ref{exp}).  Let us consider the  operators $U_1=e^{-i\varphi_1(\sigma^i_1 \otimes \sigma^j_2 \otimes \sigma^k_3 ...)}$ and $U_2=e^{-i\varphi_2(\sigma^o_1\otimes \sigma^p_2 \otimes \sigma^q_3 ... )}$ and choose the Pauli matrices $\sigma^i_1, \sigma^j_2, ...$ and $ \sigma^o_1, \sigma^p_2, ... $ such that $U_1$ and $U_2$ commute and both anticommute with the Pauli operator to be measured $\sigma^\alpha_a$. In this manner, we have that
\begin{eqnarray}
\langle \sigma^\alpha_a \rangle_{\varphi_1, \varphi_2 =\frac{\pi}{4}} &=& \langle U_1^{\dag}(\frac{\pi}{4})U_2^\dag(\frac{\pi}{4}) \ \sigma^\alpha_a \  U_1(\frac{\pi}{4}) U_2(\frac{\pi}{4}) \rangle\nonumber\\
&=& \langle \sigma^i_1 \sigma^o_1 \otimes \sigma^j_2\sigma^p_2  \otimes ... \otimes \sigma_a^\alpha \sigma_a^l \sigma_a^r... \rangle.
\end{eqnarray}
Then, a suitable choice of Pauli matrices will produce the desired correlation. Note that this protocol always results in a correlation of an odd number of Pauli matrices. In order to access a correlation of an even number of qubits, we have to measure a two-qubit correlation $\sigma^\alpha_a \otimes \sigma^\beta_b$ instead of just $\sigma_a^\alpha$. For the particular case of correlations of only Pauli matrices and no identity operators, evolution $U_2$ is not needed and no distinction between odd and even correlations has to be done. For instance, if one is interested in an even correlation like ${\sigma^y_1 \otimes \sigma^x_2 \otimes \sigma^x_3 \otimes \sigma^x_4 \otimes \mathbb{I}_5 \otimes ... \otimes \mathbb{I}_N}$, $N$ being the number of ions of the system, then one would have to measure observable
$\sigma_1^y \otimes \sigma_2^x$ after the  evolutions  $U_1=e^{-i(\sigma_1^x \otimes \sigma_2^y \otimes \sigma_3^y \otimes \sigma_4^y \otimes \sigma_5^y \otimes ... \otimes \sigma_j^y \otimes ...) \varphi}$ and ${U_2=e^{-i(\sigma_1^x \otimes \sigma_2^y \otimes \sigma_3^z \otimes \sigma_4^z \otimes \sigma_5^y \otimes ...\otimes \sigma_j^y \otimes ...)\varphi}}$. However, for the particular case of $N=4$ a single evolution $U_1=e^{-i(\sigma_1^x \otimes \sigma^x_2 \otimes \sigma^x_3
\otimes \sigma^x_4) \varphi}$ and subsequent measurement of $\langle \sigma_1^z \rangle$ is enough. Note that all the gates in the protocol, as they are of the type of Eq.~(\ref{exp}), are implementable with single qubit and MS gates.

\subsubsection{Examples} Consider the Ising Hamiltonian for two spins, ${ H=\hbar\omega_1 \sigma^y_1 + \omega_2 \sigma^y_2 + g \sigma^x_1 \otimes \sigma^x_2 }$ whose image in the enlarged space corresponds to $ \tilde H = \omega_1 \sigma^y_1 + \omega_2 \sigma^y_2 - g \sigma^y_0 \otimes \sigma^x_1 \otimes \sigma^x_1 $. The evolution operator associated to this Hamiltonian can be implemented using the Trotter method from Eq.~(\ref{Trotterexpansion}) with $(\tilde{H}_{1}, \tilde{H}_2, \tilde{H}_3) =(\omega_1 \sigma^y_1, \omega_2 \sigma^y_2, - g \sigma^y_0 \otimes \sigma^x_1 \otimes \sigma^x_2)$. While evolutions $e^{-\frac{i}{\hbar}\tilde{H}_1 t/n}$ and $e^{-\frac{i}{\hbar}\tilde{H}_2 t/n}$ can be implemented with single ion rotations, the evolution $e^{-\frac{i}{\hbar}\tilde{H}_3 t/n}$, which is of the kind described in Eq.~(\ref{exp}), is implemented with two MS gates and a single ion rotation. This simple case allows us to compute directly quantities such as the concurrence measuring $\langle \sigma_0^z \sigma_1^y \sigma_2^y \rangle$ and $\langle \sigma_0^x \sigma_1^y \sigma_2^y \rangle$. According to the measurement method introduced above, to access these correlations we first evolve the system under the gate $U=e^{-i(\sigma_0^y \otimes \sigma_1^y \otimes \sigma_2^y)\varphi}$ for a time such that $\varphi=\frac{\pi}{2}$, and then measure $\langle \sigma_0^x \rangle$ for the first correlation and $\langle \sigma_0^z \rangle$ for the second one.

\begin{figure}[h!]
\begin{center}
\vspace{0.5cm}
\hspace{-0.5cm}
\includegraphics [width= 0.9 \columnwidth]{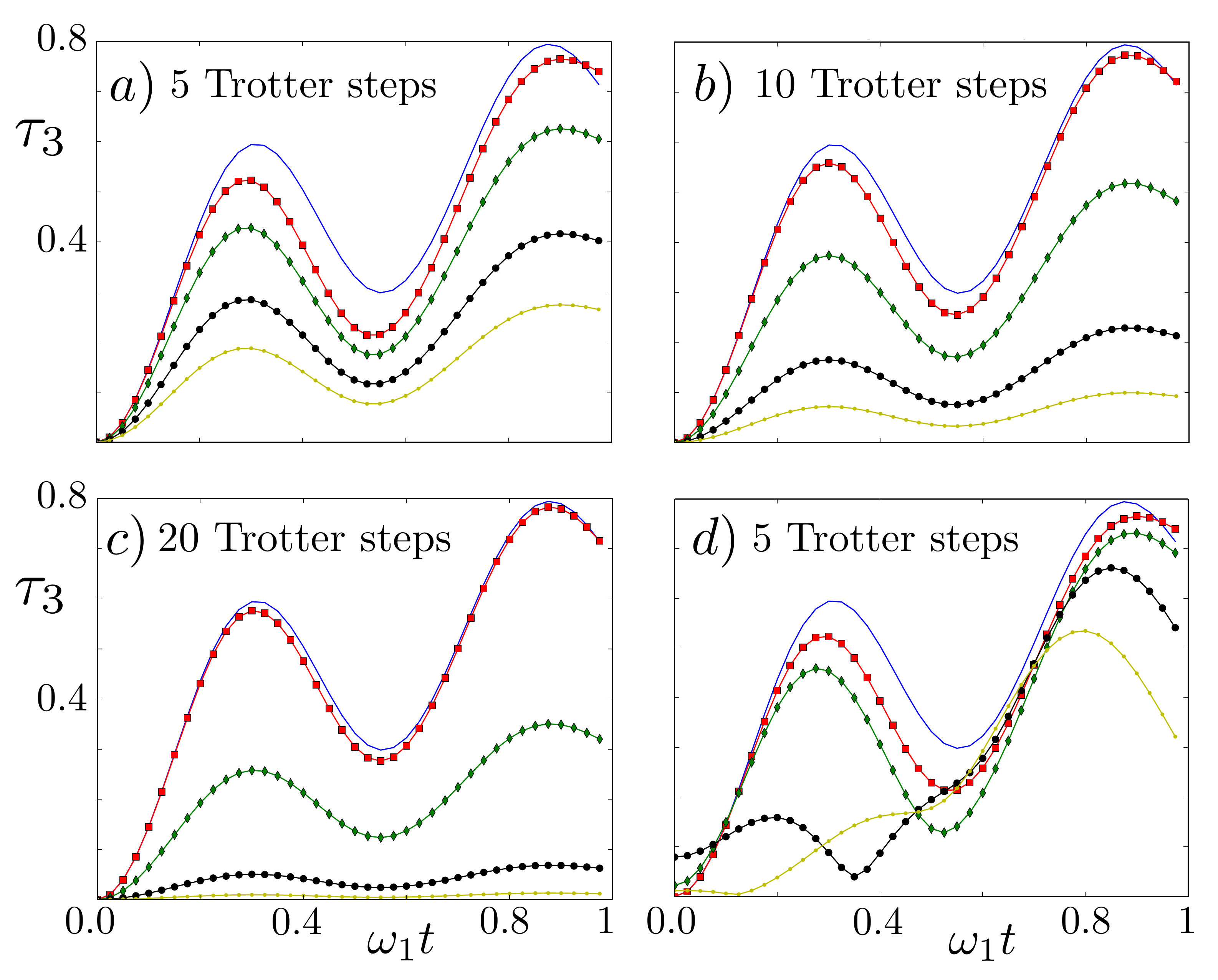}
\end{center}
\caption[Numerical simulation of the $3$-tangle evolving under Hamiltonian in Eq.~($\ref{EnlargedIsing3qubit}$) and assuming different error sources.]{\footnotesize{ {\bf Numerical simulation of the $3$-tangle evolving under Hamiltonian in Eq.~($\ref{EnlargedIsing3qubit}$) and assuming different error sources.} In all the plots the blue line shows the ideal evolution. In a), b), c) depolarizing noise is considered, with N=5,10 and 20 Trotter steps, respectively. Gate fidelities are $\epsilon =1, 0.99$, $0.97$, and $0.95$ marked by red rectangles, green diamonds, black circles and yellow dots, respectively. In d) crosstalk between ions is added with strength $\Delta_0 = 0, 0.01$, $0.03$, and $0.05$ marked by red rectangles, green diamonds, black circles and yellow dots, respectively. All the simulations in d) were performed with 5 Trotter steps. In all the plots we have used $\omega_1=\omega_2 = \omega_3=g/2=1$.}} \label{figures simulation}
\end{figure}

Based on the two-qubit example, one can think of implementing a three-qubit model as $ H_{\rm GHZ}= \omega_1 \sigma^y_1 + \omega_2 \sigma^y_2 + \omega_3 \sigma^y_3 + g \sigma^x_1 \otimes \sigma^x_2 \otimes \sigma^x_3 $, which in the enlarged space corresponds to
\begin{equation}
 \label{EnlargedIsing3qubit}
 \tilde H_{\rm GHZ} = \omega_1 \sigma^y_1 +  \omega_2 \sigma^y_2 + \omega_3 \sigma^y_3 - g \sigma^y_0 \otimes \sigma^x_1 \otimes \sigma^x_2 \otimes \sigma^x_3.
\end{equation}
This evolution results in GHZ kind states, which can be
readily detected using the 3-tangle $\tau_3$~\cite{Dur2000}. This is an entanglement monotone of the general class of  Eq.~(\ref{anti}) that can be computed in the enlarged space by measuring  $\big|-\langle\tilde\psi(t)|\sigma_z\otimes\mathbb{I}_2\otimes\sigma_y\otimes\sigma_y-i\sigma_x\otimes\mathbb{I}_2\otimes\sigma_y\otimes\sigma_y|\tilde\psi(t)\rangle^2 +\langle\tilde\psi(t)|\sigma_z\otimes\sigma_x\otimes\sigma_y\otimes\sigma_y-i\sigma_x\otimes\sigma_x\otimes\sigma_y\otimes\sigma_y|\tilde\psi(t)\rangle^2  + \langle \tilde \psi(t) | \sigma_z \otimes \sigma_z \otimes \sigma_y \otimes \sigma_y - i \sigma_x \otimes \sigma_z \otimes \sigma_y \otimes \sigma_y | \tilde \psi(t) \rangle^2 \big| $.
More complex Hamiltonians with interactions involving only three of the four particles can also be implemented.  In this case, the required entangling operations acting only on a part of the entire register can be realized with the aid of splitting the MS operations into smaller parts and
inserting refocusing pulses between them as shown in Ref.~\cite{Muller2011}. An alternative method is
to decouple the spectator ions from the laser light by shelving the quantum information into additional Zeeman substates of the ions as sketched in Fig.~(\ref{figures levelscheme}) for $^{40}$Ca$^+$ ions.
This procedure has been successfully demonstrated in Ref.~\cite{Schindler2013a}.
For systems composed of a larger number of qubits, for example $N> 10$, our method yields nontrivial results given that the standard computation of  entanglement monotones of the kind  $\langle \psi(t)| \Theta | \psi(t)^*\rangle$ requires the measurement of a number of observables that grows exponentially with $N$. For example, in the case of $\Theta = \sigma^y \otimes \sigma^y \otimes \dots \otimes \sigma^y$~\cite{Osterloh2005} our method requires the evaluation of $2$ observables while the standard procedure based on state tomography requires, in general, the measurement of $2^{2N}-1$ observables.

\subsubsection{Experimental considerations} A crucial issue of a quantum simulation algorithm is its susceptibility to experimental imperfections. In order to investigate the deviations with respect to the ideal case, the system dynamics needs to be described by completely positive maps instead of unitary dynamics. Such a map is defined by the process matrix $\chi$ acting on a density operator $\rho$ as follows: $\rho \rightarrow \sum_{i,j} \chi_{i,j} \sigma^i \rho \sigma^j,$ where $\sigma^i$ are the Pauli matrices spanning a basis of the operator space.  In complex algorithms, errors can be modeled by adding a depolarizing process  with a probability
$1 - \varepsilon$ to the ideal process $\chi^{id}$
\begin{equation}
\label{depnoisemodel}
\rho \rightarrow \varepsilon \sum_{i,j} \chi^{id}_{i,j} \sigma^i \rho \sigma^j + (1- \varepsilon) \frac{\mathbb{I}}{2^N}  .
\end{equation}
In order to perform a numerical simulation including this error model, it is required to decompose the quantum simulation into an implementable gate sequence. Numerical simulations of the Hamiltonian in Eq.~(\ref{EnlargedIsing3qubit}), including realistic values gate fidelity $\varepsilon = \{1, 0.99, 0.97, 0.95\}$ and for $\{5,10,20\}$ Trotter steps, are shown in Figs.~\ref{figures simulation} a), b), and c).
Naturally, this analysis is only valid if the noise in the real system is close to depolarizing noise. However, recent analysis of entangling operations indicates that this noise model is accurate~\cite{Navon2014, Schindler2013}.  According to Eq. (\ref{depnoisemodel}), after $n$ gate operations, we show that
\begin{equation}
\label{idealexpectation}
\langle O \rangle_{\mathcal{E}_{id}(\rho)}=\frac{\langle O\rangle_{\mathcal{E}(\rho)}}{\varepsilon^n}-\frac{1-\varepsilon^n}{\varepsilon^n}\text{Tr}\,(O),
\end{equation}
where $\langle O \rangle_{\mathcal{E}_{id}(\rho)}$ corresponds to the ideal expectation value in the absence of decoherence, and $\langle O\rangle_{\mathcal{E}(\rho)}$ is the observable measured in the
experiment. Given that we are working with observables composed of tensorial products of Pauli operators $\sigma^y_0\otimes\sigma_1^x ...$ with $\text{Tr}\,(O)=0$, Eq.~(\ref{idealexpectation}) will simplify to $\langle O \rangle_{\mathcal{E}_{id}(\rho)}=\frac{\langle O\rangle_{\mathcal{E}(\rho)}}{\varepsilon^n}$. In order to retrieve with uncertainty $k$ the expected value of an operator $O$, the experiment will need to be repeated $N_{emb}= \left( \frac{1}{k \epsilon^n} \right)^2$ times. Here, we have used $k\equiv\sigma^{\mathcal{E}(\rho)}_{\langle O \rangle}=\sigma^{\mathcal{E}(\rho)}_O/\sqrt{N}$ ( for large $N$), and that the relation between the standard deviations of the ideal and experimental expectation values is $\sigma^{\mathcal{E}(\rho)}_O=\sigma^{\mathcal{E}_{id}(\rho)}_O/\varepsilon^n$. If
we compare $N_{emb}$ with the required number of repetitions to measure the same entanglement monotone to the same accuracy $k$ in a one-to-one quantum simulator, $N_{oto} = 3^{N_{qubits}} \left(\frac{1}{k \delta^{n}}\right)^2$, we have
\begin{equation}
\label{tomoVSemb}
\frac{N_{emb}}{N_{oto}}=l\left(\frac{\delta}{ \sqrt{3}\varepsilon}\right)^{2N_{qubit}}.
\end{equation}
Here, $l$ is the number of observables corresponding to a given entanglement monotone in the enlarged space, and $\delta$ is the gate fidelity in the one-to-one approach. We are also asuming that full state tomography of $N_{qubits}$ qubits requires $3^{N_{qubit}}$ measurement settings for experiments exploiting single-qubit discrimination during the measurement process~\cite{Roos2004}. Additionally, we assume the one-to-one quantum simulator to work under the same error model but with $\delta$
fidelity per gate. Finally, we expect that the number of gates grows linearly with the number of qubits, that is $n\sim N_{qubit} $, which is a fair assumption for a nearest-neighbor interaction model. In general, we can assume that $\delta$ is always bigger than $\varepsilon$ as the embedding quantum simulator requires an additional qubit which naturally could increase the gate error rate. However, for realistic values of $\varepsilon$ and $\delta$, e.g. $\varepsilon=0.97$ and $\delta=0.98$ one can prove that $\frac{N_{emb}}{N_{oto}}\ll 1$. This condition is always fulfilled for large systems if $\frac{\delta}{ \sqrt{3}\varepsilon}<1$. The latter is a reasonable assumption given that in any quantum platform it is expected $\delta\approx\epsilon$ when the number of qubits grows, i.e. we expect the same gate fidelity for $N$ and $N+1$ qubit systems when $N$ is large.  Note that this analysis assumes that the same amount of Trotter steps is required for the embedded and the one-to-one simulator. This is a realistic assumption if one considers the relation between $H$ and $\tilde{H}$ in Eq.~(\ref{map}). A second type of imperfections are undesired unitary operations due to imperfect calibration of the applied gates or due to crosstalk between neighboring qubits. This crosstalk occurs when performing operations on a single ion due to imperfect single site illumination~\cite{Schindler2013}. Thus the operation $s_j^{z}(\theta)=\exp(-i \, \theta \, \sigma_{j}^{z}/ 2 )$ needs to be written as $s_j^{z}(\theta)=\exp(-i \, \sum_k \epsilon_{k,j} \, \theta \sigma_k^z / 2 )$ where the crosstalk is characterized by the matrix $\Delta$. For this analysis, we assume that the crosstalk affects only the nearest neighbors with strength $\Delta_0$ leading to a matrix $\Delta=\delta_{k,j} + \Delta_0 \, \delta_{k\pm1,j} $. In Fig.~\ref{figures simulation} d) simulations including crosstalk are shown. It can be seen, that simulations with increasing crosstalk show qualitatively different behavior of the 3-tangle, as in the simulation for $\Delta_0=0.05$ (yellow line) where the entire dynamics is distorted. This effect was not observed in the simulations including depolarizing noise and, therefore, we identify unitary crosstalk as a
critical error in the embedding quantum simulator. It should be noted that, if accurately characterized, the described crosstalk can be completely compensated experimentally~\cite{Schindler2013}.

In conclusion, in this section we have proposed an embedded quantum algorithm for trapped-ion systems to efficiently compute entanglement monotones for $N$ interacting qubits at any time of their evolution and without the need for full state tomography. It is noteworthy to mention that the performance of EQS would outperform similar efforts with one-to-one quantum simulators, where the case of 10 qubit may be considered already as intractable. Furthermore, we showed that the involved decoherence effects can be corrected if they are well characterized. We believe that EQS methods will prove useful as long as the Hilbert-space dimensions of quantum simulators grows in complexity in different quantum platforms.


\subsection{An experimental implementation of an EQS with linear optics} \label{sec:EQSPhotons}

In this section, we experimentally demonstrate an embedding quantum simulator, using it to efficiently measure two-qubit entanglement. Our EQS uses three polarization-encoded qubits in a circuit with two concatenated controlled-sign gates. The measurement of only 2 observables on the resulting tripartite state gives rise to the efficient measurement of bipartite concurrence, which would otherwise need 15 observables.
\\

\subsubsection{The protocol}~We consider the simulation of two-qubit entangling dynamics governed by the Hamiltonian $H{=}{-} g \sigma_z\otimes\sigma_z$, where $\sigma_z{=} | 0 \rangle \langle 0 | - | 1 \rangle \langle 1 |$ is the z-Pauli matrix written in  the computational basis, $\{ | 0 \rangle, | 1 \rangle \}$, and $g$ is a constant with units of frequency. For simplicity, we let $\hbar{=}1$. Under this Hamiltonian, the concurrence~\cite{Wootters1998} of an evolving pure state $| \psi(t) \rangle$ is calculated as $\mathcal{C}{=}\left| \langle \psi(t) | \sigma_y\otimes\sigma_yK | \psi(t) \rangle \right|$, where $K$ is the complex conjugate operator defined as $K| \psi(t) \rangle{=}| \psi(t)^* \rangle$.

\begin{figure}[h]
\centering
\includegraphics[width=0.75 \columnwidth]{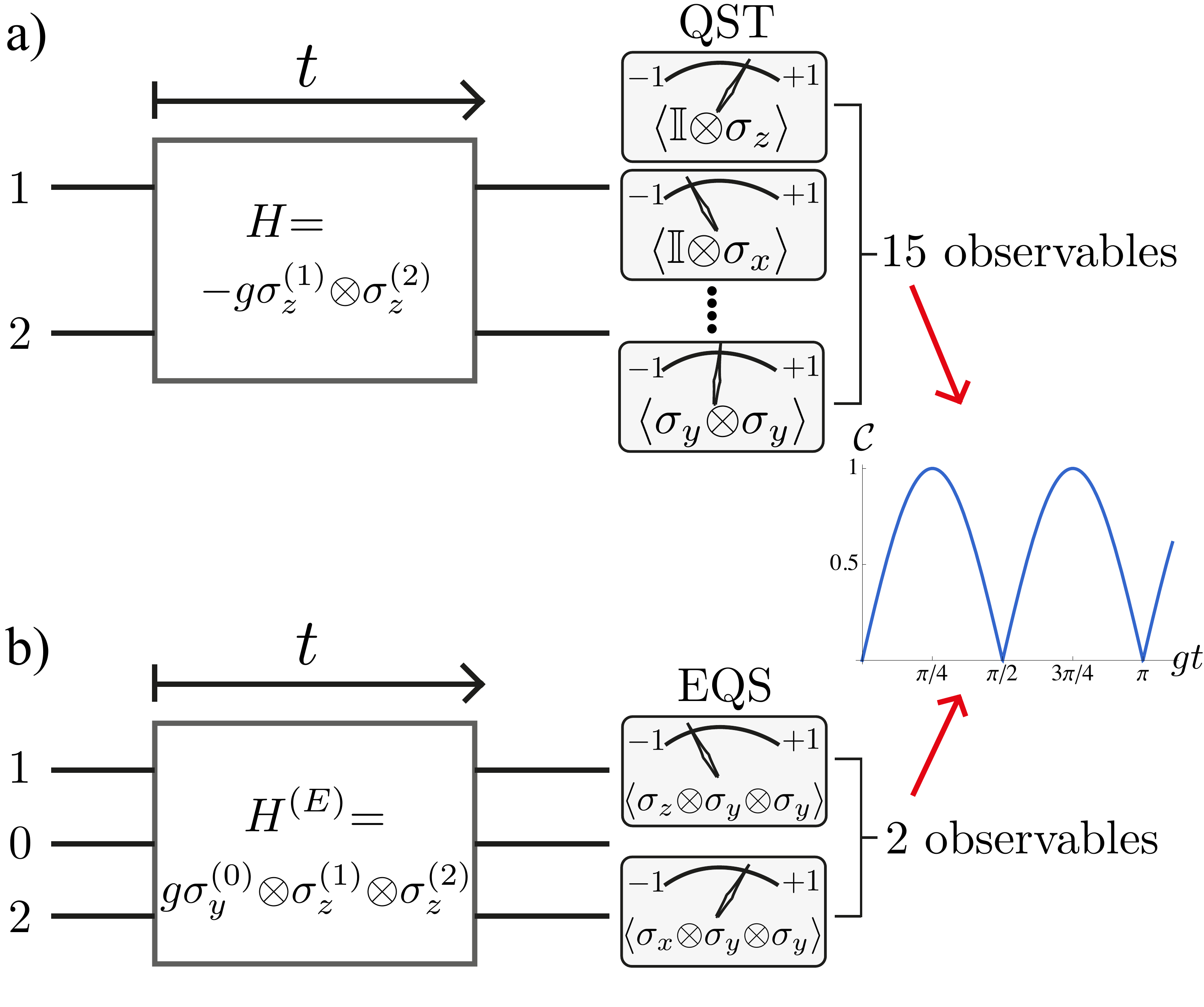}
\caption[Strategy to extract concurrence in an EQS versus a one-to-one quantum simulator.]{\footnotesize{{\bf Strategy to extract concurrence in an EQS versus a one-to-one quantum simulator.} (a) Qubits $1$ and $2$ evolve via an entangling Hamiltonian $H$ during a time interval $t$, at which point QST is performed via the measurement of $15$ observables to extract the amount of evolving concurrence. (b) An efficient alternative corresponds to adding one extra ancilla, qubit $0$, and having the enlarged system---the EQS---evolve via $H^{(E)}$. Only two observables are now required to reproduce measurements of concurrence of the system under simulation.}}
\label{Fig:1}
\end{figure}

Notice here the explicit dependance of $\mathcal C$ upon the unphysical transformation $K$. We now consider the dynamics of the initial state $| \psi(0) \rangle{=}(| 0 \rangle{+}| 1 \rangle){\otimes}(| 0 \rangle{+}| 1 \rangle)/2$. Under these conditions one can calculate the resulting concurrence at any time $t$ as
\begin{equation}
\label{eq:csim}
\mathcal{C}=|\sin(2gt)| .
\end{equation}
The target evolution, $e^{-iHt} | \psi(0) \rangle$, can be embedded in a $3$-qubit simulator. Given the state of interest $| \psi \rangle$, the transformation
\begin{equation}
\label{eq:toSIM}
| \Psi \rangle=| 0 \rangle\otimes\text{Re}| \psi \rangle+| 1 \rangle\otimes\text{Im}| \psi \rangle,
\end{equation}
gives rise to a real-valued $3$-qubit state $| \Psi \rangle$ in the corresponding embedding quantum simulator. The decoding map is, accordingly, $| \psi \rangle{=}\langle 0 | \Psi \rangle{+}i\langle 1 | \Psi \rangle$. The physical unitary gate $\sigma_z{\otimes}\mathbb{I}_4$ transforms the simulator state into $\sigma_z{\otimes}\mathbb{I}_4 | \Psi \rangle {=}| 0 \rangle{\otimes}\text{Re}| \psi \rangle{-}| 1 \rangle{\otimes}\text{Im}| \psi \rangle$, which after the decoding  becomes $\langle 0 | \Psi \rangle{-}i\langle 1 | \Psi \rangle{=}\text{Re} | \psi \rangle {-}{i}\text{Im} | \psi \rangle {=}| \psi^* \rangle$. Therefore, the action of the complex conjugate operator $K$ corresponds to the single qubit rotation $\sigma_z{\otimes}\mathbb{I}_4$. Now, following the same encoding rules: $\langle \psi |OK| \psi \rangle{=}\langle \Psi |(\sigma_z{-}i\sigma_x){\otimes}O| \Psi \rangle$, with $O$ an observable in the simulation. In the case of  $O{=}\sigma_y{\otimes}\sigma_y$, we obtain
\begin{equation}
\label{eq:C}
\mathcal{C}=|\langle\sigma_z\otimes\sigma_y\otimes\sigma_y\rangle-i\langle\sigma_x\otimes\sigma_y\otimes\sigma_y\rangle|,
\end{equation}
which relates the simulated concurrence to the expectation values of two nonlocal operators in the embedding quantum simulator. Regarding the dynamics, it can be shown that the Hamiltonian $H^{(E)}$ that governs the evolution in the simulator is $H^{(E)}{=}{-}\sigma_y{\otimes}(\text{Re}H){+}i\mathbb{I}_2{\otimes}(\text{Im}H)$. Accordingly, in our case, it will be given by $H^{(E)}{=}g \sigma_y{\otimes}\sigma_z{\otimes}\sigma_z$.

Our initial state under simulation is $| \psi(0) \rangle{=}(| 0 \rangle{+}| 1 \rangle){\otimes}(| 0 \rangle{+}| 1 \rangle)/2$, which requires, see Eq.~(\ref{eq:toSIM}), the initialization of the simulator in $|{\Psi}(0)\rangle{=}|0\rangle{\otimes}\left( |0\rangle+|1\rangle \right){\otimes}\left( |0\rangle+|1\rangle \right){/}2$. Under these conditions, the relevant simulator observables, see Eq.~(\ref{eq:C}), read $\langle\sigma_x{\otimes}\sigma_y{\otimes}\sigma_y\rangle{=}\sin{(2gt)}$ and $\langle\sigma_z{\otimes}\sigma_y{\otimes}\sigma_y\rangle{=}0$, from which the concurrence of Eq.~(\ref{eq:csim}) will be extracted. Therefore, our recipe, depicted in Fig.~\ref{Fig:1}, allows the encoding and efficient measurement of two-qubit concurrence dynamics.

\begin{figure}[h!]
\centering
\includegraphics[width=  \columnwidth]{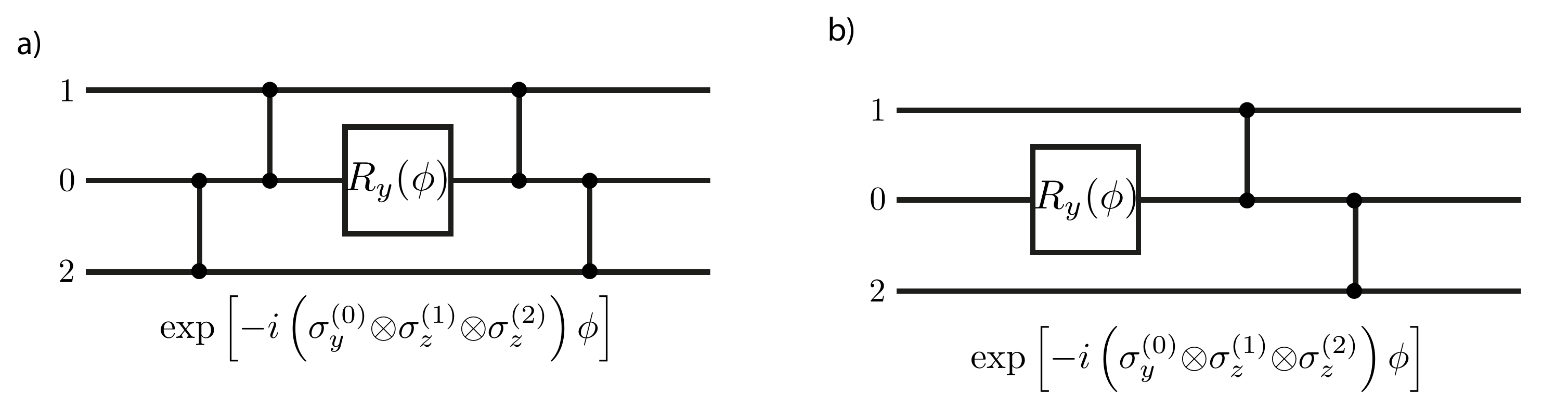}
\caption[Quantum circuit for an EQS.]{\footnotesize{{\bf Quantum circuit for an EQS.} (a) $4$ controlled-sign gates and one local rotation $R_{y}(\phi)$ implement the evolution operator $U({t}){=}\text{exp}\left(-i g \sigma_y^{(0)}{\otimes}\sigma_z^{(1)}{\otimes}\sigma_z^{(2)}t\right)$, with $\phi=gt$. (b) A reduced circuit employing only two controlled-sign gates reproduces the desired three-qubit dynamics for inputs with the ancillary qubit in $|0\rangle$.}}
\label{Fig:2}
\end{figure}

To construct the described three-qubit simulator dynamics, it can be shown\footnote{See appendix~\ref{app:EQSPCircuit} for a detailed explanation.} that a quantum circuit consisting of $4$ controlled-sign gates and one local rotation $R_y(\phi){=}\text{exp}\left({-i}\sigma_{y} \phi \right)$, as depicted in Fig.~\ref{Fig:2}(a), implements the evolution operator $U(t){=}\text{exp}\left[-ig \left(\sigma_y{\otimes}\sigma_z{\otimes}\sigma_z\right)t\right]$, reproducing the desired dynamics, with $\phi=gt$. This quantum circuit can be further reduced if we consider only inputs with the ancillary qubit in state $| 0 \rangle$, in which case, only two controlled-sign gates reproduce the same evolution, see Fig.~\ref{Fig:2}~(b). This reduced subspace of initial states corresponds to simulated input states of only real components. 

\subsubsection{Experimental implementation}
\begin{figure}[h!]
\centering
\includegraphics[width=  \columnwidth]{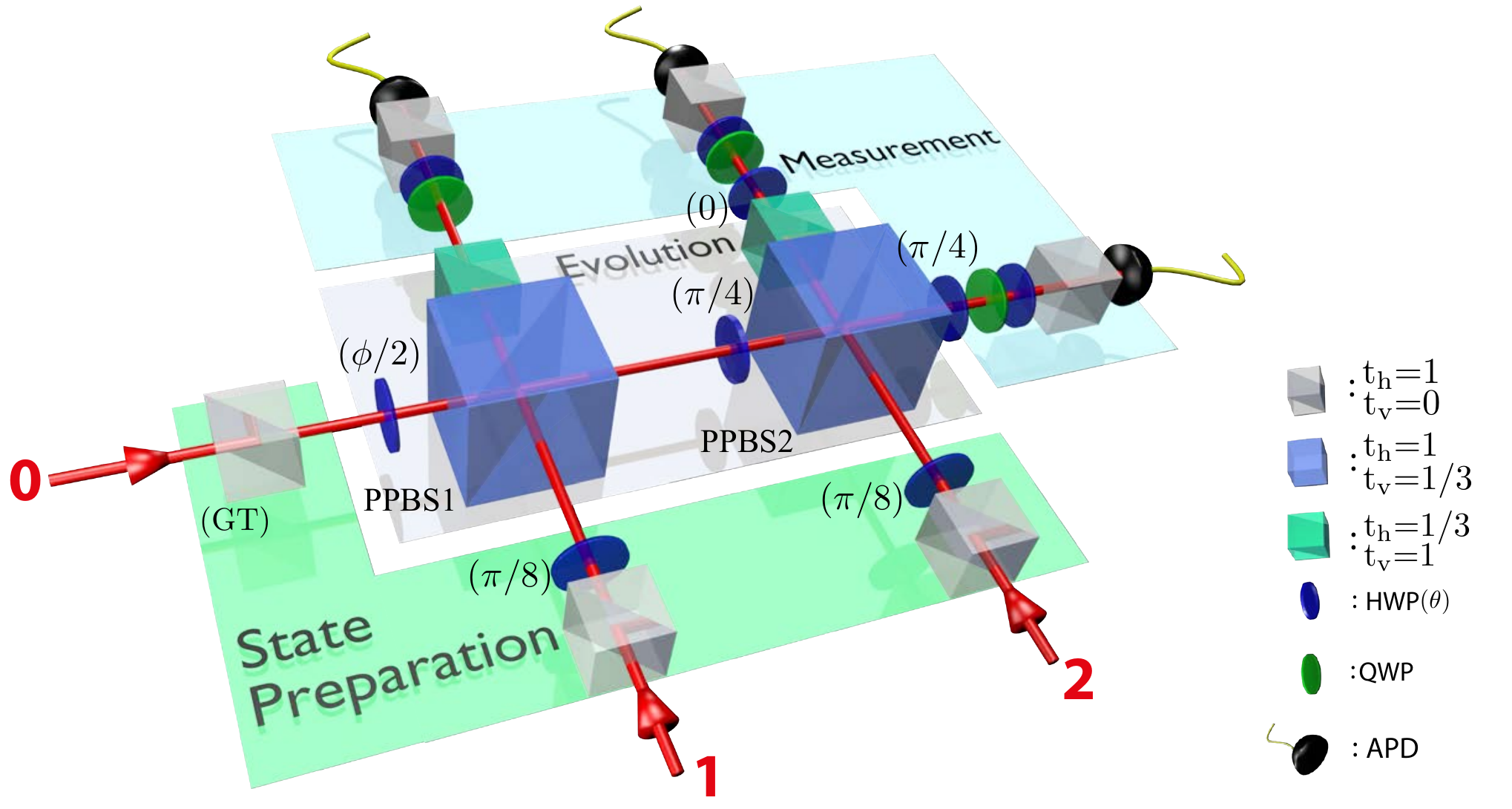}
\caption[Experimental setup for the implementation of a photonic EQS.]{\footnotesize{{\bf Experimental setup for the implementation of a photonic EQS.} Three single-photons with wavelength centered at 820~nm are injected via single-mode fibers into spatial modes $0$, $1$ and $2$. Glan-Taylor (GT's) prisms, with transmittance $t_\text{h}{=}1$ ($t_\text{v}{=}0$) for horizontal (vertical) polarization, and half-wave plates (HWP's) are employed to initialize the state. Controlled two-qubit operations are performed based on two-photon quantum interference at partially polarizing beam-splitters (PPBS's). Projective  measurements are carried out with a combination of half-wave plates, quarter-wave plates (QWP's) and Glan-Taylor prisms. The photons are collected via single-mode fibers and detected by avalanche photodiodes (APD's).}}
\label{Fig:3}
\end{figure}

~We encode a three-qubit state in the polarization of 3 single-photons. The logical basis is encoded according to $| h \rangle{\equiv}| 0 \rangle,| v \rangle{\equiv} | 1 \rangle$, where $| h \rangle$ and $| v \rangle$ denote horizontal and vertical polarization, respectively. The simulator is initialized in the state $|{\Psi}(0)\rangle{=}|h\rangle^{(0)}{\otimes}\left(|h\rangle^{(1)}+|v\rangle^{(1)}\right){\otimes}\left( |h\rangle^{(2)}+|v\rangle^{(2)}\right){/}2$ of qubits $0$, $1$ and $2$, and evolves via the optical circuit in Fig.~\ref{Fig:2}~(b). Figure~\ref{Fig:3} is the physical implementation of Fig.~\ref{Fig:2}~(b), where the dimensionless parameter $\phi{=}gt$ is controlled by the angle $\phi{/}2$ of one half-wave plate. The two concatenated controlled-sign gates are implemented by probabilistic gates based on two-photon quantum interference~\cite{Ralph2001,OBrien2003,Ralph2004}, see appendix~\ref{app:CsignGates}.

In order
to reconstruct the two three-qubit observables in Eq.~(\ref{eq:C}), one needs to collect $8$ possible tripartite correlations of the observable eigenstates. For instance, the observable $\langle\sigma_x{\otimes}\sigma_y{\otimes}\sigma_y\rangle$ is obtained from measuring the $8$ projection combinations of the $\{| d \rangle,| a \rangle \} \otimes \{| r \rangle,| l \rangle \} \otimes \{ | r \rangle, | l \rangle \}$ states, where $| d \rangle{=}(| h \rangle{+}| v \rangle)/\sqrt2$, $| r \rangle{=}(| h \rangle{+}i| v \rangle)/\sqrt2$, and $| a \rangle$ and $| l \rangle$ are their orthogonal states, respectively. To implement these polarization projections, we employed Glan-Taylor prisms due to their high extinction ratio. However, only their transmission mode is available, which required each of the $8$ different projection settings separately, extending our data-measuring time. The latter can be avoided by simultaneously registering both outputs of a projective measurement, such as at the two output ports of a polarizing beam splitter, allowing the simultaneous recording of all $8$ possible projection settings. Thus, an immediate reconstruction of each observable is possible.

Our source of single-photons consists of four-photon events collected from the forward and backward pair emission in spontaneous parametric down-conversion in a \emph{beta}-barium borate (BBO) crystal pumped by a $76$~MHz frequency-doubled mode-locked femtosecond Ti:Sapphire laser. One of the four photons is sent directly to an APD to act as a trigger, while the other $3$ photons are used in the protocol. This kind of sources are known to suffer from undesired higher-order photon events that are ultimately responsible of a non-trivial gate performance degradation~\cite{Barbieri2009,Broome2011,Pan2012}, although they can be reduced by decreasing the laser pump power. However, given the probabilistic nature and low efficiency of down-conversion processes, multi-photon experiments are importantly limited by low count-rates\footnote{See appendix~\ref{app:CountRates}.}. Therefore, increasing the simulation performance quality by lowering the pump requires much longer integration times to accumulate meaningful statistics, which ultimately limits the number of measured experimental settings.

As a result of these higher-order noise terms, a simple model can be considered to account for non-perfect input states. The experimental input $n$-qubit state $\rho_{\rm exp}$ can be regarded as consisting of the ideal state $\rho_{\rm id}$ with certain probability $\varepsilon$, and a white-noise contribution with probability $1{-}\varepsilon$, i.e.~$\rho_{\rm exp}{=}\varepsilon \rho_{\rm id}{+}(1{-}\varepsilon)\mathbb{I}_{2^n}{/}{2^n}$.

\begin{figure}[h!]
\centering
\includegraphics[width=\columnwidth]{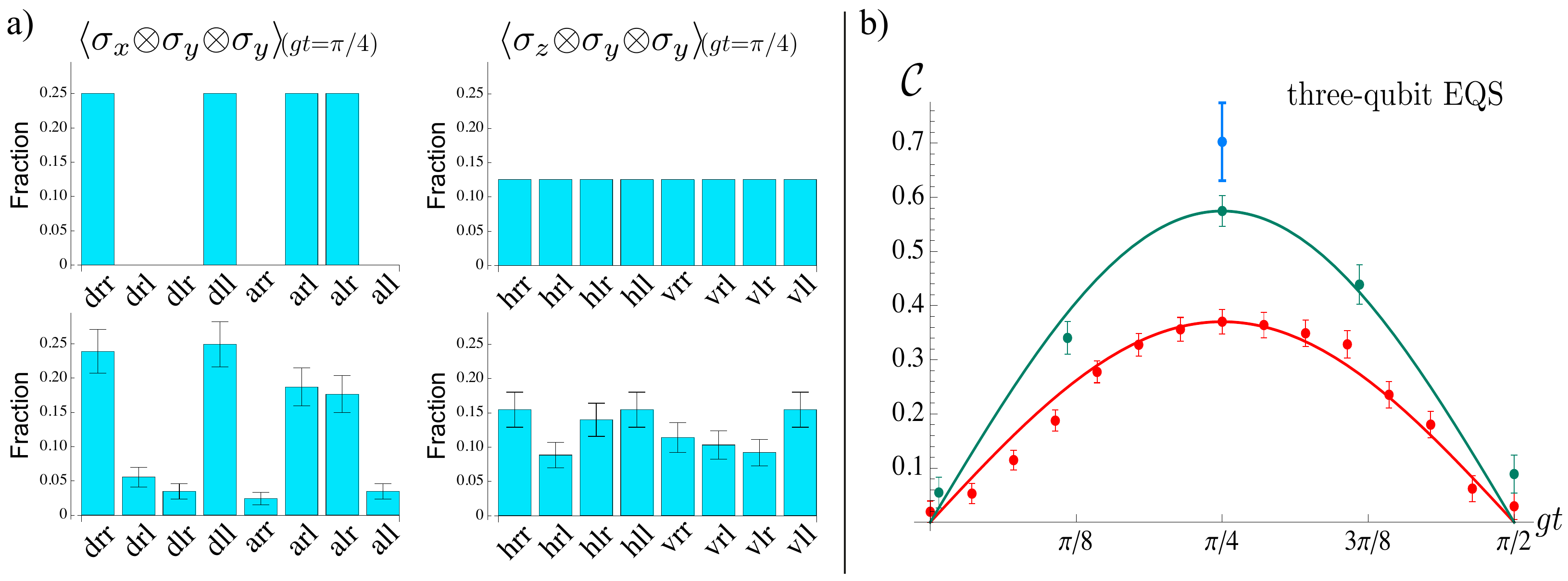} 
\caption[Measurement of the concurrence with a photonic EQS.]{\footnotesize{{\bf Measurement of the concurrence with a photonic EQS.} (a) Theoretical predictions (top) and experimentally measured (bottom) fractions involved in reconstructing $\langle\sigma_x{\otimes}\sigma_y{\otimes}\sigma_y\rangle$ (left) and $\langle\sigma_z{\otimes}\sigma_y{\otimes}\sigma_y\rangle$ (right), taken at $gt{=}\pi{/}4$ for a $10\%$ pump. (b) Extracted simulated concurrence within one evolution cycle, taken at $10\%$ (blue), $30\%$ (green), and $100\%$ (red) pump powers. Curves represent $\mathcal{C}{=}\mathcal{C}_{\text{pp}}|\sin(2gt)|$, where $\mathcal{C}_{\text{pp}}$ is the maximum concurrence extracted for a given pump power (pp): $\mathcal{C}_{10\%}{=}0.70\pm0.07$, $\mathcal{C}_{30\%}{=}0.57\pm0.03$ and $\mathcal{C}_{100\%}{=}0.37\pm0.02$. Errors are estimated from propagated Poissonian statistics. The low count-rates of the protocol limit the number of measured experimental settings, hence only one data point could be reconstructed at $10\%$ pump.}}
\label{Fig:4}
\end{figure}

Since the simulated concurrence is expressed in terms of tensorial products of Pauli matrices, the experimentally simulated concurrence becomes $\mathcal{C}_{\rm exp}{=}\varepsilon|\sin(2gt)|$.

In Fig.~\ref{Fig:4}, we show our main experimental results from our photonic embedding quantum simulator for one cycle of concurrence evolution taken at different pump powers:~$60$~mW, $180$~mW, and $600$~mW---referred as to $10\%$, $30\%$, and $100\%$ pump, respectively. Figure~\ref{Fig:4}~(a) shows theoretical predictions (for ideal pure-state inputs) and measured fractions of the different projections involved in reconstructing $\langle\sigma_x{\otimes}\sigma_y{\otimes}\sigma_y\rangle$ and $\langle\sigma_z{\otimes}\sigma_y{\otimes}\sigma_y\rangle$ for $10\%$ pump at $gt{=}\pi/4$. From measuring these two observables, see Eq.~(\ref{eq:C}), we construct the simulated concurrence produced by our EQS, shown in Fig.~\ref{Fig:4}~(b). We observe a good behavior of the simulated concurrence, which preserves the theoretically predicted sinusoidal form. The overall attenuation of the curve is in agreement with the proposed model of imperfect initial states. Together with the unwanted higher-order terms, we attribute the observed degradation to remaining spectral mismatch between photons created by independent down-conversion events and injected to inputs $0$ and $2$ of Fig.~\ref{Fig:3}---at which outputs $2$~nm band-pass filters with similar but not identical spectra were used.

We compare our measurement of concurrence via our simulator with an explicit measurement from state tomography. In the latter we inject one down-converted pair into modes $0$ and $1$ of Fig.~\ref{Fig:3}. For any value of $t$, set by the wave-plate angle $\phi$, this evolving state has the same amount of concurrence as the one from our simulation, they are equivalent in the sense that one is related to the other at most by local unitaries, which could be seen as incorporated in either the state preparation or within the tomography settings.

\begin{figure}[h!]
\centering
\includegraphics[width= 0.8 \columnwidth]{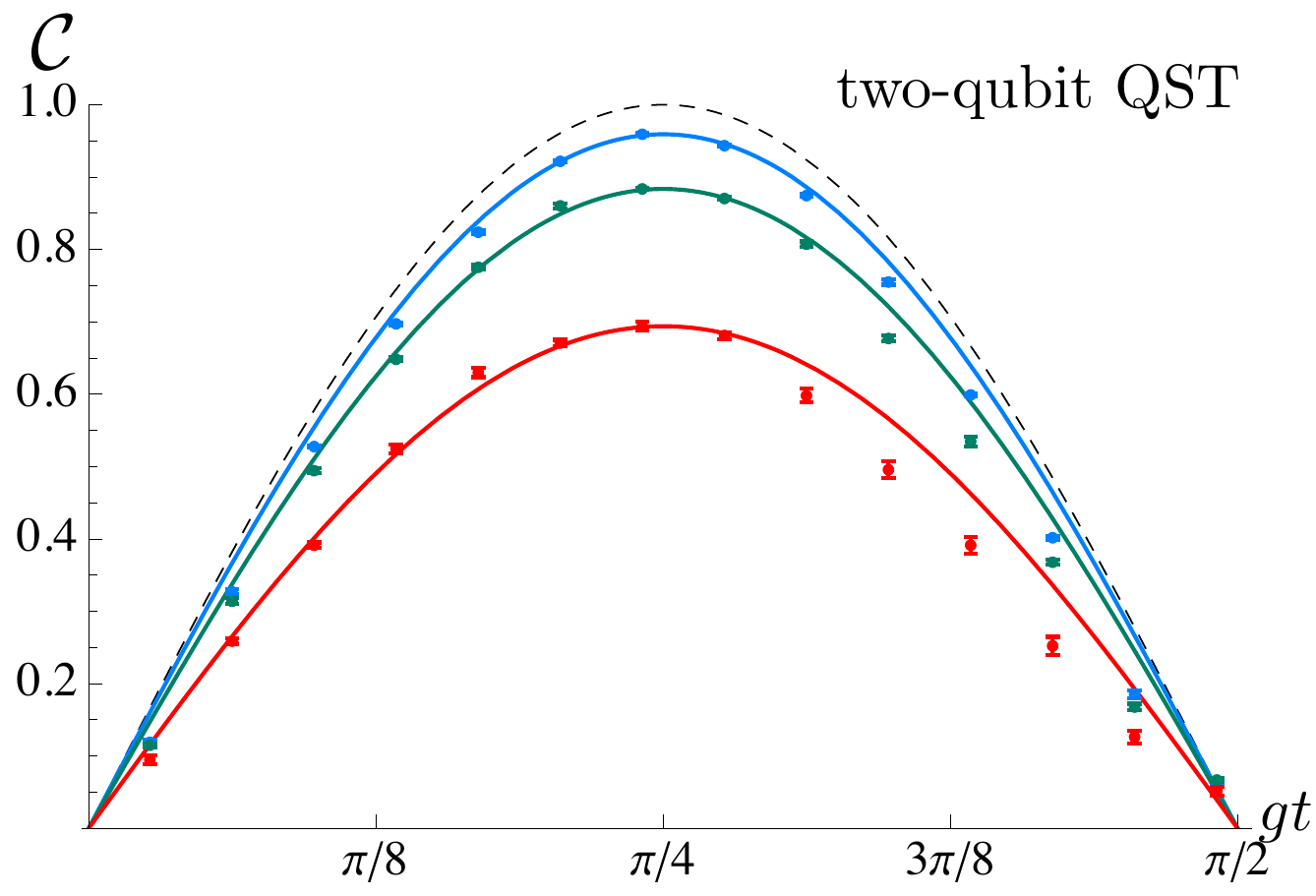}
\caption[Measurement of the concurrence in a on-to-one photonic quantum simulator.]{\footnotesize{{\bf Measurement of the concurrence in a on-to-one photonic quantum simulator.} Concurrence measured via two-qubit QST on the explicit two-photon evolution, taken at $10\%$ (blue), $30\%$ (green), and $100\%$ (red) pump powers. The corresponding curves indicate $\mathcal{C}{=}\mathcal{C}_{\text{pp}}|\sin(2gt)|$, with $\mathcal{C}_{\text{pp}}$ the maximum extracted concurrence for a given pump power (pp): $\mathcal{C}_{10\%}{=}0.959\pm0.002$, $\mathcal{C}_{30\%}{=}0.884\pm0.002$, and $\mathcal{C}_{100\%}{=}0.694\pm0.006$. Errors are estimated from Monte-Carlo simulations of Poissonian counting fluctuations.}} 
\label{Fig:5}
\end{figure}

Figure~\ref{Fig:5} shows our experimental results for the described two-photon protocol. We extracted the concurrence of the evolving two-qubit state from overcomplete measurements in quantum state tomography~\cite{James2001}.

A maximum concurrence value of $\mathcal{C}{=}1$ is predicted in the ideal case of perfect pure-state inputs. Experimentally, we measured maximum values of concurrence of $\mathcal{C}_{10\%}{=}0.959\pm0.002$, $\mathcal{C}_{30\%}{=}0.884\pm0.002$ and $\mathcal{C}_{100\%}{=}0.694\pm0.006$, for the three different pump powers, respectively. For the purpose of comparing this two-photon protocol with our embedding quantum simulator, only results for the above mentioned powers are shown. However, we performed an additional two-photon protocol run at an even lower pump power of $30$~mW ($5\%$ pump), and extracted a maximum concurrence of $\mathcal{C}_{5\%}{=}0.979\pm0.001$. A clear and pronounced decline on the extracted concurrence at higher powers is also observed in this protocol. However, a condition closer to the ideal one is reached. This observed pump power behavior and the high amount of measured concurrence suggest a high-quality gate performance, and that higher-order terms---larger for higher pump powers---are indeed the main cause of performance degradation.

While only mixed states are always involved in experiments, different degrees of mixtures are present in the $3$- and $2$-qubit protocols, resulting in different extracted concurrence from both methods. An inspection of the pump-dependence\footnote{See appendix \ref{app:PumpDep}.} reveals that both methods decrease similarly with pump power and are close to performance saturation at the $10\%$ pump level. This indicates that in the limit of low higher-order emission our $3$-qubit simulator is bounded to the observed performance. Temporal overlap between the $3$ photons was carefully matched. Therefore, we attribute the remaining discrepancy to spectral mismatch between photons originated from independent down-conversion events. This disagreement can in principle be reduced via error correction~\cite{OBrien2005,Yao2012a} and entanglement purification~\cite{Pan2003} schemes with linear optics.

We have shown experimentally that entanglement measurements in a quantum system can be efficiently done in a higher-dimensional embedding quantum simulator. The manipulation of larger Hilbert spaces for simplifying the processing of quantum information has been previously considered~\cite{Lanyon2009}. However, in the present scenario, this advantage in computing concurrence originates from higher-order quantum correlations, as it is the case of the appearance of tripartite entanglement~\cite{Lu2007,Lanyon2007}.

The efficient behavior of embedding quantum simulators resides in reducing an exponentially-growing number of observables to only a handful of them for the extraction of entanglement monotones. We note that in this non-scalable photonic platform the addition of one ancillary qubit and one entangling gate results in count rates orders of magnitude lower as compared to direct state tomography on the $2$-qubit dynamics. This means that in practice absolute integration times favor the direct $2$-qubit implementation. However, this introduced limitation escapes from the purposes of the embedding protocol and instead belongs to the specific technology employed in its current state-of-the-art performance.

This section contains the first proof-of-principle experiment showing the efficient behaviour of embedding quantum simulators for the processing of quantum information and extraction of entanglement monotones. A parallel and independent experiment was performed in Hefei following similar ideas~\cite{Chen2016}. These experiments validate an architecture-independent paradigm that, when implemented in a scalable platform, as explained in section~\ref{sec:EQSIons}, would overcome a major obstacle in the characterization of large quantum systems. The relevance of these techniques will thus become patent as quantum simulators grow in size and currently standard approaches like full tomography become utterly unfeasible. We believe that these results pave the way to the efficient measurement of entanglement in any quantum platform via embedding quantum simulators.

%% file: chap/chapter4.tex
\lettrine[lines=2, findent=3pt,nindent=0pt]{T}{he} quantum Rabi model (QRM) describes the most fundamental light-matter interaction involving quantized light and quantized matter.  It is different from the Rabi model~\cite{Rabi1936}, where light is treated classically. In general, the QRM is used to describe the dipolar coupling between a two-level system and a bosonic field mode. Although it plays a central role in the dynamics of a collection of quantum optics and condensed matter systems~\cite{Arakawa2015}, such as cavity quantum electrodynamics (CQED), quantum dots, trapped ions, or circuit QED (cQED), an analytical solution of the QRM in all coupling regimes has only recently been proposed~\cite{Braak2011,Solano2011}. In any case, standard experiments naturally happen in the realm of the Jaynes-Cummings (JC) model~\cite{Jaynes1963}, a solvable system where the rotating-wave approximation (RWA) is applied to the QRM. Typically, the RWA is valid when the ratio between the coupling strength and the mode frequency is small. In this sense, the JC model is able to correctly describe most observed effects where an effective two-level system couples to a bosonic mode, be it in more natural systems as CQED~\cite{Miller2005,Walther2006,Haroche2006}, or in simulated versions as trapped ions~\cite{Leibfried2003,Haffner2008} and cQED~\cite{Wallraff2004,Devoret2013a}. However, when the interaction grows in strength until the ultrastrong coupling (USC)~\cite{Ciuti2005,Niemczyk2010,Forn-Diaz2010} and deep strong coupling (DSC)~\cite{Casanova2010,DeLiberato2014} regimes, the RWA is no longer valid. While the USC regime happens when the coupling strength is some tenths of the mode frequency, the DSC regime requires this ratio to be larger than unity. In such extreme cases, the intriguing predictions of the full-fledged QRM emerge with less intuitive results.

Recently, several systems have been able to experimentally reach the USC regime of the QRM, although always closer to conditions where perturbative methods can be applied or dissipation has to be added. Accordingly, we can mention the case of circuit QED~\cite{Niemczyk2010, Forn-Diaz2010}, semiconductor systems coupled to metallic microcavities~\cite{Todorov2009, Askenazi2014, Kena-Cohen2013}, split-ring resonators connected to cyclotron transitions~\cite{Scalari2012}, or magnetoplasmons coupled to photons in coplanar waveguides~\cite{Muravev2011}. The advent of these impressive experimental results contrasts with the difficulty to reproduce the nonperturbative USC regime, or even approach the DSC regime with its radically different physical predictions~\cite{Werlang2008,DeLiberato2009,Casanova2010,Stassi2013,Wolf2013,Felicetti2014,DeLiberato2014}. Nevertheless, these first achievements, together with recent theoretical advances, have put the QRM back in the scientific spotlight. At the same time, while we struggle to reproduce the subtle aspects of the USC and DSC regimes, quantum simulators~\cite{Feynman1982} may become a useful tool for the exploration of the QRM and related models~\cite{Ballester2012}.

In this chapter, we will propose a technique for the experimental implementation of the quantum Rabi model in trapped ions, as well as extensions of it to more atoms, the Dicke model, or more photons, the two-photon quantum Rabi model. Our simulation protocols will allow us to explore this model in all relevant interaction regimes, which is of interest for studying the model and also for the generation of quantum correlations in a trapped ion platform, like for instance the highly-correlated ground state of the QRM in the DSC regime. 

\subsection{Quantum Rabi model in trapped ions}

Trapped ions are considered as one of the prominent platforms for building quantum simulators~\cite{Blatt2012}. In fact, the realization and thorough study of the JC model in ion traps, a model originally associated with CQED, is considered a cornerstone in physics~\cite{Haroche2013,Wineland2013}.  This is done by applying a red-sideband interaction with laser fields to a single ion~\cite{Diedrich1989,Blockley1992,Leibfried2003} and may be arguably presented as the first quantum simulation ever implemented. In this sense, the quantum simulation of all coupling regimes of the QRM in trapped ions would be a historically meaningful step forward in the study of dipolar light-matter interactions. In this section, we propose a method that allows the access to the full-fledged QRM with trapped-ion technologies by means of a suitable interaction picture associated with inhomogeneously detuned red and blue sideband excitations. Note that, in the last years, bichromatic laser fields have been successfully used for different purposes~\cite{Sorensen1999,Solano1999,Solano2001,Haljan2005}. In addition, we propose an adiabatic protocol to generate the highly-correlated ground states of the USC and DSC regimes, paving the way for a full quantum simulation of the experimentally elusive QRM.

\subsubsection{The simulation protocol}

Single atomic ions can be confined using radio-frequency Paul traps and their motional quantum state cooled down to its ground state by means of sideband cooling techniques~\cite{Haffner2008}. In this respect, two internal metastable electronic levels of the ion can play the role of a quantum bit (qubit). Driving a monochromatic laser field in the resolved-sideband limit allows for the coupling of the internal qubit and the motional mode, whose associated Hamiltonian reads ($\hbar=1$)
\begin{equation}
H  =  \frac{\omega_0}{2} \sigma_z + \nu a^\dag a + \Omega(\sigma^+ +  \sigma^-) \Big( e^{i[\eta(a + a^\dag) - \omega_lt + \phi_l]} +  e^{-i[\eta(a + a^\dag) -\omega_l t + \phi_l] } \Big) .
\end{equation}
Here, $a^\dag$ and $a$ are the creation and annihilation operators of the motional mode, $\sigma^+$ and $\sigma^-$ are the raising and lowering Pauli operators, $\nu$ is the trap frequency, $\omega_0$ is the qubit transition frequency, $\Omega$ is the Rabi coupling strength, and $\eta=k\sqrt{\frac{\hbar}{2m\nu}}$ is the Lamb-Dicke parameter, where $k$ is the component of the wave vector of the laser on the direction of the ions motion and $m$ the mass of the ion~\cite{Leibfried2003}; $\omega_l$ and $\phi_l$ are the corresponding frequency and phase of the laser field. For the case of a bichromatic laser driving, changing to an interaction picture with respect to the uncoupled Hamiltonian, ${H_0 = \frac{\omega_0}{2}\sigma_z} + \nu a^\dag a $ and applying an optical RWA, the dynamics of a single ion reads~\cite{Leibfried2003}
\begin{eqnarray}
H^{\rm I} & = & \!\! \sum_{n=r,b}\frac{\Omega_n}{2}\left[e^{i\eta[a(t)+a^\dag(t)]}e^{i(\omega_0-\omega_n)t}\sigma^++\textrm{H.c.}\right] ,
\label{trapped_ion_hamil}
\end{eqnarray}
with  $a(t)=ae^{-i\nu t}$ and $a^\dag (t)=a^\dag e^{i\nu t}$. 
We will consider the case where both fields are off-resonant, first red-sideband (r) and first blue-sideband (b) excitations, with detunings $\delta_r$ and $\delta_b$, respectively,
\begin{eqnarray}
\omega_r = \omega_0 - \nu + \delta_r , \,\,\, \omega_b = \omega_0 + \nu + \delta_b . \nonumber
\end{eqnarray}
In such a scenario, one may neglect fast oscillating terms in Eq.~(\ref{trapped_ion_hamil}) with two different vibrational RWAs. We will restrict ourselves to the Lamb-Dicke regime, that is, we require that $\eta\sqrt{\langle a^\dag a\rangle}\ll1$. This allows us to select terms that oscillate with minimum frequency, assuming that weak drivings do not excite higher-order sidebands, $\delta_n,\Omega_n\ll\nu$ for $n = r , b$. These approximations lead to the simplified time-dependent Hamiltonian
\begin{equation}
\label{RabiInteraction}
\bar{H}^{\rm I} = \frac{i\eta\Omega}{2} \sigma^+ \left( a e^{-i\delta_rt} + a^\dag e^{-i\delta_bt} \right) + \textrm{H.c.},
\end{equation}
where we consider equal coupling strengths for both sidebands, $\Omega = \Omega_r = \Omega_b$.
Equation~(\ref{RabiInteraction}) corresponds to the interaction picture Hamiltonian of the Rabi Hamiltonian with respect to the uncoupled Hamiltonian $H_0 = \frac{1}{4}(\delta_b+\delta_r)\sigma_z + \frac{1}{2}(\delta_b-\delta_r) a^\dag a $, 
\begin{equation}
H_{\rm QRM} = \frac{\omega_0^R}{2}\sigma_z + \omega^R a^\dag a + ig (\sigma^+-\sigma^-)(a+a^\dag) ,
\label{effective_hamiltonian}
\end{equation} 
with the effective model parameters given by
\begin{eqnarray}
\omega_0^R=-\frac{1}{2}(\delta_r+\delta_b), \,\, \omega^R = \frac{1}{2}(\delta_r-\delta_b), \,\, g=\frac{\eta\Omega}{2} ,
\end{eqnarray}
where the qubit and mode frequencies are represented by the sum and difference of both detunings, respectively. The tunability of these parameters permits the study of all coupling regimes of the QRM via the suitable choice of the ratio $ g / \omega^R$. It is important to note that all realized interaction-picture transformations, so far, are of the form $\alpha a^\dag a +\beta\sigma_z$. This expression commutes with the observables of interest, $\{ \sigma_z, | n \rangle \langle n |, a^{\dagger} a \}$, warranting that their experimental measurement will not be affected by the transformations.

The protocol can be naturally extended to the Dicke model,
\begin{equation}
H_{\rm D}= \frac{\omega_0}{2} \sum_{i=1}^N\sigma^z_i + \omega a^\dag a + g \sum_{i=1}^N(\sigma^+_i-\sigma^-_i)(a+a^\dag),
\end{equation}
where $N$ two-level systems interact with a single mode of the radiation field, through a Rabi kind interaction. The Dicke model has several properties that make it appealing, as its capacity to generate multi-partite entanglement and its superradiant phase transition. Following our scheme for the simulation of the Rabi model one could consider to trap several ions in a string, and repeat the simulation procedure, this time detuning the bichromatic sideband with respect to one of the collective motional modes. If the center of mass mode is selected, then the coupling of all the ions to this mode will be homogenous, however, if one were interested in simulating inhomogeneous Dicke Hamiltonians, other modes could be selected. With this procedure a string of trapped ions can reproduce de Dicke model in all relevant parameter regimes. 

\subsubsection{Accessible regimes} The quantum Rabi model in Eq.~(\ref{effective_hamiltonian}) will show distinct dynamics for different regimes, which are defined by the relation among the three Hamiltonian parameters: the mode frequency $\omega^R$, the qubit frequency $\omega_0^R$, and the coupling strength $g$. 
\begin{figure}[h!]
\centering
\includegraphics[width=0.9 \columnwidth]{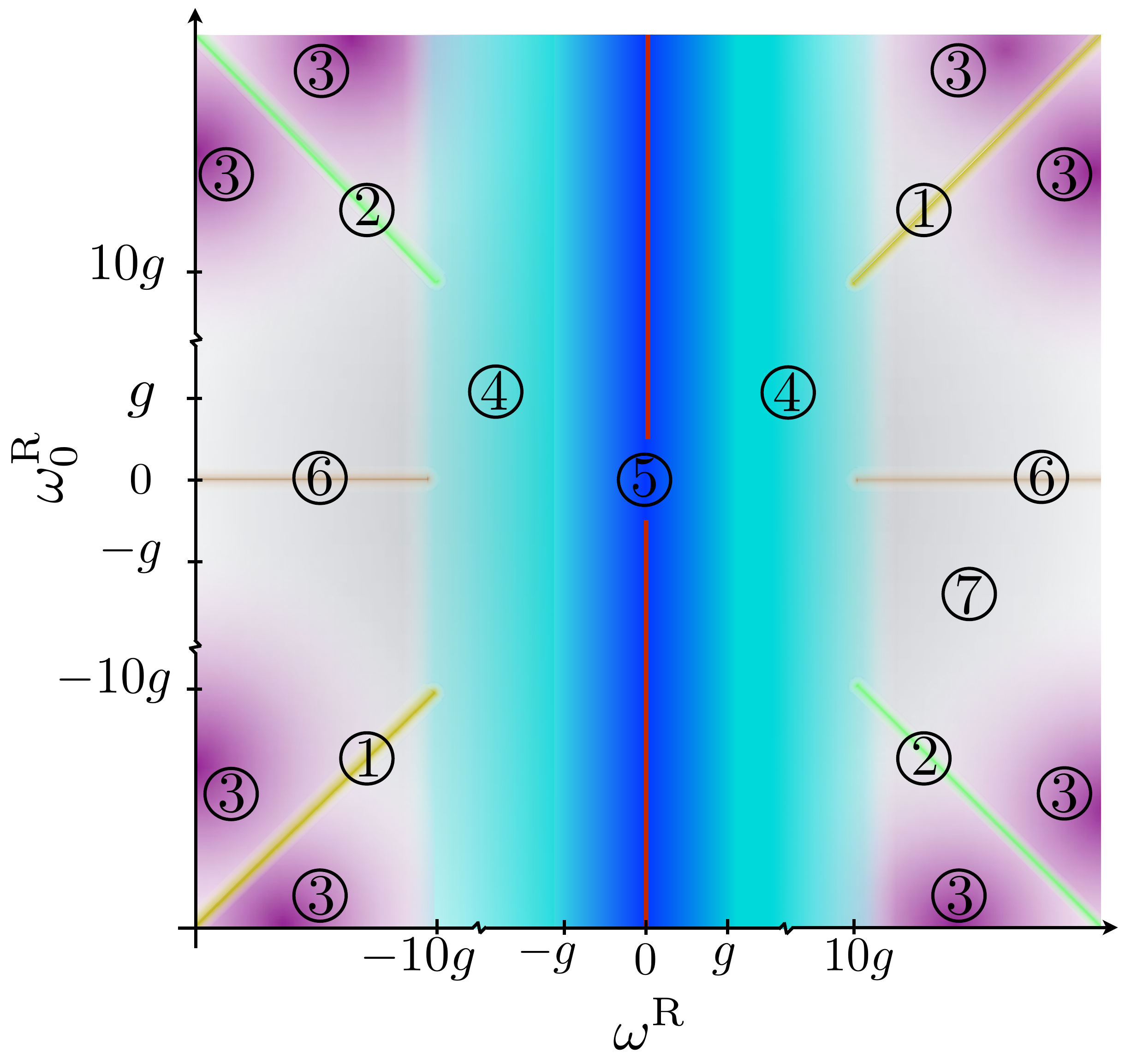}
\caption[Configuration space of the QRM.]{\footnotesize{{\bf Configuration space of the QRM.} 
(1) JC regime: $ g \ll \{|\omega^R|,|\omega_0^R| \}$ and $| \omega^R - \omega_0^R | \ll |\omega^R + \omega_0^R | $.
(2) AJC regime: $ g \ll \{|\omega^R |,|\omega_0^R| \}$ and $|\omega^R - \omega_0^R| \gg | \omega^R + \omega_0^R | $.
(3) Two-fold dispersive regime: $ g < \{ | \omega^R | , | \omega_0^R | , | \omega^R - \omega_0^R |, |\omega^R + \omega_0^R | \}$.
(4) USC regime: $| \omega^R | < 10 g $, 
(5) DSC regime: $ | \omega^R | < g $, 
(6) Decoupling regime: $ | \omega_0^R | \ll g \ll |\omega^R| $.
(7) This intermediate regime ($ |\omega_0^R| \sim g \ll |\omega^R|$) is still open to study.
The (red) central vertical line corresponds to the Dirac equation regime. Colours delimit the different regimes of the QRM, colour degradation indicates transition zones between different regions. All the areas with the same colour correspond to the same region.}}
\label{rabiregions}
\end{figure}
We first explore the regimes that arise when the coupling strength is much weaker than the mode frequency $g \ll |\omega^R| $. Under such a condition, if the qubit is close to resonance, $|\omega^R| \sim |\omega_0^R|$, and {$|\omega^R + \omega_0^R| \gg | \omega^R - \omega_0^R |$} holds, the RWA can be applied. This implies neglecting terms that in the interaction picture rotate at frequency $ \omega^R + \omega_0^R $, leading to the JC model. This is represented in Fig.~\ref{rabiregions} by the region 1 in the diagonal. Notice that these conditions are only possible if both the qubit and the mode frequency have the same sign. However, in a quantum simulation one can go beyond conventional regimes and even reach unphysical situations, as when the qubit and the mode have frequencies of opposite sign. In this case, $| \omega^R - \omega_0^R| \gg | \omega^R + \omega_0^R | $ holds, see region 2, and we will be allowed to neglect terms that rotate at frequencies $ | \omega^R - \omega_0^R |$. This possibility will give rise to the anti-Jaynes Cummings (AJC) Hamiltonian, $H^{AJC} = \frac{\omega_0^R}{2} \sigma_z + \omega^R a^\dag a + ig (\sigma^+a^\dag - \sigma^-a)$. It is noteworthy to mention that, although both JC and AJC dynamics can be directly simulated with a single tuned red or blue sideband interaction, respectively, the approach taken here is fundamentally different. Indeed, we are simulating the QRM in a regime that corresponds to such dynamics, instead of directly implementing the effective model, namely the JC or AJC model.

If we depart from the resonance condition and have all terms rotating at high frequencies $ \{ |\omega^R|, |\omega^R_0|, |\omega^R + \omega_0^R |, | \omega^R - \omega_0^R | \} \gg g$, see region 3, for any combination of frequency signs, the system experiences dispersive interactions governed by a second-order effective Hamiltonian. In the interaction picture, this Hamiltonian reads
\begin{equation}
H_{\textrm{eff}}  =  g^2\left[\frac{|e\rangle\langle e|}{\omega^R - \omega_0^R}- \frac{|g\rangle\langle g| }{\omega^R + \omega_0^R}\right. + \left.\frac{2 \omega^R }{(\omega_0^R + \omega^R)(\omega^R - \omega_0^R)}a^\dag a\sigma_z\right],
\end{equation}
inducing AC-Stark shifts of the qubit energy levels conditioned to the number of excitations in the bosonic mode.

The USC regime is defined as $0.1 \lesssim g/\omega^R \lesssim 1$, with perturbative and nonperturbative intervals, and is represented in Fig.~\ref{rabiregions} by region 4. In this regime, the RWA does not hold any more, even if the qubit is in resonance with the mode. In this case, the description of the dynamics has to be given in terms of the full quantum Rabi Hamiltonian. For $g/\omega^R \gtrsim 1$, we enter into the DSC regime, see region 5 in Fig.~\ref{rabiregions}, where the dynamics can be explained in terms of phonon number wave packets that propagate back and forth along well defined parity chains~\cite{Casanova2010}.

In the limit where $\omega^R=0$, represented by a vertical centered red line in Fig.~\ref{rabiregions}, the quantum dynamics is given by the relativistic Dirac Hamiltonian in 1+1 dimensions, 
\begin{equation}
H_D=mc^2\sigma_z+cp\sigma_x,
\end{equation}
which has been successfully implemented in trapped ions~\cite{Gerritsma2010, Lamata2007}, as well as in other platforms~\cite{Salger2011, Dreisow2010}.

Moreover, an interesting regime appears when the qubit is completely out of resonance and the coupling strength is small when compared to the mode frequency, $ \omega^R_0 \sim 0 $ and $ g \ll |\omega^R| $. In this case, the system undergoes a particular dispersive dynamics, where the effective Hamiltonian becomes a constant. Consequently,  the system does not evolve in this region that we name as decoupling regime, see region 6 in Fig.~\ref{rabiregions}. The remaining regimes correspond to region 7 in Fig.~\ref{rabiregions}, associated with the parameter condition $ |\omega_0^R| \sim g \ll |\omega^R|$.

The access to different regimes is limited by the maximal detunings allowed for the driving fields, which are given by the condition $\delta_{r,b} \ll \nu$, ensuring that higher-order sidebands are not excited. The simulations of the JC and AJC regimes, which demand detunings $|\delta_{r,b}| \le | \omega^R | + |\omega_0^R| $, are the ones that may threaten such a condition. We have numerically simulated the full Hamiltonian in Eq.~(\ref{trapped_ion_hamil}) with typical ion-trap parameters: $\nu=2\pi \times 3 {\rm MHz}$, $\Omega=2\pi \times 68 {\rm kHz}$ and $\eta=0.06$~\cite{Gerritsma2010}, while the laser detunings were $\delta_b=- 2\pi \times 102 {\rm kHz}$ and $\delta_r=0$, corresponding to a simulation of the JC regime with $g/\omega^{\rm R}=0.01$. The numerical simulations show that second-order sideband transitions are not excited and that the state evolution follows the analytical JC solution with a fidelity larger than $99\%$ for several Rabi oscillations. This confirms that the quantum simulation of these regimes is also accessible in the lab. We should also pay attention to the Lamb-Dicke condition $\eta \sqrt{\langle a^\dag a \rangle} \ll 1$, as evolutions with an increasing number of phonons may jeopardize it. However, typical values like $\eta=0.06$ may admit up to some tens of phonons, allowing for an accurate simulation of the QRM in all considered regimes.

Regarding coherence times, the characteristic timescale of the simulation will be given by $t_{\rm char}= \frac{2\pi}{g}$. In our simulator, $g=\frac{\eta \Omega}{2}$, such that $t_{\rm char}=\frac{4 \pi}{\eta \Omega}$. For typical values of $\eta=0.06-0.25$ and of $\Omega/2\pi=0-500$~kHz, the dynamical timescale of the system is of milliseconds, well below coherence times of qubits and motional degrees of freedom in trapped-ion setups~\cite{Haffner2008}. 
\begin{figure}[h!]
\centering
\includegraphics[width=\columnwidth]{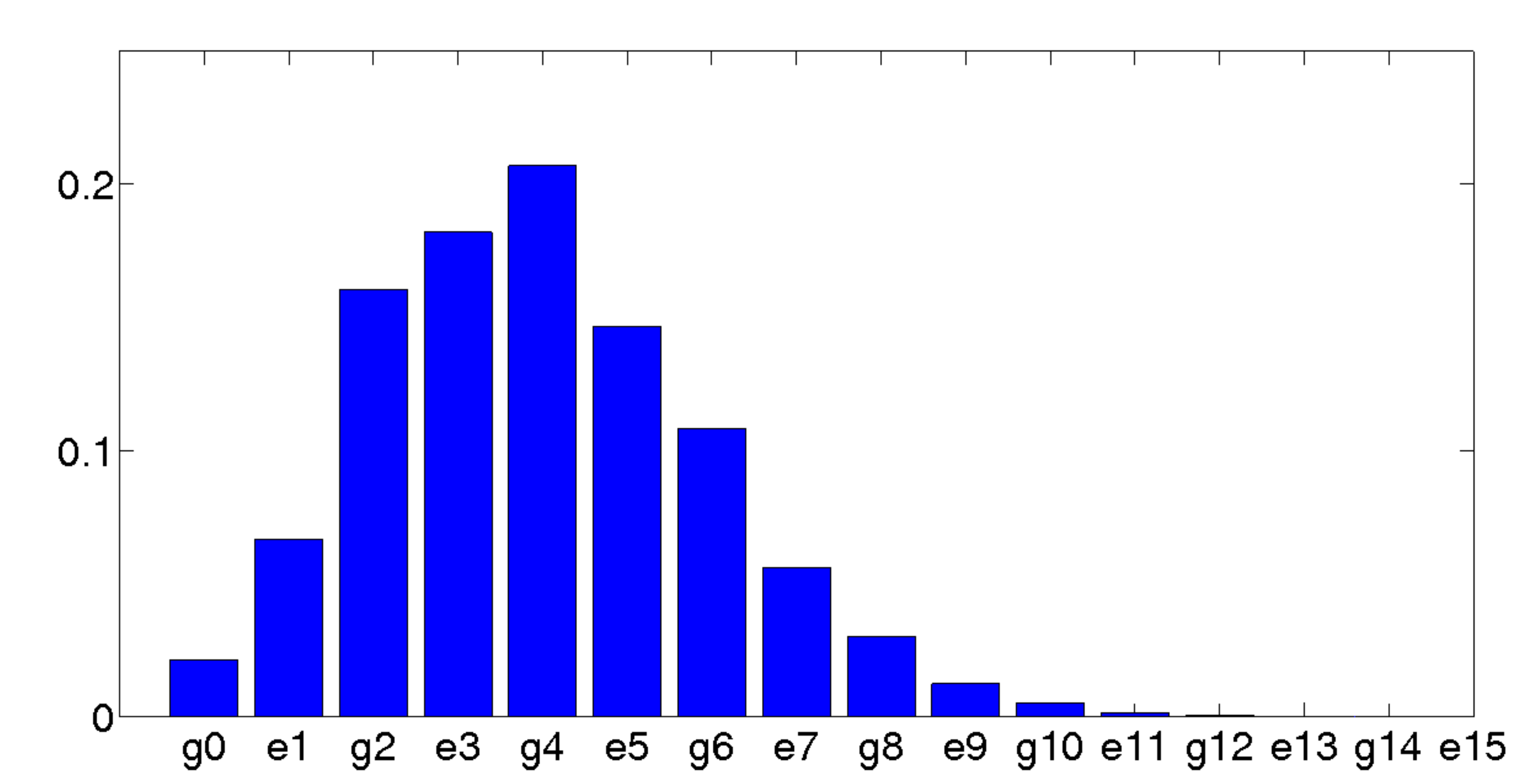}
\caption[State population of the QRM ground state.]{\footnotesize{{\bf State population of the QRM ground state.} We plot the case of $g/\omega^R=2$, parity $p=+1$, and corresponding parity basis $\lbrace |g,0\rangle,|e,1\rangle,|g,2\rangle,|e,3\rangle,\hdots\rbrace$. Here, $p$ is the expectation value of the parity operator ${P=\sigma_z e^{-i\pi a^\dag a}}$~\cite{Casanova2010}, and only states with even number of excitations are populated. We consider a resonant red-sideband excitation ($\delta_r=0$), 
a dispersive blue-sideband excitation ($\delta_b/2\pi=-11.31$kHz), and $g=-\delta_b$, leading to the values $\omega^R=\omega_0^R=-\delta_b/2$ and $g/\omega^R=2$.}}
\label{Gstate_barplot_fig}
\end{figure}
\subsubsection{Ground state preparation} The ground state $|G\rangle$  of the QRM in the JC regime ($g\ll\omega^R$) is given by the state $|g,0\rangle$, that is, the qubit ground state, $| g \rangle$, and the vacuum of the bosonic mode, $| 0 \rangle$. It is known that $|g,0\rangle$ will not be the ground state of the interacting system for larger coupling regimes, where the contribution of the counter-rotating terms becomes important~\cite{Hwang2010}. As seen in Fig.~\ref{Gstate_barplot_fig}, the ground state of the USC/DSC Hamiltonian is far from trivial~\cite{Braak2011}, essentially because it contains qubit and mode excitations, $\langle G|a^\dag a|G\rangle >0$.

\begin{figure}[h!]
\centering
\includegraphics[width=\columnwidth]{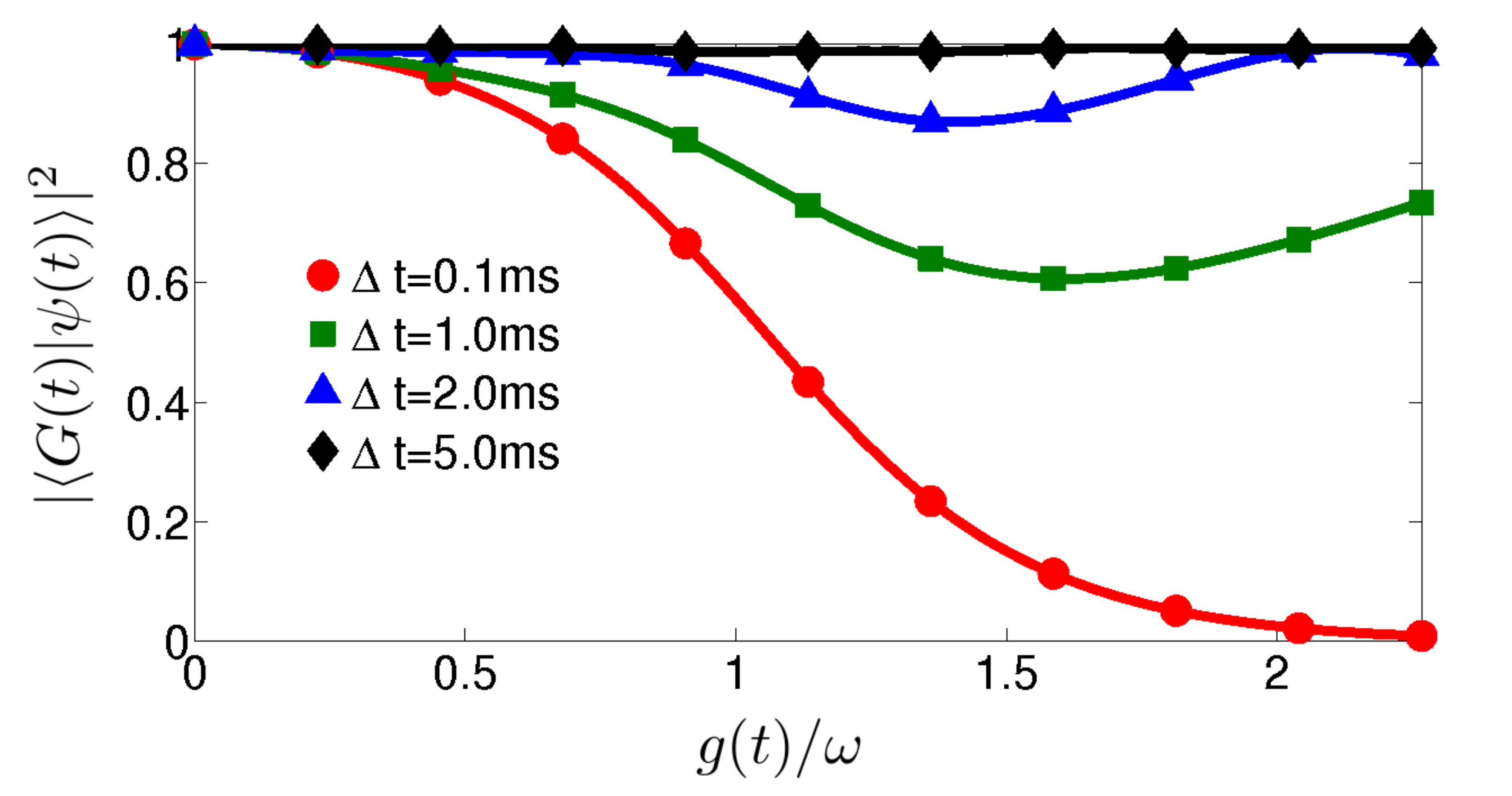}
\caption[Fidelity of the adiabatic evolution for the preparation of the GS of the QRM.]{\footnotesize{{\bf Fidelity of the adiabatic evolution for the preparation of the GS of the QRM.} Let us assume that the system is initially prepared in the JC ground state $|g,0\rangle$, that is, when $g\ll\omega^R$.  Then, the coupling is linearly chirped during an interval $\Delta t$ up to a final value $g_f$, i. e., $g(t)=g_f t/\Delta t$. For slow changes of the laser intensity, the ground state is adiabatically followed, whereas for non-adiabatic processes, the ground state is abandoned. The instantaneous ground state $|G(t)\rangle$ is computed by diagonalizing the full Hamiltonian at each time step, while the real state of the system $|\psi(t)\rangle$ is calculated by numerically integrating the time dependent Schr\"odinger equation for a time-varying coupling strength $g(t)$.
For the simulation, a $^{40}$Ca$^+$ ion has been considered with parameters: 
$\nu=2\pi\times3$MHz, $\delta_r=0$, $\delta_b=-6\times10^{-4}\nu$, $\eta=0.06$ and $\Omega_f=2\pi\times68$kHz~\cite{Gerritsma2010}.}}
\label{JC_groundstate_adiabatic_fig}
\end{figure}

Hence, preparing the qubit-mode system in its actual ground state is a rather difficult state-engineering task in most parameter regimes, except for the JC limit. The non analytically computable ground state of the QRM has never been observed in a physical system, and its generation would be of theoretical and experimental interest. We propose here to generate the ground state of the USC/DSC regimes of the QRM via adiabatic evolution. Figure \ref{JC_groundstate_adiabatic_fig} shows the fidelity of the state prepared following a linear law of variation for the coupling strength at different evolution rates. When our system is initialized in the JC region, achieved with detunings $\delta_r=0$ and $|\delta_b |\gg g$, it is described by a JC Hamiltonian with the ground state given by $|G\rangle=|g,0\rangle$. Notice that the $g/\omega^{\rm R}$ ratio can be slowly turned up, taking the system adiabatically through a straight line in the configuration space to regions 4-5~\cite{Kyaw2015}. This can be done either increasing the value of $g$ by raising the intensity of the driving, or decreasing the value of $\omega^{\rm R}$ by reducing the detuning $|\delta_b|$. The adiabatic theorem~\cite{Messiah1999} ensures that if this process is slow enough, transitions to excited states will not occur and the system will remain in its ground state.  As expected, lower rates ensure a better fidelity.  Once the GS of the QRM is generated, one can extract the populations of the different states of the characteristic parity basis shown in Fig.~(\ref{Gstate_barplot_fig}). To extract the population of a specific Fock state, one would first generate a phonon-number dependent ac-Stark shift~\cite{Solano2005}. A simultaneous transition to another electronic state will now have a frequency depending on the motional quantum number. By matching the frequency associated with Fock state $n$, we will flip the qubit with a probability proportional to the population of that specific Fock state. This will allow us to estimate such a population without the necessity of reconstructing the whole wave function.

In this section, we have proposed a method for the quantum simulation of the QRM in ion traps. Its main advantage consists in the accessibility to the USC/DSC regimes and the convenient switchability to realize full tomography, outperforming other systems where the QRM should appear more naturally, such as cQED~\cite{Peropadre2010,Felicetti2014}. In addition, we have shown how to prepare the qubit-mode system in its entangled ground state through adiabatic evolution from the known JC limit into the USC/DSC regimes. This would allow for the complete reconstruction of the QRM ground state, never realized before, in a highly controllable quantum platform as trapped ions. The present ideas are straightforwardly generalizable to many ions, opening the possibility of going from the more natural Tavis-Cummings model to the Dicke model. In our opinion, the experimental study of the QRM in trapped ions will represent a significant advance in the long history of dipolar light-matter interactions.

\subsection{Two-photon quantum Rabi model in trapped ions} \label{sec:TPQRM}



The two-photon quantum Rabi model is defined analogous to the quantum Rabi model with squared creation and annihilation operators in the interaction term,
\begin{equation}
H_\text{TPQRM} = \omega\ a^\dagger a + \frac{\omega_q}{2} \sigma_z + g\ \sigma_x \left( a^2 + {a^\dagger}^2 \right).
\end{equation}
It enjoys a spectrum with highly counterintuitive features~\cite{Ng1999, Emary2002}, which appear when the coupling strength becomes comparable with the bosonic mode frequency. In this sense, it is instructive to compare these features with the ultrastrong~\cite{Bourassa2009,Niemczyk2010,Forn-Diaz2010} and deep strong~\cite{Casanova2010} coupling regimes of the quantum Rabi model. The two-photon Rabi model has been applied as an effective model to describe second-order processes in different physical setups, such as Rydberg atoms in microwave superconducting cavities \cite{Bertet2002} and quantum dots \cite{Stufler2006, delValle2010}. However, the small second-order coupling strengths restrict the observation of a richer dynamics. As already explained, in trapped-ion systems~\cite{Leibfried2003,Haffner2008}, it is possible to control the coherent interaction between the vibrations of an ion crystal and its internal electronic states, which form effective spin degrees of freedom.  Second sidebands , where this interaction happens through terms of the form $\sigma^+ a^2 + H.c. (\sigma^- a^2 + H.c.)$, have been considered for laser cooling~\cite{Filho1994} and for generating nonclassical motional states~\cite{Meekhof1996, Filho1996, Gou1997}.

In this section, we design a trapped-ion scheme in which the two-photon Rabi and two-photon Dicke models can be realistically implemented in all relevant regimes. We theoretically show that the dynamics of the proposed system is characterized by harmonic two-phonon oscillations or by spontaneous generation of excitations, depending on the effective coupling parameter. In particular, we consider cases where complete spectral collapse---namely, the fusion of discrete energy levels into a continuous band---can be observed.

\subsubsection{The simulation protocol}
We consider a chain of $N$ qubits interacting with a single bosonic mode via two-photon interactions
\begin{equation}
\mathcal{H} = \omega a^\dagger a + \sum_n \frac{\omega_q^n}{2} \sigma_z^n + \frac{1}{N}\sum_n g_n \sigma_x^n \left( a^2 + {a^\dagger}^2 \right),
\label{2phdicke}
\end{equation}
where $\hbar=1$, $a$ and $a^\dagger$ are bosonic ladder operators; $\sigma_x^n$ and $\sigma_z^n$ are qubit Pauli operators; parameters $\omega$, $\omega_q^n$, and $g_n$, represent the mode frequency, the $n$-th qubit energy spacing and the relative coupling strength, respectively. We will explain below how to implement this model using current trapped-ion technology, considering in detail the case $N=1$ and discussing the scalability issues for $N > 1$.

As for the simulation of the linear quantum Rabi model, we will consider a setup where the qubit energy spacing, $\omega_{\rm int}$, represents an optical or hyperfine/Zeeman internal transition in a single trapped ion and the vibrational motion of the ion is described by bosonic modes $a,\ a^\dagger$, with trap frequency $\nu$. Turning on a bichromatic driving, with frequencies $\omega_r$ and $\omega_b$, an effective coupling between the internal and motional degrees of freedom is activated as described by Eq.~(\ref{trapped_ion_hamil}). Again, we will consider the system to be in the Lamb-Dicke regime, but now we will set the frequencies of the bichromatic driving to be detuned from the second sidebands, 
$\omega_r = \omega_0 - 2\nu + \delta_r$, $\omega_b = \omega_0 +2 \nu + \delta_b$.
We choose  homogeneous coupling strengths $\Omega_j = \Omega$ for both sideband excitations.  Expanding the exponential operator in Eq.~\eqref{trapped_ion_hamil} to the second order in $\eta$, and performing a RWA with $\delta_j, \Omega \ll \nu$, we can rewrite the interaction picture Hamiltonian
\begin{equation}
\mathcal{H}^I= -\frac{\eta^2\Omega}{4}\left[a^2\ e^{-i \delta_r t} + {a^\dagger}^2\ e^{-i\delta_b t} \right] \sigma_+ + \text{H.c.}
\label{rwaINT}
\end{equation}
The first-order correction to approximations made in deriving Eq.~\eqref{rwaINT} is given by $\frac{\Omega}{2}\ e^{\pm i2\nu t}\sigma_+ + \text{H.c.}$, which produce spurious excitations with negligible probability $P_e = \left( \frac{\Omega}{4\nu} \right)^2$. Further corrections are proportional to $\eta \Omega$ or $\eta^2$ and oscillate at frequency $\nu$, yielding $P_e = \left( \frac{\eta\Omega}{4\nu} \right)^2$. Hence, they are negligible in standard trapped-ion implementations. The explicit time dependence in Eq.~\eqref{rwaINT} can be removed by going to another interaction picture with $\mathcal{H}_0 = \frac{1}{4}\left(\delta_b - \delta_r \right)a^\dagger a + \frac{1}{4}\left( \delta_b + \delta_r\right) \sigma_z$, which we dub the simulation picture. Then, the system Hamiltonian resembles the two-phonon quantum Rabi Hamiltonian
\begin{equation}
\mathcal{H}_\text{eff} = \omega\ a^\dagger a + \frac{\omega_q}{2} \sigma_z - g\ \sigma_x \left( a^2 + {a^\dagger}^2 \right),
\label{implem}
\end{equation}
where the effective model parameters are linked to physical variables through $\omega = \frac{1}{4}\left( \delta_r - \delta_b \right)$, $\omega_q = -\frac{1}{2}\left( \delta_r + \delta_b \right)$, and $g=\frac{\eta^2 \Omega }{4}$. Remarkably, by tuning $\delta_r$ and $\delta_b$, the two-phonon quantum Rabi model of Eq.~\eqref{implem} can be implemented in all regimes. Moreover, the {\it N}-qubit two-phonon Dicke model of Eq.~\eqref{2phdicke} can be implemented using a chain of {\it N} ions by applying a similar method. In this case, the single bosonic mode is represented by a collective motional mode~\cite{James1998} (see Appendix~\ref{app_dicke}).

\begin{figure}[h!]
\centering
\includegraphics[width= \columnwidth]{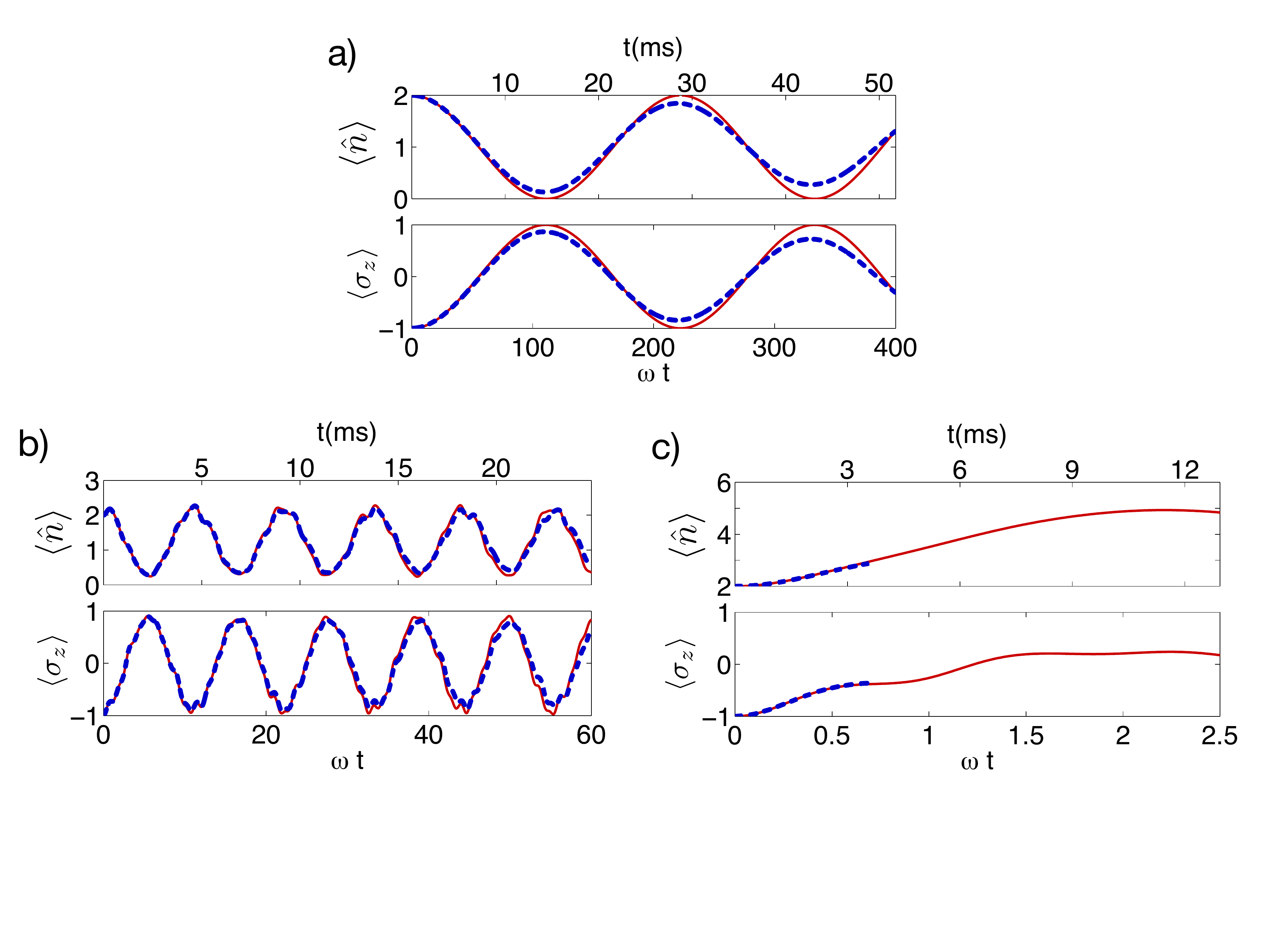}
\caption[Real-time dynamics of the two-photon QRM.]{\footnotesize{{\bf Real-time dynamics of the two-photon QRM.} The model parameters are $N=1$, resonant qubit $\omega_q = 2\omega$, and effective couplings: (a) $g=0.01\omega$, (b) $g=0.2\omega$, and (c) $g=0.4\omega$. The initial state is given by $| g, 2 \rangle$, i.e., the two-phonon Fock state and the qubit ground state. In all plots, the red solid line corresponds to numerical simulation of the exact Hamiltonian of Eq.~\eqref{2phdicke}, while the blue dashed line is obtained simulating the full model of Eq.~\eqref{trapped_ion_hamil}, including qubit decay $t_1 = 1$s, pure dephasing $t_2 = 30$ms and vibrational heating of one phonon per second. In each plot, the lower abscissa shows the time in units of $\omega$, while the upper one shows the evolution time of a realistic trapped-ion implementation. In panel (c), the full model simulation could not be performed for a longer time due to the fast growth of the Hilbert-space.}}
\label{realtime}
\end{figure}

The validity of the approximations made in deriving Eq.~\eqref{implem} has been checked comparing the simulated two-photon quantum Rabi dynamics with numerical evaluation of the simulating trapped-ion model of Eq.~\eqref{trapped_ion_hamil}, as shown in Fig.~\ref{realtime}. Standard parameters and dissipation channels of current setups have been considered. In all plots of Fig.~\ref{realtime},  the vibrational frequency is $\nu/2\pi = 1$~MHz and the coupling coefficient is $\Omega/2\pi = 100$~KHz. The Lamb-Dicke parameter is $\eta=0.04$ for Fig.~\ref{realtime}a and Fig.~\ref{realtime}b, while $\eta=0.02$ for Fig.~\ref{realtime}c. Notice that larger coupling strengths imply a more favourable ratio between dynamics and dissipation rates. Hence, the implementation accuracy improves for large values of $g/\omega$, which correspond to the most interesting coupling regimes.   

\subsubsection{Real-time dynamics}
Depending on the ratio between the normalized coupling strength $g$ and the mode frequency $\omega$, the model of  Eq.~\eqref{2phdicke} exhibits qualitatively different behaviors. Two parameter regimes can be identified accordingly. 
For the sake of simplicity, we will consider the homogeneous coupling case $g_n=g$, $\omega_q^n = \omega_q$, for every $n$, and we will focus
 in the resonant or near-resonant case $\omega_q \approx 2\omega$. 

In accordance with the quantum Rabi model, we define the strong coupling (SC) regime by the condition $g/\omega \ll1$. Under this restriction, the RWA can be applied to the coupling terms, replacing each direct interaction $ g \sum_n \sigma_x^n \left( a^2 + {a^\dagger}^2 \right)$ with  $ g \sum_n \left( \sigma_+^n a^2 + \sigma_-^n{a^\dagger}^2 \right)$, where we defined the raising/lowering single-qubit operators $\sigma^n_\pm = \left(\sigma^n_x \pm i\sigma^n_y \right)/2$. 
When the RWA is valid, the system satisfies a continuous symmetry, identified by the operator $\zeta = a^\dagger a + 2\sum_n \sigma_+^n\sigma_-^n$, which makes the model superintegrable~\cite{Braak2011}.  In the SC regime, the interaction leads to two-photon excitation transfers between the bosonic field and the qubits, as shown in Fig.~\ref{realtime}a. 
Jaynes-Cummings-like collapses and revivals of population inversion are also expected to appear~\cite{Alsing1987, Joshi2000}.

As the ratio $g/\omega$ increases, the intuitive dynamics of the SC regime disappears and excitations are not conserved  (see Fig.~\ref{realtime}b and Fig.~\ref{realtime}c).
When the normalized coupling approaches the value $g\sim 0.1\omega$, the RWA cannot be performed, and the full quantum Rabi model must be taken into account.  We define the USC regime as the parameter region for which $0.1 \lesssim g/\omega < 0.5$.
An analytical solution for the system eigenstates has  been derived in~\cite{Travenec2012, Maciejewski2015, Travenec2015}. However, this approach relies basically on a numerical instability related to the presence of dominant and minimal solutions of an associated three-term recurrence relation~\cite{Zhang2013} and gives no qualitative insight into the behavior of the spectrum close to the collapse point. While continued-fraction techniques are applicable in principle~\cite{Zhang2013}, only a few low-lying levels can be computed and the method fails again in approaching the critical coupling (see below). While the $G$-function derived in~\cite{Chen2012} allows for the desired understanding of the qualitative features of the collapse, its mathematical justification is still incomplete. On the other hand, direct numerical simulation becomes challenging close to collapse due to the large number of excitations involved. Especially the dynamics of the two-photon Dicke model is demanding for classical numerical techniques.

\begin{figure}[h!]
\centering
\includegraphics[width= \columnwidth]{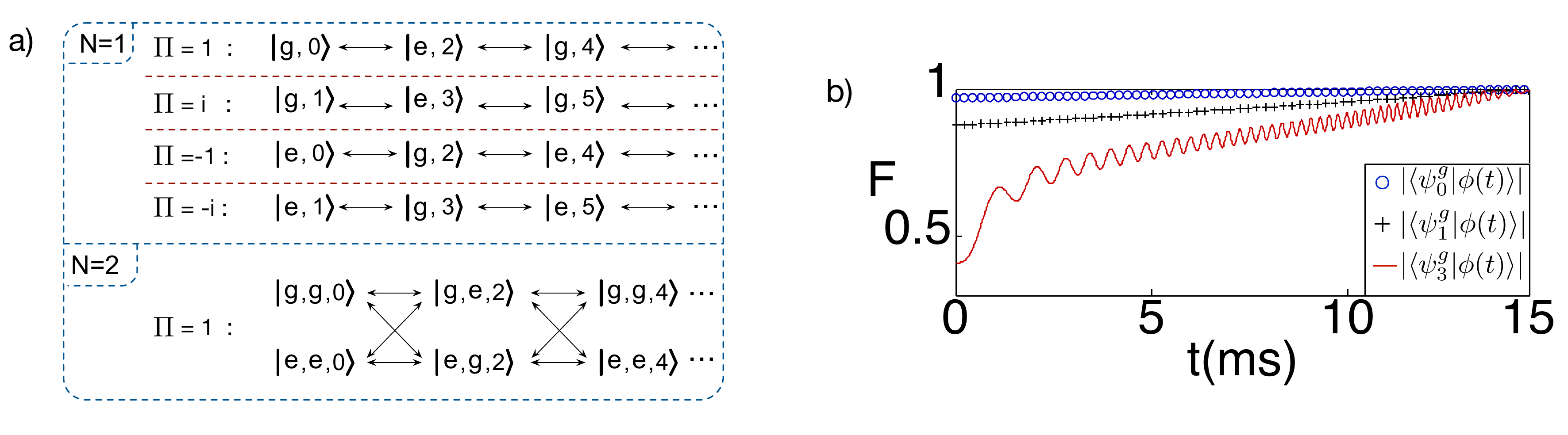}
\caption[Parity chains and adiabatic generation of eigenstates of the 2-photon QRM.]{\footnotesize{{\bf Parity chains and adiabatic generation of eigenstates of the 2-photon QRM.} (a) Generalized-parity chains for $N=1, 2$. For simplicity, for $N=2$, only one generalized-parity subspace is shown. (b) Quantum  state fidelity between the system state $ | \phi(t) \rangle$ and the target eigenstates $| \psi^g_n \rangle$ during adiabatic evolution. The Hamiltonian at $t=0$ is given by Eq.~\eqref{2phdicke} with $N=1$, $\omega_q/\omega = 1.9$ and $g=0$. During the adiabatic process, the coupling strength is linearly increased until reaching the value $g/\omega = 0.49$. For the blue circles, the initial state is given by the ground state $| \phi(t=0) = | \psi^{g=0}_0 \rangle$. For black crosses, $| \phi(t=0) \rangle = | \psi^{g=0}_1 \rangle$, while for the red solid line, $| \phi(t=0) \rangle = | \psi^{g=0}_4 \rangle $. The color code indicates  generalized parity as in Fig.~\ref{spectrum}a. Notice that, due to generalized parity conservation, the fourth excited eigenstate $ | \psi^{g=0}_4 \rangle$  of the decoupled Hamiltonian is transformed into the third one $| \psi^{g}_3 \rangle$ of the full Hamiltonian.}}
\label{paritychain}
\end{figure}

In the SC/USC transition, the continuous symmetry $\zeta$ breaks down to a $\mathbb{Z}_4$ discrete symmetry identified by the  operator 
\begin{equation}
\Pi = (-1)^N \bigotimes_{n=1}^N \sigma_z^n\ \exp\left( i \frac{\pi}{2} a^\dagger a \right).
\label{parityoperator}
\end{equation} 
We will call $\Pi$ the generalized parity operator, in analogy with the standard quantum Rabi model~\cite{Braak2011}.
Four invariant Hilbert subspaces are identified by the four eigenvalues $\lambda = \{ 1,-1, i, -i \}$ of $\Pi$.
Hence, for any coupling strength, the symmetry $\Pi$  restricts the dynamics to generalized-parity chains, shown in Fig.~\ref{paritychain}a, for $N=1,2$. 

When the normalized coupling $g$ approaches $g = \omega /2 $ (see Fig~\ref{spectrum}c), the dynamics is dominated by the interaction term and it is characterized by photon production. Finally, when  $g > \omega/2$, the Hamiltonian is not bounded from below. However, it still provides a well defined dynamics when applied for a limited time, like usual displacement or squeezing operators.

\subsubsection{The spectrum}
\begin{figure}[h!]
\centering
\includegraphics[width= \columnwidth]{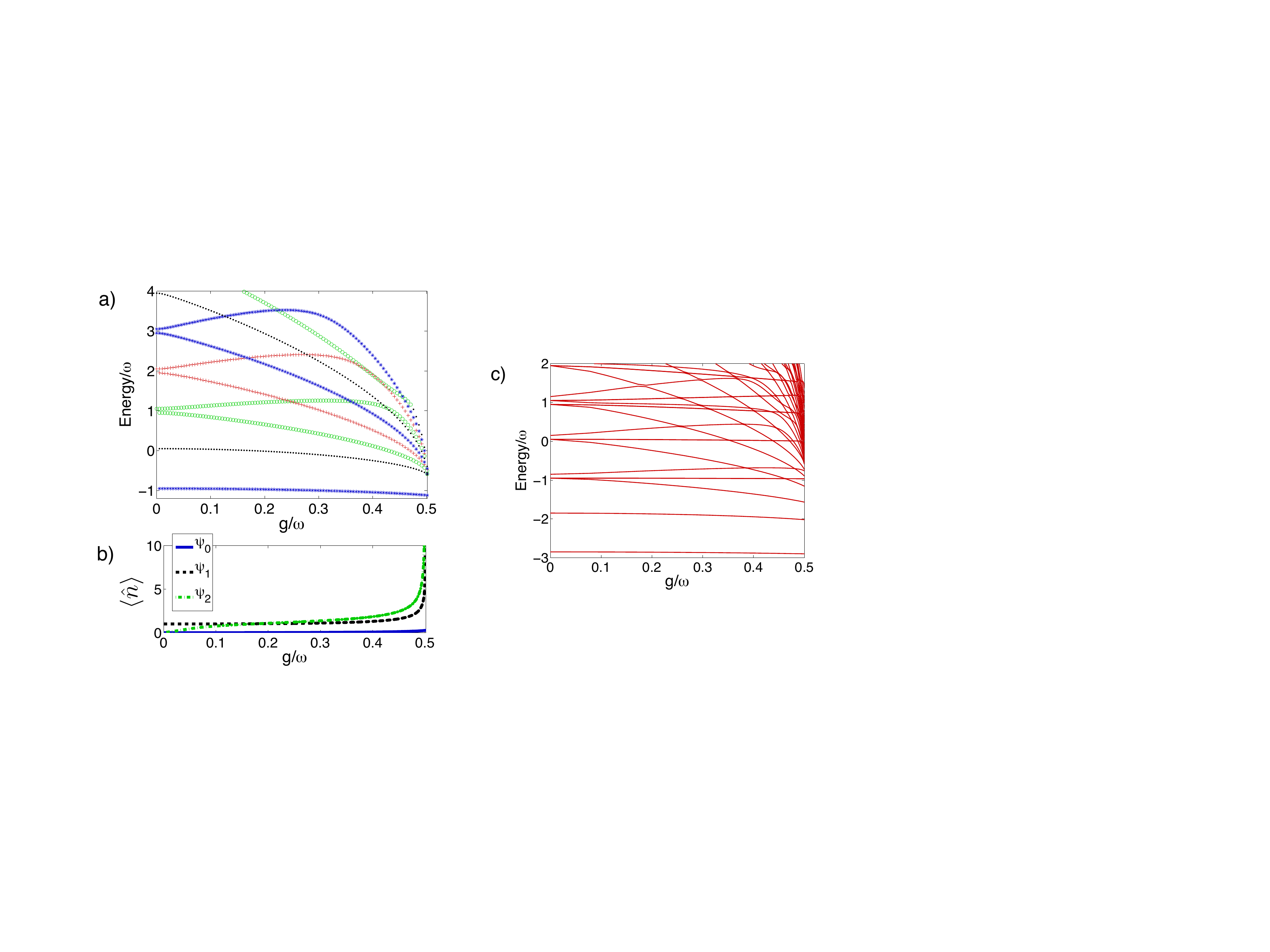}
\caption[Spectral properties of the 2-photon QRM Hamiltonian.]{\footnotesize{{\bf Spectral properties of the 2-photon QRM Hamiltonian.} The considered Hamiltonian is that of Eq.~\eqref{2phdicke}, in units of $\omega$, for $\omega_q = 1.9$, as a function of the coupling strength $g$. For $g>0.5$, the spectrum is unbounded from below. (a) Spectrum for $N=1$. Different markers identify the generalized parity of each eigenstate: green circles for $p=1$, red crosses for $p=i$, blue stars for $p=-1$, and black dots for $p=-i$. (b) Average photon number for the ground and first two excited states, for $N=1$. (c) Spectrum for $N=3$. For clarity, the generalized parity of the eigenstates is not shown.}}
\label{spectrum}
\end{figure}

The eigenspectrum of the Hamiltonian in Eq.~\eqref{2phdicke} is shown in Figs.~\ref{spectrum}a and \ref{spectrum}c for $N=1$ and $N=3$, respectively. Different markers are used to identify the generalized parity $\Pi$ of each Hamiltonian eigenvector, see Eq.~\eqref{parityoperator}. In the SC regime, the spectrum is characterized by the linear dependence of the energy splittings, observed for small values of $g$.  On the contrary, in the USC regime the spectrum is characterized by level crossings known as Juddian points, allowing for closed-form isolated solutions~\cite{Emary2002} in the single-qubit case.

The most interesting spectral features appear when the normalized coupling $g$ approaches the value $\omega/2$. In this case, the energy spacing between the system eigenenergies asymptotically vanishes and the average photon number  for the first excited eigenstates diverges (see Fig.~\ref{spectrum}b). When $g = \omega/2$, the discrete spectrum collapses into a continuous band, and its eigenfunctions are not normalizable (see Appendix~\ref{app_math}). Beyond that value, the Hamiltonian is unbounded from below~\cite{Ng1999, Emary2002}. This can be shown by rewriting the bosonic components of Hamiltonian of Eq.~\eqref{2phdicke} in terms of the effective position and momentum operators of a particle of mass $m$, defined as $\hat{x} = \sqrt{\frac{1}{2m\omega}}\left( a + a^\dagger \right)$ and $\hat{p} = i\sqrt{\frac{m\omega}{2}}\left( a - a^\dagger \right)$. Therefore, we obtain
\begin{eqnarray}
\mathcal{H} &=& \frac{m\omega}{2}\left[ (\omega - 2g\ \hat{S}_x )\frac{\hat{p}^2}{m^2 \omega^2} + (\omega + 2g\ \hat{S}_x ) \hat{x}^2 \right]  \nonumber \\
&+&  \frac{\omega_q}{2}  \sum_n \sigma_z^n,
\label{effH}
\end{eqnarray}
where $\hat{S}_x = \frac{1}{N}\sum_n \sigma_x^n$.  Notice that  $\hat{S}_x$ can take values included in the interval $\langle S_x\rangle \in [-1,1]$. Hence, the parameter $(\omega + 2g)$ establishes the shape of the effective potential. For $g < \omega/2$, the particle experiences an always  positive quadratic potential. For $g = \omega/2$, there are qubit states which turn the potential flat and the spectrum collapses, like for a free particle (see Appendix~\ref{app_math}). Finally, when $g>\omega/2$, the effective quadratic potential can be positive, for $\langle\hat{S}_x\rangle < -\omega/2g$, or negative, for $\langle\hat{S}_x\rangle >  \omega/2g$. Therefore, the Hamiltonian~\eqref{effH} has neither an upper nor a lower bound.

\subsubsection{Measurement techniques}
A key experimental signature of the spectral collapse (see Fig.~\ref{spectrum}a) can be obtained by measuring the system eigenenergies~\cite{Senko2014} when  $g$ approaches $0.5 \omega$. Such a measurement could be done via the quantum phase estimation algorithm~\cite{Abrams1999}. A more straightforward method consists of directly generating the system eigenstates~\cite{Felicetti2014} by means of the adiabatic protocol shown in Fig.~\ref{paritychain}b. When $g=0$, the eigenstates $| \psi^{g=0}_n \rangle$ of  Hamiltonian in Eq.~\eqref{2phdicke} have an analytical form and can be easily generated~\cite{Jurcevic2014}. Then, adiabatically increasing $g$, the eigenstates $| \psi^{g}_n \rangle$ of the full model can be produced. Notice that generalized-parity conservation protects the adiabatic switching at level crossings (see Fig.~\ref{paritychain}b).

Once a given eigenstate has been prepared, its energy can be inferred by measuring the expected value of the Hamiltonian in Eq.~\eqref{2phdicke}. We consider separately the measurement of each Hamiltonian term. The measurement of $\sigma^n_z$ is  standard in trapped-ion setups and is done with fluorescence techniques~\cite{Leibfried2003}. The measurement of the phonon number expectation value was already proposed in Ref.~\cite{Bastin2006}. Notice that operators $\sigma^n_z$ and $a^\dag a$ commute with all transformations performed in the derivation of the model. The expectation value of the interaction term $g\sigma^n_x \left( a^2 + {a^\dag}^2\right)$ can be mapped into the value of the first time derivative of $\langle \sigma^n_z \rangle$ at measurement time $t=0$, with the system evolving under $\mathcal{H}_m=\omega a^\dag a + \frac{\omega_q}{2} \sigma^n_z - g \sigma^n_y \left( a^2 + {a^\dag}^2\right)$. This Hamiltonian is composed of a part $A=\omega a^\dag a + \frac{\omega_q}{2} \sigma^n_z$ which commutes with $\sigma^n_z$, $[A,\sigma^n_z]=0$, and a part $B=-g \sigma^n_y \left(a^2 + {a^\dag}^2\right)$ which anti-commutes with $\sigma^n_z$, $\{B,\sigma^n_z \}=0$, yielding
$\langle e^{i(A+B)t} \sigma^n_z e^{-i(A+B)t} \rangle = \langle e^{i(A+B)t} e^{-i(A-B)t}\sigma^n_z  \rangle$.
The time derivative of this expression at $t=0$ is given by
$\langle [i (A+B) - i (A - B)] \sigma^n_z \rangle = 2 i \langle B \sigma^n_z \rangle$,
which is proportional to the expectation value of the interaction term of Hamiltonian in Eq.~(\ref{implem}),
$\partial_t \langle e^{i\mathcal{H}_m t} \sigma^n_z e^{-i \mathcal{H}_m t} \rangle|_{t=0} =  2 \langle g \sigma^n_x \left(a^2 + {a^\dag}^2\right) \rangle.$
The evolution under Hamiltonian $\mathcal{H}_m$ in the simulation picture is implemented  in the same way as the Hamiltonian in Eq.~(\ref{implem}), but selecting the laser phases $\phi_j$ to be $\frac{\pi}{2}$.
Moreover, expectation values for the generalized-parity operator $\Pi$ of Eq.~\eqref{parityoperator} can be extracted following the techniques described in Appendix~\ref{app_meas}.

In this section, we have introduced a trapped-ion scheme which allows one to experimentally investigate two-photon interactions in unexplored regimes of light-matter coupling, replacing photons in the model by trapped-ion phonons. It provides a feasible method to observe an interaction-induced spectral collapse in a two-phonon quantum Rabi model, approaching recent mathematical and physical results with current quantum technologies. Furthermore, the proposed scheme provides a scalable quantum simulator of a complex quantum system, which is difficult to approach with classical numerical simulations even for low number of qubits, due to the large number of phonons involved in the dynamics.

%% file: chap/chapter5.tex
\lettrine[lines=2, findent=3pt,nindent=0pt]{Q}{uantum} simulators are devices designed to mimic the dynamics of physical models encoded in quantum systems, enjoying high controllability and a variety of accessible regimes~\cite{Feynman1982}.  It was shown by Lloyd~\cite{Lloyd1996a} that the dynamics of any local Hamiltonian can be efficiently implemented in a universal digital quantum simulator, which employs a universal set of gates upon a register of qubits. Recent experimental demonstrations of this concept in systems like trapped ions~\cite{Lanyon2011} or superconducting circuits~\cite{Salathe2015, Barends2015, Barends2016} promise a bright future to the field. However, the simulation of nontrivial dynamics requires a considerable number of gates, threatening the overall accuracy of the simulation when gate fidelities do not allow for quantum error correction. Analog quantum simulators represent an alternative approach that is not restricted to a register of qubits, and where the dynamics is not necessarily built upon gates~\cite{Buluta2009, Georgescu2014}. Instead, a map is constructed that transfers the model of interest to the engineered dynamics of the quantum simulator. An analog quantum simulator, unlike digital versions, depends continuously on time and may not enjoy quantum error correction. In principle, analog quantum simulators provide less flexibility due to their lack of universality. 

\begin{figure}[h!]
\centering
\includegraphics[width=0.8 \columnwidth]{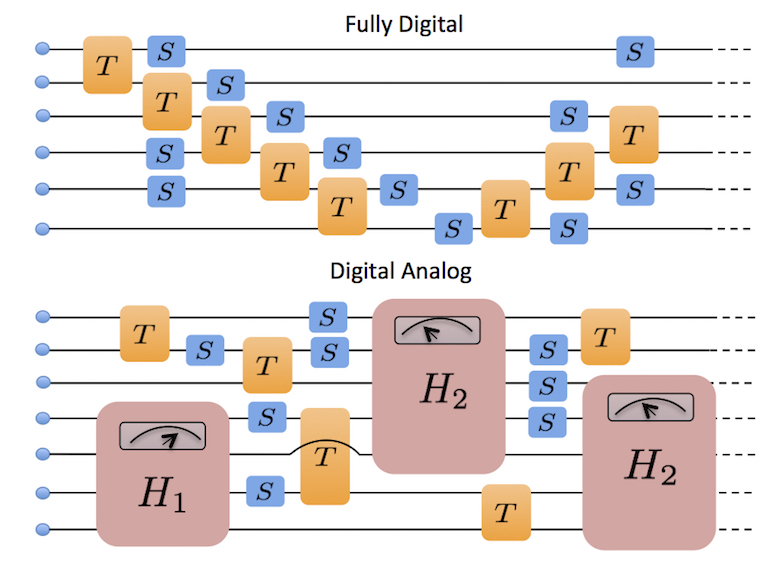}
\caption[Scheme of a fully digital versus a digital-analog protocol.]{ \footnotesize{{\bf Scheme of a fully digital versus a digital-analog protocol.} We depict the circuit representation of the digital and digital-analog approaches for quantum simulation. The fully digital approach is composed exclusively of single-qubit ($S$) and two-qubit ($T$) gates, while the digital-analog one significantly reduces the number of gates by including analog blocks. The latter, depicted in large boxes ($H_1$ and $H_2$), depend on tunable parameters, represented by an analog indicator, and constitute the analog quantum implementation of a given Hamiltonian dynamics.}}\label{fig:Fig1}
\end{figure}

In this chapter, we propose a merged approach to quantum simulation that combines digital and analog methods. We show that a sequence of analog blocks can be complemented with a sequence of digital steps to enhance the capabilities of the simulator. In this way, the larger complexity provided by analog simulations can be complemented with local operations providing flexibility to the simulated model.  We have named our approach digital-analog quantum simulation (DAQS), a concept that may be cross-linked to other quantum technologies. The proposed digital-analog quantum simulator is built out of two constitutive elements, namely, analog blocks and digital steps (see Fig.~\ref{fig:Fig1}). Digital steps consist of one- and two-qubit gates, the usual components of a universal digital quantum simulator. On the other hand, analog blocks consist in the implementation of a larger Hamiltonian dynamics, which typically involve more degrees of freedom than those involved in the digital steps. In general, analog blocks will depend on tunable parameters and will be continuous in time. In section~\ref{sec:HeisenbergIons}, we apply the concept to the simulation of spin chains following the Heisenberg model in trapped ions, while in section~\ref{sec:RabiCircuits}, we show how to simulate the Rabi and Dicke models in superconducting circuits.
%


\subsection{Digital-analog quantum simulation of spin models in trapped ions} \label{sec:HeisenbergIons}

Trapped-ion technologies represent an excellent candidate for the implementation of both, digital and analog quantum simulators~\cite{Blatt2012}. Using electromagnetic fields, a string of ions can be trapped such that  their motional modes display bosonic degrees of freedoms, and two electronic states of each atom serve as qubit systems. A wide variety of proposals for either digital or analog quantum simulations exist~\cite{Casanova2012,Mezzacapo2012,Casanova2011a,Alvarez-Rodriguez2013,Hayes2014,Cheng2015}, and several experiments have demonstrated the efficiency of these techniques in trapped ions, in the digital~\cite{Lanyon2011}, and analog cases. Examples of the latter include the quantum simulation of spin systems~\cite{Porras2004,Friedenauer2008, Kim2010,Bermudez2011,Britton2012,Islam2013,Jurcevic2014,Richerme2014} and relativistic quantum physics~\cite{Lamata2007,Gerritsma2010,Gerritsma2011}. 

In this section, we propose a method to simulate spin models in trapped ions using a digital-analog approach, consisting in a suitable gate decomposition in terms of analog blocks and digital steps. More precisely, we show that analog quantum simulations of a restricted number of spin models can be extended to more general cases, as the Heisenberg model, by the inclusion of single-qubit gates. Our proposal is exemplified and validated by numerical simulations with realistic trapped-ion dynamics. In this way, we show that the quantum dynamics of an enhanced variety of spin models could be implemented with substantially less number of gates than a fully digital approach. 


We consider a generic spin-$\!1\!/\!2$ Heisenberg model ($\hbar=1$)
\begin{equation}\label{HB}
H_{\rm{H}}=\sum_{i<j}^N J_{i\!j} \ {\vec{\sigma}}_{\!i}\cdot{\vec{\sigma}}_{\!j}=\sum_{i<j}^N J_{i\!j} (\sigma_i^x \sigma_{\!j}^x+\sigma_i^y \sigma_{\!j}^y+\sigma_i^z \sigma_{\!j}^z),
\end{equation}
where the vector of Pauli matrices ${\vec{\sigma}}_i=(\sigma_i^x,\sigma_i^y,\sigma_i^z)$ characterizes the spin of particle $i$, while $J_{i\!j}$  is the coupling strength between spins $i$ and $j$. To simulate this model, we will consider off-the-shelf interactions of ion chains~\cite{Jurcevic2014,Richerme2014}. More specifically, spin Hamiltonians
\begin{eqnarray}
H_{XX}=\sum_{i<j}J_{i\!j} \sigma_i^x \sigma_{\!j}^x, \ \ 
H_{XY}=\sum_{i<j}J_{i\!j} (\sigma_i^x \sigma_{\!j}^x +\sigma_i^y \sigma_{\!j}^y) 
\label{analogblock}
\end{eqnarray}
can be used as analog blocks, while single-qubit rotations $R_{x,y}(\theta)= \exp({-i\theta\sum_i \sigma_i^{x,y}})$ perform digital steps.

It is known that if a Hamiltonian can be decomposed into a sum of local terms, $H=\sum_{k} H_k$, its dynamics $U=e^{-iHt}$ can be approximated by discrete stepwise unitaries, according to the Trotter formula
\begin{equation}\label{Trotter}
U=(\prod_{k} e^{-iH_kt/ l})^l + O(t^2/l),
\end{equation}
where $l$ is the number of Trotter steps.  Here, the error of the approximation to the second order $O(t^2/l)$ is bounded by $\|O(t^2/l)\|_{\rm{sup}} \leq \sum_{k=2}^{\infty} l\|Ht/\hbar l\|^k_{\rm{sup}}/k!$ Thus, the digital error will decrease for a larger number of Trotter steps $l$. For the specific case of the antiferromagnetic Heisenberg Hamiltonian ($J_{i\!j}>0$), we have that $\|H\|_{\rm{sup}}=\sum_{i<j}^N J_{i\!j}$. This indicates the growth of the digital error bound with the number of spins in the chain, $N$, and with the range of the interaction between the spins. On the other hand, each particular decomposition of the Hamiltonian will show a different truncation error, which will grow linearly with the sum of the commutators of all the Hamiltonian terms~\cite{Lloyd1996a}. For the Heisenberg Hamiltonian, a suitable decomposition is given by $H_{\rm{H}}=H_{XY}+H_{ZZ}$. The dynamics of the $H_{ZZ}=\sum_{i<j}J_{i\!j} \sigma_i^z \sigma_{\!j}^z $ term can be generated with the proposed DAQS protocol, by combining the global qubit rotation $R_y(\pi/4)$ with the Ising-like dynamics $H_{XX}$. In this case, a Trotter step is given by the decomposition
\begin{equation}\label{UnitaryEvolution}
U^{\rm{H}}(t/l)=e^{-i H_{XY}t/l} R_y e^{-i H_{XX}t/l}R_y^\dagger ,
\end{equation}
where $R_y \equiv R_y(\pi/4)$. 

\begin{figure}[h!]
\centering
\includegraphics[width= 0.8 \columnwidth]{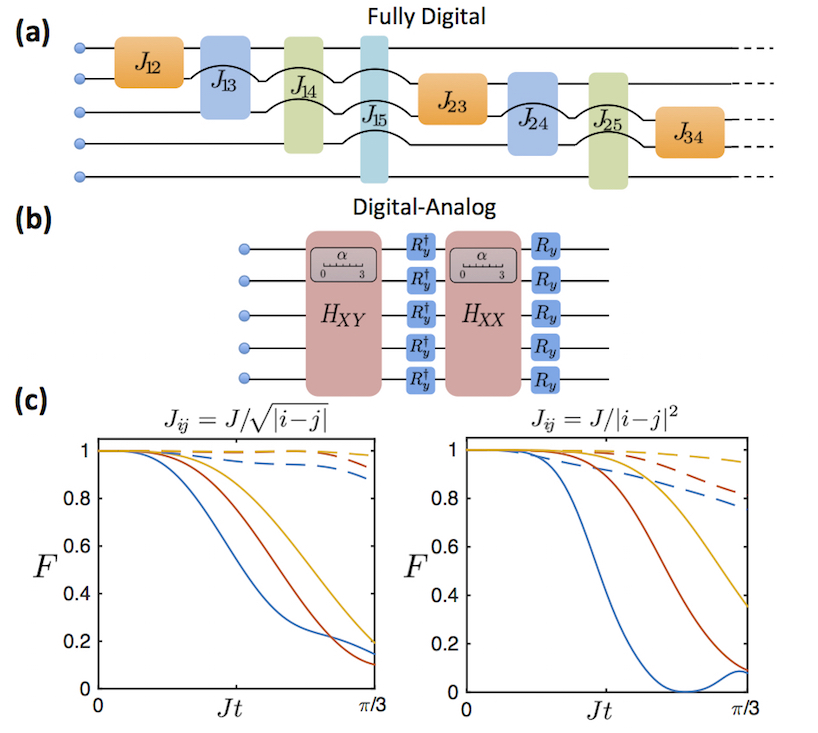}
\caption[Digitization of the Heisenberg model.]{\footnotesize{{\bf Digitization of the Heisenberg model.} (a) Scheme of a Trotter step of a purely digital quantum simulation for a generic spin dynamics with five sites. (b) Digital-Analog protocol of a Trotter step for the simulation of the Heisenberg model with tunable $\alpha$. (c) Fidelity loss obtained with the application of fully digital (solid lines) and digital-analog (dashed lines) protocols for the initial state $|\!\! \downarrow \downarrow \uparrow \downarrow \downarrow \rangle$. Blue (lower), orange (middle), and yellow (upper) colours represent one, two, and three Trotter steps, respectively. For the digital case, fidelity $F$ decays faster with $t$ for long-range interactions, while $F$ remains similar for the digital-analog protocol.}}\label{fig:Fig2}
\end{figure}
In Fig.~\ref{fig:Fig2}a and \ref{fig:Fig2}b, we show the circuit representation of the simulation algorithm following such a Trotter decomposition, as compared to its equivalent in a purely digital quantum simulator, that is, a simulator built only upon one- and two-qubit gates. The latter will need to include in the algorithm a two-qubit gate for each two-body interaction contained in the Hamiltonian. Even though these elementary gates can be realized with high control level, one needs to apply a large number of them, especially when the model has a long interaction range. The induced global fidelity loss is given not only by the imperfection of experimental gates, but also by the noncommutativity of these gates, which increases the Trotterization error. The inclusion of analog blocks like $H_{XY}$ and $H_{XX}$, accessible in trapped ions~\cite{Jurcevic2014,Richerme2014}, can become beneficial for the simulation of many-qubit spin models.

In Fig.~\ref{fig:Fig2}c, we plot the fidelity of the time evolution of two particular coupling regimes of the Heisenberg Hamiltonian for different numbers of Trotter steps, and we compare them to the fidelity of a purely digital algorithm. Numerical results show that the digital-analog approach achieves higher fidelities at all studied times for both models. Furthermore, the DAQS method represents a higher advantage with respect to the digital approach when the interaction range of the simulated model is longer. In general, a long-range Hamiltonian has more noncommuting terms that contribute to a larger digital error. DAQS takes advantage of its versatility in the Hamiltonian decompositions, as given by the sum of only two terms in the considered example. These terms do not always commute, but the associated commutator happens to be small for long-range spin interactions. Actually, for the limiting $J_{i\!j}=J$ case, DAQS produces no digital error, i.e., the analog blocks commute. In consequence, we consider this approach to represent a solid alternative for simulating generic long-range Heisenberg models. 

As we already mentioned, the digital-analog protocol  shown in Fig.~\ref{fig:Fig2}b needs two analog blocks per Trotter step, independently of the number of spins $N$. On the contrary, the number of entangling gates in a fully digital protocol grows with $N$. For generating each two-body interaction, at least a two-qubit gate is needed, and the number of two-body interactions will vary depending on the simulated model. This ranges from $N-1$, in the case of nearest neighbour interactions, to $N(N-1)/2$, in the long-range interaction case. Apart from the Trotter error, any realistic digital simulation has to deal with errors arising from the imperfection of experimental gates, which we quantify by the gate infidelity. In this respect, and in the long-range case, DAQS leads to a better result as long as the analog-block gate infidelity fulfills $\epsilon_{AB}\leq \frac{N(N-1)}{4}\epsilon_{T}$, where $\epsilon_{T}$ is the two-qubit-gate infidelity. Consequently, a purely digital proposal would need to compensate the larger number of gates with better gate fidelities~\cite{Ballance2016}. However, it is fair to assume that the two-qubit gate fidelity will decrease when we increase the number of ions in the trap~\cite{Monz2011}. The gate fidelity of the analog block, on the other hand, will also decrease with the size of the system. 

\subsubsection{Proposal for an experimental implementation}

The experimental implementation of the considered digital steps, which correspond to local spin rotations, is easily achieved in trapped ions through carrier transitions~\cite{Haffner2008}. The spin-spin interactions, corresponding to the proposed analog blocks, were first suggested in Ref.~\cite{Porras2004}, and have been implemented in several experiments~\cite{Friedenauer2008,Kim2010, Jurcevic2014}. To show their derivation, we first consider a set of $N$ two-level ions confined in a linear trap, coupled to the $2N$ radial modes of the string by a pair of non-copropagating monochromatic laser beams. These lasers are oriented orthogonally to the ion chain, with a $45$ degree angle with respect to the x and y radial directions. We work in an interaction picture with respect to the uncoupled Hamiltonian ${H_0=\frac{\omega_0}{2} \sum_j \sigma_j^z +\sum_m \nu_m a_m^\dagger a_m }$. Here, $\omega_0$ is the frequency of the electronic transition of the two-level ion, and $\nu_m$ the frequency of the transverse motional mode $m$ of the ion string, with  annihilation(creation) operator $a_m$($a_m^\dagger$). The interaction Hamiltonian for the system reads
\begin{equation}\label{IntHamil}
H^I = \sum_{j=1}^N \Omega_{\!j}\sigma_j^+  e^{-i(\epsilon t-\phi_L)} \sin\big[{\sum_{m=1}^{2N}\!\!\eta_{j\!,m}(a_m e^{-i\nu_m t} + a^\dag_m e^{i\nu_m t})\big]} + {\rm{H.c.}},
\end{equation}
where $\Omega_j$ is the Rabi frequency of the laser for the $j$th ion, $\epsilon=\omega_L-\omega_0$ is the detuning of the laser frequency with respect to the electronic transition, $\phi_L$ is the laser phase, and $\eta_{j,m}$ is the Lamb-Dicke parameter, which is proportional to the displacement of the $j$th ion in the $m$th collective mode~\cite{James1998}.

To obtain the effective spin-spin interactions, the two pairs of laser beams are tuned off-resonantly to the red and blue sidebands of the $2N$ radial modes with symmetric detunings ${\epsilon_{\pm}=\pm(\nu_{\rm{COM}}+\Delta)}$, where $\Delta \ll \nu_{\rm{COM}}$. Here, $\Delta$ denotes the detuning of the laser with respect to the first blue sideband of the motional mode with highest frequency. This corresponds to the center-of-mass (COM) mode in the radial $x$-axis, in the case where this axis has the highest trapping frequency ($\omega_x > \omega_y$). The Lamb-Dicke regime, which corresponds to keeping only the linear term in the expansion of the sine in Eq.~\ref{IntHamil}, can be considered when $|\eta_{j,m}|\sqrt{\langle a^\dag_ma_m\rangle} \ll 1$. Moreover, we can also neglect fast oscillating terms under the so called vibrational rotating-wave approximation (RWA), which holds when $|\eta_{j\!,m}\Omega_{\!j}| \ll \nu_m$. All in all, the resulting Hamiltonian is given by
\begin{equation}\label{BicHamil}
H_{\rm{bic}}=\sum_{j=1}^{N} \sum_{m=1}^{2N}\Omega_{\!j}\eta_{j\!,m} \big(\sigma_j^++\sigma_j^-\big) \big(a_me^{i\Delta_m t}+a_m^\dagger e^{-i\Delta_m t} \big),
\end{equation}
where $\Delta_m=\Delta + (\nu_{\rm{COM}}-\nu_m) > \Delta$. If $|\eta_{j\!,m}\Omega_{\!j}| \ll \Delta_m$, we can perform the adiabatic elimination of the motional modes, which are only virtually excited. As a result, a second order effective Hamiltonian with only spin-spin interaction terms arises
\begin{equation}\label{effectiveHamil}
H_{\rm{eff}}=\sum_{i<j}^N J_{i\!j} \sigma_i^x \sigma_{\!j}^x=H_{XX},
\end{equation}
where the spin-spin coupling is given by
\begin{equation}\label{effectiveCoupling}
J_{i\!j}=2\Omega_i \Omega_{\!j} \sum_{m=1}^{2N} \frac{\eta_{i\!,m}\eta_{j\!,m}}{\Delta_m}\approx \frac{J}{|i-j|^\alpha},
\end{equation}
with $J\equiv {\sum_i J_{i,i+\!1}}/{(N\!-\!1)} > 0$ and tunable $0 < \alpha < 3$~\cite{Britton2012}.
In Fig.~\ref{fig:Fig3}a, we plot the spin-spin coupling matrix obtained for five $^{40}$Ca$^+$ ions, with the values $\Delta=(2\pi)60$kHz for the detuning, ${\Omega=(2\pi)62}$kHz for the Rabi frequency, $\vec{\omega}=(2\pi)(2.65,2.63,0.65)$MHz for the trapping frequencies and $\lambda=729$nm for the laser wavelength. The coupling matrix approximately follows the power-law decay with $\alpha \approx 0.6$, which essentially can be tuned varying $\Delta$ and $\omega_z$. Here we have assumed $\Omega_j=\Omega$, which is safe for the five ion chain that we are considering~\cite{Richerme2014}. If we were to consider longer chains one would need to have into account that the laser intensity profile has a Gaussian shape and therefore that the outermost ions may have smaller Rabi frequency than the central ones. This would result in a modification of the coupling scaling law. The $XY$ Hamiltonian can be generated introducing a slight asymmetry in the detuning of the bichromatic laser $\epsilon_{\pm}=\pm(\nu_{\rm{COM}}+\Delta)+\delta$, $\delta \ll \Delta$. This introduces in the spin ladder operators a time-dependent phase factor, $\sigma^+ \rightarrow \sigma^+e^{-i\delta t}$ and $\sigma^- \rightarrow \sigma^- e^{i\delta t}$, making several terms in the effective Hamiltonian negligible under the RWA. The effective Hamiltonian, then, reads
\begin{equation}\label{effectiveHamil1}
H_{\rm{eff}}=\sum_{i<j}^N J_{i\!j} (\sigma_i^+ \sigma_{\!j}^- +\sigma_i^- \sigma_{\!j}^+) + \frac{\delta}{\Delta}\sum_{j=1}^N B_{\!j} \sigma_j^z \approx \frac{1}{2} H_{XY},
\end{equation}
where $B_{\!j}=\Delta\Omega_j^2 \sum_{m}{\!(\!\eta_{j,m}/\Delta_m)}^2 (a_m^\dagger a_m \! +\!1\!/2) $. In addition to the $XY$ interaction, terms proportional to $ a_m^\dagger a_m\sigma_{\!j}^z$ appear. However, the contribution of these terms is smaller than the spin-spin term by a factor of $\delta/\Delta$ and can be neglected in the case where a small number of phonons is excited ($\langle B_j \rangle \sim J_{i\!j}$). The $XY$ Hamiltonian can also be implemented using a single monochromatic laser field tuned off-resonantly to the first blue sidebands of the $2N$ modes. As for the bichromatic case, the vibrational modes are only virtually excited and this gives rise to an effective spin-spin Hamiltonian. Nevertheless, in this case the strength of the terms $a_m^\dagger a_m\sigma_{\!j}^z $ is of the same order of magnitude as that of the spin-spin coupling term. This makes this last approach more sensitive to the heating of the phononic degrees of freedom.

\subsubsection{Numerical simulations}

\begin{figure}[h!]
\centering
\includegraphics[width=0.8 \columnwidth]{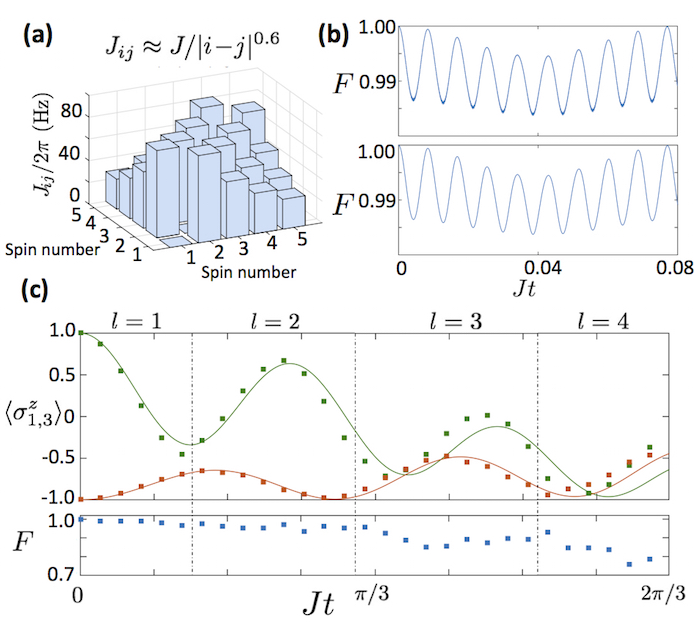}
\caption[Numerical simulations of the implementation of spin model in trapped ions.]{\footnotesize{ {\bf Numerical simulations of the implementation of spin model in trapped ions.} (a) Long-range ($\alpha \approx 0.6$) spin-spin coupling matrix $J_{ij}$ for $N=5$ spins.  (b) Fidelity of the $H_{XY}$ analog block for the state ${|\!\! \downarrow \downarrow \uparrow \downarrow \downarrow \rangle}$, with (lower plot) and without (upper plot) applying the vibrational RWA. The fidelity is periodic in time and thus we just plot one period. The numerical simulation assumes only the Lamb-Dicke regime, therefore accounting for the main sources of error that are the RWA and the adiabatic elimination. (c) Magnetization of the first (orange, lower curve) and third (green, upper curve) spins $\langle\sigma_{1,3}^z\rangle$ (upper plot) and the state fidelity of the digital-analog protocol (lower plot) versus time, for the protocol in Fig. \ref{fig:Fig2}b with initial state as in (b). Solid curves correspond to the ideal state produced by the Heisenberg Hamiltonian in Eq. (\ref{HB}), while dots correspond to the state produced by the DAQS approach. We divide the time interval into regions and simulate each time region using optimized numbers of Trotter steps, in order to maximize the state fidelity produced by our protocol.}}\label{fig:Fig3}
\end{figure}

In Fig.~\ref{fig:Fig3}b, we depict the fidelity of an $H_{XY}$ analog block for five ions as a function of time. For numerical feasibility, instead of considering the ten radial modes present in the five ion case, we have considered a single COM mode with an effective Lamb-Dicke parameter $\eta^{\rm{eff}}\equiv \Omega^{-1}\!\sqrt{J \Delta /2}$ that represents the effect of all radial modes.  This can be done as long as the chosen effective Lamb-Dicke parameter results in a coupling strength of the same order of magnitude of the one under study. This is true because the infidelity of the adiabatic approximation depends directly on the coupling strength J. Moreover, we choose the COM mode because it is the most unfavorable one for the approximation in terms of its detuning $\Delta$. In this manner we are able to give a safe fidelity estimate, overcoming the computationally demanding task of simulating the model with the ten motional modes. The analog blocks result from an effective second-order Hamiltonian, and their fidelity is subject to the degree of accuracy of the involved approximations. In the case of the $H_{XX}$ interaction, the greater the $\Delta$, the better the approximation and the gate fidelity. However, the simulation time is longer because $J_{i\!j}$ is inversely related to $\Delta$. The same is true for the $H_{XY}$ interaction, but the latter involves additional approximations that require $\delta \ll \Delta$ and $J \ll \delta$. For $\Delta=(2\pi)60$kHz and $\delta=(2\pi)3$kHz, the $H_{XY}$ gate infidelity can go up to $\epsilon_{AB}\approx 0.02$, as we can observe in Fig.~\ref{fig:Fig3}b. Obviously, the $H_{XX}$ analog block gives better results, since it is subject to fewer approximations.  We have also plotted the time evolution for the Hamiltonian in Eq. (\ref{BicHamil}), in which the vibrational RWA has been applied, and the Lamb-Dicke regime has been considered. It can be observed that there is no appreciable difference between both plots, which validates the vibrational RWA in the considered parameter regimes.

A numerical simulation of the dynamics produced by the digital-analog protocol in Fig.~\ref{fig:Fig2}b is presented in Fig. \ref{fig:Fig3}c. More precisely, we plot the magnetization of the first (orange, lower curve) and third (green, upper curve) spins, $\langle \sigma_1^z (t)\rangle$ and $\langle \sigma_3^z (t)\rangle$ (upper plot), and the fidelity associated with the digital-analog protocol (lower plot) as a function of time, in a five-ion chain. In order to maximize the fidelity of the quantum simulation, we need to reach a compromise between the number of Trotter steps, which increases the fidelity by reducing the digital error, and the total number of gates, which lowers the total fidelity by increasing the accumulated gate error. For that, we divide the time interval in regions, and we numerically simulate each region with the optimal number of Trotter steps. The fidelity of single-qubit gates is in general high~\cite{Harty2014}, so we treat them as perfect in our calculation. As can be seen in Fig. \ref{fig:Fig3}c, we reach times of $Jt=2\pi/3$ with a state fidelity of approximately $70\%$, assuming $\Delta=(2\pi)60$kHz and $\delta=(2\pi)3$kHz. As we discussed, we could lower the error coming from the analog block by taking a larger value for $\Delta$ and, thus, improve the fidelity of the quantum simulation. However, this would increase the experimental time, which is limited by the coherence time of the system. For our analysis, we have considered real time dynamics of up to 13ms, which is below coherence times in trapped ion chains~\cite{Jurcevic2014}.

Summarizing, in this section we have introduced the digital-analog approach to quantum simulation of spin models, which represents a solid alternative to universal digital quantum simulations, whenever gate fidelities are not high enough to allow for quantum error correction. Also, we have validated through numerical simulations that an implementation of our protocol is within experimental reach. With the proposed DAQS approach, we expect that a larger number of ions can be employed when compared with purely digital methods, reaching the size of analog quantum simulators~\cite{Jurcevic2014,Richerme2014}. The natural continuation of this research line is to explore how other models could benefit from the DAQS techniques in trapped ions. Under the general argument that analog blocks concentrate the complexity of the model in high fidelity analog simulations, it is reasonable to expect that plenty of models will profit from such a simulation procedure. We consider the introduced DAQS techniques to be an important ingredient enhancing the toolbox of quantum simulations in trapped ions.


\subsection{Digital-analog implementation of the Rabi and Dicke models in superconducting circuits} \label{sec:RabiCircuits}
\fancyhead[LE]{5.2 \  DA implementation of the Rabi and Dicke models in superconducting circuits}
In section~\ref{sec4}, we have already introduced the Rabi class of Hamiltonians that describe the most fundamental interaction of quantum light and quantum matter, consisting of the dipolar coupling of a two-level system with a single radiation mode~\cite{Rabi1936}. The Dicke model~\cite{Dicke1954} was later introduced to generalize this interaction to an ensemble of $N$ two-level systems. Typically, the coupling strength is small compared to the transition frequencies of the two-level system and the radiation mode, leading to effective Jaynes-Cummings and Tavis-Cummings interactions, respectively, after performing a  RWA. This introduces a $U(1)$ symmetry and integrability to the model for any $N$~\cite{Jaynes1963,Tavis1968}. Recently, analytical solutions for the generic quantum Rabi and Dicke model for $N=3$ were found~\cite{Braak2011,Braak2013}. However, the general case for arbitrary $N$ is still unsolved, while its direct study in a physical system remains an outstanding challenge. A variety of quantum platforms, such as cavity QED, trapped ions, and circuit QED, provide a natural implementation of the Jaynes-Cummings and Tavis-Cummings models, due to the weak qubit-mode coupling strength. But, an experimental observation of the full quantum Rabi and Dicke models in all parameter regimes has not yet been realized. In particular, the quantum simulation~\cite{Feynman1982} of the Dicke Hamiltonian could outperform analytical and numerical methods, while enabling the simulation of engineered super-radiant phase transitions~\cite{Hepp1973,Wang1973,Carmichael1973}.  

In this section, we propose the digital-analog quantum simulation of the quantum Rabi and Dicke models in a circuit QED setup, provided only with Jaynes-Cummings and Tavis-Cummings interactions, respectively. We show how the rotating and counter-rotating contributions to the corresponding dynamics can be effectively realized with digital techniques. By interleaved implementation of rotating and counter-rotating steps, the dynamics of the quantum Rabi and Dicke models can be simulated for all parameter regimes with negligible error. Lastly, we show how a relativistic Dirac dynamics can be retrieved in the limit where the mode frequency cancels.

\subsubsection{The digital-analog simulation protocol}
We start by considering a generic circuit QED setup consisting of a charge-like qubit, e.g. a transmon qubit~\cite{Koch2007}, coupled to a microwave resonator. The setup is well described by the Hamiltonian ($\hbar=1$)~\cite{Blais2004} 
\begin{equation}
H=\omega_r a^{\dagger}a +\frac{\omega_q}{2}\sigma^z +g(a^{\dagger}\sigma^-+a\sigma^+),\label{QubitResHam}
\end{equation}
where $\omega_r$ and $\omega_q$ are the resonator and qubit transition frequencies, $g$ is the resonator-qubit coupling strength, $a^{\dagger}$$(a)$ is the creation(annihilation) operator for the resonator mode, and $\sigma^{\pm}$ raise and lower excitations on the qubit. 
The capacitive interaction in Eq.~(\ref{QubitResHam}) excludes excitations of the higher levels of the qubit device, because typically the coupling $g$ is much smaller than other transition frequencies of the system. By trying to design setups with larger capacitive couplings, pushing them above dispersive regimes, one starts to populate the higher levels of the transmons, producing unwanted leakage. On the other hand, methods based on orthogonal drivings of the qubits~\cite{Ballester2012,Pedernales2013a} may increase the resonator population. Here, we show that the dynamics of the quantum Rabi Hamiltonian 
\begin{equation}
H_R=\omega^R_r a^{\dagger}a +\frac{\omega^R_q}{2}\sigma^z +g^R\sigma^x(a^{\dagger}+a)\label{RabiHam}
\end{equation}
can be encoded in a  superconducting setup provided with a Jaynes-Cummings interaction, as in Eq.~(\ref{QubitResHam}), using a digital expansion. 

The quantum Rabi Hamiltonian in Eq.~(\ref{RabiHam}) can be decomposed into two parts, $H_R=H_1+H_2$, where
\begin{eqnarray}
&&H_1=\frac{\omega^R_r}{2} a^{\dagger}a +\frac{\omega^1_q}{2}\sigma^z +g(a^{\dagger}\sigma^-+a\sigma^+) , \nonumber \\
&&H_2=\frac{\omega^R_r}{2} a^{\dagger}a -\frac{\omega^2_q}{2}\sigma^z +g(a^{\dagger}\sigma^++a\sigma^-) ,
\label{Ham12} 
\end{eqnarray}
and we have defined the qubit transition frequency in the two steps such that $\omega_q^1-\omega_q^2=\omega^R_q$. These two interactions can be simulated in a typical circuit QED device with  fast control of the qubit transition frequency. Starting from the qubit-resonator Hamiltonian in Eq.~(\ref{QubitResHam}), one can define a frame rotating at frequency $\tilde{\omega}$, in which the effective interaction Hamiltonian becomes 
\begin{equation}
\tilde{H}=\tilde{\Delta}_ra^{\dagger}a+\tilde{\Delta}_q\sigma^z+g(a^{\dagger}\sigma^-+a\sigma^+),\label{IntHam}
\end{equation}  
with $\tilde{\Delta}_r=(\omega_r-\tilde{\omega})$ and $\tilde{\Delta}_q=\left(\omega_q-\tilde{\omega}\right)/2$. Therefore, Eq.~(\ref{IntHam}) is equivalent to $H_1$, following a proper redefinition of the coefficients.
The counter-rotating term $H_2$ can be simulated by applying a local qubit rotation to $\tilde{H}$ and a different detuning for the qubit transition frequency,
\begin{equation}
e^{-i \pi\sigma^x/2}\tilde{H}e^{i \pi\sigma^x/2}=\tilde{\Delta}_ra^{\dagger}a-\tilde{\Delta}_q\sigma^z+g(a^{\dagger}\sigma^++a\sigma^-).\label{RotHam}
\end{equation}
\begin{figure}[h!]
\centering
\includegraphics[width=0.8 \columnwidth]{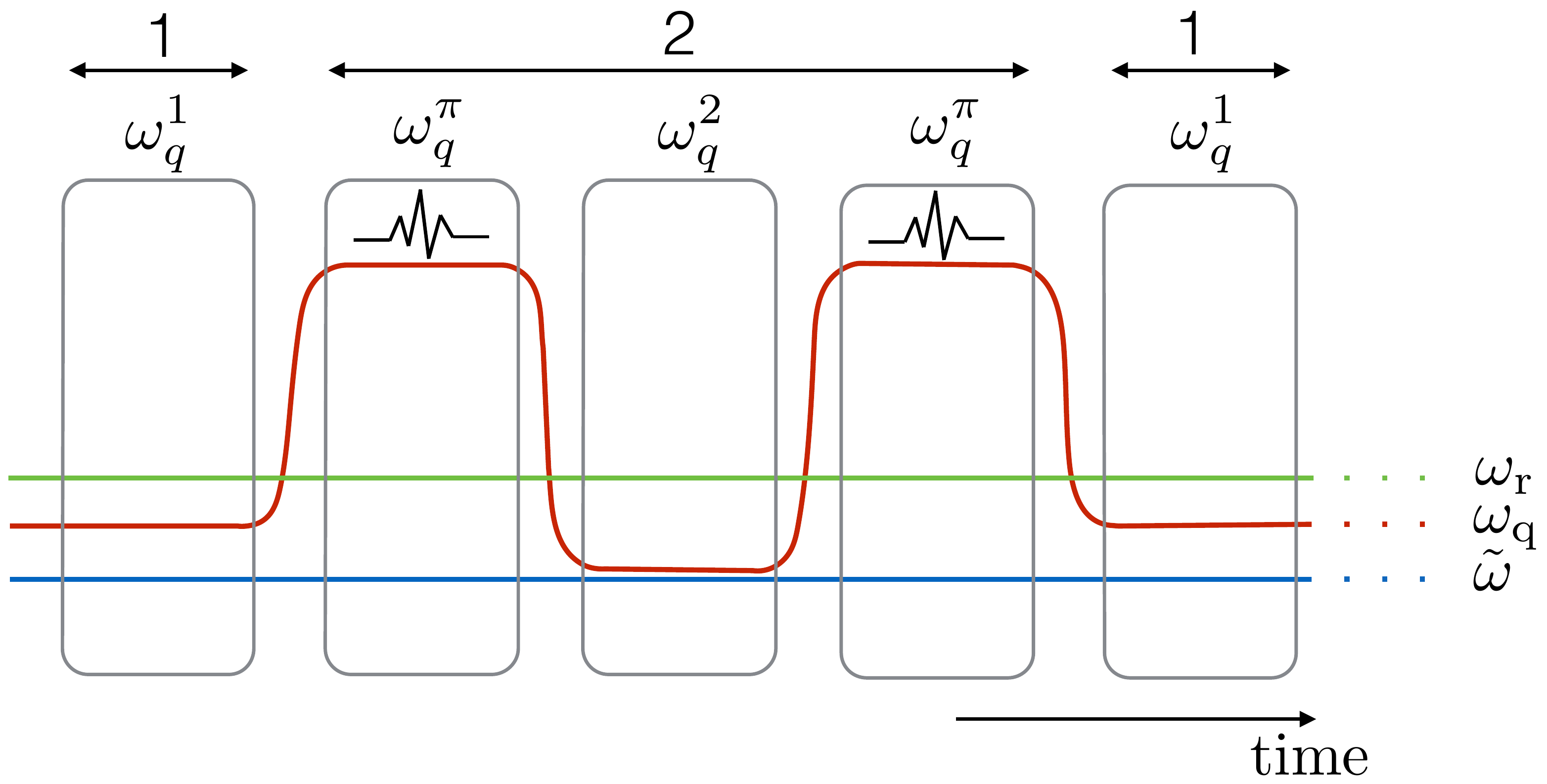}
\caption[Frequency scheme of the stepwise implementation for the QRM Hamiltonian in circuits.]{\footnotesize{{\bf Frequency scheme of the stepwise implementation for the QRM Hamiltonian in circuits.} A transmon qubit of frequency $\omega_q$ is interacting with a microwave resonator, whose transition frequency is $\omega_r$. The interactions $H_{1,2}$ in Eq.~(\ref{Ham12}) are simulated respectively with a Jaynes-Cummings interaction (step 1), and another one with different detuning, anticipated and followed by $\pi$ pulses (step 2).}} \label{FrequencyScheme}
\end{figure}
By choosing different qubit-resonator detuning for the two steps, $\tilde{\Delta}^1_q$ for the first one and $\tilde{\Delta}^2_q$ for the rotated step, one is able to simulate the quantum Rabi Hamiltonian, Eq.~(\ref{RabiHam}), via digital decomposition~\cite{Lloyd1996a}, by interleaving the simulated interactions. The frequency scheme of the protocol is shown in Fig.~\ref{FrequencyScheme}. Standard resonant Jaynes-Cummings interaction parts with different qubit transition frequencies are interrupted by microwave pulses, in order to perform customary qubit flips~\cite{Blais2007}. This sequence can be repeated according to the digital simulation scheme to obtain a better approximation of the quantum Rabi dynamics.

The simulated Rabi parameters can be obtained as a function of the physical parameters of the setup by inverting the derivation presented above. In this way, one has that the simulated bosonic frequency is related to the resonator detuning $\omega_r^R=2\tilde{\Delta}_r$, the two-level transition frequency is related to the transmon frequency in the two steps, $\omega_q^R=\tilde{\Delta}_q^1-\tilde{\Delta}_q^2$, and the coupling to the resonator remains the same, $g^R=g$. Notice that even if the simulated two-level frequency $\omega_q^R$ depends only on the frequency difference, large detunings $\tilde{\Delta}_q^{1(2)}$ will affect the total  fidelity of the simulation. In fact, since the digital error depends on the magnitude of individual commutators between the different interaction steps, using larger detunings linearly increases the latter, which results in fidelity loss of the simulation. To minimize this loss, one can choose, for example, the transmon frequency in the second step to be tuned to the rotating frame, such that $\tilde{\Delta}_q^2=0$. Nevertheless, to avoid sweeping the qubit frequency across the resonator frequency, one may choose larger detunings.
To estimate the loss of fidelity due to the digital approximation of the simulated dynamics, we consider a protocol performed with typical transmon qubit parameters~\cite{Koch2007}.
We estimate a resonator frequency of $\omega_r/2\pi=7.5$~GHz, and a transmon-resonator coupling of $g/2\pi=100$~MHz. The qubit frequency $\omega_q$ and the frequency of the rotating frame $\tilde{\omega}$ are varied to reach different parameter regimes.
\\
\\

To perform the simulation for the quantum Rabi model with ${g^R/2\pi=\omega^R_q/2\pi}={\omega^R_r/2\pi=100}$~MHz, for example, one can set $\omega^1_q/2\pi=7.55$~GHz, $\omega^2_q/2\pi=7.45$~GHz. In this way, one can define an interaction picture rotating at $\tilde{\omega}/2\pi=7.45$~GHz  to encode the dynamics of the quantum Rabi model with minimal fidelity loss.
Considering that single-qubit rotations take approximately~$\sim10$~ns, tens of Trotter steps could be comfortably performed within the coherence time. Notice that, in performing the protocol, one has to avoid populating the third level of the transmon qubit. Considering transmon anharmonicities of about $\alpha=-0.1$, for example, in this case one has third level transition frequencies of $6.795$~GHz and $6.705$~GHz. Therefore, given the large detuning with the resonator, it will not be populated.
Similarly, by choosing different qubit detunings and rotating frames, one can simulate a variety of parameter regimes, e.g. see Table~\ref{Table}. 
\begin{table}
\centering
\caption[Simulated quantum Rabi dynamics parameters versus frequencies of the system.]{\footnotesize{{\bf Simulated quantum Rabi dynamics parameters versus frequencies of the system.} For all entries in the right column, the resonator frequency is fixed to $\omega_r/2\pi=7.5$~GHz, and the coupling $g^R/2\pi=100$~MHz. Frequencies are shown up to a $2\pi$ factor.}} \label{Table}
\vspace{0.5cm}
\begin{tabular}{l|l}
\hline\hline

  $g^R=\omega_q^R/2=\omega_r^R/2$ \;\; & $\tilde{\omega}=7.4$~GHz, $\omega_q^1-\omega_q^2=200$~MHz \;\; \\
  $g^R=\omega_q^R=\omega_r^R$ \;\; & $\tilde{\omega}=7.45$~GHz, $\omega_q^1-\omega_q^2=100$~MHz \;\; \\
  $g^R=2\omega_q^R=\omega_r^R$ \;\; & $\tilde{\omega}=7.475$~GHz, $\omega_q^1-\omega_q^2=100$~MHz \;\; \\

\hline\hline
\end{tabular}
\end{table}

\subsubsection{Experimental considerations}
\begin{figure}[h]
\centering
\includegraphics[width=0.8 \columnwidth]{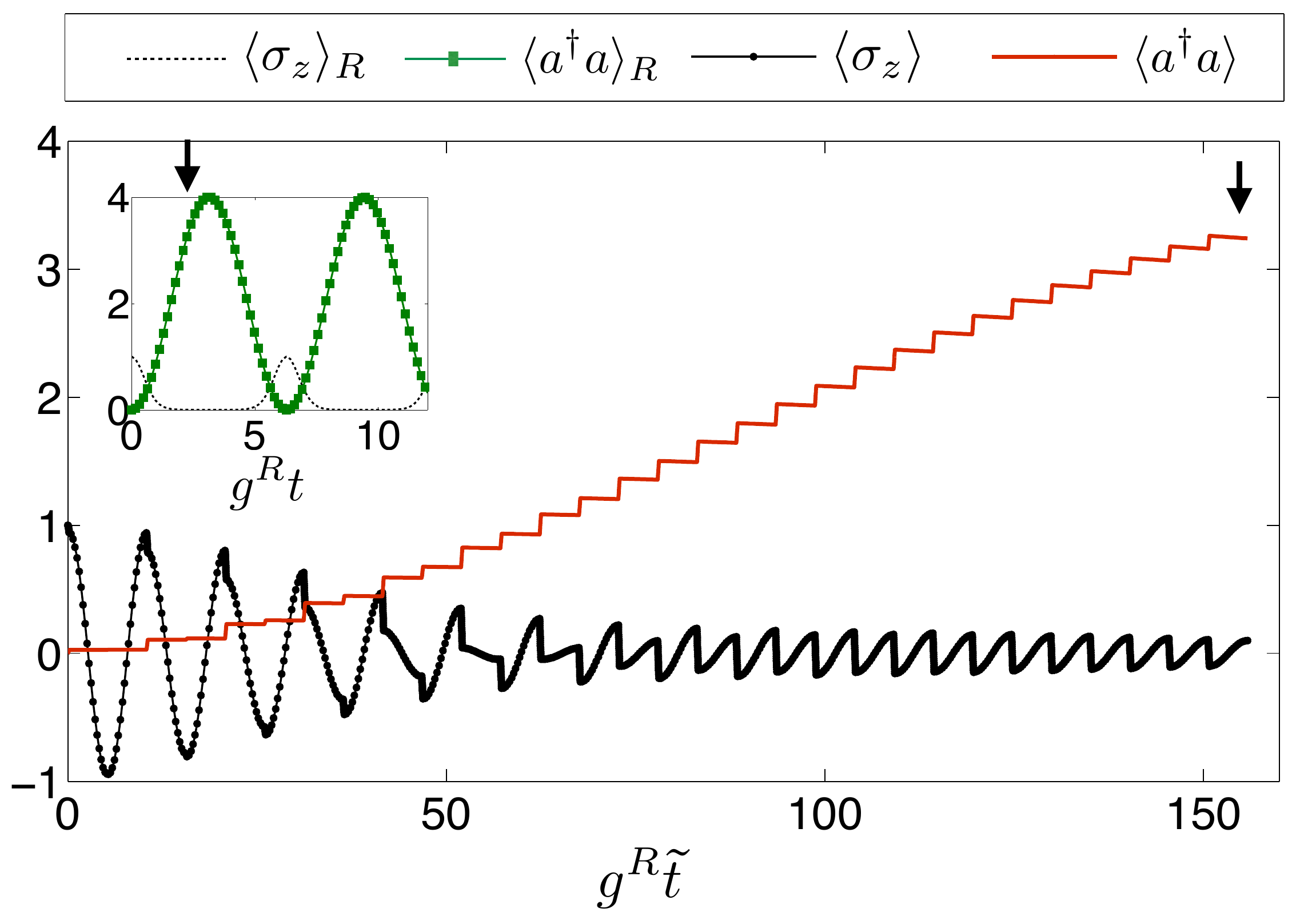}
\caption[A transmon qubit and microwave resonator simulating the quantum Rabi Hamiltonian.]{\footnotesize{{\bf A transmon qubit and microwave resonator simulating the quantum Rabi Hamiltonian.} The simulated regime corresponds to $g^R=\omega_r^R$, $\omega_q^R=0$. The ideal dynamics, plotted in the inset, shows collapses and revivals of the photon and qubit population. The latter are recovered via sequential qubit-resonator interactions and qubit flips. The photon population is pumped to the expected value at the time marked by the arrow. Note that the simulating time $\tilde{t}$ is different from the simulated time $t$.}}  \label{Dynamics}
\end{figure}
In order to capture the physical realization of the simulation, we plot in Fig.~\ref{Dynamics} the behavior of the transmon-resonator system during the simulation protocol. We numerically integrate a master equation, alternating steps of Jaynes-Cummings interaction with single-qubit flip pulses. We consider $\dot{\rho}=-i[H,\rho]+\kappa L(a)\rho+\Gamma_\phi L(\sigma^z)\rho+\Gamma_- L(\sigma^-)\rho$, with  Jaynes-Cummings terms $\tilde{H}=\tilde{\Delta}_ra^{\dagger}a+\tilde{\Delta}_q\sigma^z+g(a^{\dagger}\sigma^-+a\sigma^+)$, alternated with qubit-flip operations $H_f=f(t)\sigma^x$, where $f(t)$ is a smooth function such that $\int_0^{T_f}f(t)dt=\pi/2$, $T_f$ being the qubit bit-flip time. The quantum dynamics is affected by Lindblad superoperators $\Gamma_\phi L(\sigma^z)\rho$, $\Gamma_- L(\sigma^-)\rho$, and $\kappa L(a) \rho$ modelling qubit dephasing, qubit relaxation and resonator losses. We have defined $L(A)\rho=(2A\rho A^{\dagger}-A^{\dagger}A\rho-\rho A^{\dagger}A)/2$. We set a resonator-qubit coupling of $g/2\pi=80$~MHz, and a frame rotating at the qubit frequency, $\tilde{\Delta}_q=0$, $\tilde{\Delta}_r/2\pi=40$~MHz. We consider $\Gamma_-/2\pi=30$~kHz, $\Gamma_\phi/2\pi=60$~kHz, and $\kappa/2\pi=100$~kHz. The inset of Fig.~\ref{Dynamics} shows collapses and revivals of both the photon and spin dynamics, which are typical signatures of the regimes of the quantum Rabi dynamics dominated by the coupling strength.  We consider prototypical DSC dynamics, with $\omega_q^R=0$, and $g^R=\omega_r^R$. Notice that to encode the dynamics corresponding to a certain simulated time $t$, one needs the quantum simulator to run for a simulating time $\tilde{t}$, that depends on the specific gate times of the experiment. We choose to set the simulation at the time marked by the black arrow, close to the photon population peak in the inset. A simulation with $15$ digital steps is then performed. The time for a single qubit flip pulse is set to $T_f=10$~ns. Periodic collapses and revivals of the bosonic population of the quantum Rabi model $\langle{a^{\dagger}a}\rangle_R$ are shown as a function of time, in the inset. The ideal spin and bosonic populations $\langle\sigma_z\rangle_R$ and $\langle a^{\dagger}a\rangle_R$, evolving according to the quantum Rabi Hamiltonian, are shown to be in good agreement with the simulated ones, $\langle\sigma_z\rangle$ and $\langle a^{\dagger}a\rangle$, at the final simulated time. In fact, during the Jaynes-Cummings interaction parts, photons are pumped into the resonator. Afterwards, before the photon population starts to decrease due to excitation exchanges with the transmon qubit, a qubit flip further enhances the photon production. 

\begin{figure}[h!]
\centering
\includegraphics[width=0.8 \columnwidth]{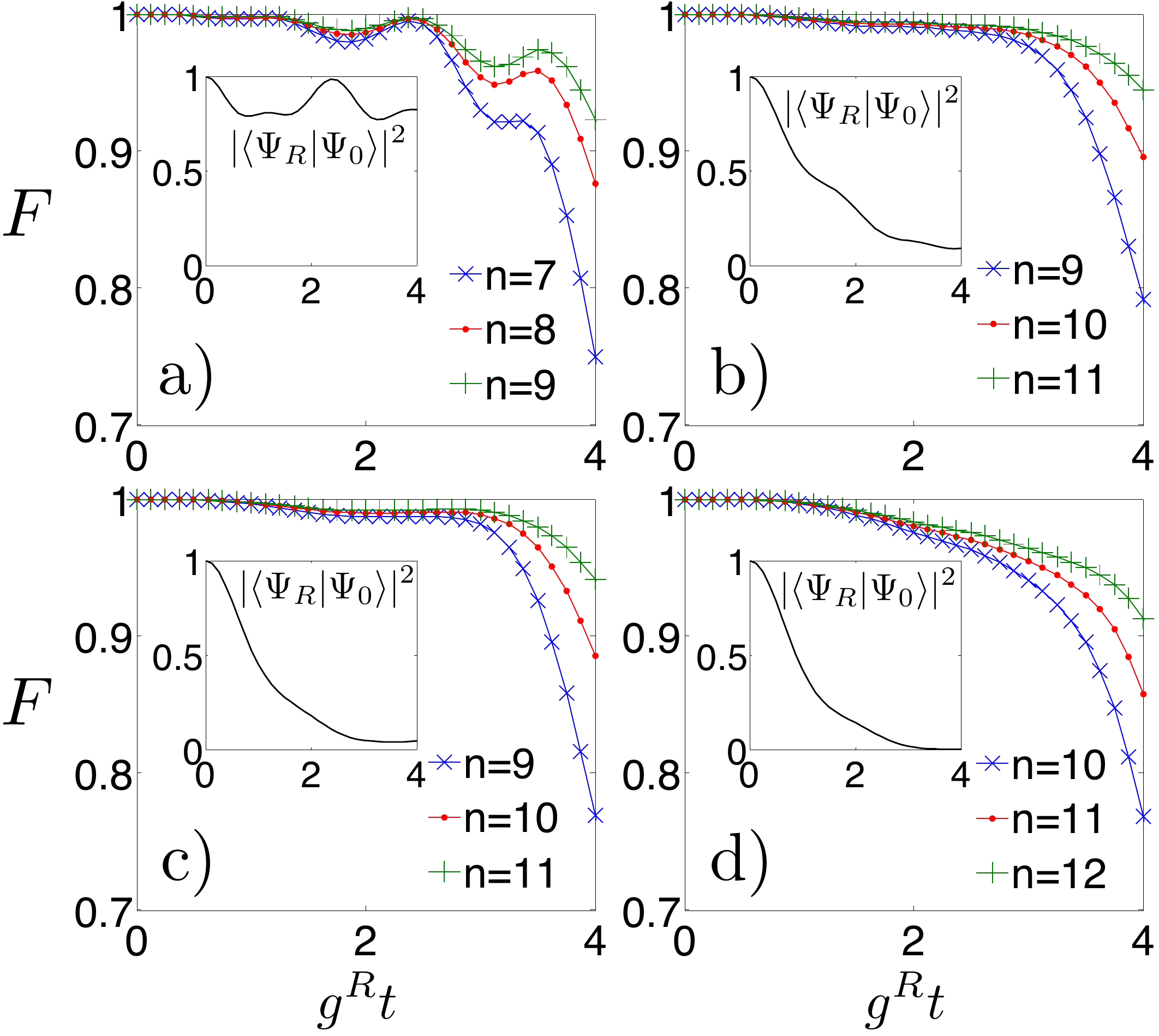}
\caption[Fidelities of the simulation of the QRM model in circuits.]{ \footnotesize{{\bf Fidelities of the simulation of the QRM model in circuits.} Time evolution of the fidelity $F=|\langle\Psi_S|\Psi_R\rangle|^2$ of state $| \Psi_S \rangle$ evolving according to the digitized protocol with respet to the ideal state $| \Psi_R \rangle$ evolving according to the quantum Rabi dynamics is plotted, with a)~$g^R=\omega^R_r/2=\omega^R_q/2$, b)~$g^R=\omega^R_r=\omega^R_q$, c)~$g^R=2\omega^R_r=\omega^R_q$, and d)~$g^R=2\omega^R_r=1.5\omega^R_q$. The simulation is performed for different number $n$ of Trotter steps. Black curves in the insets show the overlap of the ideal evolved state with the one at time $t=0$, $|\langle\Psi_R|\Psi_0\rangle|^2$, initialized with a fully excited qubit and the resonator in the vacuum state.}} \label{TrotterFidelity}
\end{figure}

The simulation protocol can be performed for every time of the dynamics, with the number of digital steps tuned to reach a satisfactory simulation fidelity. We plot in Fig.~\ref{TrotterFidelity} the fidelity $F=| \langle \Psi_S | \Psi_R \rangle |^2$ as a function of time of the simulated wavefunction $\Psi_S$, including resonator and spin degrees of freedom, versus the ideal one $\Psi_R$, evolving according to $H_R$, as defined in Eq.~(\ref{RabiHam}). The fidelity is plotted for different parameters and iteration steps. Increasing the number of steps, the fidelity grows as expected from standard Suzuki-Lie-Trotter expansions~\cite{Suzuki1990}. In principle, the whole protocol can accurately access non-analytical regimes of these models, including USC and DSC regimes.  
By adding several transmon qubits to the architecture, the presented method can be extended to simulate the Dicke Hamiltonian 
\begin{equation}
H_D=\omega^R_r a^{\dagger}a +\sum_{j=1}^N\frac{\omega^R_q}{2}\sigma_j^z +\sum_{j=1}^Ng^R\sigma^x_j(a^{\dagger}+a).
\end{equation}
This simulation can be efficiently implemented by means of collective qubit rotations. In fact, only collective Tavis-Cummings interactions and global qubit rotations are involved. In this way, the total time for the simulation does not scale with the size of the system $N$.
The Dicke model can be investigated provided enough coherence and low-enough gate errors. Notice that this kind of quantum simulation is well suited for superconducting circuits, since simultaneous single-qubit addressing is possible.
Making use of the results in Refs.~\cite{Berry2007,Wiebe2011}, we demonstrate that the quantum resources needed to approximate the Dicke Hamiltonian with an error less than $\epsilon$ scale efficiently with the number of spins $N$ and of excitations allowed in the bosonic mode $M$. In a Dicke model simulation, one can bound the number of gates $N_\epsilon$ necessary to achieve a certain error $\epsilon$ in a time $t$ by
\begin{equation}
\label{Ngates}
N_\epsilon\leq\frac{2\cdot5^{2k}\left\{2t[\omega_r^RM+N (\omega_q^R+2|g^R|\sqrt{M+1})]\right\}^{1+1/2k}}{\epsilon^{1/2k}}.
\end{equation}  
Here, we have used an upper bound for the norm of the Dicke Hamiltonian, $||H_R||\leq\omega_r^RM+N(\omega_q^R+2|g^R|\sqrt{M+1})$, where $M$ is a truncation on the number of bosonic excitations involved in the dynamics. The fractal depth is set to $k=1$ in the standard Trotter approximations. Using higher orders of fractal decompositions would be a more involved task for implementation of digital approximations in realistic devices, due to the sign inversion that appears~\cite{Suzuki1990}. Nevertheless, unitary approximants with arbitrarily high fidelity can be obtained even when $k=1$. 
The formula in Eq.~(\ref{Ngates}) gives an upper bound to the scaling of quantum resources and experimental errors in a simulation involving several qubits. In fact, if one considers a small error for each gate, the accumulated gate error grows linearly with the number of gates.  

Notice that the quantum dynamics of the Dirac Hamiltonian emerges as a specific case of the quantum Rabi dynamics. For the 1+1 dimensional case the algebra of the Dirac spinors $| \Psi \rangle$ corresponds to that of Pauli matrices, and the Dirac equation in the standard representation can be written 
\begin{equation}
\label{Dirac equation 1+1}
i  \frac{d}{d t} | \Psi \rangle = (mc^2 \sigma_z + c p \sigma_x)| \Psi \rangle,
\end{equation}
where $m$ is the mass of the particle, $c$ is the speed of light and $ p \propto (a - a^{\dagger}) / i$ is the one-dimensional momentum operator. The Dirac Hamiltonian in Eq.~(\ref{Dirac equation 1+1}), $H_{\rm D}=mc^2 \sigma_z + c p \sigma_x$, shows the same mathematical structure as the quantum Rabi Hamiltonian, Eq.~(\ref{RabiHam}), when $\omega_r^R=0$. This condition can be achieved by choosing $\tilde{\omega}=\omega_r$. The analogy is complete by relating $mc^2$ to $\omega^R_q/2$, $c$ to $g^R$, and the momentum to the quadrature of the microwave field, which can be measured with current microwave technology~\cite{DiCandia2014}. Choosing an initial state with components in both positive and negative parts of the Dirac spectrum will allow the measurement of the {\it Zitterbewegung}~\cite{Lamata2007}. By retrieving different quadratures of the microwave field, one can detect this oscillatory motion of the simulated particle in the absence of forces, and the Klein paradox, where a relativistic particle can tunnel through high-energy barriers. To detect such effects, one will be interested in measuring either the position or the momentum of the particle, standing for different quadratures of the microwave field. 

In conclusion, we have shown that the dynamics of the quantum Rabi and Dicke models can be encoded in a circuit QED setup using a digital-analog approach. These quantum simulations will contribute to the observation of quantum dynamics not accessible in current experiments and to the generation of nontrivial correlations in circuit QED setups. Indeed, a recent experiment in the lab of Prof. Leonardo DiCarlo in TU Delft, has demonstrated the ideas proposed in this section, implementing the quantum Rabi model in the deep strong coupling regime~\cite{Langford2016}.

%% file: chap/conclusions.tex
\lettrine[lines=2, findent=3pt,nindent=0pt]{T}{his} thesis shows that quantum platforms can cross the barrier of their observable magnitudes, and efficiently access magnitudes that do not correspond to Hermitian observables, like time-correlation functions, or entanglement monotones in simulation schemes. Another major conclusion of this thesis is that quantum simulators are not limited to simulating the physics of their native interactions, but that they can access a broader collection of physical regimes of the same models that describe their behavior. Our theoretical investigations were kept always close to realistic experimental considerations. As a consequence, the ideas proposed in this thesis are at the frontier of the possibilities of actual technology, guaranteeing not only that they can be implemented today or in the near future, but also that they induce a technological development. A good proof of this is that this thesis contains two experimental demonstrations of the ideas developed in it, and that another two independent experiments have been performed based on experimental proposals contained in this thesis~\cite{Chen2016, Langford2016}.

A strong emphasis is put on time-correlation functions, which have been little studied in the literature, specially in the context of quantum simulation. We have demonstrated that high order time correlations of arbitrary Hermitian operators can be efficiently extracted with the help of an ancillary qubit. We demonstrate the feasibility of our protocol with an experiment that has measured with NMR techniques time correlations of the nuclear spins in a solution of chloroform. More specifically, we have extracted time correlations of the system evolving both under time-independent and time-dependent Hamiltonian dynamics, as well as high order time-correlation functions up to a tenth order. We propose that such correlation functions can be employed in the simulation of dissipative systems following Markovian dynamics, and our results can be extended to the simulation of non-Markovian dynamics. This, we believe, shows an alternative path towards the experimental simulation of open quantum systems. Here, we have given the general framework for the simulation of open quantum systems with time-correlation functions. A natural continuation of such a research line would be to propose specific open quantum systems of interest for their simulation. Also, experimental proposals, not to simulate a reservoir, but to measure time-correlation functions of an actual open quantum system, could shed light onto the properties of such a system and help to characterize it.

On the other hand, we have entanglement monotones, which quantify the degree of entanglement of a system, and are believed to be highly inefficient to extract. We demonstrate that an EQS can be constructed to allow for an efficient extraction of such quantities from the simulated dynamics. A proof of principle experimental demonstration of an EQS is presented. The experiment, realized in the group of Prof. Andrew White in Brisbane, employs the polarizations of three single photons to encode three qubits, which are then used to extract the concurrence of the simulated dynamics of two qubits. A similar experiment was also performed in the labs of Prof. Jian-Wei Pan in Hefei~\cite{Chen2016}. We propose the implementation of the same ideas with trapped ions, arguably a platform with better scalability perspectives, and higher controllability. We are convinced that these simulation schemes, together with their experimental confirmation, will be of value for quantum simulators as well as for fundamental studies of entanglement, as an EQS shall allow for the experimental exploration of entanglement well beyond what it is possible with classical computers. The concept of EQS introduced in this thesis shall be extended to allow for other hindered magnitudes of interest, rather than just antilinear operators. For instance, does an EQS exist which allows for the extraction of the purity, which is a quadratic magnitude? If such an EQS exists, what would its overhead be? Just a single ancilla qubit, like in the case of the EQS for antilinear operators? Can EQS be classified according to the size of their ancillary system?

Besides, it is our belief that a profound investigation of the quantum correlations of a quantum platform should also contain an analysis of the models that generate this interactions, mainly the quantum optical models of light-matter interaction that govern almost every quantum platform. We have demonstrated that quantum platforms like trapped ions and circuit QED are not restricted to the regimes of these models that naturally arise, but that they can indeed simulate other physical regimes. More specifically, we have shown that trapped ions can simulate the one-photon quantum Rabi model in the ultrastrong and deep-strong coupling regimes, as well as the two-photon quantum Rabi Hamiltonian. These models have an interest from the point of view of fundamental physics and from mathematics. The ability to explore and to move through the different regimes during the simulation time enhances the capabilities of quantum simulators and could be of interest for the generation of nontrivial states with high degree of correlation. The generation of states which are correlated in ways which are more complex than the resulting from the more trivial Jaynes-Cummings interactions, suggests that more possibilities for computation and simulation should exist. The exploration of these possibilities represents one of the more straightforward directions in which this research line could grow.

In the endeavor of exploring light-matter interactions and their correlations in quantum platforms, we pay special attention to how these could be scaled. As quantum platforms grow in size, a major challenge is to keep the generation of correlations with the growth of the platform. For the goal of scalability we have introduced the concept of digital-analog simulation, which preserves the complexity and size of analog simulators and adds versatility to them by incorporating a reduced number of digital operations. In this manner, we avoid a complete digitization of the dynamics, which is unwanted when the fidelity of the individual digital steps is not accurate enough, but we abandon the restrictions of analog simulators, which can only simulate a reduced number of models. We have demonstrated that circuits can simulate Rabi and Dicke Hamiltonians in almost all the interesting regimes, and certainly beyond the the Jaynes-Cummings and Tavis-Cummings regimes that naturally arise between the artificial two-level systems and the coplanar resonators. These ideas have been experimentally demonstrated in the lab of Prof. Leonardo DiCarlo in TU Delft~\cite{Langford2016}. We also demonstrate that spin models like the Heisenberg model can be simulated in ion traps with state-of-the-art technology and for a number of spins beyond any other performed simulation. These ideas we believe should allow to scale quantum simulators with current technology, and eventually shed light on the behavior of the simulated models beyond what numerical simulations have allowed us. The concept of DAQS might be extended to the simulation of other models, and to other platforms different from the ones studied here. 

All in all, this thesis is a journey through the quantum correlations that arise in modern quantum platforms, and through the models that generate them. The understanding gained through this thesis should help enhance the capabilities of quantum platforms, for the generation of nontrivial dynamics or states, for their scalability, and for the extraction of relevant information from the system, which are the building blocks for a quantum simulation.

%% file: chap/appendix.tex
\section[Further Considerations on the n-Time Correlation Algorithm]{Further Considerations on the n-Time Correlation \\ Algorithm} \label{app:timecor} 
\fancyhead[RO]{\small \leftmark}

In this appendix, we provide additional discussions on the efficiency of the method introduced in section~\ref{sec:TimeCor} for the measurement of time-correlation functions. We also include specific calculations concerning the implementation of gates. 

\subsection{Efficiency of the method} \label{app:efficiency}

Our algorithm is conceived to be run in a setup composed of a system undergoing the evolution of interest and an ancillary qubit. Thus, the size of the setup where the algorithm is to be run is always that of the system plus one qubit, regardless of the order of the time correlation. For instance, if we are considering an $N$-qubit system then our method is performed in a setup composed of $N+1$ qubits. 

With respect to time-efficiency,  our algorithm requires the performance of $n$ controlled gates $U_c^i$ and $n-1$  time-evolution operators $U(t_{j+1}; t_{j})$, $n$ being  the order of the time-correlation function.  

If we assume that $q$ gates are needed for the implementation of the system evolution and $m$ gates are required per control operation, our algorithm employs $(m+q)*n - q$ gates. As $m$ and $q$ do not depend on the order of the time correlation we can state that our algorithm needs a number of gates that scales  as a first-order polynomial with respect to the order $n$ of the time-correlation function. 

The scaling of $q$ with respect to the  system size depends on the specific simulation  under study. However, for most relevant cases it can be shown that this scaling is polynomial. For instance, in the case of an analogue quantum simulation of unitary dynamics, what it is usually called an always-on simulation, we have $q=1$. For a model requiring digital techniques $q$ will scale polinomially if the number of terms in the Hamiltonian grows polinomially  with the number of constituents, which is a physically-reasonable assumption~\cite{Lloyd1996a, NielsenChuang, Casanova2012}. In any case, we want to point out that the way in which $q$ scales is a condition inherent to any quantum simulation process, and, hence, it is not an additional overhead introduced by our proposal.

With respect to the number $m$, and as  explained in the next section, this number does not depend on the system size, thus, from the point of view of efficiency  it amounts to  a constant factor. 

In order to provide a complete runtime analysis of our protocol we study now the number of iterations needed to achieve a certain precision $\delta$ in the measurement of the time correlations. According to Berntein's inequality~\cite{Hoeffding1963} we have that 
\begin{equation}
\Pr\left[\left|\frac{1}{L}\sum_{i=1}^LX_i-\langle X\rangle\right|>\delta\right]\leq2 \exp \left( \frac{-L \delta^2}{4 \sigma_0^2} \right),
\end{equation}
where $X_i$ are independent random variables, and $\sigma_0^2$ is a bound on their variance. Interpreting $X_i$ as a single observation of the real or imaginary part of the time-correlation function, we find the number of measurements needed to have a precision $\delta$. Indeed, we have that $\left|\frac{1}{L}\sum_{i=1}^LX_i-\langle X\rangle\right|\leq\delta$ with probability $P\geq 1-e^{-c}$, provided that $L\geq \frac{4(1+c)}{\delta^2}$, where we have set $\sigma_0^2 \le 1$, as we always measure Pauli observables. This implies that the number of gates that we need to implement to achieve a precision $\delta$ for the real or the imaginary part of the time-correlation function is $\frac{4(1+c)}{\delta^2} [(m+q)n - q]$. Again, $c$ is a constant factor which does not depend nor on the order of the time correlation neither on the size of the system.

\subsection{N-body interactions  with M\o lmer-S\o rensen gates}\label{app:MolmerSorensen}

Exponentials of tensor products of Pauli operators, $\exp[i\phi\sigma_1 \otimes \sigma_2 \otimes ... \otimes \sigma_k]$, can be systematically constructed, up to local rotations, with a M\o lmer-S\o rensen gate applied over the $k$ qubits, one local gate on one of the qubits, and the inverse M\o lmer-S\o rensen gate on the whole register. This can be schematized as follows,
\begin{equation}
U = U_{MS}(-\pi/2, 0) U_{\sigma_z} (\phi) U_{MS}(\pi/2, 0) = \exp [ i\phi \sigma_1^z \otimes \sigma_2^x \otimes ...\otimes \sigma_k^z],
\end{equation}
where $U_{MS}(\theta, \phi ) = \exp[-i\theta ( \cos \phi S_x + \sin \phi S_y)^2/4]$, $S_{x,y} = \sum_{i=1}^k \sigma_i^{x,y}$ and $U_{\sigma_z}(\phi) = \exp(i \phi' \sigma_1^z)$ for odd $k$, where $\phi ' = \phi $ for $k=4n+1$, and $\phi '= - \phi $ for $k=4n-1$, with positive integer n. For even $k$, $U_{\sigma_z}(\phi)$ is replaced by $U_{\sigma_y}(\phi) = \exp(i\phi' \sigma_1^y)$, where $\phi '=\phi$ for $k=4n$, and $ \phi ' = \phi$ for $k=4n-2$, with positive integer $n$. Subsequent local rotations will generate any combination of Pauli matrices in the exponential. 

The replacement in the previous scheme of the central gate  $U_{\sigma_z}(\phi)$ by an interaction containing a coupling with bosonic degrees of freedom, for example  $U_{\sigma_z , (a+a^{\dag})}(\phi) = \exp[i \phi' \sigma_1^z (a +a^{\dag}) ] $, will directly provide us with 
\begin{equation}\label{interaction}
U = U_{MS}(-\pi/2, 0) U_{\sigma_z, (a+a^\dag)} (\phi) U_{MS}(\pi/2, 0) = \exp [ i\phi \sigma_1^z \otimes \sigma_2^x \otimes ...\otimes \sigma_k^z (a+a^{\dag})].
\end{equation}

In order to provide a complete recipe for systems where M\o lmer-S\o rensen interactions are not directly available, we want to comment that the kind of entangling quantum gates required by our algorithm, see the right hand side of Eq.~(\ref{interaction}) above, are always decomposable in a polynomial sequence of controlled-Z  gates~\cite{NielsenChuang}. For example, in the case of a three-qubit system we have  
\begin{equation}
CZ_{1,3} CZ_{1,2} e^{-i\phi\sigma^y_1}CZ_{1,2} CZ_{1,3} = \exp{(-i\phi \sigma^y_1\otimes \sigma^z_2 \otimes \sigma^z_3)}
\end{equation}
Here, $CZ_{i,j}$ is a controlled-Z gate between the $i, j$ qubits and $e^{-i\phi\sigma^y_1}$ is local rotation applied on the first qubit. This result can be easily extended to $n$-qubit systems with the application of $2(n-1)$ controlled operations~\cite{NielsenChuang}.

Therefore, it is demonstrated the polynomial character of our algorithm, and hence its efficiency, even if M\o lmer-S\o rensen gates are not available in our setup.


\section[Further Considerations on the Simulation of Dissipative Dynamics]{Further Considerations\\ on the Simulation of \\ Dissipative Dynamics} \label{app:dissipation} \fancyhead[RO]{B \ \ CONSIDERATIONS ON THE SIMULATION OF DISSIPATIVE DYNAMICS} \fancyhead[LE]{\small \rightmark}

In this appendix, we provide explicit derivations and additional details of the results derived in section~\ref{sec:SimDiss}.

\subsection{Decomposition in Pauli operators}\label{app:decompPauli}
In this section, we discuss the decomposition of the Lindblad operators in an unitary basis. In order to implement the protocol introduced in section~\ref{sec:TimeCor} to compute a general multitime correlation function, we need to decompose a general Lindblad operator $L$ and observable $O$ in Pauli-kind orthogonal matrices $\{ Q_k\}_{k=1}^{d^2}$, where $Q_k$ are both Hermitian and unitaries and $d$ is the dimension of the system. If $d=2^l$ for some integer $l$, then a basis of this kind is the one given by the tensor product of Pauli matrices. Otherwise, it is always possible to embed the problem in a larger Hilbert space, whose dimension is the closest power of $2$ larger than $d$. Thus, we can set $\| Q_k\|_\infty=1$ and $\| Q_k\|_2=\sqrt{d}$, where $\|A\|_2\equiv\sqrt{\text{Tr}\,(A^\dag A)}$ and we have redefined $d$ as the embedding Hilbert space dimension. Here, we prove that if $\| L\|_\infty=1$ and $L=\sum_{k=1}^Mq_kQ_k$ with $M\leq d^2$, then (i) $\sum_{k=1}^M |q_k|\leq \sqrt{M}$. This relations will be useful in the proof of Eq.~(\ref{numer1}) of the main text. We first show that $\sum_{k=1}^M |q_k|^2\leq1$:
\begin{align}
\sum_{k=1}^M |q_k|^2=\frac{1}{d}\sum_{k=1}^M |q_k|^2\|Q_k\|_2^2=\frac{1}{d}\left\| \sum_{k=1}^Mq_kQ_k\right\|_2^2=\frac{1}{d}\| L\|_2^2\leq \| L\|_\infty^2=1,
\end{align}
where we have used the orthogonality of the matrices $Q_i$, i.e. $\text{Tr}\,(Q_i^{\dag}Q_j)=\text{Tr}\,(Q_i Q_j)=d\delta_{ij}$.
The relation (i) follows simply from the norm inequality for $M$-dimensional vectors $v$: $\| v\|_1\leq \sqrt{M}\|v\|_2$.

\subsection{Proof of the single-shot approach for the integration of time-correlation functions} \label{app:proofEqnumer1}
In this section, we provide a proof of Eq.~(\ref{numer1}) of the main text:
\begin{align}
\left|\sum_{[i_1,\dots,i_n]=1}^N\int dV_n\;\langle A_{[i_1,\cdots,i_n]}(\vec{s})\rangle-\frac{(Nt)^n}{n!|\Omega_n|}\sum_{\Omega_n}\tilde A_{\vec{\omega}}(\vec{t})\right|\leq\delta_n
\end{align}
with probability higher than $1-e^{-\beta}$, provided that $|\Omega_n|>\frac{36M_O^2(2+\beta)}{\delta_n^2}\frac{(2\bar{\gamma} M N t)^{2n}}{n!^2}$. Here, $\bar{\gamma}=\max_{i,s\in[0,t]}|\gamma_i(s)|$, $M=\max_i M_i$ where $M_i$ is defined by the Pauli decomposition of the Lindblad operators $L_i=\sum_{k=1}^{M_i}q^i_kQ^i_{k}$, $M_O$ is the Pauli decomposition of the observable $O$ that we will to measure, $[\vec{\omega},\vec{t}]\in\Omega_n\subset\{[\vec{\omega},\vec{t}]\;|\; \vec{\omega}=[i_1,\dots, i_n], i_k\in[1,N], \vec{t}\in V_n \}$ and $[\vec{\omega},\vec{t}]$ are sampled uniformly and independently, $|\Omega_n|$ is the size of $\Omega_n$, and $\tilde A_{\vec{\omega}}(\vec{t})$ corresponds to single-shot measurements of $A_{\vec{\omega}}(\vec{t})$. Notice that $V_n$ is the integration volume corresponding to the $n$-th order term, and $|V_n|=t^n/n!$. 

First, we write $\tilde A_{\vec{\omega}}(\vec{t})=\langle A_{\vec{\omega}}(\vec{t})\rangle+\tilde \epsilon_{[\vec{\omega},\vec{t}]}$, where $\tilde \epsilon_{[\vec{\omega},\vec{t}]}$ is the shot-noise. Note that, due to the previous identity, $\langle \epsilon_{{[\vec{\omega}, \vec{t} ]}} \rangle = 0.$  We have to bound the following quantity
\begin{align}\label{numer2}
&\left|\sum_{[i_1,\dots,i_n]=1}^N\int dV_n\;\langle A_{[i_1,\dots,i_n]}(\vec{s})\rangle-\frac{N^n|V_n|}{|\Omega_n|}\sum_{\Omega_n}\tilde A_{\vec{\omega}}(\vec{t})\right| \nonumber\\
\quad& \leq\left|\sum_{[i_1,\dots,i_n]=1}^N\int dV_n\;\langle A_{[i_1,\dots,i_n]}(\vec{s})\rangle-\frac{N^n|V_n|}{|\Omega_n|}\sum_{\Omega_n}\langle A_{\vec{\omega}}(\vec{t})\rangle\right|+\left|\frac{N^n|V_n|}{|\Omega_n|}\sum_{\Omega_n}\tilde  \epsilon_{[\vec{\omega},\vec{t}]} \right|.
\end{align}
The first term in the right side of Eq.~\eqref{numer2} is basically the error bound in a Montecarlo integration, while the second term is small as the variance of $\epsilon$ is bounded. Indeed, both quantities can be bounded using the Hoeffding1963 inequality~\cite{Hoeffding1963}:

\newtheorem*{Hoeffding1963}{Theorem}
\begin{Hoeffding1963}[Hoeffding1963 Inequality~\cite{Hoeffding1963}] Let $X_1,\dots, X_m$ be independent zero-mean random variables. Suppose ${\mathbb{E}[X_i^2]}\leq \sigma_0^2$ and $|X_i|\leq c$. Then for any $\delta>0$,
\begin{align}
\Pr\left[\left|\sum_{i=1}^mX_i\right|>\delta\right]\leq 2\exp\left({\frac{-\delta^2}{4m\sigma_0^2}}\right),
\end{align}
provided that $\delta\leq 2 m\sigma_0^2/c.$
\end{Hoeffding1963}

To compute the first term in the right-hand side of Eq.~\eqref{numer2},  we sample $[\vec{\omega},\vec{t}]$ uniformly and independently to find that $\mathbb{E}\left[\frac{N^n|V_n|}{|\Omega_n|}\langle A_{\vec{\omega}}(\vec{t})\rangle\right]=\frac{1}{|\Omega_n|}\sum_{i_1,\dots,i_n=1}^N\int dV_n\;\langle A_{[i_1,\dots,i_n]}(\vec{s})\rangle$. We define the quantity $X_{[\vec{\omega},\vec{t}]}\equiv\frac{N^n|V_n|}{|\Omega_n|}\langle A_{\vec{\omega}}(\vec{t})\rangle-\frac{1}{|\Omega_n|}\sum_{i_1,\dots,i_n=1}^N\int dV_n\;\langle A_{[i_1,\dots,i_n]}(\vec{s})\rangle $, and look for an  estimate $\left|\sum_{\Omega_n}X_{[\vec{\omega},\vec{t}]}\right|$, where $\mathbb{E}[X_{[\vec{\omega},\vec{t}]}]=0$. We have that 
\begin{eqnarray}
\nonumber \mathbb{E}[X_{[\vec{\omega},\vec{t}]}^2] =\frac{1}{|\Omega_n|^2}N^{2n}|V_n|^2\mathbb{E}[\langle A_{\vec{\omega}}(\vec{t})\rangle^2]-\frac{1}{|\Omega_n|^2}\left(\sum_{[i_1,\dots,i_n]=1}^N\int dV_n\;\langle A_{[i_1,\dots,i_n]}(\vec{s})\rangle\right)^2\\ 
\nonumber  \leq \frac{N^{n}|V_n|}{|\Omega_n|^2}\sum_{[i_1,\dots,i_n]=1}^N\int dV_n\;\langle A_{[i_1,\dots,i_n]}(\vec{s})\rangle^2\leq\frac{N^{2n}|V_n|^2}{|\Omega_n|^2}\max_{[i_1,\dots,i_n],\vec{s}}\langle A_{[i_1,\dots,i_n]}(\vec{s})\rangle^2,\\ 
\end{eqnarray}
where we have used the inequality $(\int dV\;f)^2\leq|V|\int dV\;f^2$. Moreover, we have that 
\begin{eqnarray}
|X_{[\vec{\omega},\vec{t}]}| &=& \frac{1}{|\Omega_n|}\left| N^n|V_n|\langle A_{\vec{\omega}}(\vec{t})\rangle-\sum_{[i_1,\dots,i_n]=1}^N\int dV_n\;\langle A_{[i_1,\dots,i_n]}(\vec{s})\rangle\right| \nonumber \\ &\leq& \frac{2N^n|V_n|}{|\Omega_n|}\max_{[i_1,\dots,i_n],\vec{s}}|\langle A_{[i_1,\dots,i_n]}(\vec{s})\rangle|,
\end{eqnarray}
where we have used the inequality $|\sum_{i=1}^N\int dV\; f|\leq N|V|\max |f|$.
\\
Now, recall that
\begin{align}\label{defA}
 A_{[i_1,\dots,i_n]}(\vec{s})&\equiv  e^{s_n\mathcal{L}_{H_s}^\dag}\mathcal{L}_D^{s_n,i_n\dag}\dots \mathcal{L}_D^{s_2,i_2\dag}e^{(s_1-s_2)\mathcal{L}^\dag_{H_s}} \mathcal{L}_D^{s_1,i_1\dag}e^{(t-s_1)\mathcal{L}^\dag_{H_s}}O,
\end{align}
where $\mathcal{L}_D^{s,i} \xi\equiv\gamma_{i}^{}(s)\left( L_{i}\xi L_{i}^{\dag}-\frac{1}{2}\{ L^{\dag}_{i} L_{i}, \xi\}\right)$, and $\mathcal{L}^\dag\xi\equiv(\mathcal{L}\xi)^\dag$ for a general superoperator $\mathcal{L}$. It follows that  $\max_{[i_1,\dots,i_n],\vec{s}}\langle A_{[i_1,\dots,i_n]}(\vec{s})\rangle^2\leq(2\bar{\gamma})^{2n}\| O\|_\infty^2\prod_{k=1}^n\|L_{i_k}\|_\infty^4= (2\bar{\gamma})^{2n}$, and $\max_{[i_1,\dots,i_n],\vec{s}}|\langle A_{[i_1,\dots,i_n]}(\vec{s})\rangle|\leq (2\bar{\gamma})^n$, where $\bar{\gamma}=\max_{i,s\in[0,t]}|\gamma_i(s)|$ and we have set $\| O\|_\infty=1$ and $\|L_i\|_\infty=1$. Here, we have used the fact that $\langle A_{[i_1,\dots,i_n]}(\vec{s})\rangle$ is real, the inequality $\left|\text{Tr}\,(AB)\right|^2\leq\|A\|_\infty\|B\|_1$, and the result in Eq.~\eqref{Lc} of the next section. 
Now, we can directly use the Hoeffding1963 inequality, obtaining
\begin{align}\label{pr1}
\text{Pr}\left[\left|\sum_{\Omega_n}X_{[\vec{\omega},\vec{t}]}\right|>\delta'\right]\leq 2\exp\left(-{\frac{n!^2|\Omega_n|\delta'^2}{4(2\bar{\gamma} Nt)^{2n}}}\right)\equiv p_1
\end{align}
provided that $\delta'\leq (2\bar{\gamma} Nt)^{n}/n!$, and where we have set $|V_n|=t^n/n!$.

Now, we show that the second term in the right hand side of Eq~\eqref{numer2} can be bounded for all $\Omega_n$. From the definition of $\tilde \epsilon_{[\vec{\omega},\vec{t}]}$, we note that 
\begin{align}
\mathbb{E}\left[\frac{N^n|V_n|}{|\Omega_n|}\tilde \epsilon_{[\vec{\omega},\vec{t}]}\right]=\frac{N^n|V_n|}{|\Omega_n|}\sum_{i}\tilde \epsilon^i_{[\vec{\omega},\vec{t}]}p^i_{[\vec{\omega},\vec{t}]}=\frac{N^n|V_n|}{|\Omega_n|}\left(\sum_{i}\tilde A^i_{\vec{\omega}}(\vec{t})p^i_{[\vec{\omega},\vec{t}]}-\langle A_{\vec{\omega}}(\vec{t})\rangle\right)=0,
\end{align} 
where $\tilde \epsilon^{i}_{[\vec{\omega},\vec{t}]}$ ($\tilde A^{i}_{\vec{\omega}}(\vec{t})$) is a particular value that the random variable $\tilde \epsilon_{[\vec{\omega},\vec{t}]}$ ($\tilde A_{\vec{\omega}}(\vec{t})$) can take, and $p^i_{[\vec{\omega},\vec{t}]}$ is the corresponding probability. Notice that the possible values of the random variable $\tilde \epsilon_{[\vec{\omega},\vec{t}]}$ depend on the Pauli decomposition of $A_{\vec{\omega}}(\vec{t})$. In fact, $A_{\vec{\omega}}(\vec{t})$ is a sum of $n$-time correlation functions of the Lindblad operators, and our method consists in decomposing each Lindblad operator in Pauli operators (see section I), and then measuring the real and the imaginary part of the corresponding time-correlation functions. As the final result has to be real, eventually we consider only the real part of $\tilde A_{\vec{\omega}}(\vec{t})$, so that also $\tilde \epsilon_{[\vec{\omega},\vec{t}]}$ can take only real values. In the case $n=2$, one of the terms to be measured is
\begin{eqnarray}\label{prod} 
&&{L_{\omega_2}^\dag(t_2)L_{\omega_1}^\dag(t_1)O(t)L_{\omega_1}(t_1) L_{\omega_2}(t_2)} \nonumber\\
&=&\sum_{l=1}^{M_O}\sum_{k_1,k_2,k'_1, k'_2=1}^M q_l^Oq_{k_1}^{\omega_1*} q_{k_2}^{\omega_2*}q_{k'_1}^{\omega_1} q_{k'_2}^{\omega_2}\,Q_{k_2}^{\omega_2\dag}(t_2) Q_{k_1}^{\omega_1\dag}(t_1)Q_{l}^{O}(t)Q^{\omega_1}_{k'_1}(t_1) Q^{\omega_2}_{k'_2}(t_2), \nonumber \\
\end{eqnarray}
where we have used the Pauli decompositions $L_{\omega_i}=\sum_{k_i=1}^{M_{\omega_i}} q_{k_i}^{\omega_i}Q^{\omega_i}_{k_i}$, $O=\sum_{l=1}^{M_O}q_l^OQ_l^O$, and we have defined $M\equiv \max_i M_{\omega_i}$. We will find a bound for the case $n=2$, and the general case will follow straightforwardly. For the term in Eq.~\eqref{prod}, we have that 
\begin{align}\label{bound2}
&\sum_{l=1}^{M_O}\sum_{k_1,k_2,k'_1, k'_2=1}^M |q_l^O|| \Re \;q_{k_1}^{\omega_1*} q_{k_2}^{\omega_2*}q_{k'_1}^{\omega_1} q_{k'_2}^{\omega_2}( \lambda^{\omega_1 \omega_2}_{k_2k_1lk'_1k'_2, r}+i \lambda^{\omega_1\omega_2}_{k_2k_1lk'_1k'_2, im})|\nonumber\\
&\leq 2\sum_{l=1}^{M_O}\sum_{k_1,k_2,k'_1, k'_2=1}^M |q_l^O| | q_{k_1}^{\omega_1*} q_{k_2}^{\omega_2*}q_{k'_1}^{\omega_1} q_{k'_2}^{\omega_2}|\,\| Q_{k_2}^{\omega_2\dag}(t_2) Q_{k_1}^{\omega_1\dag}(t_1)Q_l^O(t)Q^{\omega_1}_{k'_1}(t_1) Q^{\omega_2}_{k'_2}(t_2)\|_\infty \nonumber\\
&\leq2\sum_{l=1}^{M_O}|q_l^O|\sum_{k_1,k_2,k'_1, k'_2=1}^M | q_{k_1}^{\omega_1*} q_{k_2}^{\omega_2*}q_{k'_1}^{\omega_1} q_{k'_2}^{\omega_2}|\leq 2\sqrt{M_O}\,M^2,
\end{align}
where we have defined the real part ($ \lambda^{\omega_1 \omega_2}_{k_2k_1lk'_1k'_2, r}$) and the imaginary part ($\lambda^{\omega_1\omega_2}_{k_2k_1lk'_1k'_2, im}$) of the single-shot measurement of $Q_{k_2}^{\omega_2\dag}(t_2) Q_{k_1}^{\omega_1\dag}(t_1)Q_l^O(t)Q^{\omega_1}_{k'_1}(t_1) Q^{\omega_2}_{k'_2}(t_2)$, and we have used the fact that $\| Q_k^i\|_\infty=1$, $\|Q_l^O\|_\infty=1$, and relation (i) of the previous section. Eq.~\eqref{bound2} is a bound on the outcomes of $L_{\omega_2}^\dag(t_2) L_{\omega_1}^\dag(t_1)O(t)L_{\omega_1}(t_1) L_{\omega_2}(t_2)$.  Notice that the bound in Eq.~\eqref{bound2} neither depends on the particular order of the Pauli operators, nor on the times $s_i$, so it holds for a general term in the sum defining $A_{\vec{\omega}}(\vec{t})$. Thus, we  find that, in the case $n=2$, $\tilde A_{\vec{\omega}}(\vec{t})$ is upper bounded by $|\tilde A_{\vec{\omega}}(\vec{t})|\leq 2\sqrt{M_O}(2\bar{\gamma}M)^2$. In the general case of order $n$, it is easy to show that $|\tilde A_{\vec{\omega}}(\vec{t})|\leq 2\sqrt{M_O}(2\bar{\gamma}M)^n$. It follows that 
\begin{eqnarray}
\left|\frac{N^n|V_n|}{|\Omega_n|}\tilde \epsilon_{[\vec{\omega},\vec{t}]}\right|&=&\frac{N^n|V_n|}{|\Omega_n|}\left|\tilde A_{\vec{\omega}}(\vec{t})-\langle A_{\vec{\omega}}(\vec{t})\rangle\right| \\ &\leq& \frac{(2\bar\gamma N)^n|V_n|}{|\Omega_n|}(1+2\sqrt{M_O}M^n)\leq \frac{3\sqrt{M_O}(2\bar\gamma M N)^n|V_n|}{|\Omega_n|}. \nonumber
\end{eqnarray}
Regarding the bound on the variance, we have that 
\begin{eqnarray}
\mathbb{E}\left[\left(\frac{N^n|V_n|}{|\Omega_n|}\tilde \epsilon_{[\vec{\omega},\vec{t}]}\right)^2\right]&=& \sum_{i}\left(\frac{N^n|V_n|}{|\Omega_n|}\tilde \epsilon^{i}_{[\vec{\omega},\vec{t}]}\right)^2p^i_{[\vec{\omega},\vec{t}]}\leq\frac{N^{2n}|V_n|^2}{|\Omega_n|^2}\sum_i\tilde A^{i\,2}_{\vec{\omega}}(\vec{t})p^i_{[\vec{\omega},\vec{t}]} \nonumber\\
\quad&\leq&\frac{N^{2n}|V_n|^2}{|\Omega_n|^2}\max_i \tilde A^{i\,2}_{\vec{\omega}}(\vec{t}) =\frac{N^{2n}|V_n|^2}{|\Omega_n|^2}\left(\max_i |\tilde A^{i}_{\vec{\omega}}(\vec{t})|\right)^2 \nonumber \\ 
&\leq& \frac{4M_O(2\bar{\gamma}MN)^{2n}|V_n|^2}{|\Omega_n|^2}.
\end{eqnarray}
Using Hoeffding1963 inequality, we obtain
\begin{align}\label{pr2}
\text{Pr}\left[\left|\frac{N^n|V_n|}{|\Omega_n|}\sum_{\Omega_n}\tilde \epsilon_{[\vec{\omega},\vec{t}]}\right|>\delta''\right]\leq 2\exp\left(- \frac{n!^2|\Omega_n|\delta''^2}{16M_O^2(2\bar{\gamma} MNt)^{2n}}\right)\equiv p_2,
\end{align}
provided that $\delta''\leq \frac{8}{3}\sqrt{M_O}(2\bar{\gamma} MNt)^{n}/n!$, where we have set, as before, $|V_n|=t^n/n!$. Now, choosing $\delta'=\frac{1}{2M^n+1}\delta_n$, $\delta''=\frac{2M^n}{2M^n+1}\delta_n$, $|\Omega_n|>\frac{36M_O^2(2+\beta)}{\delta_n^2}\frac{(2\bar{\gamma} MN t)^{2n}}{n!^2}$, we have that $p_1, p_2\leq \frac{e^{-\beta}}{2}$. Notice that $\delta_n\leq (2\bar{\gamma}Nt)^n/n!$ always holds, so the conditions on $\delta'$, $\delta''$ are satisfied. By using the union bound, we conclude that
\begin{flalign}
&\text{Pr}\left[ \left|\sum_{[i_1,\dots,i_n]=1}^N\int dV_n\;\langle A_{[i_1,\cdots,i_n]}(\vec{s})\rangle-\frac{(Nt)^n}{n!|\Omega_n|}\sum_{\Omega_n}\tilde A_{\vec{\omega}}(\vec{t})\right|>\delta_n \right]\nonumber\\
&\leq \text{Pr}\Bigg[ \left|\sum_{[i_1,\dots,i_n]=1}^N\int dV_n\;\langle A_{[i_1,\dots,i_n]}(\vec{s})\rangle-\frac{N^n|V_n|}{|\Omega_n|}\sum_{\Omega_n}\langle A_{\vec{\omega}}(\vec{t})\rangle\right|>\frac{1}{1+2M^n}\delta_n \nonumber \\
&\hspace{6cm} \lor \left|\frac{N^n|V_n|}{|\Omega_n|}\sum_{\Omega_n}\tilde  \epsilon_{[\vec{\omega},\vec{t}]} \right| > \frac{2M^n}{1+2M^n}\delta_n\Bigg] \nonumber\\
&\leq p_1+p_2\leq e^{-\beta}.
\end{flalign}

\subsection{Proof of the bounds on the trace distance} \label{app:proofbounds}
\fancyhead[LE]{\rightmark}
In this section, we provide the proof for the bound in Eq.~(\ref{D1zero}), and the general bound in Eq.~(\ref{bounds}) of the main text. We note that 
\begin{align}\label{11}
D_1(\rho(t),\tilde{\rho}_0(t))&\leq \frac{1}{2}\int_0^tds\; \|\mathcal{L}_D^s\|_{1\rightarrow1}\|\rho(s)\|_1=\frac{1}{2}\int_0^tds\; \|\mathcal{L}_D^s\|_{1\rightarrow1},
\end{align}
where we have introduced the induced superoperator norm $\| \mathcal{A}\|_{1\rightarrow 1}\equiv\sup_\sigma\frac{\|\mathcal{A}\sigma\|_1}{\|\sigma\|_1}$~\cite{Watrous2004}. Moreover, the following bound holds 
\begin{eqnarray}\label{Lc}
\|\mathcal{L}_D^t\sigma\|_{1} &=& \left\|\sum_{i=1}^N\gamma_i(t)\left(L_i\sigma L_i^\dag-\frac{1}{2}L_i^\dag L_i\sigma-\frac{1}{2}\sigma L_i^\dag L_i\right)\right\|_1 \nonumber \\
&\leq& \sum_{i=1}^N|\gamma_i(t)|\left(\|L_i \sigma L_i^\dag\|_1+\frac{1}{2}\|L_i^\dag L_i\sigma\|_1+\frac{1}{2}\|\sigma L_i^\dag L_i\|_1\right) \nonumber \\ 
&\leq& 2\sum_{i=1}^N|\gamma_i(t)|\| L_i\|_\infty^2\|\sigma\|_1,
\end{eqnarray}
where we have used the triangle inequality and the inequality $\|AB\|_1\leq \left\{\|A\|_\infty\|B\|_1,\|A\|_1\|B\|_\infty\right\}$. Eq.~\eqref{Lc} implies that $\| \mathcal{L}_D^t\|_{1\rightarrow 1}\leq2\sum_{i=1}^N|\gamma_i(t)|\| L_i\|_\infty^2$. Inserting it into Eq.~\eqref{11}, it is found that
\begin{align}
D_1(\rho(t),\tilde{\rho}_0(t))&\leq\sum_{i=1}^N\| L_i\|_\infty^2 \int_0^tds\;|\gamma_i(s)|=\sum_{i=1}^N|\gamma_i(\epsilon_i)|\| L_i\|_\infty^2t,
\end{align}
where we have assumed that $\gamma_i(t)$ are continuous functions in order to use the mean-value theorem ($0\leq\epsilon_i\leq t$). Indeed, $|\gamma_i(\epsilon_i)|=\frac{1}{t}\int_0^tds\;|\gamma_i(s)|$, that can be directly calculated or estimated. 

The bound in Eq.~(\ref{bounds}) has to been proved by induction. Let us assume that Eq.~(\ref{bounds}) in the text holds for the order $n-1$. We have that
\begin{flalign}
D_1(\rho(t),\tilde{\rho}_n(t))&\leq\int_0^tds\;\|\mathcal{L}_D\|_{1\rightarrow1}D_1(\rho(s),\tilde{\rho}_{n-1}(s)) \nonumber \\
&\leq \prod_{k=0}^{n-1}\bigg[2\sum_{i_k=1}^N |\gamma_{i_k}(\epsilon_{i_k})|\|L_{i_k}\|_\infty^2\bigg]\sum_{i=1}^N\|L_i\|_\infty^2\frac{1}{n!}\int_0^tds\;|\gamma_i(s)|s^n,
\end{flalign}
where we need to evaluate the quantities $\int_0^tds\;|\gamma_i(s)|s^n$. By using the mean-value theorem, we have $\int_0^tds\;\gamma_i(s)s^n=|\gamma_i(\epsilon_i)|\int_0^tds\;s^n$, with $0\leq\epsilon_i\leq t$, and Eq.~(\ref{bounds}) follows straightforwardly. In any case, we can evaluate $\int_0^tds\;|\gamma_i(s)|s^n$ by solving directly the integral or we can estimate it by using H\"older's inequalities:
\begin{equation}
\int_0^tds\;|\gamma_i(s)|s^n\leq\left\{\sqrt{\int_0^tds\;\gamma_i(s)^2}\;\sqrt{\frac{t^{2n+1}}{2n+1}}\;, \;\max_{0\leq s\leq t}|\gamma_i(s)|\;\frac{t^{n+1}}{n+1}\right\}.
\end{equation}

\subsection{Error bounds for the expectation value of an observable}
 
 In this section, we find an error bound for the expectation value of a particular observable $O$. As figure of merit, we choose $D_O(\rho_1,\rho_2)\equiv \left|\text{Tr}\,(O(\rho_1-\rho_2))\right|/ (2\| O\|_\infty)$. The quantity $D_O(\rho_1,\rho_2)$ tells us how close the expectation value of $O$ on $\rho_1$ is to the expectation value of $O$ on $\rho_2$, and it is always bounded by the trace distance, i.e. $D_O(\rho_1,\rho_2)\leq D_1(\rho_1,\rho_2)$. Taking the expectation value of $O$ in both sides of Eq.~(\ref{generaldif}) of the main text, we find that 
\begin{eqnarray}\label{boundO}
D_O(\rho(t),\tilde{\rho}_n(t))&=&\frac{1}{2\|O\|_\infty}\left|\int_0^tds\;\text{Tr}\,\left(e^{(t-s)\mathcal{L}_H}\mathcal{L}_D^s(\rho(s)-\tilde{\rho}_{n-1}(s))O\right) \right| \nonumber \\
&=& \frac{1}{2\|O\|_\infty}\left|\int_0^tds\;\text{Tr}\; \left(\mathcal{L}_D^{s\dag} O(\rho(s)-\tilde{\rho}_{n-1}(s))\right)\right| \nonumber\\
\quad&\leq& \frac{1}{\|O\|_\infty}\int_0^tds\;\|\mathcal{L}_D^{s\dag} O\|_\infty D_1(\rho(s),\tilde{\rho}_{n-1}(s)) \nonumber \\
&\leq& \frac{\|\mathcal{L}_D^{s\dag} O\|_\infty}{\|O\|_\infty}\left(2\bar \gamma N\right)^{n}\frac{t^{n+1}}{2(n+1)!},
\end{eqnarray}
where  $\mathcal{L}_D^{s\dag} O=\sum_{i=1}^N\gamma_i(s)\left(L_i^\dag O L_i-\frac{1}{2}\{L_i^\dag L_i ,O\}\right)$. The bound in Eq.~\eqref{boundO} is particularly useful when $L_i$ and $O$ have a tensor product structure. In fact, in this case, the quantity $\|\mathcal{L}_D^{s\dag} O\|_\infty$ can be easily calculated or bounded. For example, consider a $2$-qubit system with $L_1=\sigma^-\otimes \mathbb{I}$, $L_2=\mathbb{I}\otimes \sigma^-$, $\gamma_i(s)=\gamma>0$ and the observable $O=\sigma_z\otimes \mathbb{I}$. Simple algebra leads to $\|\mathcal{L}_D^{s\dag} O\|_\infty=\gamma\| (\mathbb{I}+\sigma_z)\otimes \mathbb{I}\|_\infty=\gamma\|\mathbb{I}+\sigma_z\|_\infty\|\mathbb{I}\|_\infty=2\gamma$, where we have used the identity $\|A\otimes B\|_\infty=\|A\|_\infty\|B\|_\infty$. 

\subsection{Total number of measurements} \label{app:numofmeas}

In this section, we provide a magnitude for the scaling of the number of measurements needed to simulate a certain dynamics with a given error $\varepsilon$ and for a time $t$. We have proved that 
\begin{align}
\varepsilon'\equiv D_1(\rho(t),\tilde \rho_n(t))\leq\frac{(2\bar \gamma Nt)^{n+1}}{2(n+1)!},
\end{align}
where $\bar \gamma \equiv \max_i |\gamma_i|$.
We want to establish at which order $K$ we have to truncate in order to have an error $\varepsilon'$ in the trace distance. We have that, if $n\geq ex+\log\frac{1}{\tilde\varepsilon}$, with $x\geq 0$ and $\tilde \varepsilon \leq 1$, then $\frac{x^n}{n!}\leq \tilde \varepsilon$. In fact 
\begin{align}
\frac{x^n}{n!}\leq \left(\frac{ex}{n}\right)^n\leq\left(1+\frac{\log\frac{1}{\tilde \varepsilon}}{ex}\right)^{-ex-\log\frac{1}{\tilde \varepsilon}}\leq\left(1+\frac{\log\frac{1}{\tilde \varepsilon}}{ex}\right)^{-ex}\leq e^{-\log \frac{1}{\tilde \varepsilon}}= \tilde \varepsilon,
\end{align}
where we have used the Stirling inequality $n!\geq \sqrt{2\pi n} \,(n/e)^n\geq (n/e)^n$. This implies that, if we truncate at the order $K\geq 2e\bar \gamma Nt+\log\frac{1}{2\varepsilon'}-1=O(2e\bar \gamma Nt+\log\frac{1}{\varepsilon'})$, then we have an error lower than $\varepsilon'$ in the trace distance. The total number of measurements in order to apply the protocol up to an error $\varepsilon'+\sum_{n=0}^K\delta_n$ is bounded by $\sum_{n=0}^K3^n|\Omega_n|$. If we choose $\varepsilon'=c\varepsilon$, $\delta_n=(1-c)\frac{\varepsilon}{(K+1)}$ ($0< c<1$), we have that the total number of measurements to simulate the dynamics at time $t$ up to an error  $\varepsilon$ is bounded by
\begin{align}
\sum_{n=0}^K 3^n|\Omega_n|&=\frac{36M_O^2(2+\beta)(1+K)^2}{(1-c)^2\varepsilon^2}\sum_{n=0}^K\frac{(6\bar \gamma N M t)^{2n}}{n!^2} \nonumber\\
\quad&\leq\frac{36M_O^2(2+\beta)(1+K)^2}{(1-c)^2\varepsilon^2} e^{12\gamma N M t}=O\left(\left(6\bar t+\log\frac{1}{\varepsilon}\right)^2\frac{e^{12M\bar t}}{\varepsilon^2}\right),
\end{align}
where we have defined $\bar t=\bar \gamma N t$.

\subsection{Bounds for the non-Hermitian Hamiltonian case} \label{app:nonHermitian}

The previous bounds apply as well to the  simulation of a non-Hermitian Hamiltonian $J=H-i\Gamma$, with $H$ and $\Gamma$ Hermitian operators. In this case, the Schr\"odinger  equation reads
\begin{equation}\label{masterc}
\frac{d\rho}{dt}=-i[H,\rho]-\{\Gamma,\rho\}=(\mathcal{L}_H+\mathcal{L}_\Gamma)\rho,
\end{equation}
where $\mathcal{L}_\Gamma$ is defined by $\mathcal{L}_{\Gamma}\,\sigma\equiv-\{\Gamma,\sigma\}$.
Our method consists in considering $\mathcal{L}_\Gamma$ as a perturbative term. To ascertain the reliability of the method, we have to show that bounds similar to those in Eqs.~(13)-(14) of the main text hold. Indeed, after finding a bound for $\|\rho(t)\|_1$ and $\|\mathcal{L}_{\Gamma}\|_{1\rightarrow1}$, the result follows by induction, as in the previous case.

For a pure state, the Schr\"odinger equation for the projected wavefuntion reads~\cite{Muga2004}
\begin{equation}\label{fesch}
\frac{dP\psi(t)}{dt}=-iP{\bf H}P\psi(t)-\int_0^t\,dsP{\bf H}Qe^{-iQ{\bf H}Qs}Q{\bf H}P\psi(t-s),
\end{equation}
where $P+Q=\mathbb{I}$ and ${\bf H}$ is the Hamiltonian of the total system. One can expand $\psi(t-s)$ in powers of $s$, i.e. $\psi(t-s)=\sum_{n=0}^\infty\frac{(-s)^n}{n!}\psi^{(n)}(t)$, and then truncate the series to a certain order, depending on how fast $e^{-iQ{\bf H}Qs}$ changes. Finally one can find, by iterative substitution, an equation of the kind $d P\psi(t)/dt=JP\psi(t)$, and generalise it to the density matrix case, achieving the equation~\eqref{masterc}, where $\rho$ is the density matrix of the projected system. If the truncation is appropriately done, then we always have $\|\rho(t)\|_1\leq1$ $\forall t\geq0$ by construction. For instance, in the Markovian limit, the integral in Eq.~\eqref{fesch} has a contribution only for $s=0$, and we reach an effective Hamiltonian $J=P{\bf H}P-\frac{i}{2}P{\bf H}Q{\bf H}P\equiv H-i\Gamma$. Here, $\Gamma$ is positive semidefinite, and $\|\rho(t)\|_1$ can only decrease in time.
 
Now, one can easily find that 
\begin{equation}
\|\mathcal{L}_{\Gamma}\sigma\|_1\leq 2\|\Gamma\|_\infty\|\sigma\|_1.
\end{equation}
Hence, $\|\mathcal{L}_{\Gamma}\|_{1\rightarrow1}\leq2\|\Gamma\|_\infty$. 

With these two bounds, it follows that 
\begin{align}
D_1(\rho(t),\tilde{\rho}_0(t))\leq \frac{1}{2}\int_0^tds\; \|\mathcal{L}_\Gamma\|_{1\rightarrow1}\|\rho(s)\|_1\leq\frac{1}{2}\int_0^tds\; \|\mathcal{L}_\Gamma\|_{1\rightarrow1}\leq \|\Gamma\|_\infty t.
\end{align}
One can find bounds for an arbitrary perturbative order by induction, as in the dissipative case.


\section[Details of the NMR Experiment]{Details of the NMR \\ Experiment} 
\fancyhead[RO]{\leftmark} 

In this appendix, we describe the platform used for the experiments reported in section~\ref{sec:NMRExp}, as well as detailed explanation of the used sequences.

\subsection{Description of the platform} \label{app:NMRplatform}

\begin{figure*}[htb]
\begin{center}
\includegraphics[width= 0.9\columnwidth]{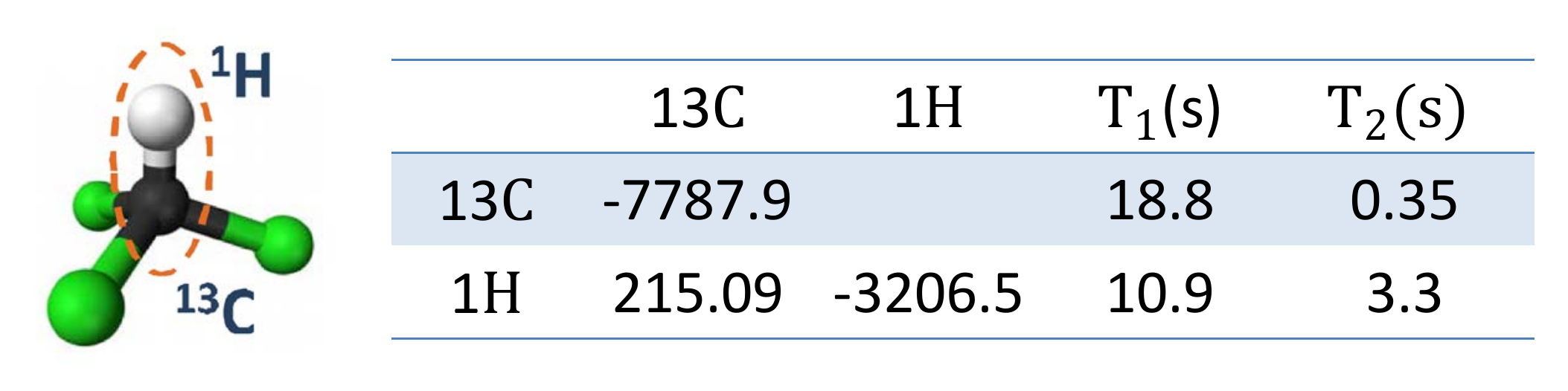}
\end{center}
\setlength{\abovecaptionskip}{-0.00cm}
\caption[Molecular structure and relevant parameters of $^{13}$C-labeled Chloroform.]{\footnotesize{\bf{Molecular structure and relevant parameters of $^{13}$C-labeled Chloroform.} Diagonal elements and off-diagonal elements in the table provide the values of the chemical shifts (Hz) and $J$-coupling constant (Hz) between $^{13}$C and $^{1}$H nuclei of the molecule. The right table also provides the longitudinal time $T_1$ and  transversal relaxation $T_2$ , which can be measured using the standard inversion recovery and Hahn echo sequences.}}\label{molecule}
\end{figure*}

\begin{figure*}[htb]
\begin{center}
\includegraphics[width= 0.9\columnwidth]{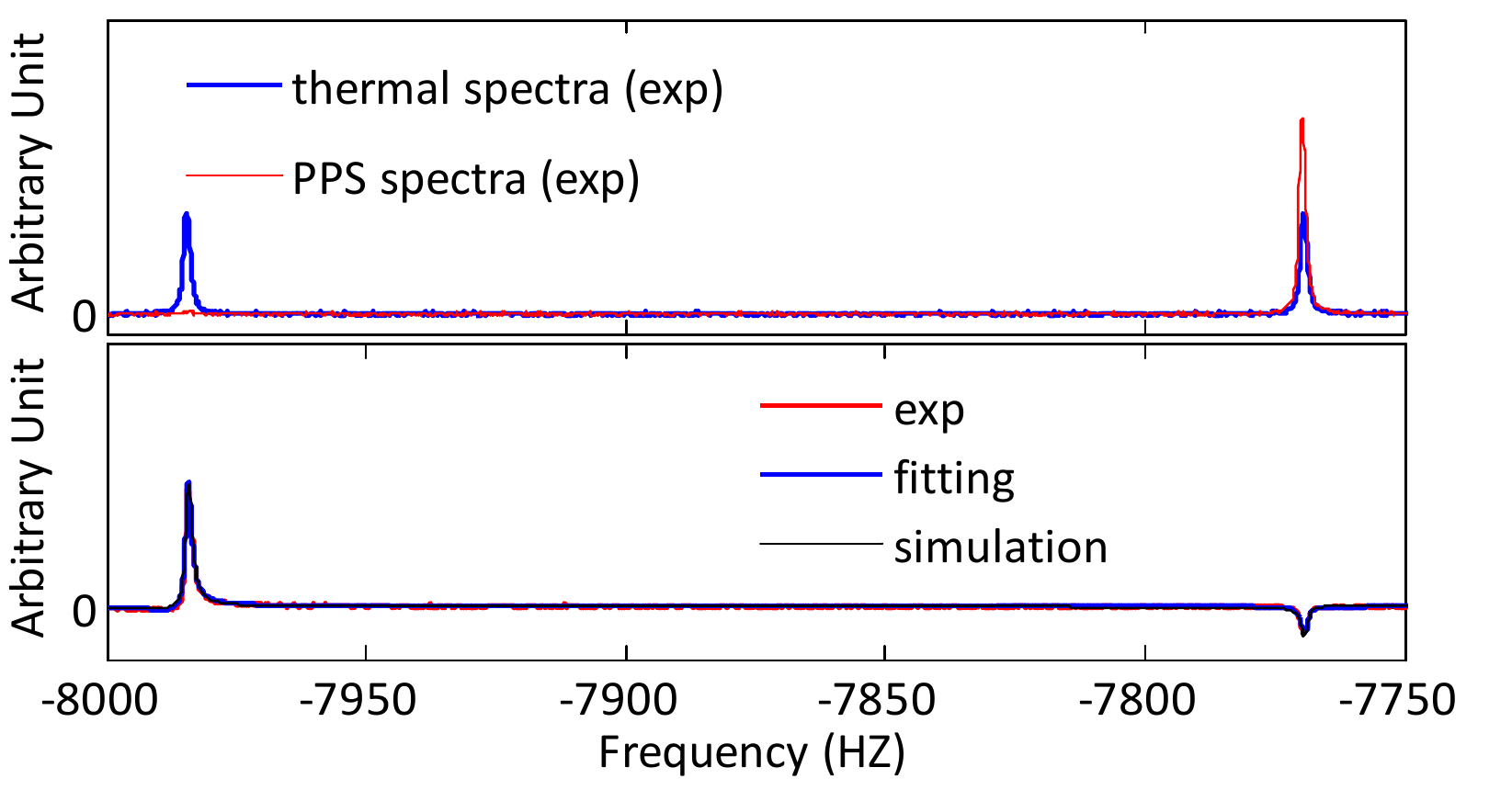}
\end{center}
\setlength{\abovecaptionskip}{-0.00cm}
\caption[Experimental spectra of $^{13}$C nuclei.]{\footnotesize{\bf{Experimental spectra of $^{13}$C nuclei.} The blue line of the top plot shows the observed spectrum after a $\pi/2$ pulse is applied on $^{13}$C nuclei in the thermal equilibrium state. The signal measured after applying a $\pi/2$ pulse following the preparation of the PPS is shown by the red line of the top plot. The bottom plot shows the spectrum when we measure $\mathcal{M}^n_{xy}$ ($n=2$) in the main text. The red, blue and black lines represent the experimental spectra, fitting results and corresponding simulations, respectively.}}\label{spectra}
\end{figure*}

Experiments are carried out using nuclear magnetic resonance (NMR), where the sample $^{13}$C-labeled Chloroform is used as our two-qubit quantum computing processor. $^{13}$C and $^{1}$H in the Chloroform act as the ancillary qubit and system qubit, respectively. Fig. \ref{molecule} shows the molecular structure and properties of the sample. The top plot in Fig. \ref{spectra} presents some experimental spectra, such as the spectra of the thermal equilibrium and pseudo pure state.  The bottom plot in Fig. \ref{spectra} shows NMR spectra which is created after we measure the 2-time correlation function $\mathcal{M}^n_{xy}$ ($n=2$) in the main text. 

\subsection{Detailed NMR sequences} \label{app:detailsequence}

\begin{figure*}[htb]
\begin{center}
\includegraphics[width= 0.9\columnwidth]{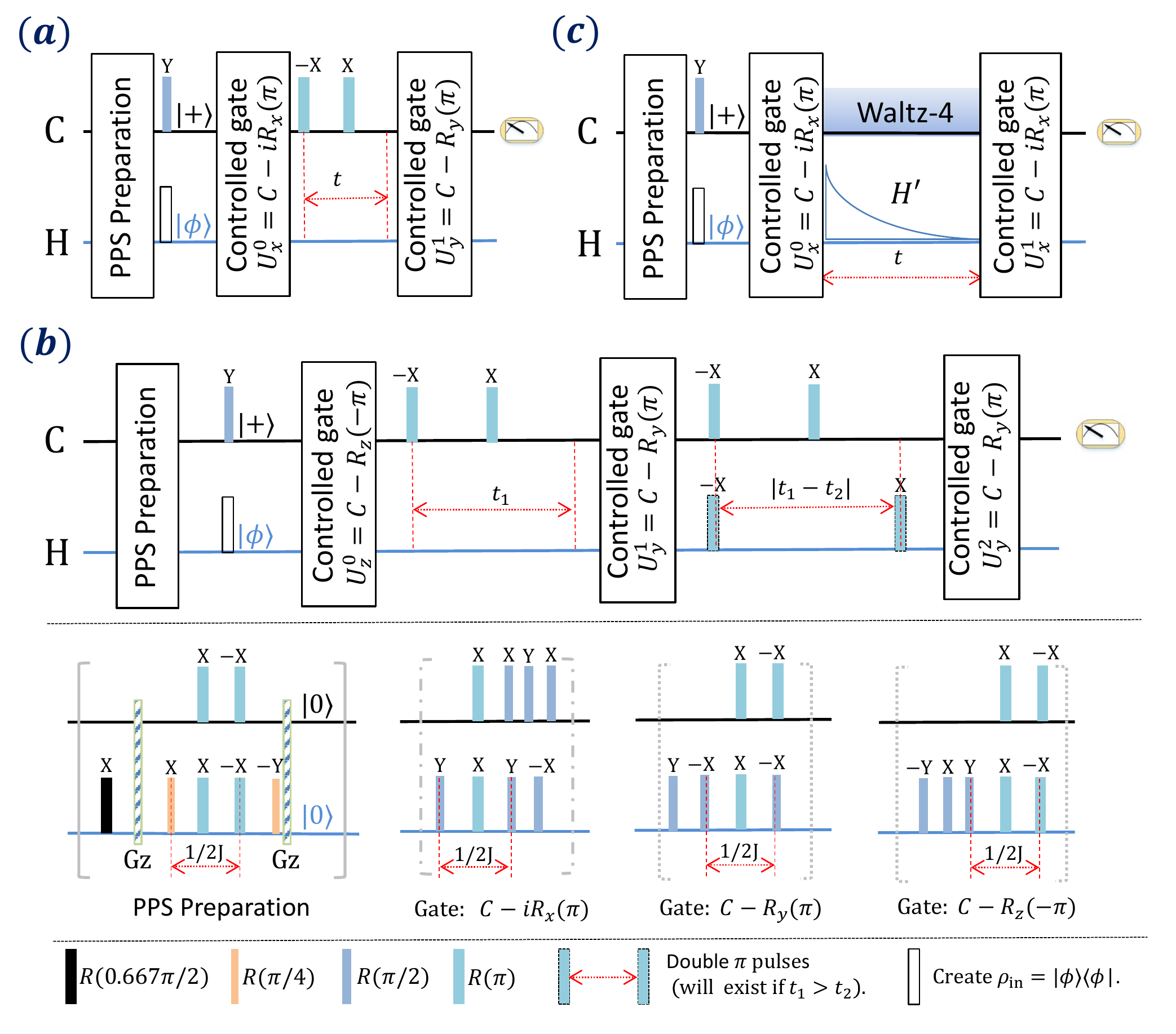}
\end{center}
\setlength{\abovecaptionskip}{-0.00cm}
\caption[NMR sequence to realize the quantum algorithm for measuring $n$-time correlation functions.]{ \footnotesize{{\bf NMR sequence to realize the quantum algorithm for measuring $n$-time correlation functions.} The black line and blue line mean the ancillary qubit (marked by $^{13}$C) and the system qubit (marked by $^{1}$H). All the controlled quantum gates $U^k_{\alpha}$ are decomposed into the following sequence in the bottom of the plot. Gz means a $z$-gradient pulse which is used to cancel the polarization in $x-y$ plane. (a) NMR sequence for measuring the 2-time correlation function $\langle\sigma_y(t)\sigma_x\rangle$. Other 2-time correlation functions can be similarly measured. (b) NMR sequence for measuring the 3-time correlation function $\langle\sigma_y(t_2)\sigma_y(t_1)\sigma_z\rangle$. We use the so-called double $\pi$ pulses which will exist if $t_1$ is longer than $t_2$, to invert the phase of the Hamiltonian $\mathcal{H}_{0}$. (c) NMR sequence for measuring the 2-time correlation function $\langle\sigma_x(t)\sigma_x\rangle$ with a time-dependent Hamiltonian $\mathcal{H'}(t)$. The method to decouple the interaction between $^{13}$C and $^{1}$H nuclei is Waltz-4 sequence. Hamiltonian $\mathcal{H'}(t)=500e^{-300t}\pi\sigma_y$ is created by using a special time-dependent radio-frequency pulse on the resonance of the second qubit in the $^1$H nuclei.}}\label{circuit_detail}
\end{figure*}

The detailed NMR sequences for measuring $n$-time correlation functions in the three types of experiments in the main text are illustrated in Figs. \ref{circuit_detail}(a), (b) and (c). As an example, Fig. \ref{circuit_detail}(a) shows the experimental sequence for measuring $\langle\sigma_y(t)\sigma_x\rangle$. Other time correlation functions $\langle\sigma_x(t)\sigma_y\rangle$ and $\langle\sigma_y(t)\sigma_y\rangle$  can be measured in a similar way by replacing the corresponding controlled quantum gates.  In the NMR sequence shown in Fig. \ref{circuit_detail}(c), we use the decoupling sequence Waltz-4, whose basic pulses are as follows,
\begin{equation}
\begin{array}{l}
[R_{-x}(\frac{3\pi}{2})R_{x}(\pi)R_{-x}(\frac{\pi}{2})]^2[R_{x}(\frac{3\pi}{2})R_{-x}(\pi)R_{x}(\frac{\pi}{2})]^2.
\end{array}
\end{equation}
This efficiently cancels the coupling between the $^{13}$C and $^{1}$H nuclei such that the $^{1}$H nucleus is independently governed by the external Hamiltonian $\mathcal{H'}(t)$. 


\section[Details of the Photonic Experiment]{Details of the \\ Photonic Experiment}

In this appendix, we elaborate on the technicalities of the experiment described in section~\ref{sec:EQSPhotons}. We give a detailed description of the used quantum circuit and of its implementation with probabilistic methods. We also comment on the photon count-rates of our experiment and on the dependence of the results on the pump power.

\subsection{Quantum circuit of the embedding quantum simulator} \label{app:EQSPCircuit}

Following the main text, the evolution operator associated with the embedding Hamiltonian $H^{(E)}{=}g\sigma_y{\otimes}\sigma_z {\otimes}\sigma_z$ can be implemented via 4 control-$Z$ gates ($CZ$), and a single qubit rotation $R_y(t)$. These gates act as
\begin{eqnarray}
CZ^{ij}&=&|0\rangle\langle0|^{(i)}\otimes \mathbb{I}^{(j)}+|1\rangle\langle1|^{(i)}\otimes \sigma^{(j)}_z,\\
R^i_y(t)&=&e^{-i\sigma_y^{(i)}gt}\equiv (\cos(gt)\mathbb{I}^{(i)}-i\sin(gt)\sigma_y^{(i)}),
\end{eqnarray}
with $\sigma_z{=}|0\rangle\langle0|{-}|1\rangle\langle1|$, and $\sigma_y{=}-i|0\rangle\langle1|{+}i|1\rangle\langle0|$. The indices $i$ and $j$ indicate on which particle the operators act. The circuit for the embedding quantum simulator consists of a sequence of gates applied in the following order:
\begin{equation}
U(t)=CZ^{02}CZ^{01}R^0_y(t)CZ^{01}CZ^{02}.
\end{equation}
Simple algebra shows that this expression can be recast as
\begin{eqnarray}\nonumber
U(t)&=& \cos (gt) \mathbb{I}^{(0)} \otimes \mathbb{I}^{(1)} \otimes \mathbb{I}^{(2)} - i \sin (gt) \sigma_y^{(0)} \otimes \sigma_z^{(1)}\otimes \sigma_z^{(2)}\\ 
&=& \text{exp}\left({-i g\sigma_y^{(0)} \otimes \sigma_z^{(1)}\otimes \sigma_z^{(2)} t}\right),
\end{eqnarray} 
explicitly exhibiting the equivalence between the gate sequence and the evolution under the Hamiltonian of interest. 

\subsection{Linear optics implementation} \label{app:CsignGates}
\begin{figure}[htp]
\begin{center}
\includegraphics [width=0.9 \columnwidth]{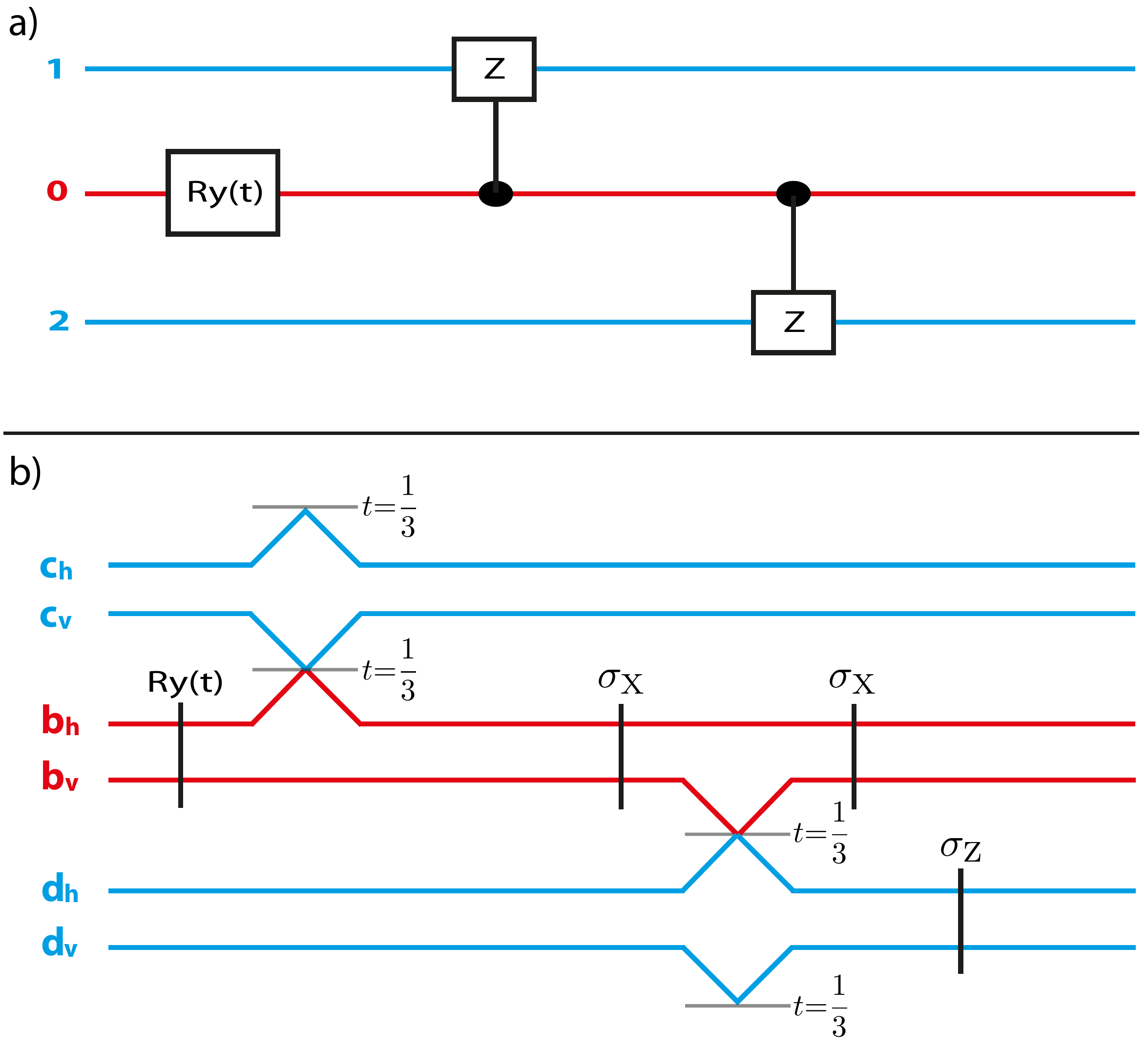}
\end{center}
\caption[Quantum circuit for the photonic implementation of an EQS]{\footnotesize{{\bf Quantum circuit for the photonic implementation of an EQS}(a)~Circuit implementing the evolution operator $U(t){=}\text{exp}\left({-i g\sigma_y^{(0)}{\otimes}\sigma_z^{(1)}{\otimes}\sigma_z^{(2)} t}\right)$, if the initial state is $|{\Psi}(0)\rangle{=}|0\rangle^{(0)}{\otimes}\left( |0\rangle^{(1)}+|1\rangle^{(1)} \right){\otimes}\left( |0\rangle^{(2)}+|1\rangle^{(2)} \right){/}2$. (b)~Dual-rail representation of the circuit implemented with linear-optics. Red (blue) lines represent trajectories undertaken by the control qubit (target qubits).}}\label{fig:SM1}
\end{figure}

The evolution of the reduced circuit is given by a $R_y(t)$ rotation of qubit $0$, followed by two consecutive control-Z gates on qubits $1$ and $2$, both controlled on qubit $0$, see Fig.~\ref{fig:SM1}~(a). These logic operations are experimentally implemented by devices that change the polarization of the photons, where the qubits are encoded, with transformations as depicted in Fig.~\ref{fig:SM1}~(b). For single qubit rotations, we make use of half-wave plates (HWP's), which shift the linear polarization of photons. For the two-qubit gates, we make use of two kinds of partially-polarizing beam splitters (PPBS's). PPBS's of type $1$ have transmittances $t_h{=}1$ and $t_v{=}1{/}3$ for horizontal and vertical polarizations, respectively. PPBS's of type $2$, on the other hand, have transmittances $t_h{=}1{/}3$ and $t_v{=}1$. Their effect can be expressed in terms of polarization dependant input-output relations---with the transmitted mode corresponding to the output mode---of the bosonic creation operators as
\begin{align}\label{eq:bsaout}
a_{p,out}^{\dag(i)}&=\sqrt{t_{p}}a_{p,in}^{\dag(i)}+\sqrt{1-t_p}a_{p,in}^{\dag(j)}\\ \label{eq:bsbout}
a_{p,out}^{\dag(j)}&=\sqrt{1-t_p}a_{p,in}^{\dag(i)}-\sqrt{t_p}a_{p,in}^{\dag(j)},
\end{align}
where $a_{p,in}^{\dag(i)}$ ($a_{p,out}^{\dag(i)}$) stands for the $i$-th input (output) port of a PPBS with transmittance $t_p$ for $p$-polarized photons. 

Our circuit is implemented as follows: the first $R_y(t)$ rotation is implemented via a HWP oriented at an angle $\theta=gt{/}2$ with respect to its optical axis. The rest of the target circuit, corresponding to the sequence of two control-Z gates, can be expressed in terms of the transformation of the input to output creation operators as
\begin{eqnarray}\label{tr:C1}\nonumber
{b_h} {c_h} {d_h} &\to& {b_h} {c_h} {d_h} \\ \nonumber
{b_h} {c_h} {d_v} &\to&{b_h} {c_h} {d_v} \\  \nonumber
{b_h} {c_v} {d_h}&\to&{b_h} {c_v} {d_h}\\  \nonumber
{b_h} {c_v} {d_v}&\to&{b_h} {c_v} {d_v}\\  \nonumber
{b_v} {c_h} {d_h}&\to&{b_v} {c_h} {d_h}\\  \nonumber
{b_v} {c_h} {d_v}&\to&-{b_v} {c_h} {d_v}\\  \nonumber
{b_v} {c_v} {d_h}&\to&-{b_v} {c_v} {d_h}\\ 
{b_v} {c_v} {d_v}&\to&{b_v} {c_v} {d_v},
\end{eqnarray}
where $b{\equiv}a^{\dag(0)}$, $c{\equiv}a^{\dag(1)}$, and $d{\equiv}a^{\dag(2)}$ denote the creation operators acting on qubits $0$, $1$, and $2$, respectively. These polarization transformations can be implemented with a probability of $1{/}27$ via a $3$-fold coincidence detection in the circuit depicted in Fig.~\ref{fig:SM1}~(b). In this dual-rail representation of the circuit, interactions of modes $c$ and $d$ with vacuum modes are left implicit.

The $\sigma_x$ and $\sigma_z$ single qubit gates in Fig.~\ref{fig:SM1}~(b) are implemented by HWP's with angles $\pi{/}4$ and $0$, respectively. In terms of bosonic operators, these gates imply the following transformations,
\begin{eqnarray}
\sigma_x:&& \ {b_h}\to{b_v},\quad  \ {b_v}\to{b_h}\\
\sigma_z:&& \  \ {d_h}\to{d_h},\quad {d_v}\to-{d_v}.
\end{eqnarray}

According to all the input-output relations involved, it can be calculated that the optical elements in Fig.~\ref{fig:SM1}~(b) implement the following transformations
\begin{eqnarray}\label{tr:C1-2}\nonumber
{b_h} {c_h} {d_h} &\to& {b_h} {c_h} {d_h}/(3\sqrt3) \\ \nonumber
{b_h} {c_h} {d_v} &\to&{b_h} {c_h} {d_v}/(3\sqrt3) \\  \nonumber
{b_h} {c_v} {d_h}&\to&{b_h} {c_v} {d_h}/(3\sqrt3)\\  \nonumber
{b_h} {c_v} {d_v}&\to&{b_h} {c_v} {d_v}/(3\sqrt3)\\  \nonumber
{b_v} {c_h} {d_h}&\to&{b_v} {c_h} {d_h}/(3\sqrt3)\\  \nonumber
{b_v} {c_h} {d_v}&\to&-{b_v} {c_h} {d_v}/(3\sqrt3)\\  \nonumber
{b_v} {c_v} {d_h}&\to&-{b_v} {c_v} {d_h}/(3\sqrt3)\\ 
{b_v} {c_v} {d_v}&\to&{b_v} {c_v} {d_v}/(3\sqrt3),
\end{eqnarray}
if events with $0$ photons in some of the three output lines of the circuit are discarded. Thus, this linear optics implementation corresponds to the evolution of interest with success probability $P=(1/(3\sqrt3))^2=1/27$.

\subsection{Photon count-rates} \label{app:CountRates}

Given the probabilistic nature and low efficiency of down-conversion processes, multi-photon experiments are importantly limited by low count-rates. In our case, typical two-photon rates from source are around ${150}$~kHz at $100\%$ pump (two-photon rates are approx. linear with pump power), which after setup transmission (${\sim}80\%$) and $1{/}9$ success probability of one controlled-sign gate, are reduced to about $13$~kHz ($1$~kHz) at $100\%$ ($10\%$) pump. These count-rates make it possible to run the two-photon protocol, described in the main text, at low powers in a reasonable amount of time. However, this situation is drastically different in the three-photon protocol, where we start with $500$~Hz of $4$-fold events from the source, in which case after setup transmission, $1{/}27$ success probability of two gates, and $50\%$ transmission in each of two $2$~nm filters used for this case, we are left with as few as ${\sim}100$~mHz (${\sim}1$~mHz) at $100\%$ ($10\%$) pump ($4$-fold events reduce quadratically with pump). Consequently, long integration times are needed to accumulate meaningful statistics, imposing a limit in the number of measured experimental settings.

\subsection{Pump-dependence} \label{app:PumpDep}
\begin{figure}[htp]
\begin{center}
\includegraphics [width= 0.9 \columnwidth]{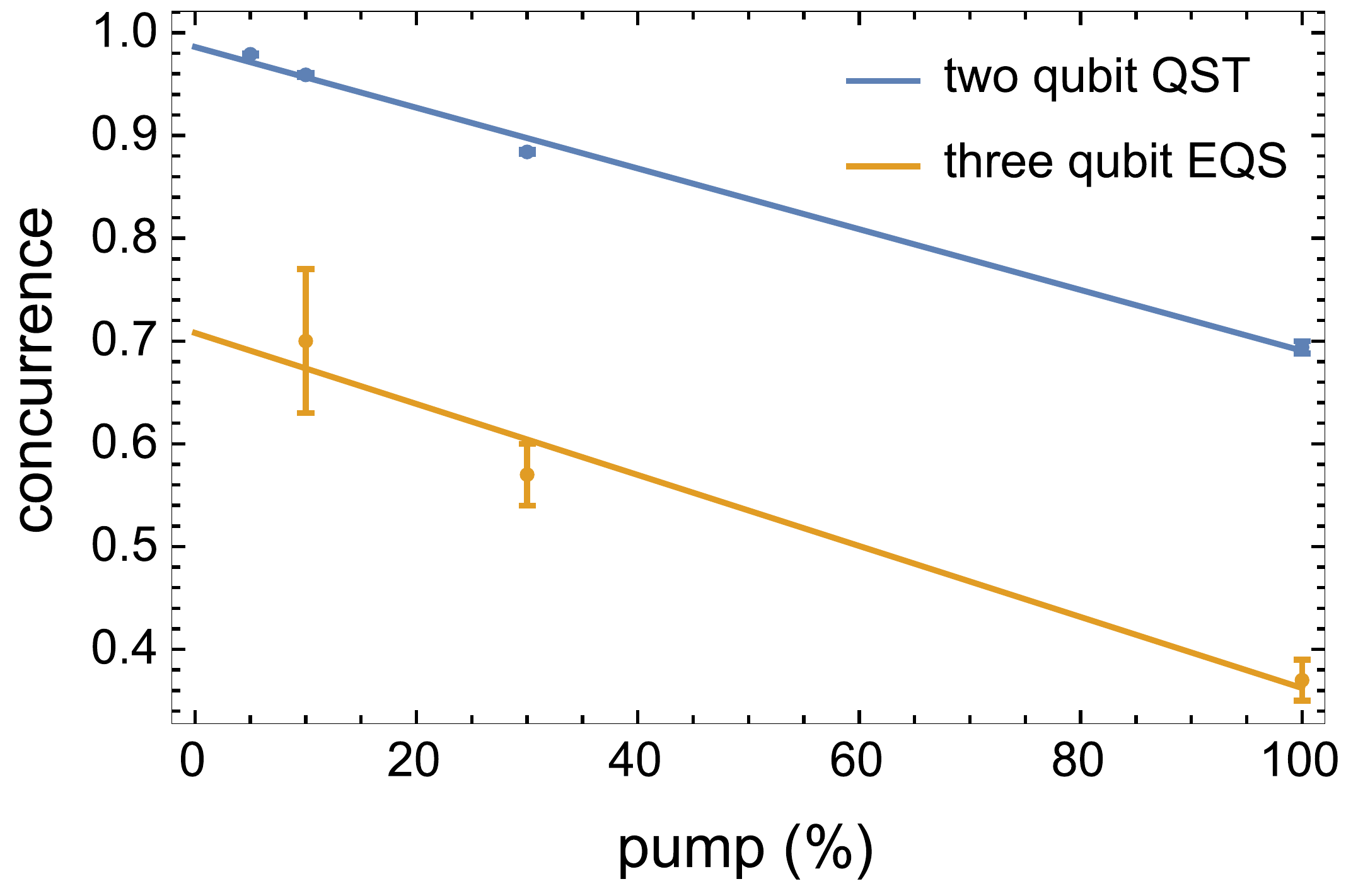}
\end{center}
\caption[Measured concurrence vs. pump power in the photonic implementation of an EQS]{\footnotesize{{\bf Measured concurrence vs. pump power.} The concurrence is extracted from both two-qubit quantum state tomography (QST) and the three-qubit embedding quantum simulator (EQS). Straight lines are linear fits to the data. Slopes overlapping within error, namely $-0.0030\pm0.0001$ from QST and $-0.0035\pm0.0007$ from EQS, show that both methods are affected by higher-order terms at the same rate.}}\label{fig:SM2}
\end{figure}
To estimate the effect of power-dependent higher-order terms in the performance of our protocols, we inspect the pump power dependence of extracted concurrence from both methods. Fig.~\ref{fig:SM2} shows that the performances of both protocols decrease at roughly the same rate with increasing pump power, indicating that in both methods the extracted concurrence at $10\%$ pump is close to performance saturation.

\begin{figure}[htp]
\begin{center}
\includegraphics [width= 0.9 \columnwidth]{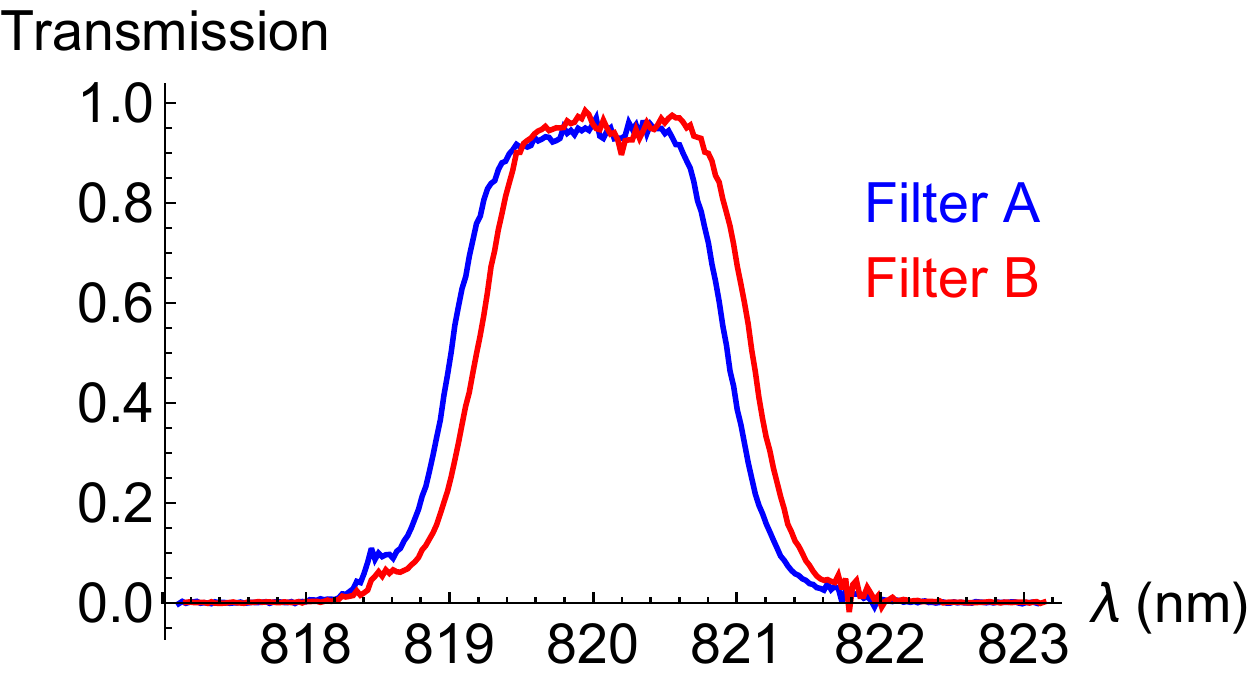}
\end{center}
\caption[Spectral filtering of photons.]{\footnotesize{{\bf Spectral filtering of photons.} The measured transmission for both filters, used in our three-qubit protocol, qualitatively reveals the remaining spectral mismatch.}}\label{fig:SM3}
\end{figure}

The principal difference between the two methods is that in the three-qubit protocol one of the photons originates from an independent down-conversion event and as such will present a somewhat different spectral shape. To reduce this spectral mismatch, we used two $2$~nm filters at the output of the two spatial modes where interference from independent events occurs, see Fig.~\ref{fig:SM3}. Note that not identical spectra are observed. This limitation would be avoided with a source that presented simultaneous high indistinguishability between all interfering photons.

%


\section{Details on the Derivation of the two-Photon Rabi Model}

In this appendix, we give further details on the trapped ion implementation of the two-photon quantum Rabi model that was introduced in section~\ref{sec:TPQRM}. We comment on the extension of the proposal to several ions, and in the behaviour of the simulation around the collapse point. We also describe a technique for the measurement of the parity operator.

\subsection{Implementation of two-photon Dicke model with collective motion of $N$ trapped ions}
\label{app_dicke}
In the main text, we showed how a two-photon Rabi model can be implemented using the vibrational degree of freedom of a single ion, coupled to one of its internal electronic transitions by means of laser-induced interactions. Here, we show how the N-qubit two-photon Dicke model
\begin{equation}
\mathcal{H} = \omega a^\dagger a + \sum_n \frac{\omega_q^n}{2} \sigma_z^n + \frac{1}{N}\sum_n g_n \sigma_x^n \left( a^2 + {a^\dagger}^2 \right),
\label{sup_2phdicke}
\end{equation}
can be implemented in a chain of $N$ ions, generalizing such a method. The $N$ qubits are represented by an internal electronic transition of each ion, while the bosonic mode is given by a collective motional mode of the ion chain. The two-phonon interactions are induced by a bichromatic laser driving with the same frequency-matching conditions used for the single-qubit case. The drivings can be implemented by shining two longitudinal lasers coupled to the whole chain, or by addressing the ions individually with transversal beams. The former solution is much less demanding, but it may introduce inhomogeneities in the coupling for very large ion chains; the latter allows complete control over individual coupling strengths.

In order to guarantee that the model of Eq.~\eqref{sup_2phdicke} is faithfully implemented, the bichromatic driving must not excite unwanted motional modes. In our proposal, the frequency of the red/blue drivings $\omega_{r/b}$ satisfy the relation $|\omega_{r/b} - \omega_{int} | = 2\nu + \delta_{r/b}$, where $\delta_{r/b}$ are small detunings that can be neglected for the present discussion. We recall that $\nu$ is the bosonic mode frequency and $\omega_{int}$ the qubit energy spacing. To be definite, we take the motion of the center of mass of the ion chain as the relevant bosonic mode. Then, the closest collective motional mode is the breathing mode~\cite{James1998}, with frequency $\nu_2 = \sqrt{3}\nu$. An undesired interaction between the internal electronic transitions and the breathing mode could appear if $|\omega_{r/b} - \omega_{int} |$ is close to $\nu_2$ or $2\nu_2$, corresponding to the first and second sidebands, respectively. In our case, the drivings are detuned by $\Delta_1 = |\omega_{r/b} - \omega_{int} | - \nu_2 \approx 0.27 \nu$  from the first and $\Delta_2 = |\omega_{r/b} - \omega_{int} | - 2\nu_2 \approx 1.46 \nu$ from the second sideband. Given that the frequency $\nu$ is much larger than the coupling strength $\Omega$, such detunings make those unwanted processes safely negligible.

\subsection{Properties of the wavefunctions below and above the collapse point}
\label{app_math}
The presence of the collapse point at $g=\omega/2$ can be inferred rigorously by studying  the asymptotic behavior of the formal solutions to the time-independent Schr\"odinger equation $\mathcal{H}\psi=E\psi$. We consider now the simplest case $N=1$. Using the representation of the model in the Bargmann space $\mathcal{B}$ of analytic functions \cite{Bargmann1961}, the Schr\"odinger equation for $\psi(z)$ in the invariant subspace with generalized-parity eigenvalue $\Pi=+1$ reads
\begin{equation}
g\psi''(z)+\omega z\psi'(z)+gz^2\psi(z) 
+ \frac{\omega_q}{2}\psi(iz)=E\psi(z),
\label{schroed-nonloc}
\end{equation}
where the prime denotes differentiation with respect to the complex variable $z$. This nonlocal linear differential equation of the second order, connecting the values of $\psi$ at the points $z$ and $iz$, may be transformed to a local equation of the fourth order,
\begin{equation}
\psi^{(4)}(z)+[(2-\bar{\omega}^2)z^2+2\bar{\omega}]\psi''(z)
+[4+2\bar{\omega}\bar{E}-\bar{\omega}^2]z\psi'(z)+
[z^4-2\bar{\omega}z^2+2-\bar{E}^2+\Delta^2]\psi(z)=0,
\label{schroed-loc}
\end{equation}
where we have used the abbreviations $\bar{\omega}=\omega/g$, $\Delta=\omega_q/(2g)$,
$\bar{E}=E/g$. Equation ~\eqref{schroed-loc} has no singular points in the complex plane except at $z=\infty$, where it exhibits an unramified irregular singular point of s-rank three \cite{Slavyanov2000}. That means that the so-called {\it normal solutions} have the asymptotic expansion
\begin{equation}
\psi(z)=e^{\frac{\gamma}{2}z^2+\alpha z}z^\rho(c_0+c_1z^{-1}+c_2z^{-2}+\ldots),
 \label{asym}
\end{equation}
for $z\rightarrow\infty$. 
Functions of this type are only normalizable (and belong therefore to $\mathcal{B}$) if the complex parameter $\gamma$, a characteristic exponent of the second kind, satisfies
$|\gamma|<1$. In our case, the possible $\gamma$'s are the solutions of the biquadratic 
equation
\begin{equation}
x^4+x^2(2-\bar{\omega}^2)+1=0.
\label{biquad}
\end{equation}
It follows
\begin{equation}
\gamma_{1,2}=\frac{\bar{\omega}}{2}\pm\sqrt{\frac{\bar{\omega}^2}{4}-1}, \quad
\gamma_{3,4}=-\frac{\bar{\omega}}{2}\pm\sqrt{\frac{\bar{\omega}^2}{4}-1}.
\label{sols}
\end{equation}
For $\bar{\omega}/2 >1$, all solutions are real. For $|\gamma_1|=|\gamma_4|>1$, we have $|\gamma_2|=|\gamma_3|<1$. In this case, there exist normalizable solutions if $\gamma_2$ or $\gamma_3$ appears in Eq.~\eqref{asym}. The condition for absence of the other characteristic exponents $\gamma_{1,4}$ in the formal solution of Eq.~\eqref{schroed-loc} is the spectral condition determining the parameter $E$ in the eigenvalue problem $\mathcal{H}\psi=E\psi$. It follows that for $g<\omega/2$, a discrete series of normalizable solutions to Eq.~\eqref{schroed-nonloc} may be found and the spectrum is therefore a pure point spectrum.

On the other hand, for $\bar{\omega}/2 <1$, all $\gamma_j$ are located on the unit circle with $\gamma_1=\gamma_2^\ast, \gamma_3=\gamma_4^\ast$. Because, then, no normalizable solutions of Eq.~\eqref{schroed-loc} exist, the spectrum of the (probably self-adjoint) operator $\mathcal{H}$ must be continuous for $g>\omega/2$, i.e. above the collapse point. The exponents $\gamma_1$ and $\gamma_2$ ($\gamma_3$ and $\gamma_4$) join at 1 (-1) for $g=\omega/2$. The exponent $\gamma=1$ belongs to the Bargmann representation of plane waves. Indeed, the plane wave states $\phi_q(x)=(2\pi)^{-1/2}\exp(iqx)$ in the rigged extension of $L^2(\mathbb{R})$ \cite{Gelfand1964}, satisfying the othogonality relation $\langle\phi_q|\phi_{q'}\rangle=\delta(q-q')$, are mapped by the isomorphism $\mathcal{I}$ between $L^2(\mathbb{R})$ and $\mathcal{B}$ onto the functions
\begin{equation}
\mathcal{I}[\phi_q](z)=\pi^{-1/4}e^{-\frac{1}{2}q^2+\frac{1}{2}z^2+i\sqrt{2}qz}, 
\end{equation}   
they correspond therefore to $\gamma=1$. It is yet unknown whether at the collapse point $g=\omega/2$, the generalized eigenfunctions of $\mathcal{H}$ have plane wave characteristics for $\omega_q\neq 0$ or which properties of these functions appear above this point, where the spectrum is unbounded from below.

\subsection{Generalized-parity measurement}
\label{app_meas}
The generalized-parity operator, defined as $\Pi=(-1)^N\bigotimes_{n=1}^N \sigma_z^n {\rm exp}\{i\frac{\pi}{2} n \}$, with $n=a^\dag a$, is a non-Hermitian operator that can be explicitly written as the sum of its real and imaginary parts,
\begin{eqnarray}
\Pi &=& (-1)^N \bigotimes_{n=1}^N \sigma_z^n \cos(\frac{\pi}{2} a^\dag a) \\
&+& i (-1)^N \bigotimes_{n=1}^N \sigma_z^n \sin ({\frac{\pi}{2}a^\dag a}). \nonumber
\end{eqnarray}
For simplicity, we will focus on the $N=1$ case, but the procedure is straightforwardly extendible to any $N$. We will show how to measure the expectation value of operators of the form 
\begin{equation}
\exp \{ \pm i n \  \sigma_i \  \phi \} \sigma_j,
\label{accesible operator}
\end{equation} 
where $\sigma_{i,j}$ are a pair of anti-commuting Pauli matrices, $\{ \sigma_i, \sigma_j \}=0$, and $\phi$ is a continuous real parameter. One can then reconstruct the real and imaginary parts of the generalized-parity operator, as a composition of observables in Eq.~(\ref{accesible operator}) for different signs and values of $i, j$,
\begin{eqnarray}
\Re (\Pi)&=& -\frac{1}{2} \{ \exp ( i n \sigma_{x} \frac{\pi}{2} ) \sigma_z + \exp ( - i n \sigma_{x} \frac{\pi}{2} ) \sigma_z \}, \\
\Im (\Pi) &=& \frac{1}{2} \{ \exp ( i n \sigma_{x} \frac{\pi}{2} ) \sigma_{y} - \exp ( - i n \sigma_{x} \frac{\pi}{2} ) \sigma_{y} \}.
\end{eqnarray}

The strategy to retrieve the expectation value of observables in Eq.~(\ref{accesible operator}) will be based on the following property of anti-commuting matrices $A$ and $B$: $e^{A}Be^{-A}=e^{2A}B=Be^{-2A}$. Based on this, the expectation value of the observables in Eq.~(\ref{accesible operator}) can be mapped onto the expectation value of $\sigma_j$ when the system has previously evolved under Hamiltonian $H=\pm n \sigma_i $ for a time $t^*=\phi/2$,
\begin{equation}
\langle \psi | \exp \{ \pm i n \  \sigma_i \  \phi \} \sigma_j | \psi \rangle = \langle \psi (t^*) | \sigma_j | \psi (t^*) \rangle,
\end{equation}
where $ | \psi(t) \rangle = e^{-i n \sigma_i t} | \psi \rangle$. The expectation value of any Pauli matrix is accesible in trapped-ion setups, $\sigma_z$ by fluorescence techniques and $\sigma_{x,y}$ by applying rotations prior to the measurement of $\sigma_z$. The point then is how to generate the dynamics of Hamiltonian $H=\pm n \sigma_i $. For that, we propose to implement a highly detuned simultaneous red and blue sideband interaction,
\begin{equation}
H= \frac{ \Omega_0 \eta}{2} (a + a^\dag) \sigma^+ e^{i \delta t} e^{i\varphi}+ {\rm H.c.},
\end{equation}
where $\varphi$ is the phase of the laser with respect to the dipole moment of the ion. This Hamiltonian can be effectively approximated to the second-order Hamiltonian,
\begin{equation}
H_{\rm eff}= \frac{1}{\delta}\Big( \frac{ \Omega_0 \eta}{2} \Big)^2 (2n + 1) \sigma_z e^{i \varphi} ,
\label{Effective Hamiltonian}
\end{equation}
when $\delta \gg \eta \Omega_0/2$. The laser phase will allow us to select the sign of the Hamiltonian. Of course, one would need to be careful and maintain $\delta$ in a regime where $\delta \ll \nu$, $\nu$ being the trapping frequency, to guarantee that higher-order resonances are not excited. Finally, in order to get rid of the undesired extra term $\sigma_z$ in Hamiltonian Eq.~\eqref{Effective Hamiltonian}, one needs to implement one more evolution under the Hamiltonian $H= - (1/2) \Omega_0 \eta \sigma_z$. This evolution can be generated by means of a highly detuned carrier transition. So far, we have given a protocol to generate the Hamiltonian $H=\pm n \sigma_z$. In order to generate Hamiltonians $H=\pm n \sigma_{y}$, one would need to modify the evolution with two local qubit rotations, 
\begin{equation}
e^{ \pm i  n \sigma_{y}t}= e^{i \sigma_{x} \pi/4} e^{ \pm i n \sigma_z t} e^{-i \sigma_{x} \pi/4}.
\end{equation}
Similarly, for Hamiltonian $H=\pm n \sigma_x$ one would need to perform rotations around $\sigma_y$.